\documentclass[aps,prb,twocolumn,superscriptaddress,longbibliography]{revtex4-2}

\usepackage{amsmath}
\usepackage{amssymb,amsfonts,textcomp}
\usepackage[utf8]{inputenc}

\usepackage{threeparttable}
\usepackage{graphicx}
\usepackage{hyperref}

\usepackage[capitalise,noabbrev]{cleveref}

\makeatletter
\def\maketitle{
\@author@finish
\title@column\titleblock@produce
\suppressfloats[t]}
\makeatother

\hypersetup{
    colorlinks = true,
    linkcolor = blue,
    citecolor = blue
}

\usepackage{lmodern}
\usepackage[version=4]{mhchem}
\usepackage[dvipsnames]{xcolor}

\usepackage{float} 
\usepackage{tabularx}   
\newcolumntype{Y}{>{\centering\arraybackslash}X}  

\usepackage{array}
\newcolumntype{M}[1]{>{\centering\arraybackslash}m{#1}}

\usepackage{booktabs}

\usepackage{tabularray}

\makeatletter
\newcommand*{\rom}[1]{\expandafter\@slowromancap\romannumeral #1@}
\makeatother

\usepackage{colortbl}
\usepackage{multirow}

\definecolor{Gray}{gray}{0.9}

\usepackage{setspace}
\crefname{equation}{Eq.}{Eqs.}
\Crefname{equation}{Eq.}{Eqs.}

\makeatletter
\renewcommand{\thesubsection}{\Alph{subsection}}
\renewcommand{\thesubsubsection}{\arabic{subsubsection}}

\renewcommand{\p@subsection}{\thesection.}
\renewcommand{\p@subsubsection}{\thesection.\thesubsection.}
\makeatother

\usepackage{bibunits}

\newcommand{\beginsupplement}{%

        \setcounter{table}{0}
        \renewcommand{\thetable}{S\arabic{table}}%
        \setcounter{figure}{0}
        \renewcommand{\thefigure}{\arabic{figure}}
        \setcounter{equation}{0}
        \renewcommand{\theequation}{S\arabic{equation}}%
        \setcounter{page}{0}
        \renewcommand{\thepage}{S\arabic{page}}

        \setlength{\parskip}{5pt}
        \everypar{\setlength{\parindent}{24pt}}
}

\usepackage[T1]{fontenc}
\usepackage{newtxtext,newtxmath}
\usepackage[english]{babel}

\usepackage{siunitx}
\renewcommand{\AA}{\text{\r{A}}}   
\DeclareSIUnit{\angstrom}{\text{\AA}}

\makeatletter
\@ifundefined{contentsname}{%
}{}
\newcommand{\sitablesofcontents}{%
  \begingroup
    \renewcommand{\addcontentsline}[3]{}%
    \section*{Table of Contents}%
  \endgroup
  \begingroup
    \hypersetup{linkcolor=black}%
    \@starttoc{stoc}%
  \endgroup
}
\newcommand{\usestocfile}{%
  \let\si@saved@addcontentsline\addcontentsline
  \renewcommand{\addcontentsline}[3]{%
    \def\si@tocfile{##1}%
    \def\si@tocname{toc}%
    \ifx\si@tocfile\si@tocname
      \si@saved@addcontentsline{stoc}{##2}{##3}%
    \else
      \si@saved@addcontentsline{##1}{##2}{##3}%
    \fi
  }%
}
\makeatother

\begin{document}
\title{
Charge-transfer mechanisms at Li/Ga-doped $\mathbf{Li_7La_3Zr_2O_{12}}$ (LLZO) interfaces from machine-learning-assisted molecular dynamics
}

\author{Arseniy S. Burov}
\affiliation{Skolkovo Institute of Science and Technology, 121205 Moscow, Russian Federation}
\author{Artem M. Abakumov}
\affiliation{Skolkovo Institute of Science and Technology, 121205 Moscow, Russian Federation}
\author{Dmitry A. Aksyonov}
\affiliation{Skolkovo Institute of Science and Technology, 121205 Moscow, Russian Federation}

\date{\today}

\begin{abstract}

Interfacial charge transfer between solid electrolytes (SEs) and Li metal is a key factor limiting all-solid-state battery performance. Conventional density functional theory and nudged elastic band calculations are performed at 0~K along a single minimum-energy path and therefore neglect concerted multi-ion motion and finite-temperature effects, which can lead to inaccurate activation barriers. Here, we trained moment tensor potentials (MTPs) for garnet LLZO systems (t-LLZO, c-LLZO, and Ga-LLZO) and Li metal, enabling machine-learning molecular dynamics (MLMD) simulations of $\ce{Li+}$ diffusion in the bulk and across Li/SE interfaces. We also introduce a residence-time window method that filters out ion rattling at the interface and isolates genuine charge-transfer events. The resulting charge-transfer activation energy at the Li/Ga-LLZO interface is only $167 \pm 18$~meV, below the 200~meV barrier for vacancy-mediated Li migration in bulk Ga-LLZO. The corresponding intrinsic charge-transfer resistance is $\sim 10^{-5}~\Omega~\mathrm{cm}^{2}$, almost four orders of magnitude below the lowest experimentally reported $R_{\mathrm{ct}}$.
These results indicate that intrinsic charge transfer across the Li/Ga-LLZO interface is not rate-limiting, and that Li transport within the electrolyte instead governs the overall kinetics. Taken together, our findings clarify the fast interfacial kinetics in Li/LLZO systems, and the proposed methodology can aid further interface optimization in solid-state batteries.

\end{abstract}

\maketitle

\begin{bibunit}[unsrt]

\section{INTRODUCTION}

All-solid-state batteries (ASSBs) are a promising energy-storage technology because they can potentially overcome the energy-density limitations of conventional lithium-ion batteries with liquid electrolytes~\cite{abraham2015prospects}. The ongoing advances of Li-rich cathodes are hindered by their chemical instability with liquid organic-based electrolytes, which often leads to parasitic reactions and significantly increases the risk of short circuits~\cite{li2020new}. In contrast, replacing liquid electrolytes with solid-state electrolytes (SEs) offers a viable solution by providing enhanced thermodynamic stability and no gas release, as well as improved compatibility with lithium metal anodes~\cite{wu2021progress, cheng2019recent}. Among the most promising SE candidates for commercialization is the garnet-type lithium lanthanum zirconate, $\ce{Li_{7-x}La_{3}Zr_{2}O_{12}}$ (LLZO), which exhibits high lithium-ion conductivity in its cubic phase  $>1$~mS~cm$^{-1}$~\cite{yang2017ionic,tao2023preparation,nasir2025excess,mishra2025stabilization,kanai2021low}, and a wide electrochemical stability window~\cite{smetaczek2021investigating}. Despite these advantages, LLZO still faces a critical challenge for practical application due to poor interfacial contact with electrodes, which can result in high interfacial resistance ($R_{\mathrm{int}}$) caused by suboptimal synthesis methods~\cite{sharafi2017impact} or degradation during cycling~\cite{krauskopf2019toward}.

Optimization of the electrode/SE interface is often a long and iterative process. For example, Buschmann~\textit{et al.} first reported a high Li/LLZO interfacial resistance of $\sim 2800~\Omega~\mathrm{cm}^2$ in 2011~\cite{buschmann2011structure}, which was later attributed mainly to poor interfacial contact and surface contamination~\cite{cheng2014origin}. Through improved surface cleaning and reduced air exposure, Sharafi~\textit{et al.} lowered $R_{\mathrm{int}}$ to $\sim 54~\Omega~\mathrm{cm}^2$~\cite{sharafi2017impact}, and Krauskopf~\textit{et al.} further reduced it to $\sim 0.1~\Omega~\mathrm{cm}^2$ immediately after synthesis~\cite{krauskopf2020fast}. This progression shows that high $R_{\mathrm{int}}$ often arises from extrinsic processing issues rather than intrinsic material limitations. Therefore, computational methods that identify material pairs capable of forming low-resistance interfaces could accelerate the development of ASSBs.

Rettenwander~\textit{et al.}~\cite{rettenwander2018interface} showed, using Raman spectroscopy and nanosecond laser-induced breakdown spectroscopy, that a tetragonal LLZO (t-LLZO) layer can form at the Fe-doped LLZO interface due to a Li deficiency. This interpretation is consistent with earlier \textit{in situ} measurements by Ma~\textit{et al.}~\cite{ma2016interfacial}. In such cases, the measured charge-transfer (CT) resistance reflects the apparent resistance of the artificial solid-electrolyte interphase rather than the intrinsic CT resistance of the Li/solid-electrolyte interface. However, special doping strategies can preserve the cubic phase at the interface~\cite{connell2020kinetic}, highlighting the importance of calculations for predicting the intrinsic interfacial resistance even when ideal interfacial stabilization cannot be achieved experimentally because of synthesis or cell-assembly imperfections.

Furthermore, the true interfacial contact area during battery operation is generally inaccessible because voids, pores, and surface roughness reduce the effective Li/solid-electrolyte contact area. As a result, electrochemical measurements typically use the projected geometric area, causing the exchange current density extracted from the Butler--Volmer equation to appear artificially lower on an area-normalized basis~\cite{krauskopf2019toward,krauskopf2020fast}. In contrast, atomistic calculations provide intrinsic area-normalized interfacial resistances and exchange-current rate constants. Therefore, comparison between calculated and experimental exchange currents can be used to estimate the true electrochemically active contact area, which cannot be measured directly during cycling or after cell disassembly.

Common computational approaches for predicting interface resistance include molecular dynamics (MD), density functional theory (DFT), and \textit{ab initio} molecular dynamics (AIMD). Classical MD is fast, but its interatomic potentials are often too limited to describe electrode, electrolyte, and interface regions with the same accuracy~\cite{mishin2021machine, zuo2020performance}. DFT-based methods commonly estimate migration barriers from the minimum-energy path using nudged elastic band (NEB) calculations~\cite{jonsson1998nudged}, but these calculations are usually performed at 0~K and may miss concerted multi-ion motion and thermal effects, which may lead to overestimated migration barriers~\cite{He2017,burov2024mechanism,gao2019ab}. AIMD naturally includes finite-temperature vibrations and cooperative effects, but it is restricted by short timescales and small system sizes, which increases statistical uncertainty~\cite{he2018statistical}. In contrast, machine-learning molecular dynamics (MLMD) offers a practical compromise by combining near-DFT accuracy with much lower computational cost~\cite{deringer2020modelling, fu2023review, li2025machine}. In this work, we use MLMD to evaluate Li/LLZO interfacial resistance for the first time.

Here, we employ a combined computational approach integrating DFT calculations, AIMD and MLMD simulations to investigate lithium-ion migration in various bulk LLZO polymorphs, including tetragonal LLZO (t-LLZO), cubic LLZO (c-LLZO), and Ga-doped LLZO (Ga-LLZO). We extend this study to supercells containing interfaces between these phases and lithium metal, specifically Li/t-LLZO and Li/Ga-LLZO interfaces.  We demonstrate that MLIPs trained on DFT data accurately reproduce lithium site occupancies, vacancy formation energies, activation energies and migration pathways in both bulk phases and interfacial supercells. In this work, we show that the intrinsic charge-transfer resistance at Li/LLZO interfaces is much lower than previously predicted~\cite{burov2024mechanism} or measured experimentally~\cite{krauskopf2020fast}, indicating that poor interfacial contact and interphase formation dominate $R_{\mathrm{int}}$.

\section{RESULTS}
\begin{figure*}[!htb]
    \includegraphics[width=0.99\textwidth]{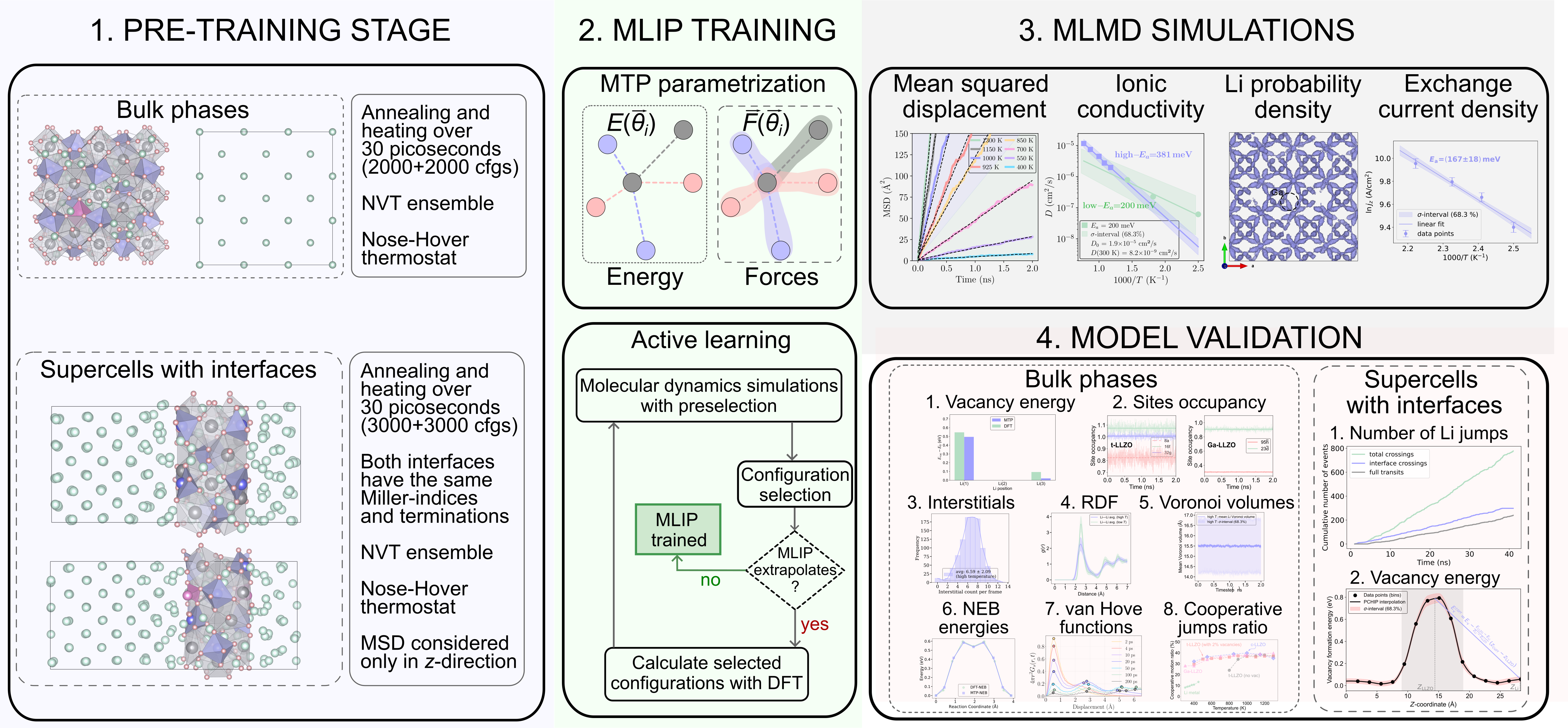}
\caption{Schematic illustration of Li-ion self-diffusion study in the metallic Li anode, \ce{Li7La3Zr2O12} (LLZO) electrolyte, and supercells with Li/LLZO interfaces. The study includes four stages: 1. Initial generation of the training configurations, using the \textit{ab initio} molecular dynamics (AIMD) simulations; 2. The training of MLIPs, using molecular dynamics (MD) simulations and active learning; 3. Large-scale MD simulations with MLIPs; 4. Validation of MTPs.}
\label{fig:methods_blocks}
\end{figure*}

The Li-ion diffusion study comprised four stages: (i) generation of an initial dataset of configurations from \textit{ab initio} molecular dynamics (AIMD) simulations; (ii) training of machine-learning interatomic potentials (MLIPs); (iii) large-scale molecular dynamics (MD) simulations; and (iv) subsequent validation. The workflow is illustrated in~\cref{fig:methods_blocks}.

\subsection{The choice of simulation systems}

In this work, we considered three bulk LLZO models: tetragonal LLZO (t-LLZO, with and without Li vacancies), cubic LLZO (c-LLZO), and Ga-doped LLZO (Ga-LLZO). The t-LLZO phase was included as a benchmark, since most computational data on activation energies and vacancy formation energies are available for this phase, and to assess whether vacancy-containing t-LLZO can approximate c-LLZO in diffusion and charge-transfer calculations. This is important because undoped c-LLZO is unstable at room temperature and may therefore be an unsuitable reference model~\cite{bernstein2012origin}.

Although pure c-LLZO is unstable, it represents an ideal, undistorted garnet framework and thus provides a useful reference for intrinsic Li-ion transport. Comparison with doped LLZO helps isolate the effect of dopant-induced vacancies and vacancy--dopant binding, which can reduce Li mobility~\cite{chen2020microstructural,npj_coatings}.

Ga-LLZO was selected because Ga substitution on Li sites ($\mathrm{Ga}_{\mathrm{Li}}$) creates two Li vacancies, stabilizes the cubic phase~\cite{chen2020manufacturing}, and yields higher ionic conductivity than Al doping~\cite{chen2020microstructural}. We therefore use Ga-LLZO to examine how Ga affects diffusion-channel blocking, activation energies, concerted Li diffusion, and we use this phase for charge-transfer barrier calculations across Li/LLZO interfaces.

The present study is limited to the Li(100)/LLZO(100) interface, whereas polycrystalline LLZO pellets expose a wide range of crystallographic orientations~\cite{sharafi2017controlling,fu2023grain}, in agreement with DFT predictions of low surface energies for multiple facets~\cite{Gao:2020,Canepa:2018,burov2024mechanism,gao2019ab}. The LLZO orientation may significantly affect the local structure and charge-transfer barrier and therefore represents an important source of variability in experimental $R_{\mathrm{ct}}$. In contrast, the initial Li orientation is expected to be less important because Li undergoes substantial structural disordering near the LLZO interface~\cite{burov2024mechanism,gao2019ab,yang2021interfacial,you2025grain}. A systematic comparison of different LLZO orientations is beyond the scope of the present work but represents an important direction for future studies. In this context, single-crystalline LLZO, which can be prepared by epitaxial growth~\cite{kim2013epitaxial} or the Czochralski method~\cite{philipp2020electronic}, provides an opportunity to experimentally investigate orientation-specific Li/LLZO interfaces and directly compare their $R_{\mathrm{ct}}$ with atomistic predictions.

\subsection{Moment tensor potential training}

We successfully trained separate MTPs for each bulk structure and interface-containing supercell. The numbers of training and validation configurations are summarized in~Table~S6 in~Section~S6.2. Their performance was assessed using the maximum absolute error (MaxAE), mean absolute error (MAE), and root mean square error (RMSE) for energies, forces, and stresses, with training and validation errors summarized in~Section~S7. For bulk phases, typical RMSE values are about 1~meV/atom for energies, 0.06~eV/\AA~for forces, and 0.1~GPa for stresses, while the corresponding values for interface supercells are 3~meV/atom, 0.07~eV/\AA, and 0.8~GPa, respectively.

We attempted to construct a unified MTP for all Li--LLZO systems, as described in~Section~S8. Bulk LLZO systems exhibit similar La--O and Zr--O bond lengths (see Figure~S11 and Table~S13) and comparable chemical diversity within the cut-off sphere (see Figure~S12 and Table~S14). To assess the transferability of the potential, we recalculated energies, forces, and stresses (EFS) for t-LLZO using the MTP trained on c-LLZO configurations (Section~S8.1) and for c-LLZO using the MTP trained on Ga-phases (Section~S8.2). In both cases, the RMSEs of predicted forces increased slightly by approximately 0.01~eV/\AA\ and stresses remained almost the same. More critically, we observed a systematic energy shift of up to 355~meV/atom, rendering the unified MTP unsuitable for predicting thermodynamic properties. Nevertheless, the comparatively small increase in force errors offers some optimism for property predictions where forces are the primary driving factor, such as kinetic barriers or mean-squared displacements. A similar trend was observed for the EFS predictions of the Li/t-LLZO system using an MTP trained on Li/Ga-LLZO configurations (see Section~S8.3).

\subsection{Validation of MTPs}

Since MTPs have already been shown to accurately reproduce elastic properties~\cite{novikov2022ai_mech} and phonon dispersion relations~\cite{rybin2024moment_phonons}, these aspects are not considered here. Instead, we focus on more challenging and less-explored properties, including average site occupancies, vacancy formation energies, and nudged elastic band (NEB) calculations, which are known to be difficult to reproduce reliably~\cite{kruglov2024surface}.

\subsubsection{Evaluation of site occupancy in LLZO phases}

\begin{figure}[!htbp]
\begin{center}\includegraphics[width=0.99\columnwidth]{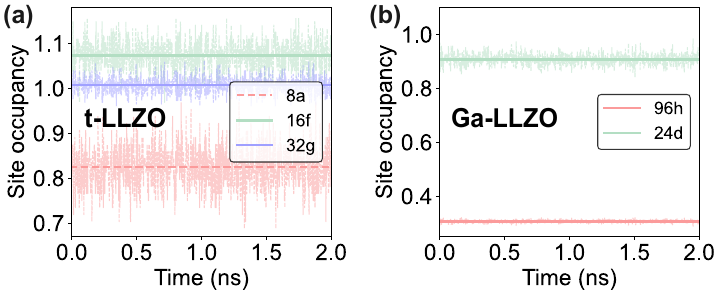}
\end{center}
\caption{Mean occupancy of Li sites over simulation time for the bulk phases: (a) t-LLZO; (b) Ga-LLZO.  The Ga-LLZO values were normalized in order to compare with reference c-LLZO occupancies by Awaka~\textit{et al.}~\cite{Awaka:2011}. }
\label{fig:occ_both_main}
\end{figure}

As part of the validation, we calculated the average Li site occupancies during MLMD simulations of t-LLZO, c-LLZO, and Ga-LLZO. The results for t-LLZO and Ga-LLZO, used for interface construction, are shown in~\cref{fig:occ_both_main}, while the c-LLZO data and additional analysis are provided in~Figure~S18.

Our MLMD results for c-LLZO (space group $Ia\overline{3}d$) closely reproduce the experimental findings of Holland~\textit{et al.}~\cite{holland2023workflow}, yielding a $24d/96h$ occupancy ratio of approximately 0.7, in good agreement with the experimental value of 0.67~\cite{Awaka:2011}. In contrast, most previous DFT and AIMD studies report significantly lower ratios of about 0.2~\cite{holland2023workflow}.

\subsubsection{Vacancy formation energy in Li--LLZO systems}

Another key property reproduced by MTPs is the Li vacancy formation energy ($E_{\mathrm{f}}$). To avoid explicit evaluation of the Li chemical potential, vacancy formation energies were calculated at $\mu_{\mathrm{Li}} = 0~\mathrm{eV}$. As summarized in~Section~S11.1, the MTP reproduces the relative energy differences in t-LLZO with an accuracy of about 23~meV and correctly captures the energetic ordering of Li sites: $E_{\mathrm{f}}(\mathrm{Li1}) > E_{\mathrm{f}}(\mathrm{Li3}) > E_{\mathrm{f}}(\mathrm{Li2})$.

In addition, vacancy formation energies were calculated for all Li sites in both Li/t-LLZO and Li/Ga-LLZO interface supercells, using single-point (unrelaxed) and fully relaxed ionic configurations. In both systems, the minimum vacancy formation energy is consistently located at the interface (see Figure~S21 and Figure~S22). The absolute energy difference between the two interfaces amounts to approximately 1~eV for the unrelaxed calculations, but reduces to only about 0.15~eV upon full ionic relaxation. This reduction is attributed to the more pronounced atomic rearrangements in Li/t-LLZO compared to Li/Ga-LLZO, as quantified by the mean atomic displacements (see Figure~S23 and Section~S11.3 for a detailed analysis). Furthermore, ionic relaxation substantially lowers the energy maxima at the center of the LLZO slab, from roughly 0.9 to 0.5~eV, highlighting the critical role of structural relaxation, in full agreement with previous DFT results~\cite{burov2024mechanism}.

\subsubsection{Nudged elastic band calculations of bulk and interfacial migration barriers}

\begin{figure}[!htbp]
\begin{center}\includegraphics[width=1.0\columnwidth]{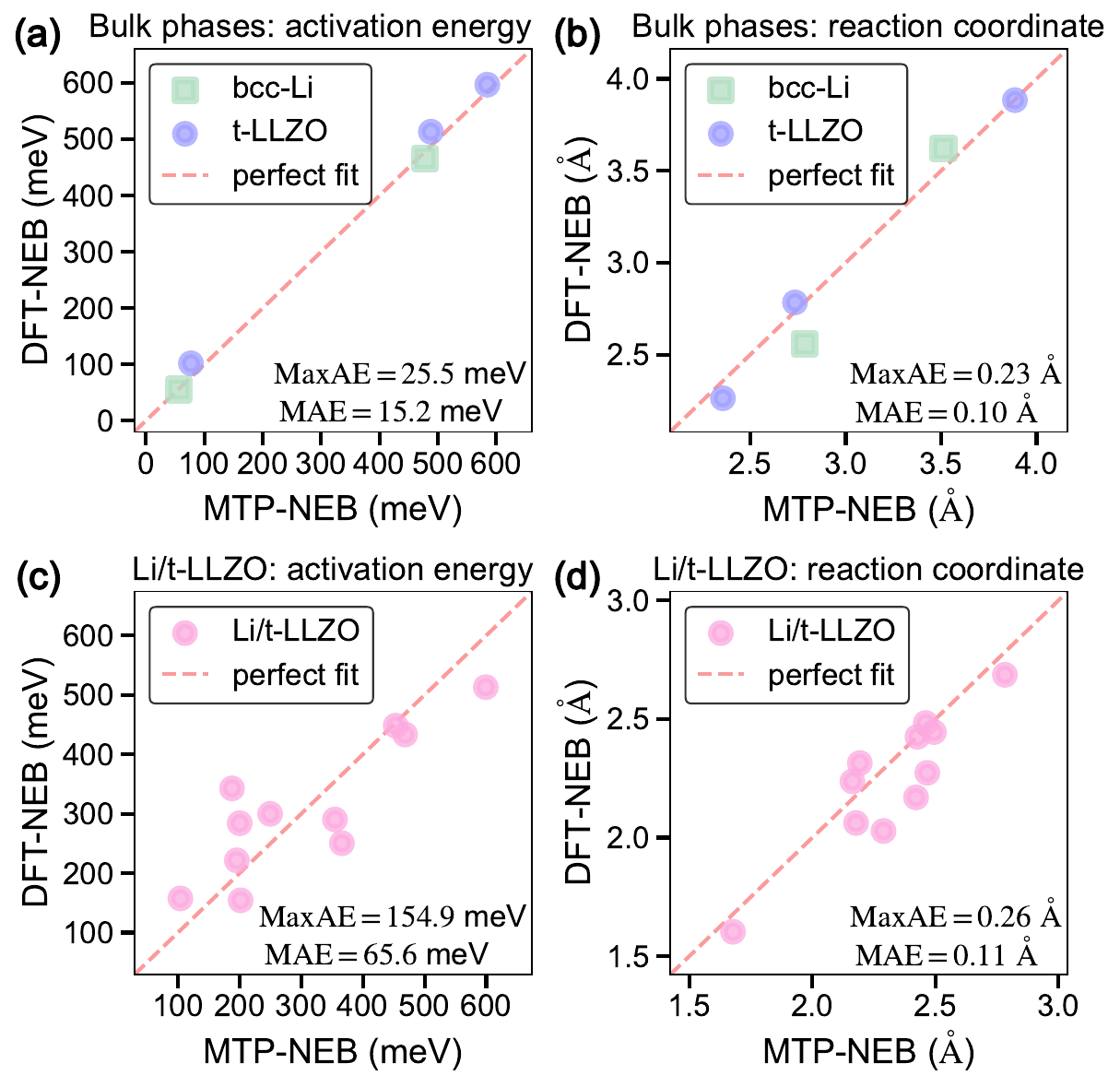}
\end{center}
\caption{Comparison of migration pathways and activation energies obtained with the DFT-NEB and MTP-NEB approaches: (a), (b) bulk t-LLZO; (c), (d) supercell with Li/t-LLZO interface.  }
\label{fig:neb_bulk_compare}
\end{figure}

Using MTP-NEB, we reproduced the low-energy $\ce{Li+}$ migration pathways in both t-LLZO and the Li/t-LLZO supercell, consistent with our previous DFT-NEB results~\cite{burov2024mechanism}. Additional energy profiles and supporting data are provided in~Section~S12, while a quantitative comparison of the DFT-NEB and MTP-NEB energy barriers is summarized in~\cref{fig:neb_bulk_compare}.

The MAEs in activation energies are 15~meV for bulk phases and 66~meV for interfacial supercells, while the MAEs in migration-path lengths are approximately 0.1~\AA~in both cases. The larger errors for interface systems are attributed to their structural complexity and incomplete sampling of vacancy configurations during MTP-NEB active learning, consistent with the higher extrapolation grades (>30) observed for these cases.

\subsubsection{MTPs training-stage timings}

To evaluate the efficiency of the two-stage training procedure proposed in~\cref{sec:mtp_method}, we analyzed the normalized computational times of the MLMD, DFT, and retraining stages for t-LLZO and Li/t-LLZO systems. The number of training configurations and relative timings across active-learning iterations are shown in~Figure~S16. Most of the computational cost originates from DFT calculations, which is advantageous because these calculations generate new configurations for improving the potential. In contrast, the MLMD and retraining stages mainly optimize the existing potential without producing additional data. For interface systems, the initial and final active-learning iterations sample only a few new configurations, increasing the relative cost of MLMD and retraining at these stages.

The total computational times are summarized in~Figure~S17 and Table~S18. For both t-LLZO and Li/t-LLZO systems, DFT calculations dominate the computational cost, accounting for more than 50\% of the total runtime. This demonstrates the efficiency of the proposed two-stage MTP training workflow, in which most of the computational effort is spent on generating new training data.

\subsection{MLMD bulk migration barriers and interfacial charge transfer}

After validating the high accuracy and reproducibility of key properties relevant to solid-state electrolytes, we conducted MLMD simulations on the bulk phases of t-LLZO, c-LLZO, Ga-LLZO, Li metal and supercells with interfaces: Li/t-LLZO and Li/Ga-LLZO.

\subsubsection{Li metal}

The current consensus is that Li diffusion in bcc Li proceeds via a vacancy mechanism~\cite{mali19886li,frank1996first,messer1975nuclear}. This is consistent with the lower vacancy formation energy compared to the lowest self-interstitial formation energy along the [111] direction (0.5 vs 0.6~eV, respectively)~\cite{yang2021interfacial}. Accordingly, Li diffusion in bcc Li was modeled using one vacancy in a 432-atom supercell.

The MLMD results for Li metal are shown in~\cref{fig:bulk_panel_main}.i--l. The calculated activation energy is 50~meV, in good agreement with the MD value of 44~meV reported by Sergeev~\textit{et al.}~\cite{sergeev2023anomalous} and experimentally measured nearest-neighbor (NN) migration barrier of 60~meV~\cite{frank1996first}. The experimental activation energy of 0.52--0.56~eV~\cite{mali19886li,lodding1970isotope_li_diffusion,messer1975nuclear} is consistent with the sum of the vacancy formation energy (515~meV) and the NN hopping barrier (50~meV), giving 565~meV, close to the experimental values. Above 425~K, the activation energy increases to 99~meV, likely due to strong anharmonic lattice vibrations and partial loss of crystalline order near the melting temperature of $\sim$450~K~\cite{boehler1983melting} as supported by the RDF and Wigner--Seitz analyses (Figure~S33 and Figure~S34). These effects distort the vacancy migration pathways and increase the effective diffusion barrier. No Li interstitials are observed up to 450~K, because the simulation time is insufficient for Frenkel pair formation given the high defect formation energy of $\sim$1.1~eV~\cite{yang2021interfacial}. Vacancy-free structures exhibit nearly zero MSD (Figure~S32.b), further supporting a vacancy-mediated diffusion mechanism.

Our calculated vacancy self-diffusion coefficient is $D_{\mathrm{vac}} = 1.5 \times 10^{-4}~\mathrm{cm}^2~\mathrm{s}^{-1}$, in good agreement with the experimental estimates of $(2.6$--$3.0) \times 10^{-4}~\mathrm{cm}^2~\mathrm{s}^{-1}$~\cite{lodding1970isotope_li_diffusion,messer1975nuclear,mali19886li}. Additional details and data are provided in~Table~S21 and Section~S14. These results suggest that, although the concentration of Li vacancies in bcc Li is extremely low, the vacancies themselves are highly mobile.

\subsubsection{LLZO phases}

\begin{figure*}[!htp]
\begin{center}
    \includegraphics[width=0.99\textwidth]{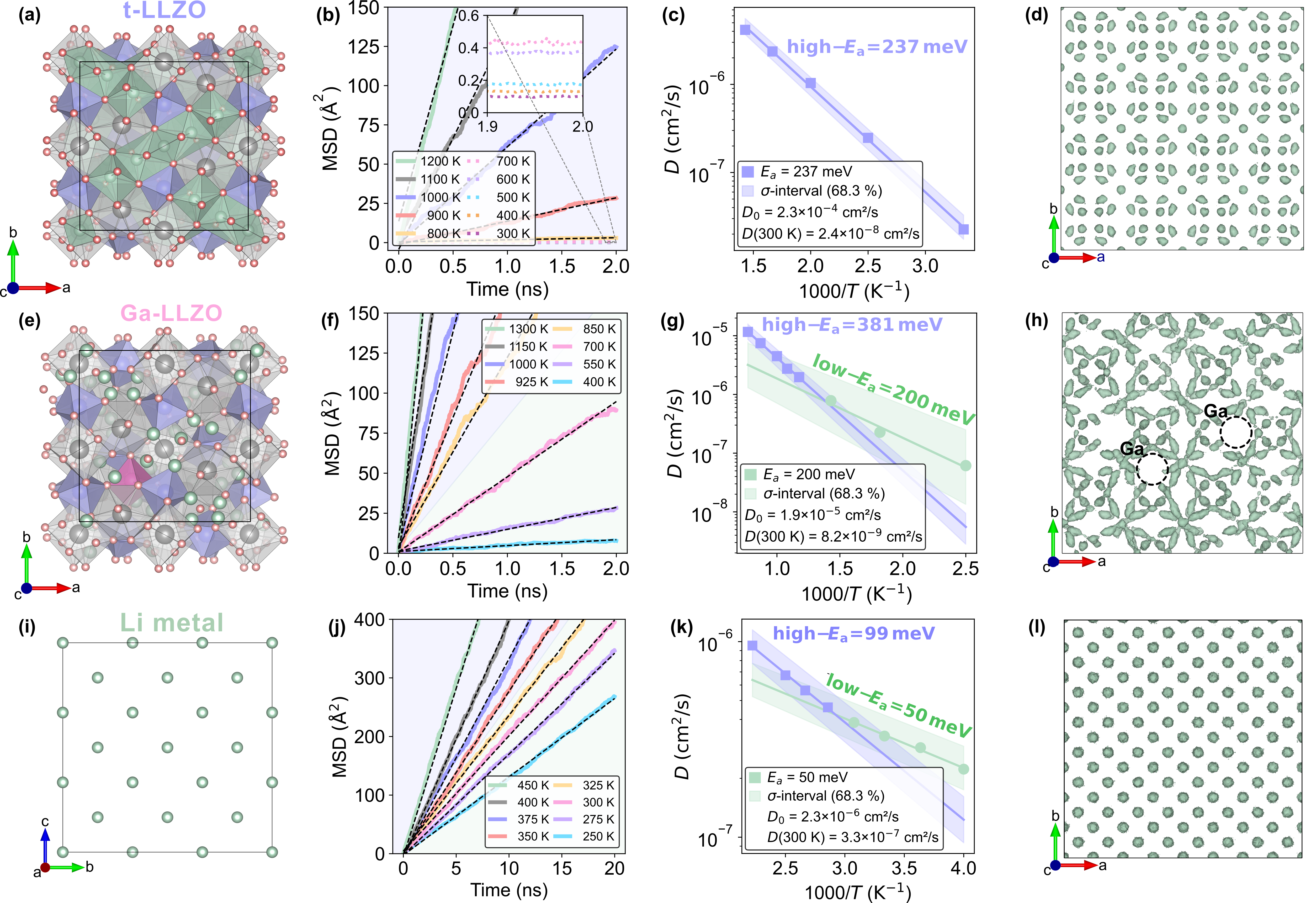}
\end{center}
\caption{Li-ion diffusion study in both the low-temperature vacancy-mediated regime and the high-temperature Frenkel-defect-mediated regime. The figure presents the crystal structure, the mean-squared displacement (MSD) as a function of temperature, diffusion coefficients with Arrhenius fits, and the Li-ion probability density map for the low-temperature regime at an isosurface level of $5 \cdot 10^{-4}\ r_{\mathrm{Bohr}}^{-3}$. The following bulk phases were analyzed: (a)--(d) tetragonal LLZO (t-LLZO) without vacancies, (e)--(h) Ga-doped LLZO (Ga-LLZO), and (i)--(l) body-centered cubic lithium (Li metal). To avoid trajectories with poor statistics, we included only those for which MSD exceeded the squared typical jump length in the bulk phases, that is, $MSD > 9~\mathrm{\AA}^2$ (dotted lines on plots). }
\label{fig:bulk_panel_main}
\end{figure*}

\textit{Li diffusion.}
The MLMD results for bulk t-LLZO without vacancies and Ga-LLZO are shown in~\cref{fig:bulk_panel_main}.a--h. Additional data for t-LLZO with 1.8\% vacancies (Figure~S35) and c-LLZO (Figure~S36) are provided in~Section~S14, while the low-temperature vacancy-mediated diffusion parameters are summarized in~\cref{tab:act_energies_our} and those for the interstitial-mediated mechanism are given in~Table~S22.

Stoichiometric t-LLZO exhibits nearly zero MSD up to 800~K due to the absence of Li vacancies. Diffusion onset corresponds to a transition from vacancy- to interstitial-mediated diffusion (see Figure~S48 in~Section~S22). Ga-LLZO shows $E_{\mathrm{a}} = 200$~meV and $D_{\mathrm{vac}} = 1.4\times10^{-8}$~cm$^2$\,s$^{-1}$, close to c-LLZO ($E_{\mathrm{a}} = 243$~meV, $D_{\mathrm{vac}} = 1.7\times10^{-8}$~cm$^2$\,s$^{-1}$). The lower activation energy arises because Ga occupies both tetrahedral and octahedral sites (see Section~S3), suppressing high-barrier octahedral--tetrahedral hops~\cite{burov2024mechanism}. At the same time, Ga partially blocks diffusion channels, reducing the vacancy diffusivity relative to c-LLZO. As a result, Ga-LLZO retains about $80\%$ of the ideal c-LLZO conductivity, consistent with experimental reports of high ionic conductivity~\cite{yang2017ionic,tao2023preparation,nasir2025excess,mishra2025stabilization,kanai2021low}. In contrast, vacancy-containing t-LLZO exhibits a much lower $E_{\mathrm{a}} = 116$~meV and higher $D_{\mathrm{vac}} = 8.6\times10^{-6}$~cm$^2$\,s$^{-1}$, demonstrating that it is not a suitable surrogate model for c-LLZO.

\textit{Anisotropy and framework stability.}
We next evaluated the diffusion anisotropy factor, $A_{\mathrm{aniso}}$, as described in~Section~S16 and shown in~Figure~S41. c-LLZO and Ga-LLZO exhibit nearly isotropic diffusion with $A_{\mathrm{aniso}} \approx 1$ (Figure~S41), whereas t-LLZO shows pronounced anisotropy with $A_{\mathrm{aniso}} \approx 0.75$, corresponding to slower diffusion along the $z$ direction, consistent with previous MD simulations by Chen~\textit{et al.}~\cite{Chen2018}.

To assess the stability of the non-Li framework, we calculated MSDs of La, Zr, Ga, and O atoms (Section~S13.1). La, Zr, and O remain essentially immobile, with MSD values below 0.5~\AA$^{2}$~throughout the simulations, indicating preservation of the La--Zr--O framework. In contrast, Ga becomes mobile above 1000~K, suggesting its easy redistribution within the LLZO lattice during high-temperature annealing~\cite{timusheva2025chemical}.

\textit{The contribution of cooperative diffusion.}
To understand the origin of the different diffusion properties of the LLZO phases, we analyzed the contributions of distinct $\ce{Li^{+}}$ transport mechanisms. Multi-ion cooperative migration has previously been identified as an important diffusion mechanism in solid-state electrolytes by He~\textit{et al.}~\cite{He2017}. Following this approach, we quantified cooperative Li jumps in bulk t-LLZO, c-LLZO, Ga-LLZO, and Li metal.

To characterize the dominant self-diffusion mechanisms, we calculated Li--Li and Li--O radial distribution functions, average Li Voronoi volumes, and performed Wigner--Seitz analyses to identify vacancy and interstitial defects. Additional details are provided in~Section~S23 and visualized in~Figure~S49. We also analyzed Li-ion probability densities (Section~S22 and Figure~S48), which show that Li ions predominantly occupy lattice sites at low temperatures, whereas interstitial occupation increases at elevated temperatures.

\begin{figure}[!htbp]
\begin{center}\includegraphics[width=0.90\columnwidth]{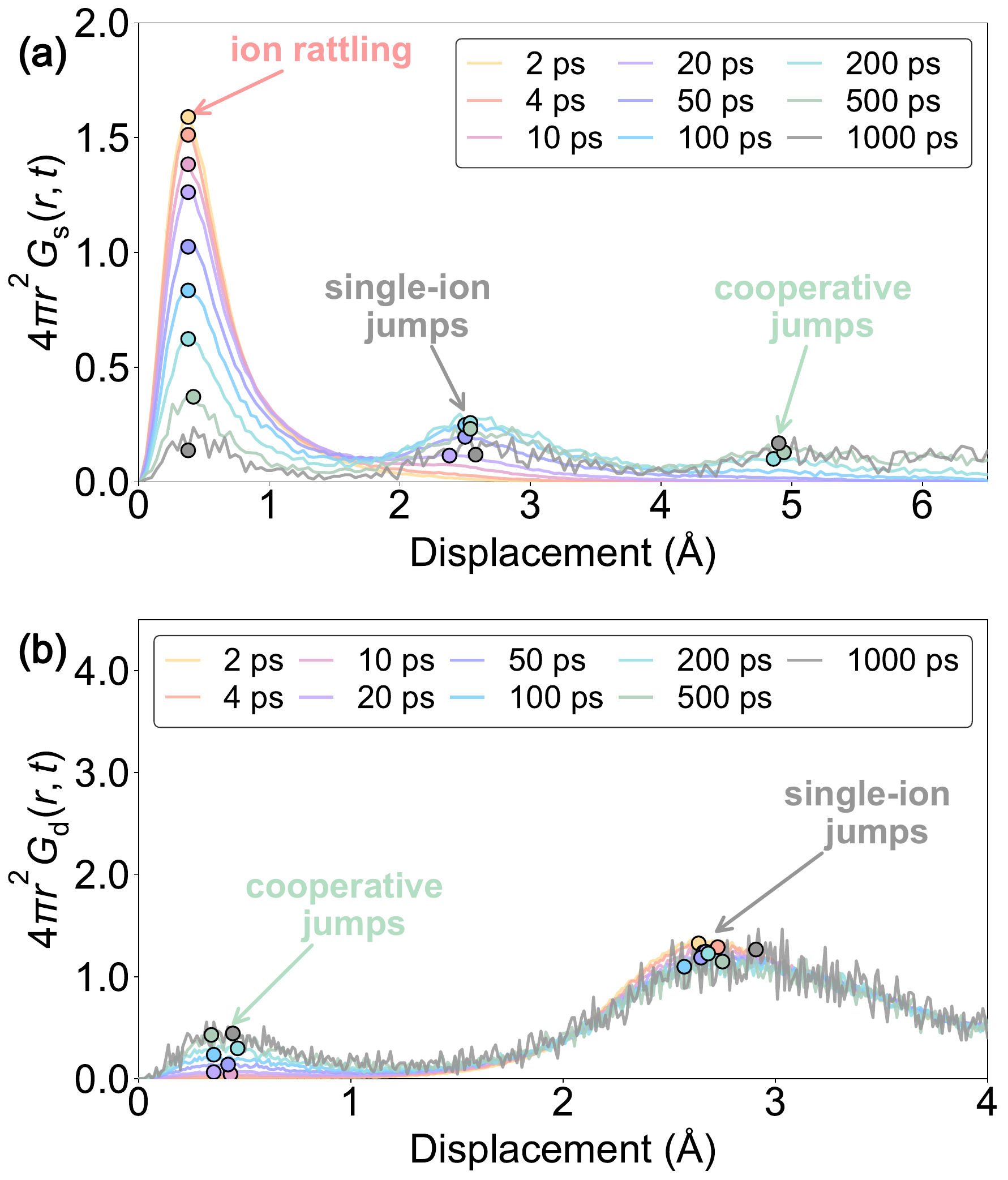}
\end{center}
\caption{(a) Self part ($4\pi r^2 G_{\mathrm{s}}(r,t)$) and (b) distinct part ($4\pi r^2 G_{\mathrm{d}}(r,t)$) of the radial one-dimensional van Hove correlation function for Ga-doped LLZO at $T = 700$~K, shown for lag times $t$ ranging from 2 to 1000~ps. Circles mark the local maxima (peak positions) for each curve. In the self part, the first peak ($0 < r \leq 1.8$~\AA) corresponds to ion rattling, the second peak ($1.8 < r \leq 4$~\AA) arises from single-ion jumps, and the third peak ($4 < r \leq 6$~\AA) is attributed to cooperative jump events. In the distinct part, the first peak ($0 < r \leq 1.5$~\AA) reflects cooperative hopping events, while the second peak ($1.5 < r \leq 4$~\AA) originates from single-ion jumps.} 
\label{fig:van_hove_main}
\end{figure}

\begin{figure}[!htbp]
\begin{center}\includegraphics[width=0.80\columnwidth]{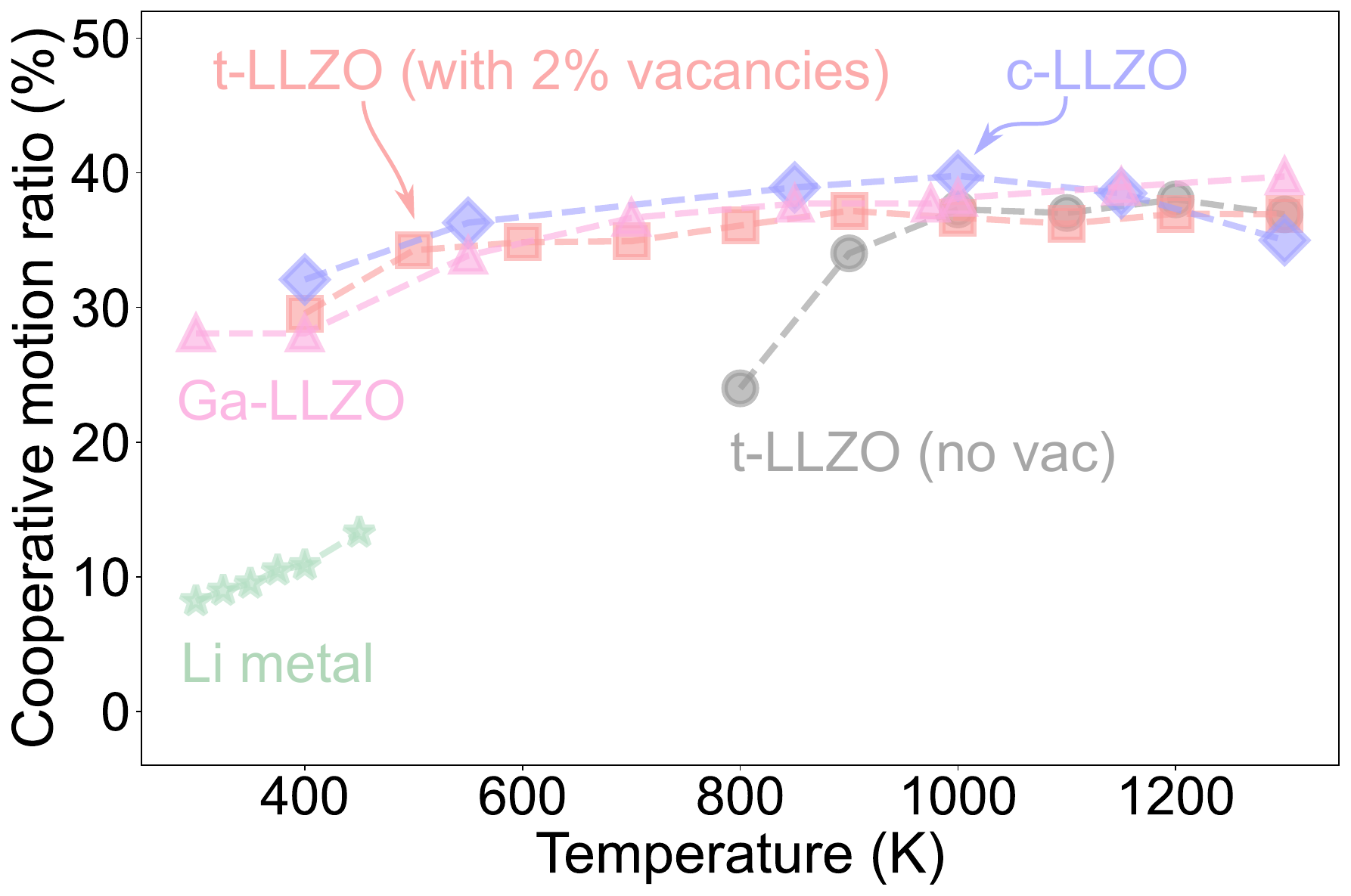}
\end{center}
\caption{Ratio of cooperative hops ($\eta_{\mathrm{norm}}$), calculated using \cref{eq:eta_norm}, for the considered bulk phases: t-LLZO (without vacancies), t-LLZO (with 1.8\% vacancies), c-LLZO, Ga-LLZO, and Li metal. } 
\label{fig:eta_both}
\end{figure}

The degree of cooperative motion was quantified from the areas corresponding to single-ion and cooperative jumps in the distinct part of the one-dimensional van Hove correlation function, as defined in~\cref{eq:eta_norm}. The methodology is described in~Section~S24 and illustrated in~Figure~S52. A representative example of the self ($G_{\mathrm{s}}$) and distinct ($G_{\mathrm{d}}$) parts of the radial one-dimensional van Hove correlation function for Ga-doped LLZO is shown in~\cref{fig:van_hove_main}. In the self part (\cref{fig:van_hove_main}.a), the second and third peaks correspond to single-ion and cooperative jumps, respectively. As the lag time increases, the height of the first peak decreases, while the second and third peaks grow. This evolution indicates the onset of ion jumps to neighboring sites, eventually reaching a steady-state regime in which the ratio of the second and third peak heights becomes constant (see Figure~S59). Accordingly, in the distinct part (\cref{fig:van_hove_main}.b), the second peak arises not only from immobile Li atoms but also from contributions associated with single-ion hopping events. The resulting cooperative fractions (\cref{fig:eta_both}) remain consistently high ($>27\%$) and increase with temperature. These findings are also consistent with the graph-based analysis used by Dembitskiy~\textit{et al.}~\cite{dembitskiy2025new}, where cooperative hopping events are identified from connected components formed by migrating ions (see Figure~S50 and Figure~S51).

\subsubsection{Li/LLZO interfaces and charge-transfer resistance}

\begin{figure*}[!htbp]
\begin{center}
    \includegraphics[width=0.99\textwidth]{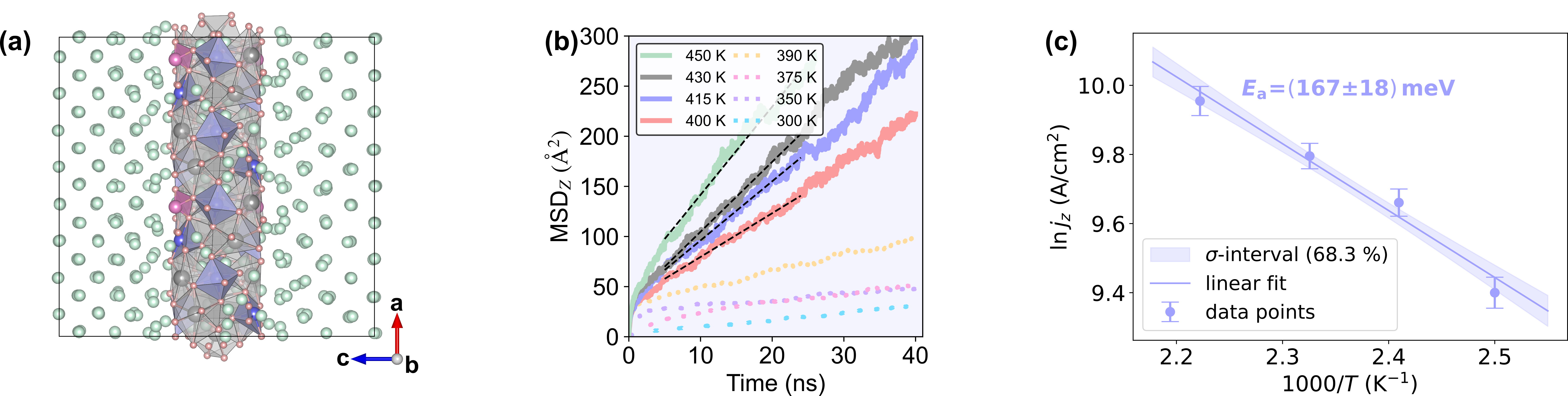}
\end{center}
\caption{(a) Crystal structure of supercell with Li/Ga-LLZO interface. (b) The mean-squared displacement in direction perpendicular to the interface plane ($MSD_{Z}$) as a function of temperature. To avoid trajectories with poor statistics, we included only those for which the $MSD_{Z}$ exceeded the squared half-size of the supercell, that is, $MSD_Z > 100~\mathrm{\AA}^2$ (dotted lines on plots). (c) Exchange-current of $\ce{Li^{+}}$ in $z$-direction as a function of temperature and its Arrhenius fit. Minimal residence time between $\ce{Li^{+}}$ jumps across interface was set to 2.8~ns.}
\label{fig:int_main_jumps_arr}
\end{figure*}

The primary interface of interest is Li/Ga-LLZO. Its structure is shown in~\cref{fig:int_main_jumps_arr}.a, while the out-of-plane component of MSD ($MSD_Z$), normal to the interface, is presented in~\cref{fig:int_main_jumps_arr}.b (and Figure~S38) for the 300--450~K temperature range. This range was selected to ensure that the Li slab remains in the solid state. 

The La--Zr--O framework remains structurally intact across the entire temperature range studied (see Section~S13.2). In contrast, Ga atoms become noticeably mobile above $\approx 400\ \mathrm{K}$, indicating a propensity for Ga loss from Ga-LLZO and dissolution into deposited Li during pellet heat treatment, which is typically performed near $500\ \mathrm{K}$~\cite{li2023excellent}. This observation agrees with X-ray photoelectron spectroscopy (XPS) and DFT results that report Ga migration across the Li/Ga-LLZO interface and subsequent Ga--Li alloy formation~\cite{klenk2024comparative}. A related effect was reported in AIMD simulations of Al-doped LLZO~\cite{Haarmann2021}. Dopant loss at the interface can promote a transition to t-LLZO and thereby hinder Li transfer.

Returning to Li diffusion, we observe a sharp increase in the mean-squared displacement (MSD) between 390 and 400~K, which we attribute to the activation of Li hops from Ga-LLZO into interstitial sites within the metallic Li layer. Because adequate statistical sampling is achieved only above 400~K, we restrict subsequent analysis to the 400--450~K temperature window.

Diffusion is strongly anisotropic ($A_{\mathrm{aniso}} < 0.03$, see Figure~S41) due to the dominant contribution of the metallic Li layer in the $xy$ plane. However, the out-of-plane component $MSD_Z$, normal to the interface, is comparable for Ga-LLZO and Li/Ga-LLZO (Figure~S39.d and Figure~S40.c), initially suggesting low interfacial resistance. This is in contrast to Li/t-LLZO (Figure~S37) where $MSD_Z$ is minor due to the absence of vacancies in t-LLZO preventing charge-transfer.

Direct use of $MSD_{Z}$ from \cref{fig:int_main_jumps_arr}.b is inappropriate for estimating the charge-transfer resistance, as it includes contributions from bulk diffusion in both the Li metal and LLZO regions. Therefore, we directly calculated the number of Li$\rightarrow$LLZO and LLZO$\rightarrow$Li jumps. To distinguish interfacial jumps from bulk transport, a minimum residence time, $t_{\mathrm{residence}}$, was introduced. Ions that spend less than $t_{\mathrm{residence}}$ in either the Li or LLZO region after crossing the interface are classified as interfacial rattling events and excluded from the statistics. In contrast, full transits across the supercell are not subject to this criterion and are counted as double-interface crossings. Additional details of the method are provided in~Section~S17.1 and Section~S17.2, while an example trajectory illustrating how interface jumps and full transits are distinguished, is shown in~Figure~S42. The current density was then calculated from the number of jumps using \cref{eq:current_density}.  We note that our residence-time window method provides reliable estimates of charge-transfer resistance even at elevated temperatures, where Li may become disordered or partially molten.

\begin{figure}[!htbp]
\begin{center}
    \includegraphics[width=0.9\columnwidth]{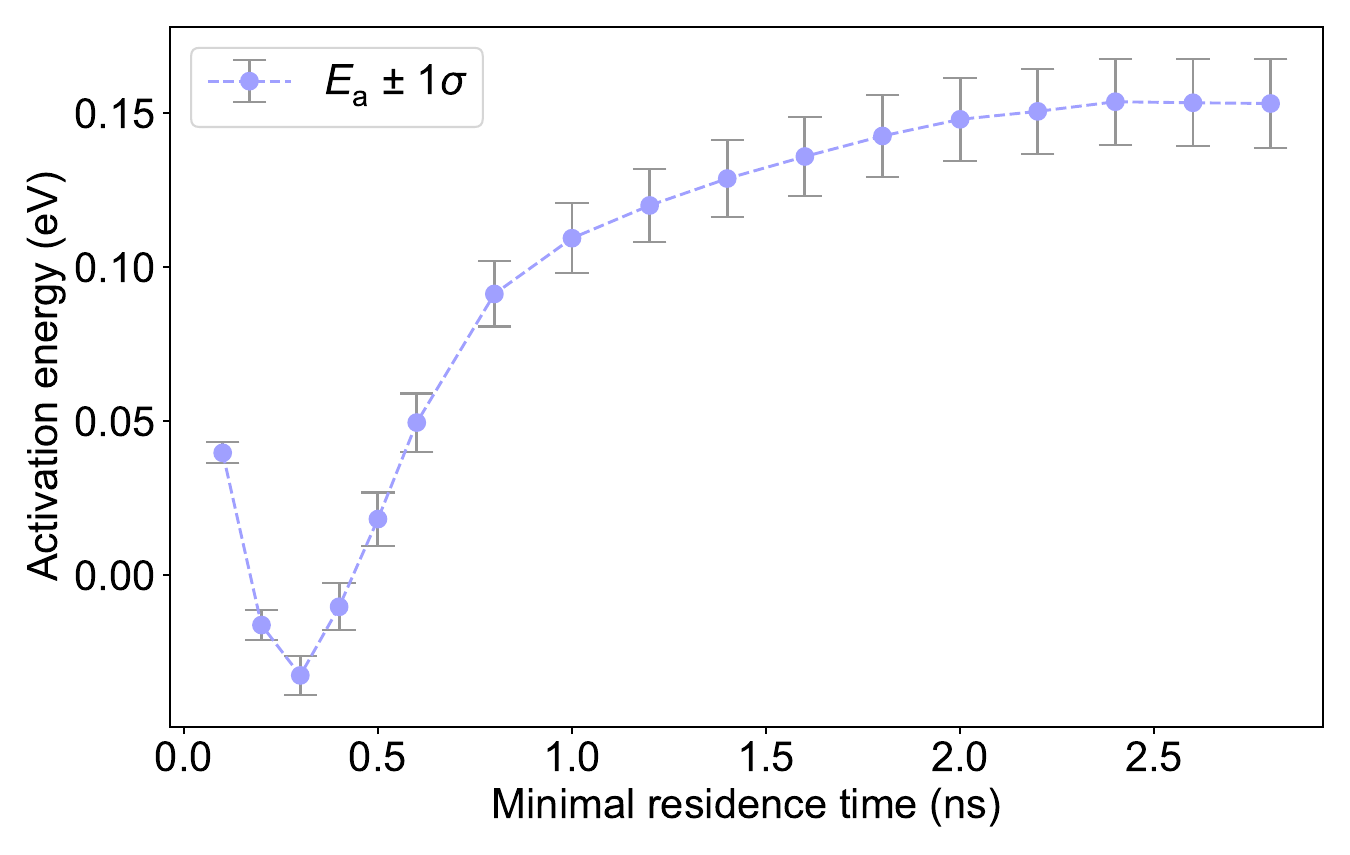}
\end{center}
\caption{Activation energy of charge transfer for Li/Ga-LLZO interface as a function of the minimal residence time $t_\mathrm{residence}$. The converged value is $(167\pm 18)$~meV.}
\label{fig:min_residence_conv}
\end{figure}

The minimum residence time, $t_{\mathrm{residence}}$, was systematically increased up to 2.8~ns until convergence of the activation energy was achieved, as shown in~\cref{fig:min_residence_conv}. This procedure yielded a charge-transfer activation energy of $167 \pm 18$~meV as shown in~\cref{fig:int_main_jumps_arr}.

Using the exchange-current density, we derived the rate constant $k_{00}^{\mathrm{calc}} = 339$~cm~s$^{-1}$ from the low-overpotential limit of the Butler--Volmer equation (\cref{eq:current_density_si}; \cref{sec:k00_from_calc}). Remarkably, this value is of the same order as the experimental estimate, $k_{00}^{\mathrm{exp}} = 102$~cm~s$^{-1}$, derived from the works of Krauskopf~\textit{et al.}~\cite{krauskopf2019toward,krauskopf2020fast} using their reported activation energies and interfacial resistances (see \cref{sec:k00_from_experiment}). The ratio $k_{00}^{\mathrm{calc}}/k_{00}^{\mathrm{exp}} \approx 3.3$ provides an estimate of the ratio between the planar and real interfacial areas. This suggests the presence of interfacial voids in the experimental systems, consistent with the interpretation proposed by Krauskopf~\textit{et al}.
The compiled charge-transfer parameters for \ce{Li^{+}} at the Li/Ga-LLZO interface are summarized in~\cref{tab:act_energies_our}.

\begin{table*}[!ht]
\centering
\setlength{\tabcolsep}{7pt}
\renewcommand{\arraystretch}{1.12} 
\caption{Diffusion parameters for the vacancy-mediated mechanism: activation energies ($E_{\mathrm{a}}$), pre-exponential factors ($D^0_{\mathrm{vac}}$), room-temperature vacancy diffusion coefficients ($D_{\mathrm{vac}}~(300\ \text{K})$), and room-temperature ionic conductivity ($\sigma~(300~\text{K})$) for the bulk phases t-LLZO with 1.8\% vacancies, c-LLZO, Ga-LLZO, and Li metal, as well as for the Li/Ga-LLZO interfacial supercell.}
\label{tab:act_energies_our}
\begin{tabularx}{\textwidth}{l c c c c c }
\hline\hline
 \multicolumn{6}{c}{Bulk phases} \\ \hline
Structure  &  Reference   &  $E_{\mathrm{a}}$ (meV) &   $D^0_{\mathrm{vac}}$ (cm$^2$/s) & $D_{\mathrm{vac}}(300~\mathrm{K})$ (cm$^2$/s) & $\sigma(300~\mathrm{K})$ (mS/cm) \\ \hline
t-LLZO (1.8\% vac.) & this work   & $116 \pm 12$ & $(7.5 \pm 1.0) \times 10^{-4}$ & $(8.6 \pm 1.2) \times 10^{-6}$ & $23.4\pm0.8$  \\
c-LLZO     &    this work    & $243 \pm 33$ & $(2.1\pm0.8) \times 10^{-4}$ & $(1.7\pm0.7) \times 10^{-8}$ & $1.4\pm0.1$  \\ 
\multirow{2}{*}{Ga-LLZO}    &   this work     & $200 \pm 55$ & $(3.3\pm2.1) \times 10^{-5}$ & $(1.4\pm0.9) \times 10^{-8}$ & $1.1\pm0.2$  \\
    &   exp.~\cite{kanai2021low}     & $200$ & --- & --- & 1.0  \\

\multirow{2}{*}{bcc-Li}     &   this work     & $50 \pm 10$ & $(1.0\pm0.1)\times 10^{-3}$ & $(1.5\pm0.3)\times 10^{-4}$   & --- \\ 
 &   exp.~\cite{mali19886li}   & $561 \pm 2$ & --- & $(2.6\pm0.3)\times 10^{-4}$ & --- \\ \hline
 \multicolumn{6}{c}{Supercells with interfaces} \\ \hline
 Structure &  Reference   &   $E_{\mathrm{a}}$ (meV) & $k_{00}$ (cm/s) & $j_{z}(300~\mathrm{K})$ (A/cm$^2$) & $R_{\mathrm{ct}}$ ($\Omega$~cm$^2$) \\ \hline
\multirow{2}{*}{Li/Ga-LLZO}   &   this work    & $167 \pm 18$ & $339 \pm 166$ & $2572\pm1261$ & $(1.0\pm0.5)\times 10^{-5}$  \\
 & exp.~\cite{krauskopf2019toward, krauskopf2020fast}  &  $370$ & $102$ & $0.3$ & $8 \times 10^{-2}$  \\
\hline\hline
\end{tabularx}
\end{table*}

The charge-transfer activation energy, $E_{\mathrm{a}} = 167~\mathrm{meV}$, is higher than the effective activation energy derived from $MSD_Z$ in~Figure~S38, because it reflects only the charge-transfer process rather than the effective diffusion of \ce{Li^{+}} within the LLZO and Li-metal regions. Using the Butler--Volmer equation (see \cref{eq:bv_methods}), we further obtained a charge-transfer resistance of $R_{\mathrm{ct}} = 1.01 \times 10^{-5}~\Omega~\mathrm{cm}^2$, which is far below the experimental value of $0.08~\Omega~\mathrm{cm}^2$ reported by Krauskopf~\textit{et al.}~\cite{krauskopf2020fast}. This indicates that the intrinsic charge-transfer resistance at the Li/Ga-LLZO interface is extremely small and does not limit Li transport in Li/LLZO solid-state batteries. This conclusion is consistent with the interpretation of Krauskopf~\textit{et al.}, who suggested that the experimentally measured resistance of $0.08~\Omega~\mathrm{cm}^2$ represents only an upper bound due to current constriction effects.

\section{DISCUSSION}

Classical MD potentials are typically fitted to limited experimental data and therefore often fail to capture changes in Li-ion transport mechanisms or to represent every local atomic environment at the anode, electrolyte, and their interface. By contrast, ML-driven models (MLMD) are trained on first-principles data, reproduce DFT energetics and forces more accurately across diverse environments, and yield diffusion and transport behavior that aligns better with experiments.

\subsection{Performance of MTPs and validation tests}

In this work, we demonstrated that moment tensor potentials (MTPs) accurately reproduce energies, forces, and stresses for both bulk materials and supercells containing interfaces. Importantly, MTPs reliably capture vacancy formation energies across these structural types, enabling precise identification of the most favorable defect sites and the relative energy differences among vacancy configurations. This capability provides a solid foundation for mapping percolating diffusion pathways in interfacial systems. In addition, MTPs correctly describe the occupation numbers on Li lattice sites. 

Moreover, MTP-NEB predictions closely align with DFT-NEB results for both bulk and interfacial structures. The mean absolute errors (MAEs) for activation energies are 15~meV for bulk phases and 66~meV for supercells containing interfaces, while the MAEs in migration-path lengths are approximately 0.1~\AA~for both cases. These results demonstrate that MTPs trained for machine-learning molecular dynamics (MLMD) simulations offer an efficient and accurate alternative to conventional DFT-NEB calculations.

\subsection{Diffusion mechanisms in LLZO systems}

One important question is whether t-LLZO with vacancies can serve as a model for c-LLZO, given the similar La--Zr--O framework and the tetragonality of t-LLZO. Our analysis shows that c-LLZO and vacancy-containing t-LLZO share a similar cooperative diffusion mechanism, with roughly 30\% of hops being cooperative at room temperature (see \cref{fig:eta_both}).  

Nevertheless, the two systems differ markedly in activation energy and room-temperature ionic conductivity (see \cref{tb:activation_en_discussion}). For t-LLZO with 1.8\% vacancies the activation energy is only 0.12~eV, versus 0.24~eV for c-LLZO. Moreover, the vacancy diffusion coefficient at room temperature in vacancy-containing t-LLZO is about two to three orders of magnitude higher than in c-LLZO: $8.6\times10^{-6}~\mathrm{cm}^2$~s$^{-1}$ vs. $1.7\times10^{-8}~\mathrm{cm}^2$~s$^{-1}$. This trend agrees with Yan~\textit{et al.}, who reported decreasing activation energies with increasing non-intrinsic vacancy concentration in t-LLZO~\cite{yan2024impact}. Consequently, t-LLZO with vacancies is not an appropriate surrogate for c-LLZO because it substantially overestimates diffusivity and ionic conductivity.

Our MLMD simulations for Ga-LLZO yield an activation energy of $\sim$0.20~eV, in good agreement with experimental values of 0.20--0.32~eV~\cite{tao2023preparation,chen2020microstructural,mishra2025stabilization,yang2017ionic,kanai2021low}. The calculated vacancy diffusion coefficient, $1.4\times10^{-8}~\mathrm{cm}^2$~s$^{-1}$, also matches reported experimental ranges $(1.2-1.8)\times10^{-8}\ \mathrm{cm}^2$~s$^{-1}$. The modest reduction in $E_{\mathrm{a}}$ for Ga-LLZO relative to c-LLZO stems from suppression of high-barrier octahedral--tetrahedral hops. At the same time, Ga occupying Li sites partially blocks diffusion channels, producing a vacancy diffusivity somewhat lower than ideal c-LLZO; overall, Ga-LLZO attains roughly 80\% of the ideal c-LLZO ionic conductivity.

\subsection{Analysis of literature Li-ion diffusion in LLZO systems}

To obtain sufficient statistics on Li-ion hops, MD and AIMD simulations are often run at elevated temperatures; although this avoids enhanced-sampling techniques (e.g., metadynamics), it can produce significant errors when the high-temperature dynamics do not reflect the room-temperature diffusion mechanism. Here we compile and systematize published Li-self-diffusion data for LLZO from MD, AIMD and MLMD studies; the extracted activation energies and room-temperature conductivities are summarized in~\cref{tb:activation_en_discussion} and compared with experiment for t-LLZO and Ga-LLZO.

\begin{table*}[!htb]
\centering
\setlength{\tabcolsep}{9pt} 
\renewcommand{\arraystretch}{1.3} 
\caption{Calculated Li vacancy self-diffusion parameters for bulk t-LLZO, c-LLZO, and Ga-LLZO: vacancy fraction ($n_{\mathrm{vac}}$, decimal value in parentheses), activation energy ($E_{\mathrm{a}}$, eV), room-temperature ionic conductivity ($\sigma(300~\mathrm{K})$, mS/cm), room-temperature diffusion coefficient ($D_{\mathrm{vac}}(300~\mathrm{K})$~cm$^2$/s), computational method, minimum simulation temperature ($T_{\mathrm{min}}$, K), and reference work. In addition, the switch temperature ($T_{\mathrm{switch}}$, K), at which the vacancy-mediated mechanism gives way to a Frenkel-pair-mediated mechanism (see Sections~S22 and~S23) is provided.}
\label{tb:activation_en_discussion}
\begin{tabularx}{\textwidth}{%
  l  
  l  
  c  
  c  
  c 
  c  
  c  
  c  
  l  
}

\hline \hline
Structure  & Method & $n_{\mathrm{vac}}$ & $T_{\mathrm{min}}$ & $T_{\mathrm{switch}}$  &  $E_{\mathrm{a}}$ &  $\sigma(300~\mathrm{K})$ &  $D_{\mathrm{vac}}(300~\mathrm{K})$ & Ref. \\ \hline
\multirow{7}{*}{t-LLZO (no vac)} 
 & MLMD & $0$       & 400       & \multirow{7}{*}{800}  & ---      & 0   &   ---    &  Our        \\ 
 & MLMD & $0$      & $\sim$700  &   & 1.23    & 7.2$\times10^{-11}$ & --- &  \cite{yan2024impact} \\ 
 & MD   & $0$      & $\sim$600        &   & $^{\dagger} 0.60$    & $^{\dagger} \sim 0$     & --- &   \cite{Chen2017_mining}          \\ 
 & AIMD & $0$      & $\sim$600  &   & 0.43    & 0.001   & --- &  \cite{miara2013effect}         \\ 
 & AIMD & $0$  & 1350         &   & 0.36    & $^{\dagger} 0.18$ & ---    &  \cite{andriyevsky2017ab}          \\ 
 & Exp. & $0$  & 297         &   & 0.41    & NA & ---    &  \cite{wang2015phase}          \\ 
 & Exp. & $0$  & 297         &   & 0.41    & 0.02 & ---    &  \cite{Wolfenstine2012}          \\ \hline

\multirow{2}{*}{t-LLZO (vac)} 
 & MLMD & $\frac{1}{56}~(0.018)$  & 550       &   \multirow{2}{*}{800} & 0.12   & 23.4  &  $8.6\times10^{-6}$ &  Our          \\ 
 & AIMD & $\frac{3}{56}~(0.054)$  & 1150        &   & 0.19    & $^{\dagger}6.3$ & $^{\dagger}3.4\times10^{-6}$    &  \cite{andriyevsky2017ab}          \\ \hline

\multirow{5}{*}{c-LLZO} 
 & MLMD & $\frac{120-56}{120}~(0.53)$      & $400$      &   \multirow{5}{*}{800}  & 0.24   & 1.4  &  $1.7\times10^{-8}$  &  Our            \\ 
 & MLMD & $\frac{120-54}{120}~(0.55)$      & $\sim$650  &  & 0.26 &  1.2 &  $^{\dagger}1.4\times10^{-8}$    &  \cite{yan2024impact} \\ 
  & MD   & $\frac{120-56}{120}~(0.53)$     & 300         &   & 0.31    & $^{\dagger}0.3$ & $^{\dagger}3.6\times10^{-9}$  &  \cite{Chen2017_mining}  \\ 
 & AIMD & $\frac{120-56}{120}~(0.53)$      & $\sim$600  &   & 0.24    &  2.0  & $^{\dagger}2.4\times10^{-8}$  &  \cite{miara2013effect}           \\ 
 & AIMD & $\frac{120-56}{120}~(0.53)$      & 1273       &   & 0.30    & $^{\dagger}0.2$ &  $^{\dagger}2.7\times10^{-9}$   &   \cite{verduzco2023atomistic}          \\  \hline

\multirow{7}{*}{Ga-LLZO} 
 & MLMD & $\frac{118-50}{118}~(0.58)$   & 400   & \multirow{7}{*}{800}   & 0.20    &  1.1 &  $1.4\times10^{-8}$ &  Our           \\ 
 & MD   & $\frac{119-53}{119}~(0.55)$  & 600  &   &  0.24   & 6.1 &  $^{\dagger}7.5\times10^{-8}$  &   \cite{jalem2015effects}        \\ 
 & Exp.   & $\frac{118-50}{118}~(0.58)$  & 297  &   &  NA   & 1.5 & $^{\dagger}1.8\times10^{-8}$   &   \cite{tao2023preparation}        \\ 
 & Exp.   & $\frac{118-50}{118}~(0.58)$  & 297  &   & 0.32  &  1.2   & $^{\dagger}1.5\times10^{-8}$   &   \cite{chen2020microstructural}        \\ 
 & Exp. & $\frac{119-53}{119}~(0.55)$  & 253  &  & 0.25  &  1.1  & $^{\dagger}1.4\times10^{-8}$ &   \cite{mishra2025stabilization}        \\ 
 & Exp. & $\frac{118-50}{118}~(0.58)$  & 297  &   & 0.25  &  1.5   & $^{\dagger}1.8\times10^{-8}$   &   \cite{yang2017ionic}        \\ 
 & Exp. & $\frac{119-53}{119}~(0.55)$  & 297  &   & 0.20  &  1.0   & $^{\dagger}1.2\times10^{-8}$   &   \cite{kanai2021low}        \\ 
 
\hline
\hline
\end{tabularx}

\begin{tablenotes}
      \small
      \item Here, the fraction of vacancies ($n_{\mathrm{vac}}$) for c-LLZO is derived from the number of vacant sites. The total available Li sites is 120, whereas only 56 are occupied (see Table~S2). $n_{\mathrm{vac}}$ for Ga-LLZO is derived with the condition that Ga occupies Li positions and blocks vacant site. \\
      The $\dagger$ means that the value was not provided in the article, but there was enough data to calculate it. \\
      The abbreviation ``NA'' is used if the data were not provided and could not be calculated from others.
\end{tablenotes}

\end{table*}

AIMD is typically limited to small simulation cells for computational reasons, which can miss collective phenomena such as cooperative migration that require larger length scales. Introducing vacancies into small cells improves hop statistics, but most solid electrolytes lack intrinsic vacancies and require doping to generate them; consequently, reported activation energies depend strongly on whether vacancies are present. For example, Andriyevsky \textit{et al.}~\cite{andriyevsky2017ab} report $E_{\mathrm{a}} \approx 0.19$~eV for vacancy-containing structures, in contrast to $0.36$--$0.43$~eV for vacancy-free cells~\cite{andriyevsky2017ab,miara2013effect}. Large-cell simulations (thousands of atoms) such as Yan~\textit{et al.} further show that self-diffusion in t-LLZO increases substantially when both vacancy concentration and system size are treated properly~\cite{yan2024impact}.

In Section~S22 and Section~S23 we show that each LLZO phase (t-LLZO, c-LLZO, Ga-LLZO) exhibits distinct low- and high-temperature diffusion regimes. At low temperature, Li transport is dominated by vacancy-mediated hops between lattice sites; at high temperature, interstitial-mediated pathways become prevalent. Because activation energies extracted from these two regimes can differ markedly, high-temperature simulations used solely to increase jump counts may yield misleading extrapolations. For example, extrapolating from the high-temperature regime predicts a 300~K diffusion coefficient for Ga-LLZO that is almost 4 times lower than the value inferred from low-temperature data (see Table~S22). Therefore, simulations intended to model ionic diffusion at 300~K should sample the low-temperature regime or otherwise account explicitly for the regime change.

These issues also explain large discrepancies in reported ionic conductivities from different simulation temperatures. Miara~\textit{et al.} predicted $\approx$ 0.001~mS~cm$^{-1}$ for t-LLZO from AIMD above 600~K, whereas Andriyevsky~\textit{et al.} reported data only above 1350~K; extrapolating the latter high-temperature limit to 300~K yields $\approx$ 0.18~mS~cm$^{-1}$ and produces the apparent mismatch. Such differences highlight the importance of identifying the dominant conduction mechanism active under battery operating conditions. For LLZO at ambient conditions, vacancy-mediated diffusion (with minimal interstitial contribution) appears to be the primary pathway, distinct from the interstitial-dominated mechanism that emerges at elevated temperatures.

\subsection{Charge transfer from MLMD vs static NEB calculations}

In this work, we proposed using a residence-time window to count only the events that lead to interfacial charge transfer. This approach filters out ion rattling, where ions remain near the interface plane and rapidly hop back and forth between Li$\rightarrow$LLZO and LLZO$\rightarrow$Li. Otherwise, such rattling can overestimate the number of interfacial jumps if they are not excluded from the statistics~\cite{Haarmann2021}. Within this framework, we can extract the real charge-transfer activation energy of 167~meV. By contrast, this energy cannot be reliably obtained from the MSD in MLMD simulations for a bcc-Li anode, where the activation barrier is only 50~meV. In that case, because diffusion in bcc-Li is much faster than at the interface, the effective barrier reflects diffusion in the anode region rather than in the interface region. As a result, LLZO becomes the rate-limiting step, with a higher activation energy of about 200~meV.

In several previous studies, including ours, the charge-transfer barrier was estimated from static NEB calculations, yielding a larger value of 0.9~eV~\cite{burov2024mechanism} compared to that of 0.17~eV, calculated within the residence-time approach. In contrast to MLMD, DFT-NEB is limited to a single minimum-energy path (MEP) corresponding to the lowest activation barrier, and it neglects thermal lattice vibrations and entropy, which can reduce the effective barrier. Moreover, AIMD sampling of local environments allows the electron density to redistribute and partially mitigates the constant-charge constraint inherent in DFT-NEB. Additionally, MLMD explores multiple diffusion pathways, sometimes uncovering lower-energy routes missed by the initial NEB guess. However, the close agreement between DFT-NEB and MTP-NEB energies confirms that these differences are intrinsic to these methods and cannot be attributed to an improperly trained MLIP (see Section~S12).

In this work, we also used long MD trajectories to extract the relative Li chemical potential from the ensemble-averaged Li vacancy formation energies. This approach gives an activation energy of 0.4~eV, which is substantially higher than the 0.17~eV obtained from the MLMD exchange-current-density analysis. The discrepancy highlights the limitations of the static ensemble-averaged chemical-potential framework, which neglects electron redistribution and interfacial electronic effects, and therefore describes Li removal as neutral vacancy formation rather than as \ce{Li^{+}} extraction coupled to electron transfer.

\subsection{Limitations of constant-charge conditions}

We showed that machine-learning interatomic potentials trained on DFT and AIMD data under constant-charge conditions are able to reproduce electrochemical potential alignment between the LLZO and Li regions. Although the calculations are performed at constant charge, electron redistribution at the interface is captured in the reference data, enabling MLMD simulations to correctly describe interfacial charge transfer and evaluate the exchange current.

The MLMD simulations in this work were performed at zero overpotential. Accessing finite-potential effects would require retraining the MTP on configurations sampled under non-zero overpotential, ideally within a grand-canonical (constant-potential) DFT framework that mimics a potentiostatically controlled electrode~\cite{hormann2024converging, melander2024constant}. This avoids the artifacts of fixed-charge calculations, where manually changing the electron count can cause spurious charge localization and unphysical potential drift. Such a constant-potential MTP would enable simulations of cyclic voltammetry and other overpotential-driven phenomena. However, this requires additional DFT data and retraining of the MTP for each specific overpotential, as the system's electron count must be varied accordingly. Previous DFT-NEB studies show that finite overpotential can shift migration barriers by tens to hundreds of meV: reductions of 25--70~meV (Campbell~\textit{et al.}~\cite{campbell2024voltage}) or increases up to 170~meV (Tran~\textit{et al.}~\cite{tran2024electric}). These results highlight the need for constant-potential MLMD to capture overpotential-dependent kinetics at Li/LLZO interfaces.

The charge transfer resistance calculated under constant-charge conditions represents a lower bound for the intrinsic resistance. To assess the possible correction associated with constant-potential boundary conditions at zero overpotential, we calculated the potential of mean force (PMF) from the Li density distribution within the interface supercell, using Eq.~(S14) to convert density deviations (relative to bulk Li and LLZO) to an energy scale (see Section~S21 for details). The PMF varies by only approximately 0.03~eV (Figure~S47) across the relevant interfacial region, indicating that the equilibrium Li chemical potential is nearly uniform.  We therefore expect the correction to the zero-overpotential activation free energy arising from enforcing constant-potential conditions to be small compared with the effective 0.17~eV barrier obtained from the residence-time analysis. At finite applied bias, an additional approximately linear electrostatic contribution to the barrier is expected and is not captured by this estimate. Accordingly, the calculated $R_{\mathrm{ct}}$ should be interpreted as a lower-bound estimate of the intrinsic interfacial resistance under equilibrium conditions, rather than as a prediction applicable to high-rate or strongly non-equilibrium operating conditions.

Another limitation of classical DFT calculations is the adiabatic approximation, as nonadiabatic effects can be crucial at interfaces because ionic motion and electronic degrees of freedom become coupled, which may lead to the transient electronic excitations during $\mathrm{Li}^0\rightarrow\mathrm{Li}^{+}$ transit. This may alter the effective barriers and open new migration pathways that are energetically inaccessible within the adiabatic Born--Oppenheimer approximation. Consequently, adiabatic AIMD may under- or overestimate interfacial hop rates and give incorrect temperature dependencies. Nonadiabatic calculations would yield more reliable kinetics and enable more accurate predictions of interfacial charge-transfer resistance and plating/stripping behavior.

\section{CONCLUSIONS}

In this work, we demonstrated that moment tensor potentials (MTPs) accurately reproduce DFT-level energies, forces, and stresses for bulk and interface-containing supercells across LLZO phases: tetragonal (t-LLZO), cubic (c-LLZO), and Ga-doped (Ga-LLZO). These MTPs also reliably predict key solid-state electrolyte properties, including vacancy formation energies, NEB activation barriers and pathways, and lattice site occupancies in both bulk and interfacial structures (Li/t-LLZO, Li/Ga-LLZO).

The MLMD simulations revealed that all LLZO systems (t-LLZO with 1.8\% vacancies, c-LLZO, Ga-LLZO) exhibit vacancy-mediated diffusion below 800~K. However, additional vacancies in t-LLZO dramatically lower activation energies and overestimate diffusion coefficients, making it unsuitable as a c-LLZO surrogate.

Our MTPs, trained on AIMD and DFT data, correctly capture interfacial charge transfer and electron density redistribution. We introduced a concept of residence-time window filtering to calculate exchange-current density, filtering non-contributory events like ion rattling. The resulting charge-transfer activation energy is only 167~meV (vs. 200~meV for Ga-LLZO), yielding a resistance of $\sim10^{-5}~\Omega~\mathrm{cm}^2$ that does not limit Li$^+$ diffusion at Li/Ga-LLZO interfaces.

These calculations enable estimation of real geometrical surface area and rate constant $k_{00}$, facilitating determination of geometrical-to-planar surface area ratios. Our approach serves as a rapid screening tool to assess interfacial resistance and interpret electrochemical measurements, accelerating interface optimization for solid-state batteries.

\section{METHODS}\label{sc:methods}

\subsection{Density functional theory}\label{sec:methods_dft}

Density functional theory (DFT) calculations were performed using the projector augmented wave (PAW) method as implemented in the Vienna \textit{Ab initio} Simulation Package (VASP)~\cite{Kresse:1996,Kresse:1999}. The generalized gradient approximation (GGA) in the Perdew--Burke--Ernzerhof (PBE) form~\cite{Perdew:PRL:1996,Perdew:PRL:1997} was employed for the exchange-correlation functional, and electron--ion interactions were described using PAW pseudopotentials (version 54)~\cite{Blochl:1994}.

The plane-wave kinetic energy cut-off ($E_{\mathrm{cut}}$) was selected at 400~eV, which is sufficient for the accurate evaluation of second derivatives of the total energy, as demonstrated in our previous work~\cite{burov2024mechanism}. A $\Gamma$-centered $k$-mesh was generated with $k$-spacing less than 0.2~\AA$^{-1}$ for bcc-Li and 0.5~\AA$^{-1}$ for other phases. For the relaxation calculations of atomic positions, the parameter of maximum force in the structure was set to be less than 50~meV/\AA. Second-order Methfessel--Paxton smearing was used for Brillouin-zone integration in Li metal, whereas Gaussian smearing was applied to the other structures. The smearing width was set to 0.1~eV. Nudged elastic band (NEB) calculations~\cite{jonsson1998nudged} were performed using the implementation developed by the Henkelman group~\cite{henkelman2000_neb_improved}, with the same setup both for the DFT and MLIP calculations.

\subsection{\textit{Ab initio} molecular dynamics}\label{sec:methods_aimd}

\textit{Ab-initio} molecular dynamics was used to model Li-ion diffusion, as implemented in the VASP code. Prior to the AIMD simulations, we performed lattice optimization for several contracted and expanded cells to mitigate Pulay errors. The simulation timestep was set to 2~fs. Each AIMD trajectory included 2,000 snapshots for bulk phases and Li/t-LLZO interface supercell (4~ps) and 4,000 snapshots for Li/Ga-LLZO interface supercell (8~ps).

For each structure, two AIMD trajectories were generated: one by heating from 0~K to $T$, and one by simulated annealing at $T$. The simulation temperature $T$ was set to 420~K for bcc Li and interface supercells, and to 1000~K for all other structures.

\subsection{Machine-learning interatomic potentials}\label{sec:mtp_method}

To describe interatomic interactions with near-DFT accuracy in the MD simulations, we employed machine-learning Moment Tensor Potentials (MTPs), which represent the local atomic environment using inertia tensors~\cite{shapeev2016moment}. The cut-off radius for the local atomic environment was set to 6.3~\AA. The selection and break-threshold parameters were set to 2 and 20, respectively. The weights for energies, forces, and stresses were set to 1, 0.1, and 0.01, respectively. Further details on MTPs can be found in the original papers by the method developers~\cite{shapeev2016moment, podryabinkin2017active}.

For each Li--LLZO system, we trained a separate MTP.  Initially, we pretrained the potential on configurations generated by AIMD simulations and then refined it using active learning~\cite{novikov2020mlip}. For this purpose, we used conventional cells containing more than $10^2$ atoms, with lattice constants exceeding 10~\AA~in each direction. The timestep was 1.0~fs. Each trajectory comprised three steps: heating from 0~K to $T$ over $10^{5}$ timesteps (0.1~ns), followed by simulated annealing at $T$ for $4 \times 10^{5}$ timesteps (0.4~ns) in the isothermal--isobaric (NPT) ensemble using the Nos\'e--Hoover thermostat and barostat. This was followed by self-diffusion for $10^{6}$ timesteps (1~ns) in the canonical (NVT) ensemble with the Nos\'e--Hoover thermostat. The simulation was repeated until no configurations exceeded the selection criterion.

The training was performed in two stages. In the first stage, the 10th-level MTP was used to collect the majority of configurations for the training set. In the second stage, the procedure was repeated with the 20th-level potential, starting from the pretraining stage and augmenting the initial training dataset with the configurations obtained in the previous stage.

\subsection{Machine-learning-assisted molecular dynamics}\label{sec:md_mlip}

For large-scale simulations involving more than $10^{3}$ atoms, we performed machine-learning molecular dynamics (MLMD) using the LAMMPS package~\cite{lammps} with the trained MTPs.

To study ionic conductivity, we increased the MLMD timestep to 2~fs. The lattice constants and atomic positions were first optimized in the NPT ensemble. The system was then heated from 0~K to the target temperature $T$ over $10^{5}$ timesteps (0.2~ns), followed by annealing at $T$ for $4 \times 10^{5}$ timesteps (0.8~ns). Li-ion self-diffusion was subsequently simulated in the NVT ensemble for $10^{6}$ timesteps (2~ns) in the bulk and up to $20 \times 10^{6}$ timesteps (40~ns) in interfacial supercells. Only the linear regime of this final stage was used in the subsequent analysis.

\subsection{Simulation of crystal structures}

To simulate bulk structures for MTP training, we used crystal structures from the Inorganic Crystal Structure Database (ICSD)~\cite{bergerhoff1987crystallographic_icsd}. The structural differences between the tetragonal (t-LLZO, sg. 142) and cubic (c-LLZO, sg. 230) phases of LLZO are described in~Section~S1. Because c-LLZO has only 56 of 120 Li sites occupied, we constructed an optimized simulation cell using the Ewald summation method~\cite{toukmaji1996ewald} combined with AIMD calculations, as described in~Section~S2. In this work, we examined several LLZO phases and assessed whether c-LLZO can be modeled using a surrogate t-LLZO structure with a finite concentration of non-intrinsic vacancies.

The Ga-doping strategy for LLZO is outlined in~Section~S3. We evaluated feasible Ga substitution sites on Li lattice sites ($\mathrm{Ga}_{\mathrm{Li}}$) and averaged total energies from AIMD simulations to identify the most stable configuration. Our results indicate that the lowest energy state corresponds to configurations where Ga occupies both tetrahedral and octahedral Li sites. This doped structure is hereafter referred to as Ga-LLZO.

Following AIMD simulations, each structure was further optimized via ionic relaxation while keeping lattice constants fixed. The relaxed lattice parameters and their comparison with computational and experimental data are provided in~Section~S5. Visualizations of the simulated cell structures are shown in~Figure~S1.

To model supercells with interfaces, we employed the methodology from our previous work~\cite{burov2024mechanism}. We first optimized the free surfaces of both electrode and electrolyte crystals using DFT calculations. Subsequently, surface orientations with minimal lattice mismatch were identified. As calculated for Li metal (see Section~S11.2), this mismatch contributes only about 0.04~eV to the Li vacancy formation energy between the two supercells and therefore does not significantly affect the thermodynamic results. To refine local morphology and cell volume, rigid-body displacements supported by Ewald summation~\cite{toukmaji1996ewald} were performed to determine the optimal separation distance and relative positioning of the two phases within the interfacial plane. We also assessed wettability and thermodynamic stability by calculating adhesion and interface energies.

The numbers of atoms in the simulation cells used for AIMD, potential training, and MLMD calculations are given in~Table~S5. The numbers of configurations sampled for training, both from AIMD and during the active-learning stages, are summarized in~Table~S6. Additional details and parameters are provided in~Section~S6.

Despite the extremely low intrinsic vacancy concentration in Li metal (on the order of $\sim10^{-7}$~\cite{frank1996first}), we deliberately introduced 3\% Li vacancies in the anode region, as Li/t-LLZO exhibits negligible diffusion without vacancies. To maintain consistency across our simulations, the same vacancy concentration was also introduced in the Li/Ga-LLZO system. Additionally, 0.2\% vacancies were incorporated into bulk bcc-Li (1 vacancy per 432-atom supercell). In this study, we focused on the Li(100) and LLZO(100) surface orientations, although the methodology is applicable to other orientations as well. The resulting interface morphology is consistent for both t-LLZO and Ga-LLZO interfaces (see Figure~S4). Hereafter, the supercells with such interfaces are referred to as Li/t-LLZO and Li/Ga-LLZO. More details on interface modeling are provided in~Section~S4.

Structure generation and calculations were performed using the SIMAN \cite{AKSYONOV2018449}, Pymatgen \cite{ong2013python_pymatgen}, and ASE \cite{larsen2017atomic_ase} Python packages. Structural visualizations were done with VESTA \cite{momma2011vesta} and OVITO \cite{stukowski2009visualization} software programs.

\subsection{Ionic conductivity}\label{sec:ionic_conductivity}

The mean squared displacement (MSD) was calculated as follows:

\begin{equation}
    MSD(t) = \frac{1}{N} \sum_{i=1}^{N} \left|\mathbf{r_{i}}(t) - \mathbf{r_{i}}(0) \right|^{2},
    \label{eq:msd}
\end{equation}
where $t$ is the simulation time, $N$ is the number of mobile ions, $\mathbf{r_{i}}(t)$ is the atomic position of the atom $i$ at the timestep $t$.

The self-diffusivity ($D_{\mathrm{Li}}$) of Li-ions was calculated from the slope of the mean squared displacement curve, ($MSD( t)$), using the Einstein relation:
\begin{equation}
    D_{\mathrm{Li}} = \frac{1}{2d} \frac{\mathrm{d}}{\mathrm{d}t} \left\langle \mathrm{MSD}(t) \right\rangle,
    \label{eq:diffusivity}
\end{equation}
where $t$ is the simulation time, $d$ is the dimensionality of diffusion. $d = 3$ for the bulk structures; and $d = 1$ for the supercells with interfaces as we consider the direction perpendicular to the interfacial plane.

The pre-exponential factor ($D^{0}_{\mathrm{Li}}$), and activation energy ($E_{\mathrm{a}}$) of Li-ion self-diffusion were calculated by fitting the Arrhenius equation: $D_{\mathrm{Li}} = D_{\mathrm{Li}}^0 \exp\left(-\frac{E_{\mathrm{a}}}{k_{\mathrm{B}} T}\right)$ and calculating diffusivities using the least squares method. 

To eliminate the concentration dependence of the Li-ion diffusion coefficient, we additionally evaluated the vacancy diffusion coefficient, which is expected to be independent of vacancy concentration in the dilute solution limit:
\begin{equation}
    D_{\mathrm{vac}} \approx \frac{D_{\mathrm{Li}}}{n_{\mathrm{vac}}},
    \label{eq:vac_diffusion}
\end{equation}
where $D_{\mathrm{vac}}$ is the vacancy diffusion coefficient and $n_{\mathrm{vac}}$ is the vacancy fraction, defined as the ratio of vacant Li sites to the total number of Li sites ($N_{\mathrm{vac}}/N_{\mathrm{site}}$).

The conductivity was calculated from the Nernst--Einstein equation as follows:

\begin{equation}
    \sigma = \frac{e^{2} Z^{2} c_{\mathrm{Li}}}{k_{\mathrm{B}} T}D_{\mathrm{Li}},
    \label{eq:ionic_conductivity}
\end{equation}
where $c_{\mathrm{Li}}$ is the diffusing particle density (number of ions per volume of the system), $e$ is the elementary charge, $Z$ is the ionic charge of the particle, $k_{\mathrm{B}}$ is Boltzmann's constant, $T$ is the temperature.

\subsection{Diffusion analysis}

To identify Li-site occupancies, we performed the following analysis. At each timestep $t_i$, every Li ion was assigned to its nearest initial position under periodic boundary conditions (PBCs). We then computed the fraction of occupied sites relative to the total number of available Wyckoff positions, i.e., their multiplicity.

To elucidate the dominant diffusion mechanism, we calculated radial distribution functions (RDFs) for Li--Li and Li--O pairs using a cut-off radius of 7~\AA. We also evaluated Voronoi volumes using polydisperse tessellation to characterize the local atomic environment. In addition, we performed Wigner--Seitz defect analysis, as implemented in OVITO \cite{stukowski2009visualization}, to quantify interstitial defects. In this analysis, if a reference site is unoccupied, it is counted as a vacancy; if a site is occupied by more than one atom, it is classified as an interstitial defect.

To quantify the concerted migration, we calculated the self-part ($G_{\mathrm{s}}$) and distinct part ($G_{\mathrm{d}}$) of the $1D$ van Hove correlation function, given by

\begin{equation}
G_{\mathrm{s}}(\mathbf{r}, t) = \frac{1}{N} \left\langle \sum_{i=1}^N \delta[ \mathbf{r} + \mathbf{r}_i(0) - \mathbf{r}_i(t)] \right\rangle 
\end{equation}

\begin{equation}
G_{\mathrm{d}}(\mathbf{r}, t) = \frac{1}{N} \left\langle \sum_{i=1}^N \sum_{j \neq i}^N \delta[\mathbf{r} + \mathbf{r}_i(0) - \mathbf{r}_j(t)] \right\rangle ,
\end{equation}

where $\delta$ is the Dirac delta function. $G_{\mathrm{s}}(\mathbf{r}, t)$ represents the fraction of particles which have performed a given displacement $\mathbf{r}_i(0) - \mathbf{r}_i(t) = \mathbf{r}$ in a time $t$.

To calculate the percent of cooperative hops, we employed the methodology described in~Section~S24.1. Here, we calculated the normalized ratio between cooperative hops and single-ion hops as follows:

\begin{equation}
\label{eq:eta_norm}
\eta_\text{norm}(t) = \frac{ \int_{r_{\mathrm{coop}}^{\mathrm{low}}}^{r_{\mathrm{coop}}^{\mathrm{up}}}  r^2 G_\text{d}(r,t)  dr }{ 
                       \int_{r_{\mathrm{single}}^{\mathrm{low}}}^{r_{\mathrm{single}}^{\mathrm{up}}} r^2 G_\text{d}(r,t)  dr + 
                       \displaystyle \int_{r_{\mathrm{coop}}^{\mathrm{low}}}^{r_{\mathrm{coop}}^{\mathrm{up}}} r^2 G_\text{d}(r,t)  dr }
\end{equation}
where $t$ is the lag time, $G_{\mathrm{d}}(r,t)$ is the distinct part of van Hove function, $r$ is the radial displacement distance, $r_{i}^{j}$ is distance threshold for the integration. Here, $i=\mathrm{single,~coop}$ is the mechanism of ion migration (single or cooperative); and $j=\mathrm{low,~up}$ is the boundary of integration (lower or upper). The values for $r_{i}^{j}$ were estimated from the NEB calculation.

\subsection{Number of jumps in supercells with interfaces}

The methodology for calculating the total number of interface crossings is described in~Section~S17.1. We use a residence time, $t_{\mathrm{residence}}$, to avoid double-counting ion hops. An ion is counted as having crossed the interface only if it remains in the LLZO or Li region for at least $t_{\mathrm{residence}}$; otherwise, such rapid back-and-forth rattling around the same positions is not counted as contributing to the ionic flux. The total number of ion jumps is then calculated using the equation below:

\begin{equation}
    \label{eq:number_of_jumps}
    N_{\Sigma} = N_{\mathrm{int}} + 2 N_{\mathrm{transit}},
\end{equation}
where $N_{\Sigma}$ is the total crossings without double-counting, $N_{\mathrm{int}}$ is number of crossings that cross one interface only, $N_{\mathrm{transit}}$
is the number of through-slab transits, each contributes two crossings.

Exchange-current density can be calculated from \cref{eq:number_of_jumps} as follows: 

\begin{equation}
\label{eq:current_density}
j_z = \frac{e N_{\Sigma}}{2 A n_{\mathrm{int}} \Delta t},
\end{equation}
where $e$ is the elementary charge, $A$ is the interface cross-sectional area, $n_{\mathrm{int}}$ is the number of interfaces included in the symmetry factor (2 in this work), and $\Delta t$ is the simulation time over which $N_{\Sigma}$ was accumulated.

\subsection{Charge transfer resistance}

To estimate the charge transfer resistance, we employed the Butler--Volmer equation using the expression derived in~Section~S18. The resulting equation is as follows:

\begin{equation}
R_\mathrm{ct} = \frac{RT}{ z F j_0(T)} = \frac{RT}{ z F^2 k_{00} C_\mathrm{oxy}^{1-\beta} C_\mathrm{red}^\beta} \exp\left(\frac{E_\mathrm{a}}{k_\mathrm{B} T}\right) ,
\label{eq:bv_methods}
\end{equation}
where $R$ is the universal gas constant, $z$ is the number of electrons transferred in the electrode reaction, $F$ is the Faraday constant, $C_{\mathrm{oxy}}$ and $C_{\mathrm{red}}$ are the concentrations of oxidizing and reducing ions in the supercell, respectively, $\beta$ is the symmetry factor, $k_{00}$ is the reaction rate constant, and $E_{\mathrm{a}}$ is the activation energy for charge transfer at the interface in units of eV.

\section{ACKNOWLEDGMENTS}
The authors acknowledge the financial support of Russian Science Foundation project No. 23-73-30003.

\clearpage
\def\bibsection{\section*{References}} 
\putbib[lib]
\end{bibunit}


\clearpage
\begin{bibunit}[unsrt]
\renewcommand{\thesection}{S\arabic{section}}

\clearpage

\onecolumngrid 
\beginsupplement
\section*{SUPPLEMENTARY MATERIALS FOR}\label{supplementary}
\maketitle
\onecolumngrid 
\date{\today}

\clearpage

\setcounter{page}{1}
\setcounter{figure}{0}
\renewcommand{\thefigure}{S\arabic{figure}}%
\setcounter{table}{0}
\renewcommand{\thetable}{S\arabic{table}}%

\setcounter{section}{0}  
\setcounter{subsection}{0}
\setcounter{subsubsection}{0}
\makeatletter
\renewcommand{\thesection}{S\arabic{section}}
\renewcommand{\thesubsection}{\thesection.\arabic{subsection}}
\renewcommand{\thesubsubsection}{\thesubsection.\arabic{subsubsection}}
\setcounter{secnumdepth}{3}
\makeatother

\makeatletter
\renewcommand{\p@subsection}{}      
\renewcommand{\p@subsubsection}{}   
\makeatother

\sitablesofcontents
\usestocfile
\clearpage

\section{Bulk crystal structures}\label{seq:s_cryst_bulk} 

\begin{figure}[H]
\begin{center}
    \includegraphics[width=0.99\columnwidth]{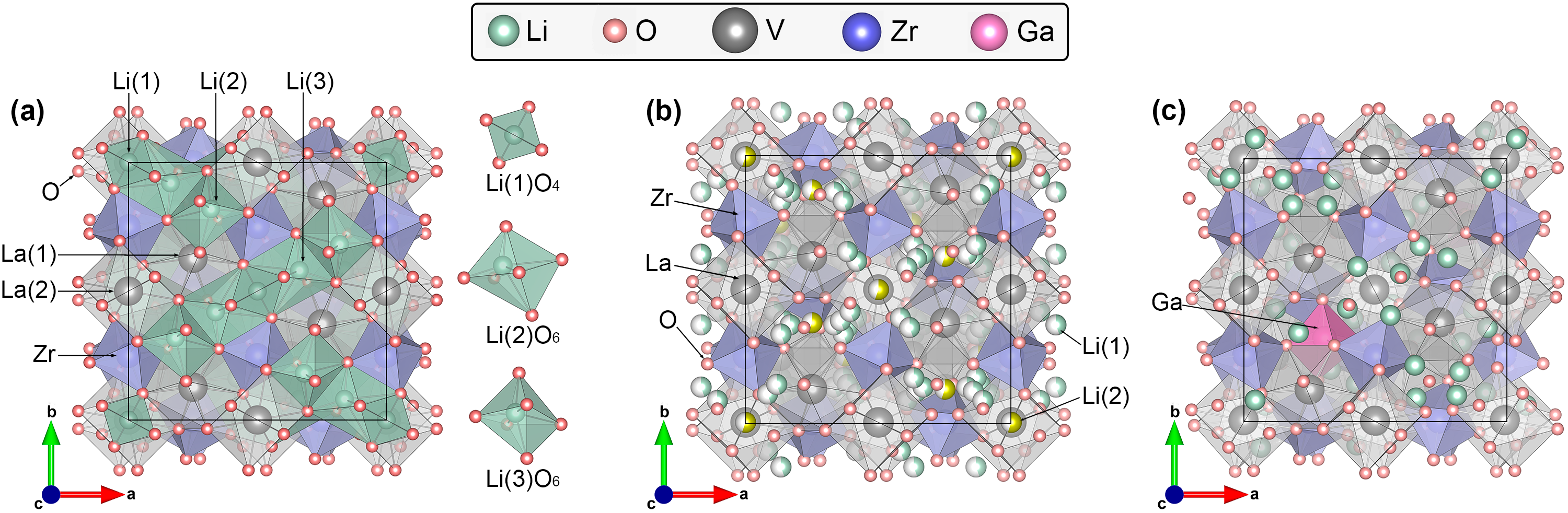}
\end{center}
\caption{Crystal structures of (a) tetragonal (sg. 142) and (b) cubic (sg. 230) $\mathrm{Li}_{7}\mathrm{La}_{3}\mathrm{Zr}_{2}\mathrm{O}_{12}$ (LLZO) phases; (c) Ga-doped LLZO with the conventional cell $\ce{Ga2Li50La24Zr16O96} $ (four vacancies per conventional cell). Separate polyhedra show different Wyckoff positions of Li in t-LLZO. The fill percentage of Li corresponds to the occupation numbers.
Ga, Li, La, Zr, and O are shown as pink, green, gray, blue, and orange spheres, respectively. }
\label{fig:crystal_structure_bulk}
\end{figure}

At room temperature, LLZO has tetragonal symmetry (\cref{fig:crystal_structure_bulk}.a,  space group $I4_1/acd$, No. 142) with $a=13.134$~\AA~and $c=12.663$~\AA~lattice parameters, which are close to each other~\cite{Awaka:2009}. The Wyckoff positions, their occupancy and notations are provided in~\cref{tb:tet_atomic_positions}. 

As the temperature increases to $\sim 900$~K~\cite{Chen2018, bernstein2012origin}, the tetragonal phase (t-LLZO) transforms into a higher symmetry cubic phase (c-LLZO) (\cref{fig:crystal_structure_bulk}.b, space group $Ia\overline{3}d$, No. 230). 
The lattice constants $a$ and $b$ merge into one with an averaged value of $12.983$~\AA~\cite{Awaka:2011}. The initial approximation of Li positions in c-LLZO was obtained from AIMD simulations, as detailed in~\cref{seq:s_model_cubic}.

To preserve cubic structure of LLZO after the annealing and at interfacial region, doping technique is widely employed~\cite{krauskopf2019toward,sharafi2017impact,wang2015phase}. In this work, we focused on Ga-doped cubic LLZO (Ga-LLZO), as we did in our previous experimental work~\cite{timusheva2025chemical}. The Ga positions were determined via AIMD simulations, as detailed in~\cref{seq:s_model_ga_doped}.

The Wyckoff positions and their occupation numbers for t-LLZO and c-LLZO are provided in~\cref{tb:tet_atomic_positions} and \cref{tb:cub_atomic_positions}, respectively. Hereinafter, we adopt the Li atomic position notation from \cref{tb:tet_atomic_positions}, where Li(1) denotes tetrahedrally coordinated Li, Li(2) and Li(3) denote octahedrally coordinated Li.

\begin{spacing}{1.0}
\begin{table}[H]
\caption{Sites, occupancy values, fractional coordinates for tetragonal $\mathrm{Li}_{7} \mathrm{La}_{3} \mathrm{Zr}_{2} \mathrm{O}_{12}$ (space group $I4_1/acd$, No. 142). The experimental data for atomic positions is taken from the work by Awaka \textit{et al.}~\cite{Awaka:2009}}

\begin{center}
\begin{tabular*}{1.0\textwidth}{c @{\extracolsep{\fill}} llllll}

\hline \hline
Atom & Site & $g$ & $x$ & $y$ & $z$ \\ \hline
$\mathrm{La}(1)$ & $8 b$ & 1 & 0 & $1 / 4$ & $1 / 8$ \\
$\mathrm{La}(2)$ & $16 e$ & 1 & $0.12716(5)$ & 0 & $1 / 4$ \\
$\mathrm{Zr}$ & $16 c$ & 1 & 0 & 0 & 0 \\
$\mathrm{Li}(1)$ & $8 a$ & 1 & 0 & $1 / 4$ & $3 / 8$ \\
$\mathrm{Li}(2)$ & $16f$ & 1 & $0.1813(13)$ & $0.4313(13)$ & $1 / 8$ \\
$\mathrm{Li}(3)$ & $32 g$ & 1 & $0.0796(12)$ & $0.0863(11)$ & $0.8099(12)$ \\
$\mathrm{O}(1)$ & $32 g$ & 1 & $-0.0335(3)$ & $0.0546(3)$ & $0.1528(3)$ \\
$\mathrm{O}(2)$ & $32 g$ & 1 & $0.0534(3)$ & $0.8525(3)$ & $0.5366(4)$ \\
$\mathrm{O}(3)$ & $32g$ & 1 & $0.1499(3)$ & $0.0273(3)$ & $0.4454(3)$ \\
\hline \hline
\end{tabular*}
\label{tb:tet_atomic_positions}
\end{center}
\end{table}
\end{spacing}

\begin{spacing}{1.0}
\begin{table}[H]
\caption{Sites, occupancy values, fractional coordinates for cubic $\mathrm{Li}_{7} \mathrm{La}_{3} \mathrm{Zr}_{2} \mathrm{O}_{12}$ (space group $Ia\overline{3}d$, No. 230). The experimental data for atomic positions is taken from the work by Awaka \textit{et al.}~\cite{Awaka:2011}}

\begin{center}
\begin{tabular*}{\textwidth}{c @{\extracolsep{\fill}} llllll}
\hline \hline
Atom &  Site & $g$ & $x$ & $y$ & $z$ \\ \hline 
$\mathrm{Li} (1) $ & $24 d$ & $0.94(7)$ & $3 / 8$ & 0 & $1 / 4$ \\
$\mathrm{Li} (2)$ & $96 h$ & $0.349$ & $0.0959(15)$ & $0.6922(14)$ & $0.5731(15)$ \\
$\mathrm{La}$ & $24 c$ & 1 & $1 / 8$ & 0 & $1 / 4$ \\
$\mathrm{Zr}$ & $16 a$ & 1 & 0 & 0 & 0 \\
$\mathrm{O}$ & $96 h$ & 1 & $-0.03163(18)$ & $0.0538(2)$ & $0.1501(2)$ \\

\hline\hline
\end{tabular*}
\label{tb:cub_atomic_positions}
\end{center}
\end{table}
\end{spacing}

\section{Modeling of cubic-LLZO}\label{seq:s_model_cubic} 

Stoichiometric, undoped cubic LLZO is unstable at room temperature~\cite{Miara:2015,bernstein2012origin}. Nevertheless, we constructed a c-LLZO simulation cell to examine its ionic conductivity.

\begin{figure}[H]
\begin{center}
\includegraphics[width=0.65\columnwidth]{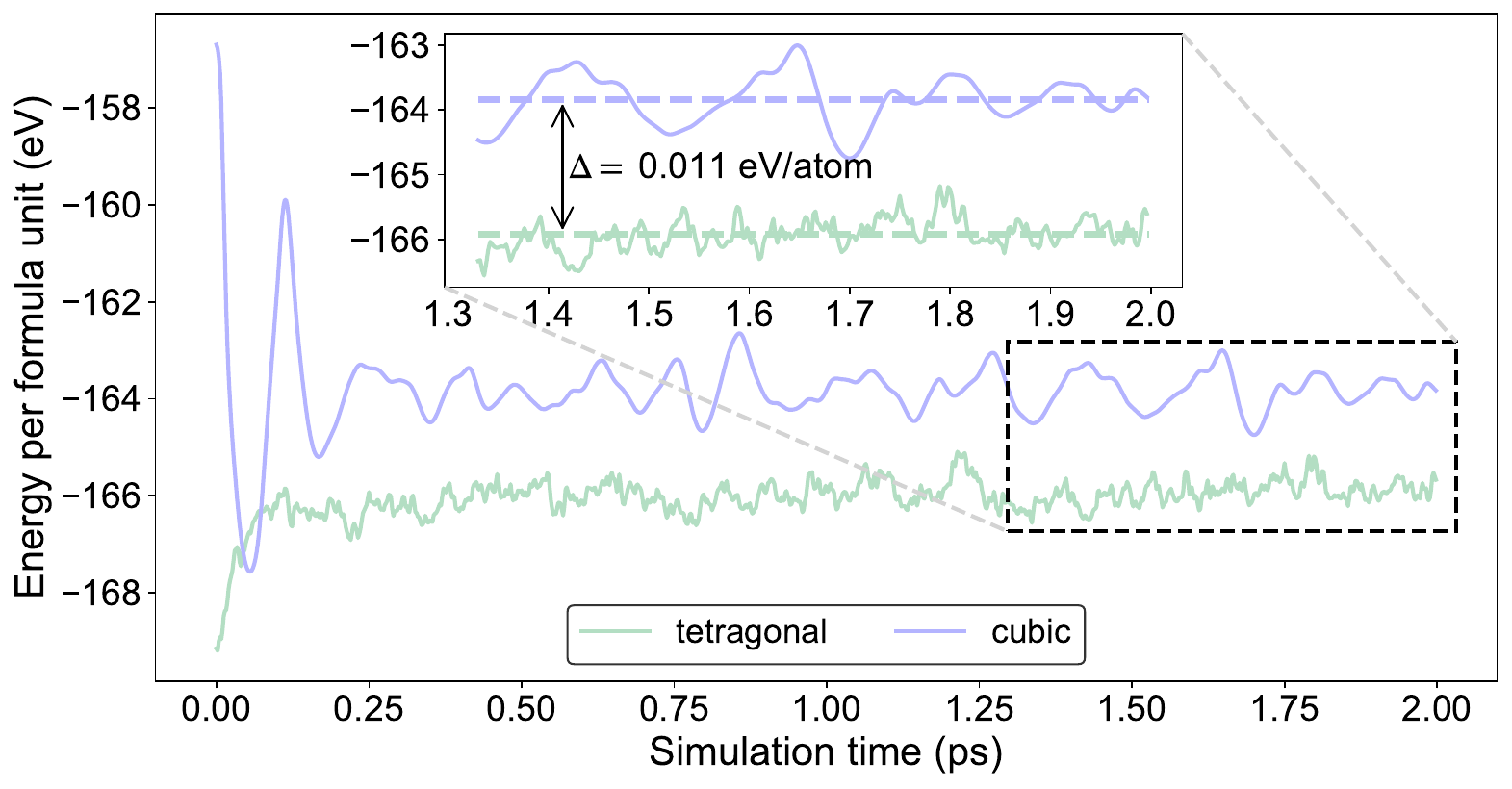}
\end{center}
\caption{AIMD trajectories for stoichiometric t-LLZO and c-LLZO were obtained at 1000~K using the NVT ensemble. The average energy difference per atom between c-LLZO and t-LLZO is 11~meV/atom. }
\label{fig:tet_cub_en_diff}
\end{figure}

We first took fully lithiated cubic LLZO from the Inorganic Crystal Structure Database (ICSD)~\cite{bergerhoff1987crystallographic_icsd} and removed the nearest atoms until the stoichiometric composition \ce{Li7La3Zr2O12} was obtained. We then performed AIMD simulated annealing at $T = 1000$~K. The results, together with a comparison to analogous simulations for t-LLZO, are shown in~\cref{fig:tet_cub_en_diff}.

At each AIMD step, stoichiometric c-LLZO has a higher energy per atom than stoichiometric t-LLZO. This indicates that the tetragonal phase remains thermodynamically favored even at high temperature when vacancies are absent. The energy difference between c-LLZO and t-LLZO is 11~meV/atom, consistent with the 4~meV/atom reported by Bernstein~\textit{et al.}~\cite{bernstein2012origin}. They also showed that the cubic phase becomes more stable than the tetragonal one only in the presence of Li vacancies.

\section{Modeling of G\lowercase{a}-doped LLZO}\label{seq:s_model_ga_doped} 

\begin{figure}[H]
\begin{center}
    \includegraphics[width=0.65\columnwidth]{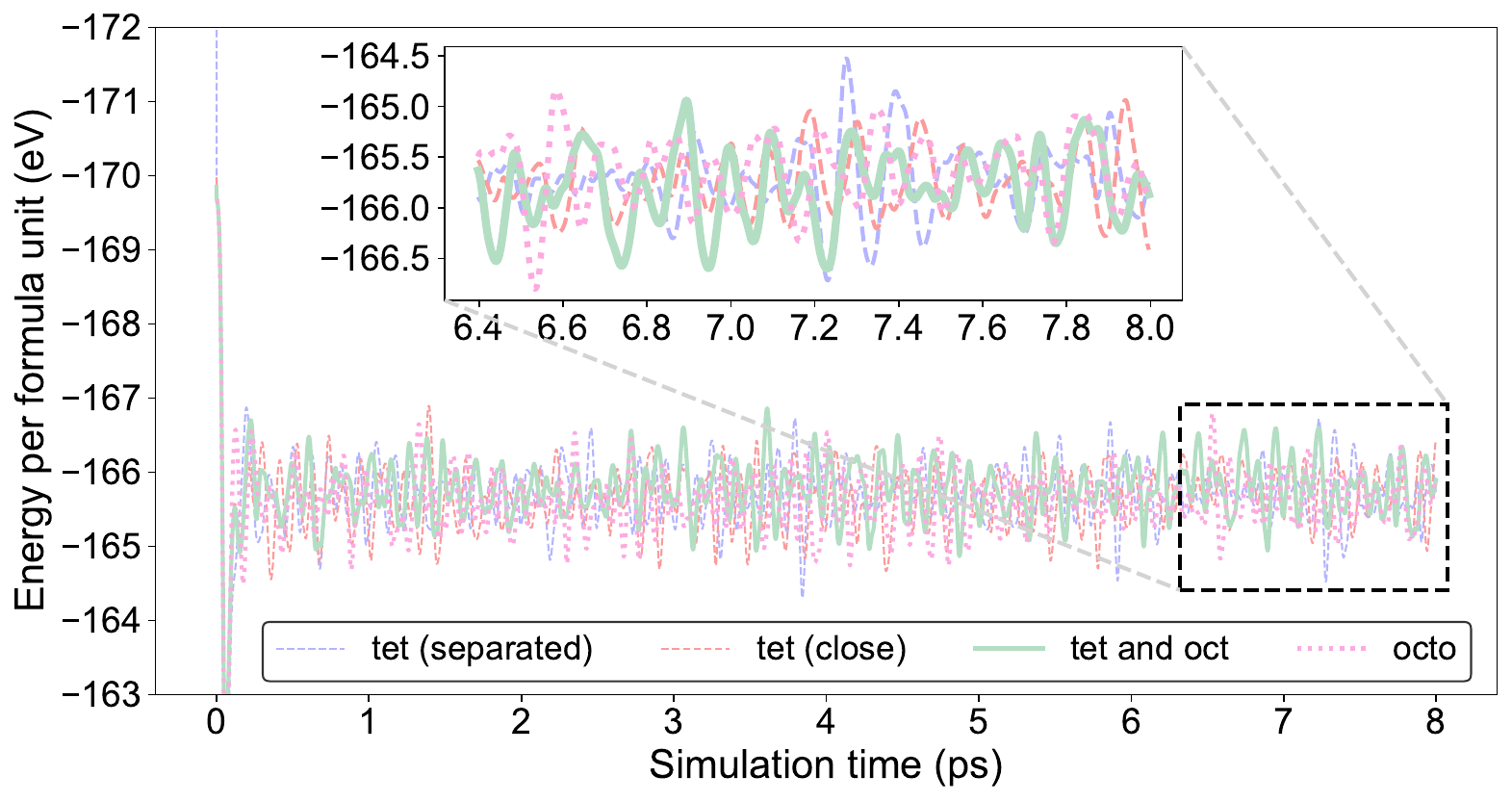}
\end{center}
\caption{Total energy of four configurations of Ga-LLZO. Blue dashed line is the configuration, where both Ga are located in tetrahedral position and close to each other (Li1, nearest neighbors); orange dashed line is the configuration, where both Ga are located in tetrahedral position and separated from each other (Li1, separated); pink dotted line is the configuration, where both Ga are located in octahedral positions (Li2); green solid line is the configuration, where two Ga are located both in tetrahedral (Li1) and octahedral positions (Li3). The inset shows the last 330 snapshots ($\sim1.5$~ps). }
\label{fig:ga_position}
\end{figure}

To identify the optimal positions of Ga in \ce{Ga2Li50La24Zr16O96} (Ga-LLZO), we modeled four combinations of Ga positions in previously modeled c-LLZO: both Ga in octahedral environment, both Ga in tetrahedral environment (where Ga has the closest and furthest possible positions within the framework), and one Ga in tetrahedral and one Ga in octahedral environments. Next, we performed AIMD simulations over 4,000 snapshots (8~ps) and calculated average total energy for each combination after the systems reached the equilibrium. The results are shown in~\cref{fig:ga_position}.

Our simulations identify the Ga-LLZO configuration with one Ga atom in an octahedral site and one in a tetrahedral site as having the lowest total energy among the structures examined. In addition, because of configurational entropy, such a configuration should be the most stable and we should not observe Ga ordering, which is consistent with the experimental data from micrographs~\cite{timusheva2025chemical}. The relaxed structure is shown in~\cref{fig:crystal_structure_bulk}.c.

\section{Modeling supercells with interfaces}\label{seq:s_model_interfaces} 

The Li(100)/t-LLZO(100) supercell was inherited from our previous work; the construction procedure is described in Section~S8 of the Supplemental Materials of Ref.~\cite{burov2024mechanism}. The resulting interfaces exhibited strong wettability and thermodynamic stability, with low adhesion and interface energies of $-0.67$~J/m$^2$ and $0.95$~J/m$^2$, respectively.

\begin{figure}[H]
\begin{center}
    \includegraphics[width=0.99\columnwidth]{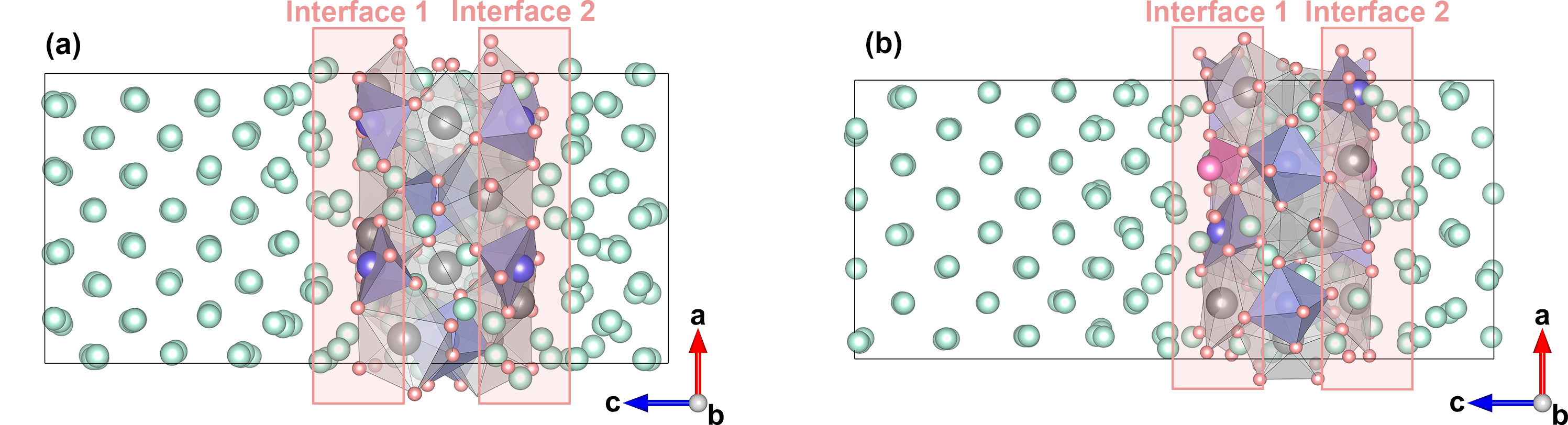}
\end{center}
\caption{The AIMD-relaxed crystal structures for Li(100)/t-LLZO(100) and Li(100)/Ga-LLZO(100) interface supercells. Ga, Li, La, Zr, and O are shown as pink, green, gray, blue, and orange spheres, respectively. }
\label{fig:crystal_structure_interface}
\end{figure}

To eliminate differences arising from surface orientation and termination, we constructed the Li(100)/Ga-LLZO(100) supercell using the same procedure, following the same sequence of steps as for t-LLZO. This yielded comparable bond lengths and local interfacial morphologies in both systems. The final Li(100)/t-LLZO(100) and Li(100)/Ga-LLZO(100) supercells are shown in~\cref{fig:crystal_structure_interface}. The stability of the Li/Ga-LLZO interface was confirmed by AIMD simulations, which showed that the interfacial region remained structurally intact with no observable morphological changes.

In the following, we refer to these interfacial supercells as Li/t-LLZO and Li/Ga-LLZO.

\section{Lattice constants}\label{seq:s_lattice_constants} 

The lattice constants of bulk phases and supercells with them are compiled in~\cref{tb:lat_const}. The lattice mismatches between LLZO slabs and Li metal slab are provided in~\cref{tb:cell_lattice_mismatch}.

\begin{spacing}{1.0}
\begin{table}[H]
\caption{Lattice constants of the bulk phases t-LLZO, c-LLZO, Ga-LLZO, and bcc Li are reported, together with data for supercells containing Li/t-LLZO and Li/Ga-LLZO interfaces. The results are compared with available calculated and experimental values, and relative deviations from literature data are provided.}

\begin{center}
\begin{tabular*}{1.0\textwidth}{c @{\extracolsep{\fill}} llllll}
\hline\hline  
 & $a$ (\AA) & $\Delta a$ (\%) & $b$ (\AA) & $\Delta b$ (\%) & $c$ (\AA) & $\Delta c$ (\%) \\ \midrule

\rowcolor[gray]{0.95}[0.425cm]
\multicolumn{7}{c}{Bulk phase: t-LLZO} \\ \midrule
Our work & 13.236 & 0  & 13.236 & 0  & 12.701  & 0   \\
Canepa, 2018 (calculations)~\cite{Canepa:2018} & 13.204 &  0.24  &  13.204  &  0.24  & 12.704 & $-0.02$   \\
Gao, 2020 (calculations)~\cite{Gao:2020} & 13.205 & 0.23   & 13.205  & 0.23 & 12.675  & 0.21   \\
Awaka, 2009 (experiment)~\cite{Awaka:2009}   & 13.134 & 0.78   & 13.134  & 0.78 &  12.663  & 0.30   \\ \midrule

\rowcolor[gray]{0.95}[0.425cm]
\multicolumn{7}{c}{Bulk phase: c-LLZO} \\ \midrule

Our work & 13.098 & 0   & 13.098  & 0   & 13.098 & 0  \\ 

Awaka, 2011 (experiment)~\cite{Awaka:2011}  & 12.983 &  0.89  & 12.983  & 0.89  & 12.983 & 0.89  \\ \midrule

\rowcolor[gray]{0.95}[0.425cm]
\multicolumn{7}{c}{Bulk phase: Ga-LLZO} \\ \midrule

Our work & 13.069 &  0  & 13.069  &  0 & 13.069  &  0 \\

Timusheva, 2025 (experiment)~\cite{timusheva2025chemical}  & 12.974 &  0.73  & 12.974  & 0.73  & 12.974 & 0.73  \\ 

Chen, 2020 (experiment)~\cite{chen2020microstructural}  & 12.977 &  0.71  & 12.977  & 0.71  & 12.977 & 0.71  \\ \midrule

\rowcolor[gray]{0.95}[0.425cm]
\multicolumn{7}{c}{Bulk phase: Li metal} \\ \midrule

Our work & 3.598 &  0  & 3.598  & 0 &  3.598  & 0   \\ 

Feder, 1970 (experiment)~\cite{feder1970equilibrium}  & 3.509    & 2.54  & 3.509   &  2.54  & 3.509 & 2.54   \\ \midrule

\rowcolor[gray]{0.95}[0.425cm]
\multicolumn{7}{c}{Supercell with interface: Li/t-LLZO} \\ \midrule

Our work & 13.236 & 0 & 12.701 & 0 & 27.285  & 0 \\ \midrule

\rowcolor[gray]{0.95}[0.425cm]
\multicolumn{7}{c}{Supercell with interface: Li/Ga-LLZO} \\ \midrule

Our work & 13.069 & 0  & 13.069 & 0  &  28.318  & 0 \\ \hline\hline
\end{tabular*}
\label{tb:lat_const}
\end{center}
\end{table}
\end{spacing}

\begin{spacing}{1.0}
\begin{table}[H]
\caption{The lattice-matched superlattices for LLZO and Li metal. $h$ and $k$ represent the lengths of vectors of created superlattice in the interface plane. $\gamma$ is the angle between two surface vectors. $\varepsilon_{x}$ and $\varepsilon_{y}$ are lattice mismatches in $x$ ($h$) and $y$ ($k$) directions.  }
    
\begin{center}
\begin{tabular*}{1.0\textwidth}{l @{\extracolsep{\fill}} ccccc}
\hline \hline
& $h$ (\AA) & $k$ (\AA) & $\gamma$  (deg) & $\varepsilon_{x}$ (\%)  & $\varepsilon_{y}$ (\%) \\ \hline \hline
\rowcolor[gray]{0.95}[0.425cm]
\multicolumn{6}{c}{Li(100)/t-LLZO(100) supercell}  \\ \hline
LLZO(100) & 13.236 & 12.701 &  90 & \multirow{ 2}{*}{3.6} & \multirow{ 2}{*}{7.9} \\
$4 \times 4$ Li(100) & 13.707 & 13.707 & 90 & & \\ \hline
\rowcolor[gray]{0.95}[0.425cm]
\multicolumn{6}{c}{Li(100)/Ga-LLZO(100) supercell}  \\ \hline
LLZO(100) & 13.069 & 13.069 &  90 & \multirow{ 2}{*}{4.9} & \multirow{ 2}{*}{4.9} \\
$4 \times 4$ Li(100) & 13.707 & 13.707 & 90 & & \\ \hline \hline
\end{tabular*}
\label{tb:cell_lattice_mismatch}
\end{center}
\end{table}
\end{spacing}

\section{Number of atoms in calculations and configurations for MTP\lowercase{s} training}\label{seq:s_number_of_cfg} 

\subsection{Number of atoms in structures} 
\begin{table}[H]
\centering
\caption{Number of atoms and vacancies ($N_{\mathrm{vac}}$) in the structures used for training and MLMD calculations, including bulk t-LLZO, c-LLZO, Ga-LLZO, bcc Li, and supercells with Li/t-LLZO and Li/Ga-LLZO interfaces.}
\label{tab:train_atom_num}
\begin{tabular}{lcccc}
\hline \hline
\multicolumn{1}{c}{Structure}  & 
Atoms (training) & $N_{\mathrm{vac}}$
(training) & Atoms (MLMD) & 
$N_{\mathrm{vac}}$
(MLMD) \\ \hline 

\rowcolor[gray]{0.95}[0.425cm]
\multicolumn{5}{c}{Bulk phases} \\ \hline

t-LLZO (without vac) & 192 & 0 & 1536   & 0        \\
t-LLZO (with vac) & 190 & 2 & 1528 & 8        \\
c-LLZO   & 192 & 0 & 1536 & 0         \\
Ga-LLZO    & 188 & 0 & 1504 & 0         \\
bcc Li     & 52 & 2 & 431 & 1         \\ \hline

\rowcolor[gray]{0.95}[0.425cm]
\multicolumn{5}{c}{Supercells with interfaces} \\ \hline

Li/t-LLZO     & 297 & 3 & 1188  & 12      \\
Li/Ga-LLZO    & 286 & 3 & 1144  & 12    \\ \hline\hline

\end{tabular}%
\end{table}

\subsection{Number of configurations in active learning stages}\label{sec:num_cfgs_al}

\begin{table}[H]
\centering
\caption{Number of configurations generated after two active learning stages (10th- and 20th- levels MTPs) and at the validation stage for t-LLZO, c-LLZO, Ga-LLZO, bcc Li, Li/t-LLZO, and Li/Ga-LLZO systems.}
\label{tab:train_cfg_num}
\begin{tabular}{lccc}
\hline \hline
\multicolumn{1}{c}{Structure}  & \begin{tabular}[c]{@{}c@{}}Active learning stage 1  \\ (MTP level 10)\end{tabular} & \begin{tabular}[c]{@{}c@{}}Active learning stage 2\\ (MTP level 20)\end{tabular} & Validation \\ \hline 
\rowcolor[gray]{0.95}[0.425cm]
\multicolumn{4}{c}{Bulk phases}
\\ \hline
t-LLZO  & 677  & 1209  & 3000        \\
c-LLZO  & 570  & 1339 & 2000         \\
Ga-LLZO    & 847  & 1595 & 4000         \\
bcc Li     & 163  & 302 & 6000         \\ \hline
\rowcolor[gray]{0.95}[0.425cm]
\multicolumn{4}{c}{Supercells with interfaces} \\ \hline
Li/t-LLZO     & 1320  & 1664  & 6000      \\
Li/Ga-LLZO    & 1246  & 1641  & 6000    \\ \hline\hline

\end{tabular}%
\end{table}

\section{Training and validation errors}\label{seq:s_train_errors} 

To evaluate the accuracy of MTPs, we calculated energies per atom (meV/atom), forces (eV/\AA), and stresses (GPa) for configurations at training and validation stages. Maximum absolute error (MaxAE), mean absolute error (MAE), and root mean square error (RMSE) are calculated.

\subsection{Bulk phases}\label{sec:errors_bulk_phases}

For the bulk phases, we trained MTPs for t-LLZO, c-LLZO, Ga-LLZO, and bcc Li. The errors of training and validation stages are provided below.

\subsubsection{t-LLZO}\label{sec:errors_bulk_tet}

\begin{table}[H]
\centering
\caption{Training and validation errors for the t-LLZO. MaxAE, MAE and RMSE errors for energies (meV/atom), forces (eV/\AA), and stresses (GPa) are provided. }
\label{tab:errors_bulk_tet}
\begin{tabular}{l|ccc}
\hline\hline
\multicolumn{1}{c|}{\bf{Train errors}} & \bf{MaxAE} & \bf{MAE} & \bf{RMSE} \\ \hline
Errors in energies (meV/atom)                                   & 2.03     & 0.40   & 0.51   \\
Errors in forces (eV/\AA)                                         & 1.28     & 0.05   & 0.07   \\
Errors in stresses (GPa)                                        & 0.43     & 0.08   & 0.10   \\ \hline
\multicolumn{1}{c|}{\bf{Validation errors}} & \bf{MaxAE} & \bf{MAE} & \bf{RMSE} \\ \hline
Errors in energies (meV/atom)                                   & 2.04     & 0.36   & 0.48   \\
Errors in forces (eV/\AA)                                         & 1.28     & 0.05   & 0.07   \\
Errors in stresses (GPa)                                        & 0.43     & 0.08   & 0.11   \\ \hline\hline
\end{tabular}%
\end{table}

\begin{figure}[H]
\begin{center}
    \includegraphics[width=0.92\columnwidth]{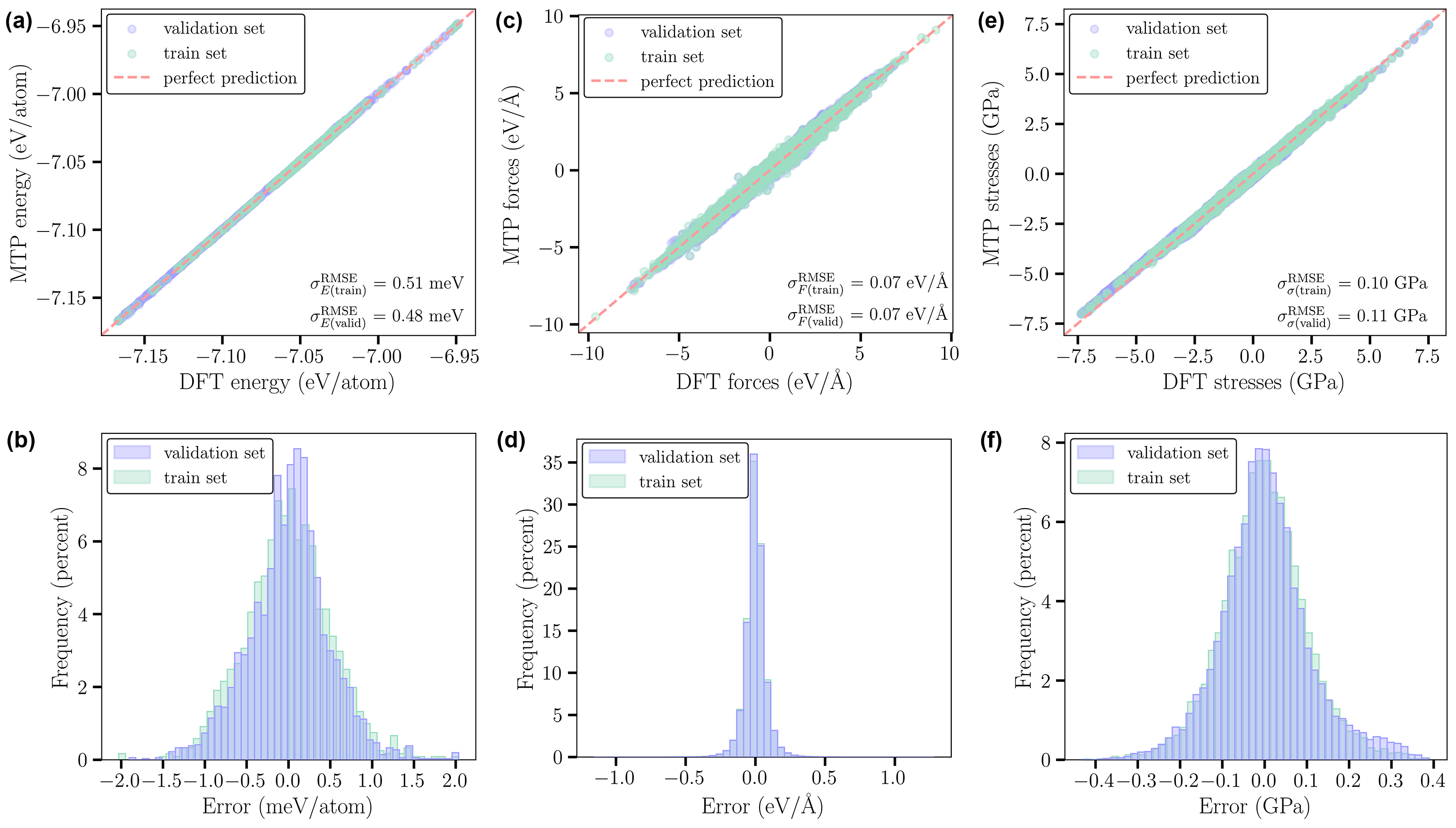}
\end{center}
\caption{Calculated training and validation errors with their distribution, obtained with the DFT method and predicted with MTP for t-LLZO: (a), (b) energies (meV/atom); (c), (d) forces, (eV/\AA); (e), (f) stresses (GPa).  }
\end{figure}

\subsubsection{c-LLZO}

\begin{table}[H]
\centering
\caption{Training and validation errors for the c-LLZO. MaxAE, MAE and RMSE errors for energies (meV/atom), forces (eV/\AA), and stresses (GPa) are provided. }
\label{tab:errors_bulk_cub}
\begin{tabular}{l|ccc}
\hline\hline
\multicolumn{1}{c|}{\bf{Train errors}} & \bf{MaxAE} & \bf{MAE} & \bf{RMSE} \\ \hline
Errors in energies (meV/atom)                                   & 30.73     & 3.08   & 5.39  \\
Errors in forces (eV/\AA)                                         & 0.47     & 0.03   & 0.05   \\
Errors in stresses (GPa)                                        & 1.12     & 0.10   & 0.14   \\ \hline
\multicolumn{1}{c|}{\bf{Validation errors}} & \bf{MaxAE} & \bf{MAE} & \bf{RMSE} \\ \hline
Errors in energies (meV/atom)                                   & 2.81     & 0.53   & 0.66   \\
Errors in forces (eV/\AA)                                         & 0.49     & 0.04   & 0.05   \\
Errors in stresses (GPa)                                        & 0.91     & 0.09   & 0.13   \\ \hline\hline
\end{tabular}%
\end{table}

\begin{figure}[H]
\begin{center}
    \includegraphics[width=0.99\columnwidth]{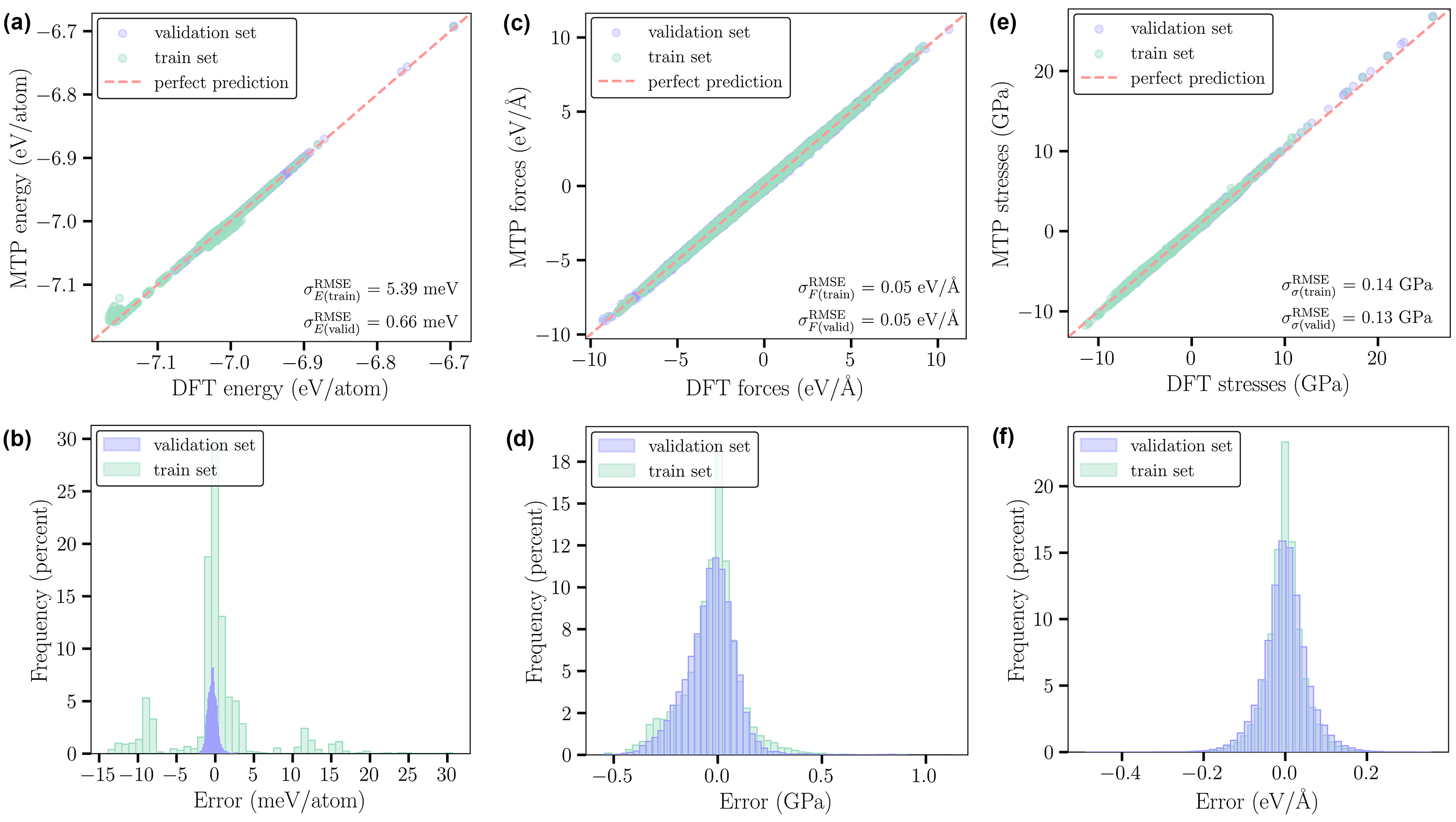}
\end{center}
\caption{Calculated training and validation errors with their distribution, obtained with the DFT method and predicted with MTP for c-LLZO: (a), (b) energies (meV/atom); (c), (d) forces, (eV/\AA); (e), (f) stresses (GPa).   }
\end{figure}

\subsubsection{Ga-LLZO}

\begin{table}[H]
\centering
\caption{Training and validation errors for the Ga-LLZO. MaxAE, MAE and RMSE errors for energies (meV/atom), forces (eV/\AA), and stresses (GPa) are provided. }
\label{tab:errors_bulk_ga_cub}
\begin{tabular}{l|ccc}
\hline\hline
\multicolumn{1}{c|}{\bf{Train errors}} & \bf{MaxAE} & \bf{MAE} & \bf{RMSE} \\ \hline
Errors in energies (meV/atom)                                   & 10.50     & 0.83   & 1.28   \\
Errors in forces (eV/\AA)                                         & 0.58     & 0.04   & 0.06   \\
Errors in stresses (GPa)                                        & 0.65     & 0.09   & 0.12   \\ \hline
\multicolumn{1}{c|}{\bf{Validation errors}} & \bf{MaxAE} & \bf{MAE} & \bf{RMSE} \\ \hline
Errors in energies (meV/atom)                                   & 2.34     & 0.51   & 0.65   \\
Errors in forces (eV/\AA)                                         & 0.60     & 0.05   & 0.06   \\
Errors in stresses (GPa)                                        & 0.52     & 0.09   & 0.12   \\ \hline\hline
\end{tabular}%
\end{table}

\begin{figure}[H]
\begin{center}
    \includegraphics[width=0.99\columnwidth]{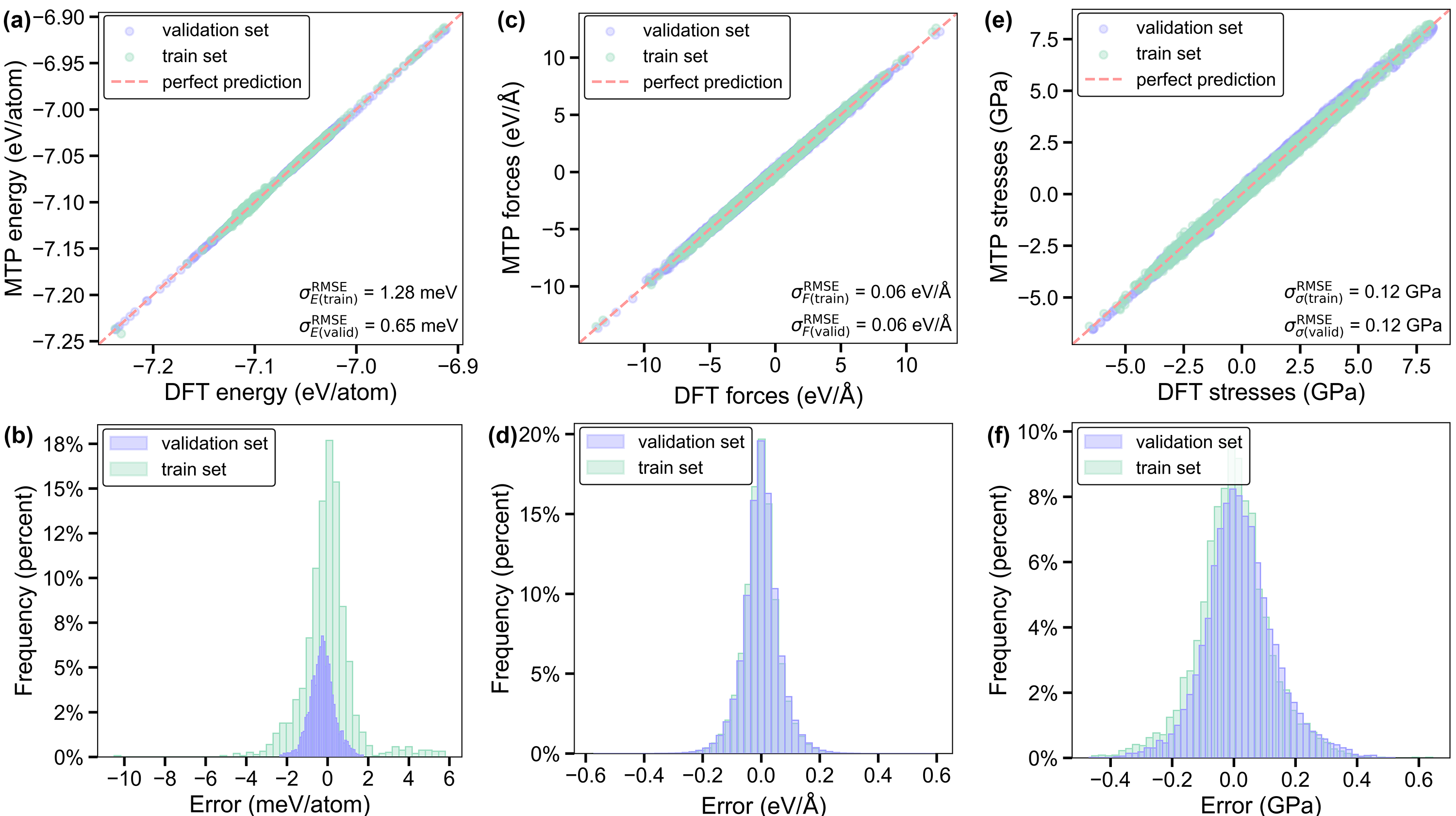}
\end{center}
\caption{Calculated training and validation errors with their distribution, obtained with the DFT method and predicted with MTP for Ga-LLZO: (a), (b) energies (meV/atom); (c), (d) forces, (eV/\AA); (e), (f) stresses (GPa).  }
\label{fig:errors_bulk_ga_cub}
\end{figure}

\subsubsection{Body-centered cubic Li}

\begin{table}[H]
\centering
\caption{Training and validation errors for the bcc Li. MaxAE, MAE and RMSE errors for energies (meV/atom), forces (eV/\AA), and stresses (GPa) are provided. }
\label{tab:errors_bulk_li_metal}
\begin{tabular}{l|ccc}
\hline\hline
\multicolumn{1}{c|}{\bf{Train errors}} & \bf{MaxAE} & \bf{MAE} & \bf{RMSE} \\ \hline
Errors in energies (meV/atom)                                   & 2.59     & 0.47   & 0.65   \\
Errors in forces (eV/\AA)                                         & 0.05     & 0.01   & 0.01   \\
Errors in stresses (GPa)                                        & 0.12     & 0.02   & 0.03   \\ \hline
\multicolumn{1}{c|}{\bf{Validation errors}} & \bf{MaxAE} & \bf{MAE} & \bf{RMSE} \\ \hline
Errors in energies (meV/atom)                                   & 2.91     & 0.37   & 0.53   \\
Errors in forces (eV/\AA)                                         & 0.04     & 0.01   & 0.01   \\
Errors in stresses (GPa)                                        & 0.12     & 0.02   & 0.03  \\ \hline\hline
\end{tabular}%
\end{table}

\begin{figure}[H]
\begin{center}
    \includegraphics[width=0.99\columnwidth]{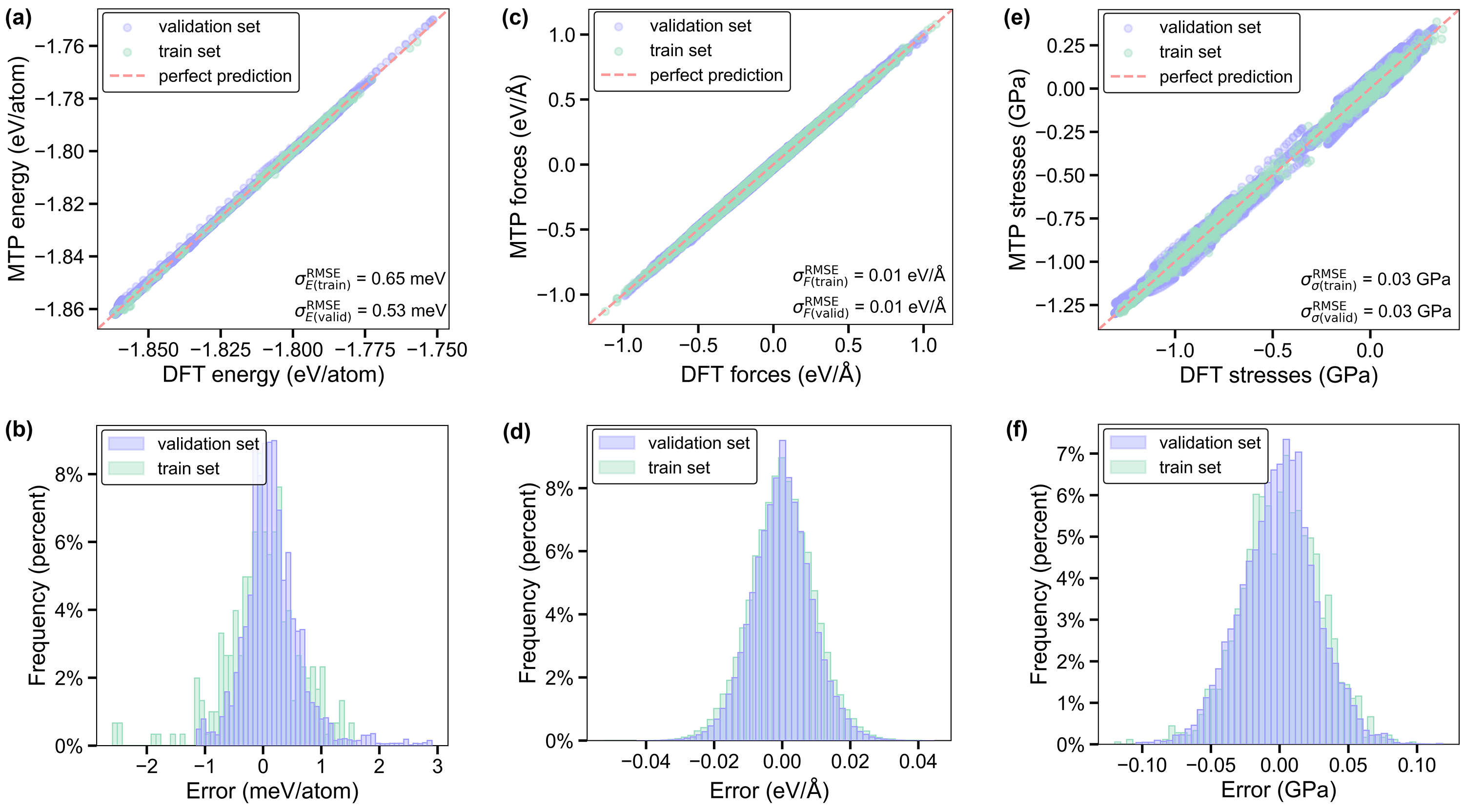}
\end{center}
\caption{Calculated training and validation errors with their distribution, obtained with the DFT method and predicted with MTP for bcc Li: (a), (b) energies (meV/atom); (c), (d) forces, (eV/\AA); (e), (f) stresses (GPa).  }
\end{figure}

\clearpage
\subsection{Supercells with interfaces}

For the supercells with interfaces, we trained MTPs for Li/t-LLZO and Li/Ga-LLZO. The errors of training and validation are provided below.

\subsubsection{Supercell with Li/t-LLZO interface}

\begin{table}[H]
\centering
\caption{Training and validation errors for the Li/t-LLZO. MaxAE, MAE and RMSE errors for energies (meV/atom), forces (eV/\AA), and stresses (GPa) are provided. }
\label{tab:errors_int_tet}
\begin{tabular}{l|ccc}
\hline\hline
\multicolumn{1}{c|}{\bf{Train errors}} & \bf{MaxAE} & \bf{MAE} & \bf{RMSE} \\ \hline
Errors in energies (meV/atom)                                   & 6.93     & 0.75   & 1.03   \\
Errors in forces (eV/\AA)                                         & 0.83     & 0.04   & 0.06   \\
Errors in stresses (GPa)                                        & 3.44     & 0.55   & 0.79   \\ \hline
\multicolumn{1}{c|}{\bf{Validation errors}} & \bf{MaxAE} & \bf{MAE} & \bf{RMSE} \\ \hline
Errors in energies (meV/atom)                                   & 6.03     & 0.98   & 1.23   \\
Errors in forces (eV/\AA)                                         & 0.70     & 0.04   & 0.06   \\
Errors in stresses (GPa)                                        & 2.34     & 0.62   & 0.88   \\ \hline\hline
\end{tabular}%
\end{table}

\begin{figure}[H]
\begin{center}
    \includegraphics[width=0.99\columnwidth]{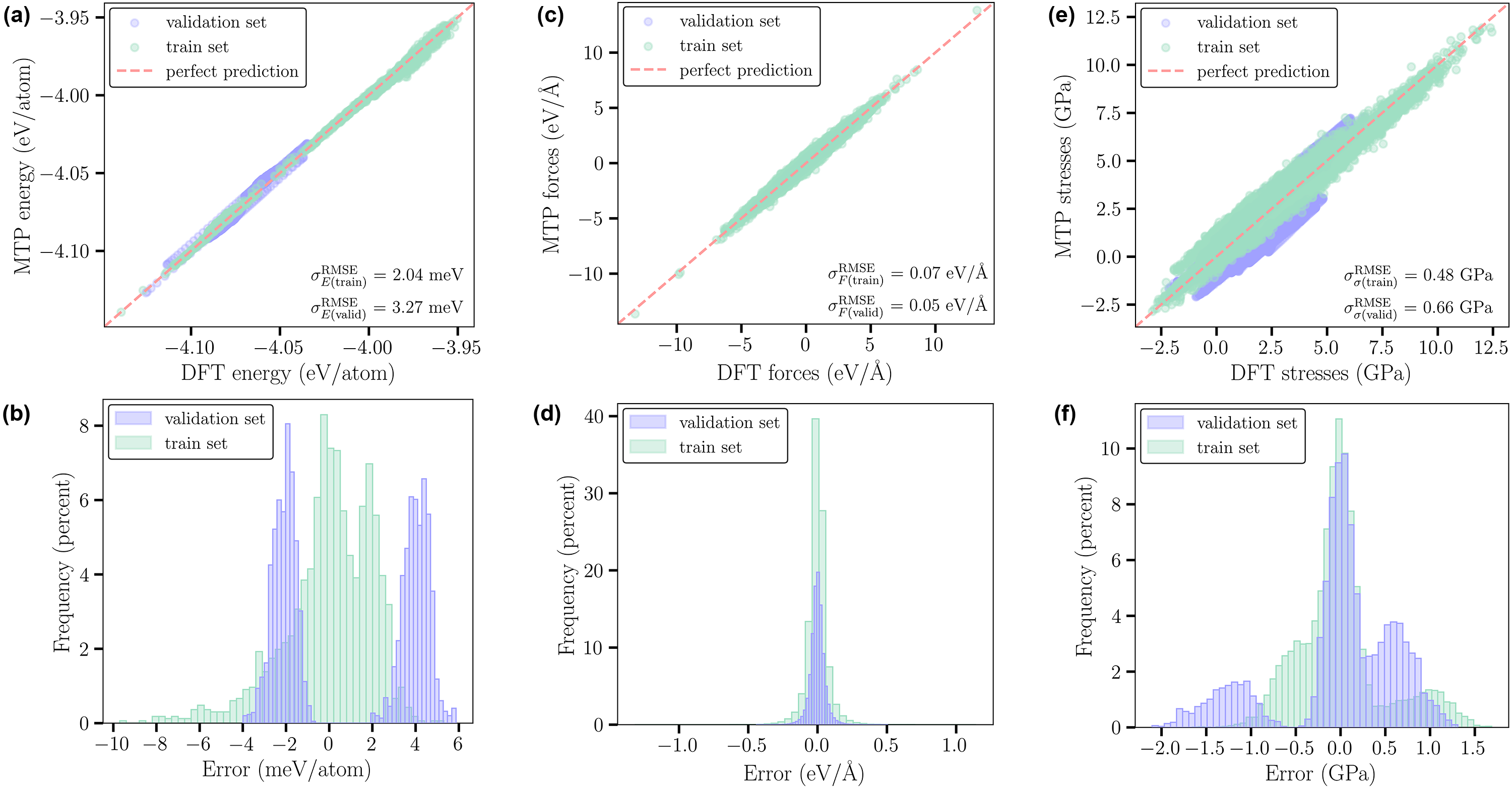}
\end{center}
\caption{Calculated training and validation errors with their distribution, obtained with the DFT method and predicted with MTP for the Li/t-LLZO: (a), (b) energies (meV/atom); (c), (d) forces, (eV/\AA); (e), (f) stresses (GPa).  }
\label{fig:errors_int_tet}
\end{figure}

\subsubsection{Supercell with Li/Ga-LLZO interface}

\begin{table}[H]
\centering
\caption{Training and validation errors for the Li/Ga-LLZO. MaxAE, MAE and RMSE errors for energies (meV/atom), forces (eV/\AA), and stresses (GPa) are provided. }
\label{tab:errors_int_cub}
\begin{tabular}{l|ccc}
\hline\hline
\multicolumn{1}{c|}{\bf{Train errors}} & \bf{MaxAE} & \bf{MAE} & \bf{RMSE} \\ \hline
Errors in energies (meV/atom)                                   & 9.69     & 1.54   & 2.04   \\
Errors in forces (eV/\AA)                                         & 1.32      & 0.05   & 0.07   \\
Errors in stresses (GPa)                                        & 1.69     & 0.34   & 0.48   \\ \hline
\multicolumn{1}{c|}{\bf{Validation errors}} & \bf{MaxAE} & \bf{MAE} & \bf{RMSE} \\ \hline
Errors in energies (meV/atom)                                   & 5.88     & 3.07   & 3.27   \\
Errors in forces (eV/\AA)                                         & 0.51     & 0.04   & 0.05   \\
Errors in stresses (GPa)                                        & 2.10     & 0.47   & 0.66   \\ \hline\hline
\end{tabular}%
\end{table}

\begin{figure}[H]
\begin{center}
    \includegraphics[width=0.99\columnwidth]{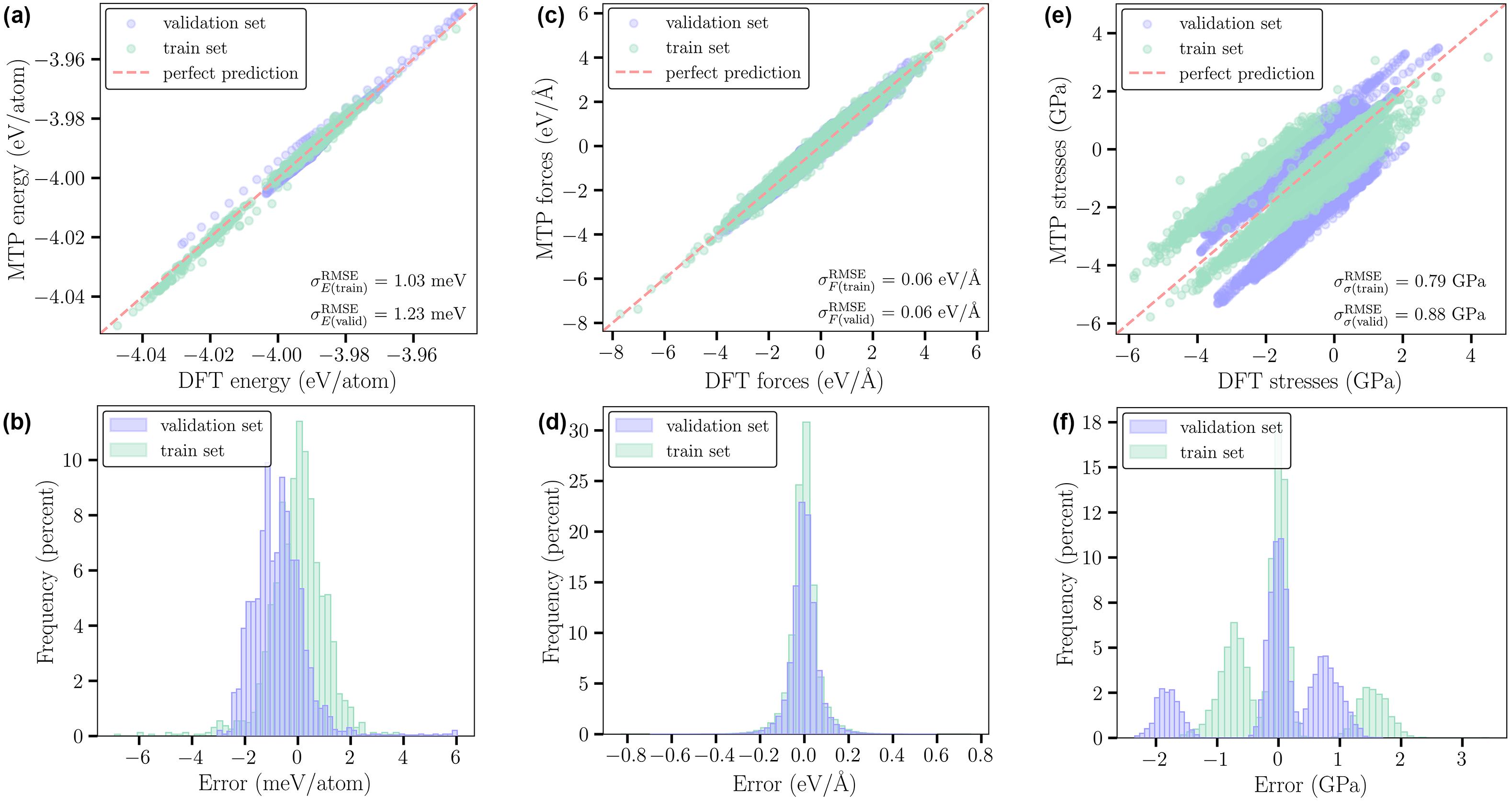}
\end{center}
\caption{Calculated training and validation errors with their distribution, obtained with the DFT method and predicted with MTP for the Li/Ga-LLZO: (a), (b) energies (meV/atom); (c), (d) forces, (eV/\AA); (e), (f) stresses (GPa).  }
\label{fig:errors_int_cub}
\end{figure}

\section{Constructing a unified MTP for L\lowercase{i}--LLZO structures }\label{seq:unified_mtps}

\begin{figure}[H]
\begin{center}
    \includegraphics[width=0.99\columnwidth]{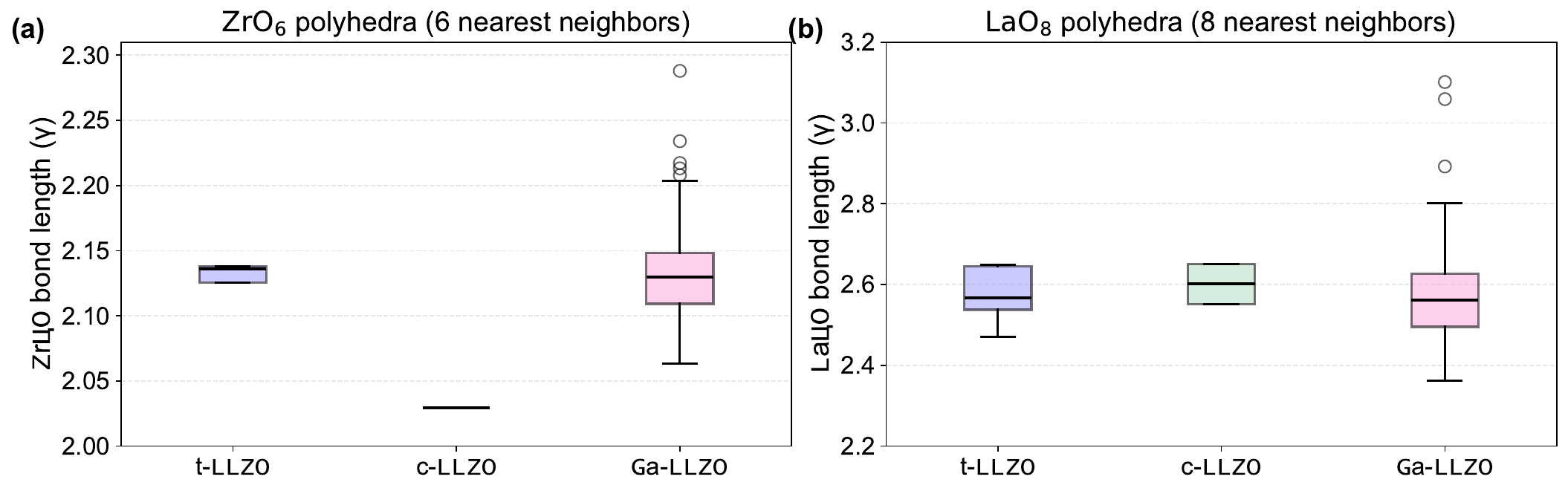}
\end{center}
\caption{Bond length distributions in LLZO systems (t-LLZO, c-LLZO, and Ga-LLZO): 
(a) Zr--O bond lengths in $\ce{ZrO6}$ octahedra; (b) La--O bond lengths 
in $\ce{LaO8}$ polyhedra.}
\label{fig:bond_distances_bulk_la_zr}
\end{figure}

\begin{figure}[H]
\begin{center}
    \includegraphics[width=0.53\columnwidth]{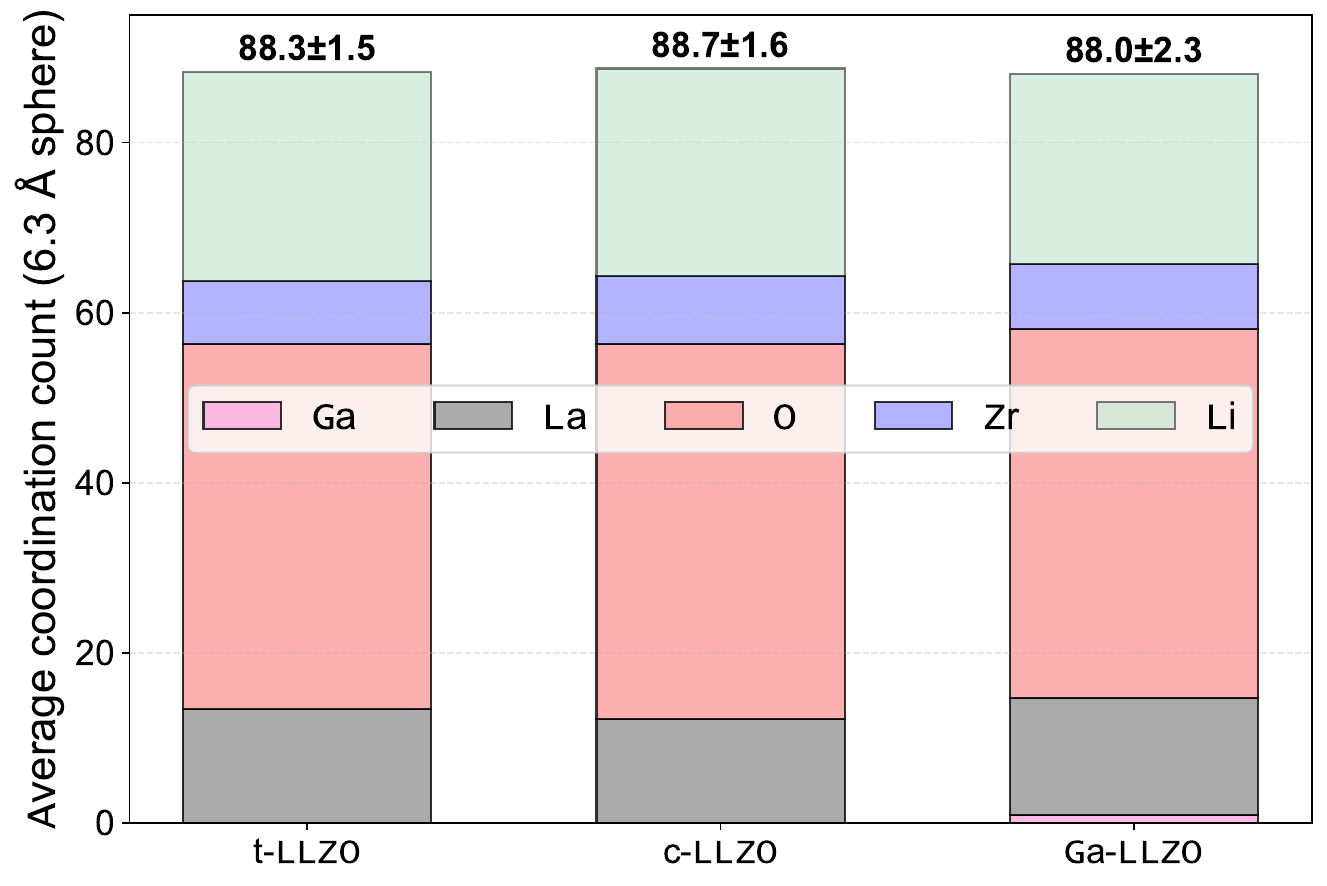}
\end{center}
\caption{Average number of atoms within a 6.3~\AA\ cut-off sphere around Li atoms in t-LLZO, c-LLZO, and Ga-LLZO. The contributions of Ga, Li, La, Zr, and O are shown in pink,  green, gray, blue, and orange, respectively. The values above each bar indicate 
the mean total number of atoms and the corresponding standard deviation.  }
\label{fig:coordination_stacked}
\end{figure}

Given that Ga-LLZO and c-LLZO have similar lattice constants (see \cref{tb:lat_const}) and t-LLZO exhibits only a slight tetragonal distortion, we explored the possibility of constructing a single unified MTP for the entire Li--LLZO family. The local atomic environment—the central concept of MTP and other MLIPs—is defined within a sphere of a chosen cut-off radius. All three LLZO systems share similar La--O and Zr--O distances (see \cref{fig:bond_distances_bulk_la_zr} and \cref{tab:bond_statistics}) and comparable chemical compositions within the 6.3~\AA\ cut-off sphere (see \cref{fig:coordination_stacked} and \cref{tab:coordination_sphere}). Despite minor differences in lattice constants (less than $0.3$~\AA), we anticipated that a unified potential could be effectively applied across all Li--LLZO systems, accurately predicting energies, forces, and stresses (EFS).

\begin{table}[H]
\centering
\caption{Bond length statistics for Zr--O and La--O bonds in the three 
LLZO phases investigated (t-LLZO, c-LLZO, and Ga-LLZO). For 
each structure, we report the minimum, maximum, mean, and 
standard deviation of the bond distances (units of \AA).}
\label{tab:bond_statistics}
\begin{tabular}{llccccc}
\hline\hline
Bond & Structure & Count & Min & Mean & Max & Std  \\
\hline
Zr--O & t-LLZO & 96 & 2.126 & 2.133 & 2.138 & 0.006 \\
Zr--O & c-LLZO & 96 & 2.030 & 2.030 & 2.030 & 0.000 \\
Zr--O & Ga-LLZO & 96 & 2.063 & 2.134 & 2.288 & 0.038 \\
\hline
La--O & t-LLZO & 192 & 2.470 & 2.572 & 2.649 & 0.063 \\
La--O & c-LLZO & 192 & 2.552 & 2.601 & 2.651 & 0.050 \\
La--O & Ga-LLZO & 192 & 2.362 & 2.571 & 3.101 & 0.106 \\
\hline\hline
\end{tabular}
\end{table}

\begin{table}[H]
\centering
\caption{Average number of atoms within a 6.3~\AA\ cut-off sphere around Li atoms in t-LLZO, c-LLZO, and Ga-LLZO. The mean total number of atoms and the corresponding standard deviation are provided. }
\label{tab:coordination_sphere}
\begin{tabular}{lcccccc}
\hline\hline
Structure & Li & La & Zr & O & Ga & Total \\
\hline
t-LLZO  & $24.57 \pm 0.90$ & $13.43 \pm 0.49$ & $7.43 \pm 0.90$ & $42.86 \pm 0.64$ & --- & $88.29$ \\
c-LLZO  & $24.43 \pm 1.45$ & $12.29 \pm 0.70$ & $8.00 \pm 0.00$ & $44.00 \pm 0.00$ & --- & $88.71$ \\
Ga-LLZO & $22.28 \pm 1.58$ & $13.66 \pm 0.51$ & $7.62 \pm 0.49$ & $43.48 \pm 1.50$ & $1.00 \pm 0.53$ & $88.04$ \\
\hline\hline
\end{tabular}
\end{table}

\subsection{Performance of t-LLZO with c-LLZO-trained MTP}\label{sec:t_llzo_with_c_mtp}

\begin{figure}[H]
\begin{center}
    \includegraphics[width=0.99\columnwidth]{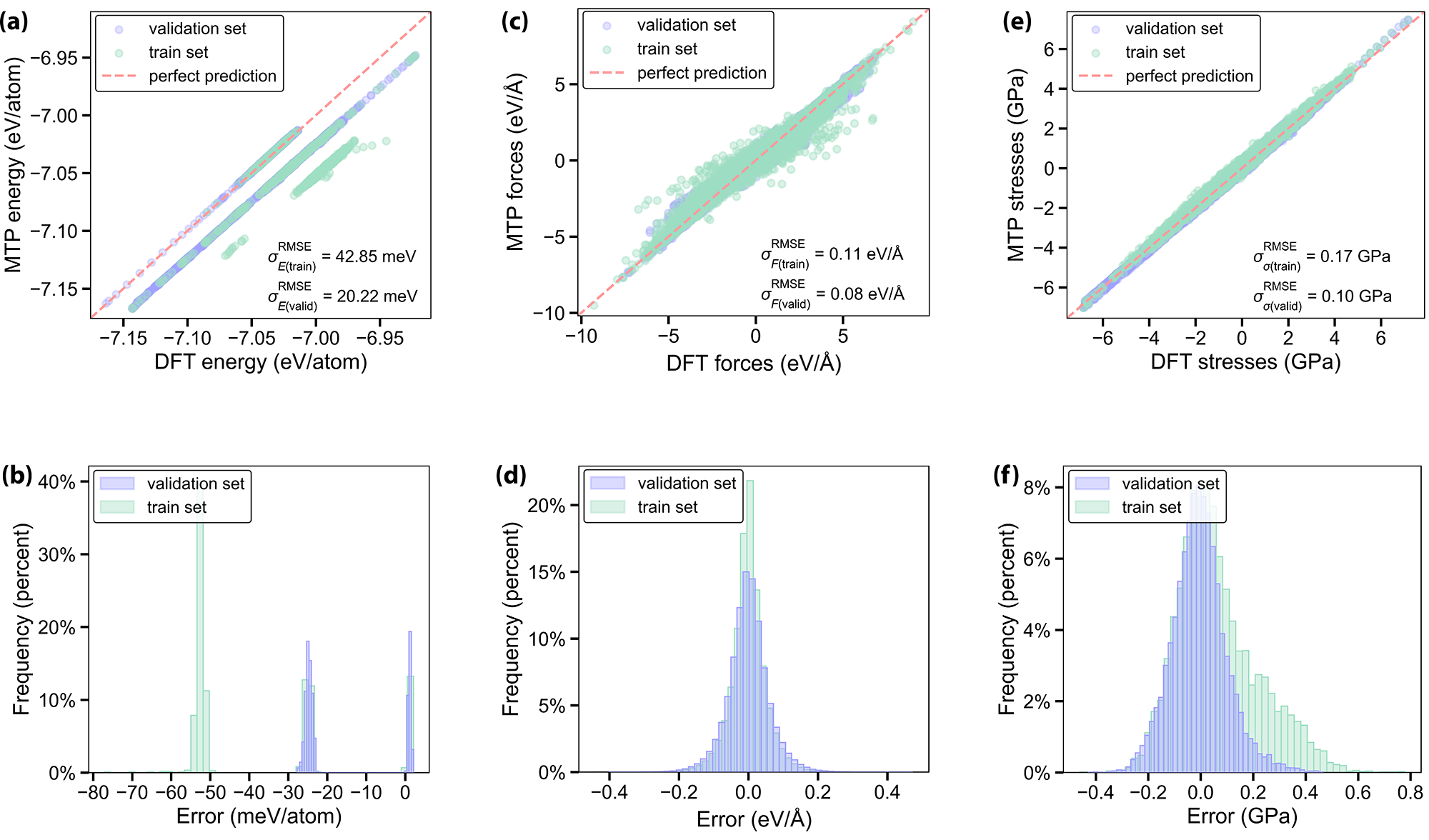}
\end{center}
\caption{Assessment of the MTP transferability from c-LLZO to t-LLZO structures, with the model trained on c-LLZO configurations and validated on t-LLZO data. Calculated training and validation errors with their distribution, obtained with the DFT method and predicted with MTP for t-LLZO: (a), (b) energies (meV/atom); (c), (d) forces, (eV/\AA); (e), (f) stresses (GPa).  }
\label{fig:errors_unified_t_with_c_pot}
\end{figure}

We first evaluated the transferability of the c-LLZO-trained MTP to t-LLZO configurations. The resulting energy, force, and stress (EFS) predictions are visualized in~\cref{fig:errors_unified_t_with_c_pot} and summarized in~\cref{tab:unified_errors_t_with_cub}. Compared to the native t-LLZO MTP validation performance, the RMSE for energies increased from 0.5 to 20.2~meV/atom, while the force RMSE rose marginally from 0.07 to 0.08~eV/\AA. Stress predictions remained largely unaffected, with only a minor variation from 0.11 to 0.10~GPa (see \cref{tab:errors_bulk_tet}).

\begin{table}[H]
\centering
\caption{Assessment of the MTP transferability from c-LLZO to t-LLZO structures, with the model trained on c-LLZO configurations and validated on t-LLZO data. Training and validation errors for the t-LLZO. MaxAE, MAE and RMSE errors for energies (meV/atom), forces (eV/\AA), and stresses (GPa) are provided.}
\label{tab:unified_errors_t_with_cub}
\begin{tabular}{l|ccc}
\hline\hline
\multicolumn{1}{c|}{\bf{Train errors}} & \bf{MaxAE} & \bf{MAE} & \bf{RMSE} \\ \hline
Errors in energies (meV/atom)                                   & 77.18    & 38.37  & 42.85  \\
Errors in forces (eV/\AA)                                       & 4.06     & 0.07   & 0.11   \\
Errors in stresses (GPa)                                        & 0.78     & 0.12   & 0.17   \\ \hline
\multicolumn{1}{c|}{\bf{Validation errors}} & \bf{MaxAE} & \bf{MAE} & \bf{RMSE} \\ \hline
Errors in energies (meV/atom)                                   & 27.66    & 16.87  & 20.22  \\
Errors in forces (eV/\AA)                                       & 1.81     & 0.05   & 0.08   \\
Errors in stresses (GPa)                                        & 0.47     & 0.08   & 0.10   \\ \hline\hline
\end{tabular}%
\end{table}

\subsection{Performance of c-LLZO with Ga-LLZO-trained MTP}\label{sec:c_llzo_with_ga_mtp}

\begin{figure}[H]
\begin{center}
    \includegraphics[width=0.99\columnwidth]{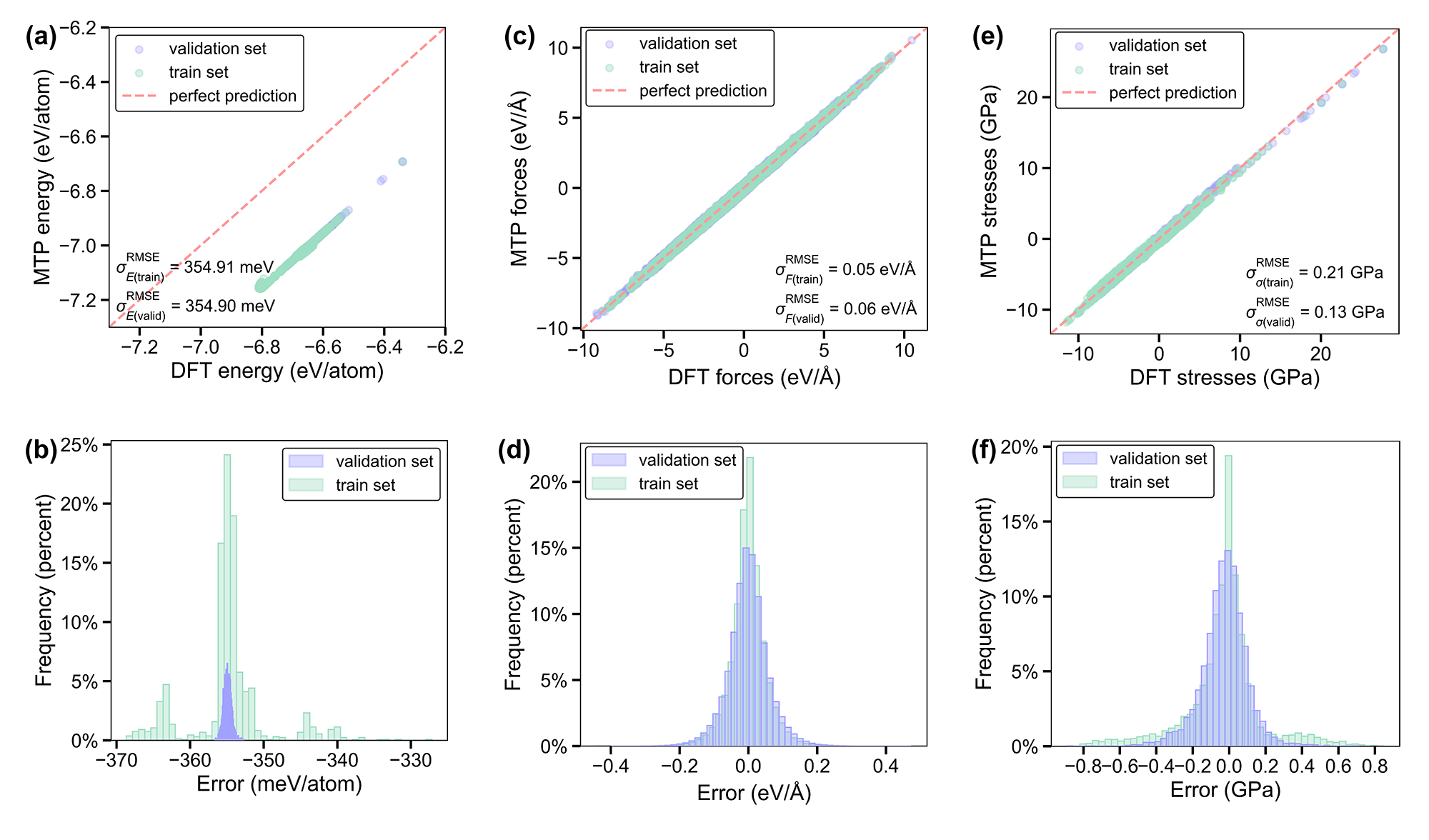}
\end{center}
\caption{Assessment of the MTP transferability from Ga-LLZO to c-LLZO structures, with the model trained on Ga-LLZO configurations and validated on c-LLZO data. Calculated training and validation errors with their distribution, obtained with the DFT method and predicted with MTP for c-LLZO: (a), (b) energies (meV/atom); (c), (d) forces, (eV/\AA); (e), (f) stresses (GPa).  }
\label{fig:errors_unified_c_with_ga_pot}
\end{figure}

Second, we evaluated the transferability of the Ga-LLZO-trained MTP to c-LLZO configurations. The EFS predictions are shown in~\cref{fig:errors_unified_c_with_ga_pot}, with corresponding errors compiled in~\cref{tab:unified_errors_c_with_ga}. Compared to the native c-LLZO MTP validation performance, the energy RMSE increased substantially from 0.7 to 354.9~meV/atom, while the force RMSE rose slightly from 0.05 to 0.06~eV/\AA\ and the stress RMSE remained unchanged at 0.13~GPa (see~\cref{tab:errors_bulk_cub}).

\begin{table}[H]
\centering
\caption{Assessment of the MTP transferability from Ga-LLZO to c-LLZO structures, with the model trained on Ga-LLZO configurations and validated on c-LLZO data. Training and validation errors for the c-LLZO. MaxAE, MAE and RMSE errors for energies (meV/atom), forces (eV/\AA), and stresses (GPa) are provided.}
\label{tab:unified_errors_c_with_ga}
\begin{tabular}{l|ccc}
\hline\hline
\multicolumn{1}{c|}{\bf{Train errors}} & \bf{MaxAE} & \bf{MAE} & \bf{RMSE} \\ \hline
Errors in energies (meV/atom)                                   & 368.65   & 354.87  & 354.91  \\
Errors in forces (eV/\AA)                                       & 0.47     & 0.04    & 0.05    \\
Errors in stresses (GPa)                                        & 0.89     & 0.13    & 0.21    \\ \hline
\multicolumn{1}{c|}{\bf{Validation errors}} & \bf{MaxAE} & \bf{MAE} & \bf{RMSE} \\ \hline
Errors in energies (meV/atom)                                   & 356.62   & 354.90  & 354.90  \\
Errors in forces (eV/\AA)                                       & 0.47     & 0.04    & 0.06    \\
Errors in stresses (GPa)    & 0.89     & 0.09    & 0.13 \\ \hline\hline
\end{tabular}%
\end{table}

\subsection{Performance of Li/t-LLZO with Li/Ga-LLZO-trained MTP}\label{sec:int_t_llzo_with_int_c_mtp}

\begin{figure}[H]
\begin{center}
    \includegraphics[width=0.99\columnwidth]{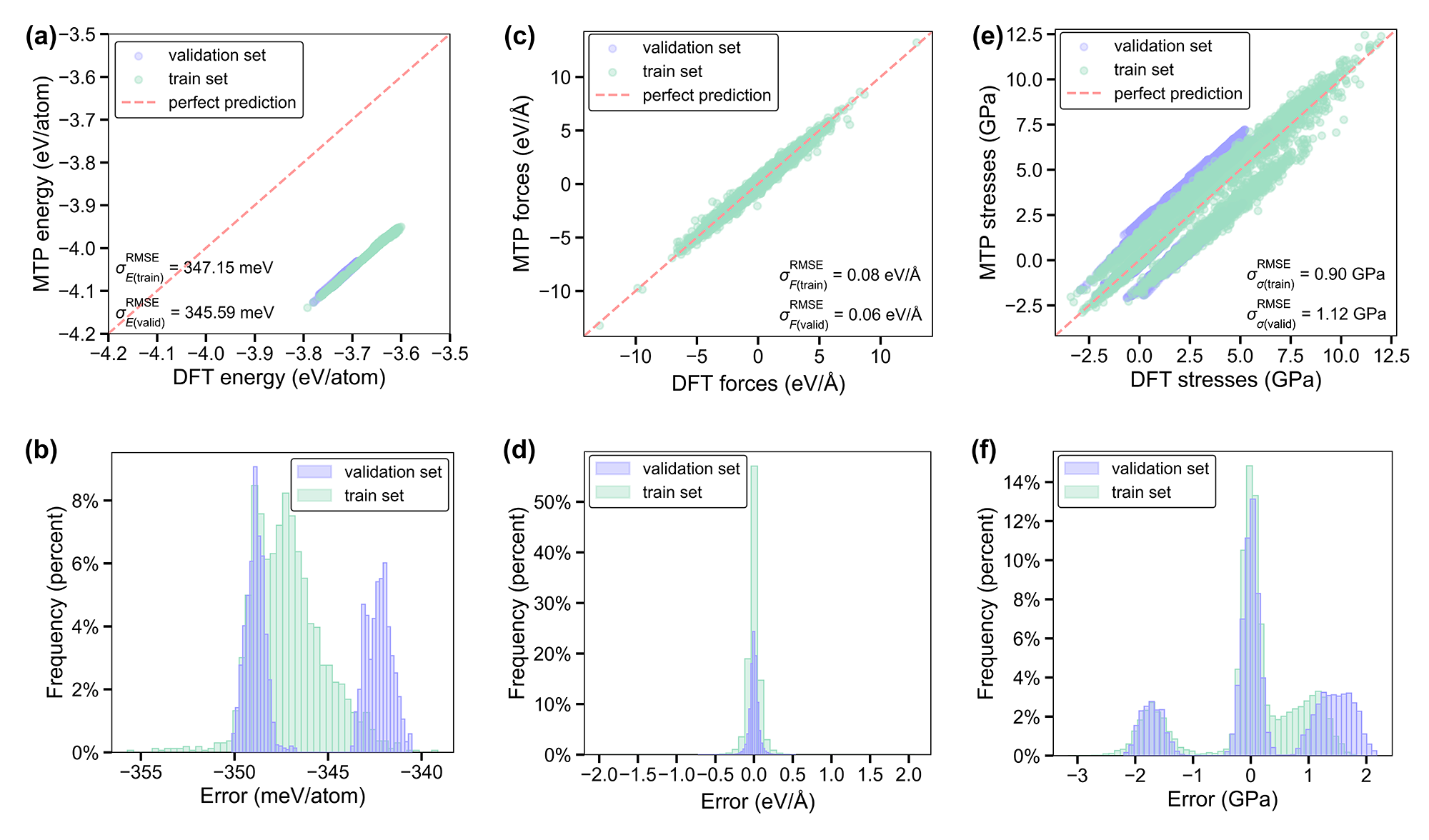}
\end{center}
\caption{Assessment of the MTP transferability from Li/Ga-LLZO to Li/t-LLZO structures, with the model trained on Li/Ga-LLZO configurations and validated on Li/t-LLZO data. Calculated training and validation errors with their distribution, obtained with the DFT method and predicted with MTP for Li/t-LLZO: (a), (b) energies (meV/atom); (c), (d) forces, (eV/\AA); (e), (f) stresses (GPa).  }
\label{fig:errors_unified_int_t_with_int_c_pot}
\end{figure}

As a final test of transferability, we applied the MTP trained on Li/Ga-LLZO configurations to predict energies, forces, and stresses for Li/t-LLZO structures. The results are visualized in~\cref{fig:errors_unified_int_t_with_int_c_pot} and the error statistics are summarized in~\cref{tab:unified_errors_int_t_with_int_ga}. Relative to the native Li/t-LLZO MTP validation performance, the energy RMSE exhibited a dramatic increase from 1.2 to 345.6~meV/atom. The stress RMSE also increased considerably from 0.88 to 1.12~GPa, whereas the force RMSE remained unchanged at 0.06~eV/\AA\ (see \cref{tab:errors_int_tet}).

\begin{table}[H]
\centering
\caption{Assessment of the MTP transferability from Li/Ga-LLZO to Li/t-LLZO structures, with the model trained on Li/Ga-LLZO configurations and validated on Li/t-LLZO data. Training and validation errors for the Li/t-LLZO interface. MaxAE, MAE and RMSE errors for energies (meV/atom), forces (eV/\AA), and stresses (GPa) are provided.}
\label{tab:unified_errors_int_t_with_int_ga}
\begin{tabular}{l|ccc}
\hline\hline
\multicolumn{1}{c|}{\bf{Train errors}} & \bf{MaxAE} & \bf{MAE} & \bf{RMSE} \\ \hline
Errors in energies (meV/atom)                                   & 355.72   & 347.14  & 347.15  \\
Errors in forces (eV/\AA)                                       & 2.08     & 0.05    & 0.08    \\
Errors in stresses (GPa)                                        & 3.15     & 0.62    & 0.90    \\ \hline
\multicolumn{1}{c|}{\bf{Validation errors}} & \bf{MaxAE} & \bf{MAE} & \bf{RMSE} \\ \hline
Errors in energies (meV/atom)                                   & 350.14   & 345.57  & 345.59  \\
Errors in forces (eV/\AA)                                       & 0.73     & 0.04    & 0.06    \\
Errors in stresses (GPa)                                        & 2.19     & 0.83    & 1.12    \\ \hline\hline
\end{tabular}%
\end{table}

\section{Training timings of MTP\lowercase{s}}\label{sec:training_timings_mtps}

\begin{figure}[H]
\begin{center}
\includegraphics[width=0.99\columnwidth]{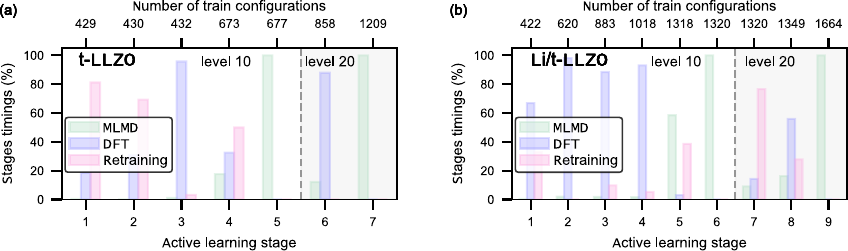}
\caption{Number of training configurations and relative timings across the active-learning stages for (a) bulk t-LLZO and (b) supercell with the Li/t-LLZO interface. Data are shown for MTP levels 10 and 20. All timings are normalized to the total duration of each active-learning stage. MLMD, DFT, and MTP retraining stages are indicated in green, blue, and pink, respectively.}
\label{fig:mtp_iter_time}

\end{center}
\end{figure}

\begin{figure}[H]
\begin{center}\includegraphics[width=0.55\columnwidth]{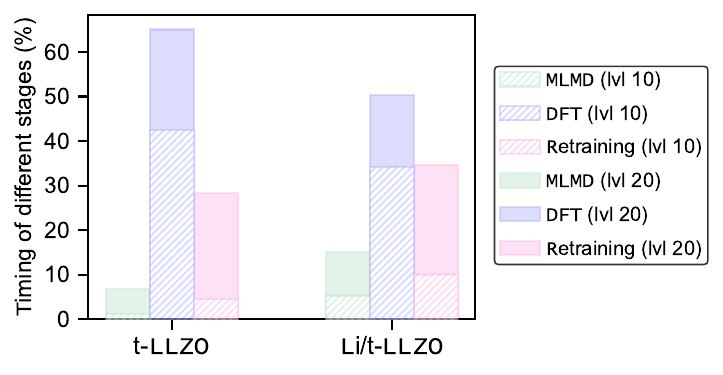}
\end{center}
\caption{Computational timing comparison of the MLMD, DFT, and retraining stages in (a) bulk t-LLZO and (b) the Li/t-LLZO interface supercell, normalized to the total time for each case. Diagonal hatching denotes MTP level 10, and solid fill denotes MTP level 20. Green, blue, and pink represent the MLMD, DFT, and retraining stages, respectively. } 
\label{fig:mtp_total_time}
\end{figure}

\begin{table}[H]
\centering
\caption{The normalized training timings for MTPs on the t-LLZO and Li/t-LLZO supercells are presented for the machine-learning molecular dynamics (MLMD), DFT calculations, and retraining stages.}
\label{tab:timings}
\renewcommand{\arraystretch}{1.2}
\begin{tabular}{l|lll|lll}
\hline \hline
          & \multicolumn{3}{c|}{MTP level 10} & \multicolumn{3}{c}{MTP level 20} \\ \hline
Timings   & MLMD    & DFT    & Retrain    &  MLMD    & DFT    & Retrain   \\ \hline
t-LLZO    &   1.2    &  42.4      &    4.6        &   5.5    &   22.6     &   23.6        \\
Li/t-LLZO &    5.3   &   34.1     &    10.0     &  9.8     &   16.1     &     24.5      \\ 
\hline \hline
\end{tabular}
\end{table}

\clearpage

\section{Occupation of L\lowercase{i} positions}\label{seq:s_occupation_sites} 

\begin{figure}[H]
\begin{center}
    \includegraphics[width=0.9\columnwidth]{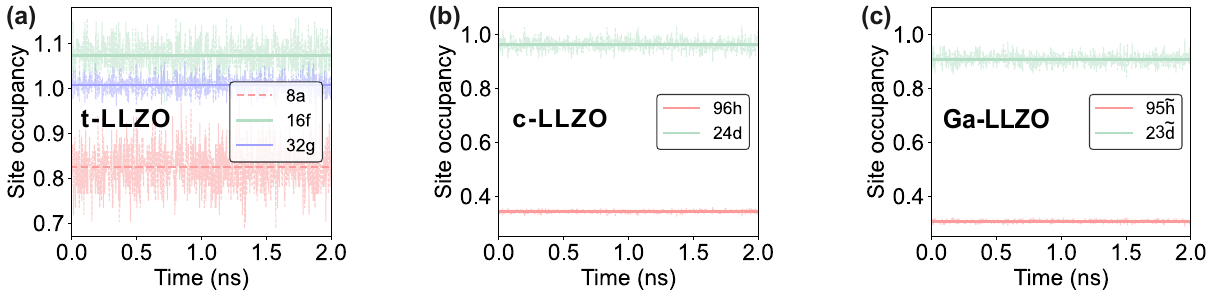}
\end{center}
\caption{Mean occupancy of Li sites over simulation time for (a) t-LLZO; (b) c-LLZO; (c) Ga-LLZO. }
\label{fig:occ_both}
\end{figure}

To identify the occupation of Li sites, we did the following analysis. At a given timestep $t_i$, for each Li-ion we found the closest initial position, considering periodic boundary conditions (PBCs). Next, we normalized the number of occupied sites by the total number of available Wyckoff positions (i.e., their multiplicities). The Wyckoff positions for t-LLZO are provided in~\cref{tb:tet_atomic_positions} and c-LLZO \cref{tb:cub_atomic_positions}. The calculated results are provided in~\cref{fig:occ_both} and \cref{tab:occ_bulk_llzo}. 

As we did not find data on site occupancies in Ga-LLZO, we also provide values, which are normalized per available sites (one $24 d$ and one $96h$ sites are occupied with Ga atoms). 

Our results are consistent with the workflow of Holland~\textit{et al.}~\cite{holland2023workflow}, who identified viable crystal structures for c-LLZO. In their Supporting Information, they compiled the ratios of $24d$ to $96h$ site occupancies calculated using various methods, finding values in the range of 0.17--0.75. Our calculated ratio of 0.7 falls within this range.

\begin{table}[H]
\centering
\caption{Mean occupancy of Li sites nearest to their equilibrium positions, over all simulation time for t-LLZO, c-LLZO, and Ga-LLZO. All data were calculated at $T=700$~K, where the low-temperature vacancy-mediated regime is dominant. The normalization of occupancies for Ga-LLZO accounts for the presence of Ga dopants and associated vacancies. Specifically, the normalized occupation factors are $\frac{24}{23}\times\frac{96}{90}$ for the $24d$ site and $\frac{96}{95}\times\frac{96}{90}$ for the $96h$ site. Reference values for comparison were taken from the works~\cite{Awaka:2011,Awaka:2009}.}
\label{tab:occ_bulk_llzo}
\renewcommand{\arraystretch}{1.15}
\begin{tabular}{ccc}
\hline\hline
\bf{Wyckoff positions} & \bf{Occupation} & \bf{Reference value}            \\ \hline
    \multicolumn{3}{c}{t-LLZO}  \\ \hline
$8a$    &  $0.824 \pm 0.046$  & 1~(Awaka, 2009~\cite{Awaka:2009}) \\
$16f$    &  $1.073 \pm 0.030$ & 1~(Awaka, 2009~\cite{Awaka:2009})    \\ 
$32g$    &  $1.007 \pm 0.018$ & 1~(Awaka, 2009~\cite{Awaka:2009})   \\ \hline
      \multicolumn{3}{c}{c-LLZO}  \\ \hline
$24d$    &  $0.963 \pm 0.020$  & 0.94(7)~(Awaka, 2011~\cite{Awaka:2011}) \\
$96h$    &  $0.343 \pm 0.005$ & 0.349~(Awaka, 2011~\cite{Awaka:2011})   \\ \hline
     \multicolumn{3}{c}{Ga-LLZO} \\ \hline 
$23\widetilde{d}$    & $0.908 \pm 0.005$ & --- \\
$24d$ (normalized)    & $1.010 \pm 0.006$ & --- \\
$95\widetilde{h}$    & $0.307 \pm 0.032$ & --- \\    
$96h$ (normalized)   & $0.331 \pm 0.032$ & --- \\ \hline\hline
\end{tabular}
\end{table}

\section{Vacancy formation energies}\label{seq:s_vac_energy} 

Vacancy formation energies were calculated as a descriptor for estimating a lower bound on the activation energy. In the case of single-ion migration, the activation energy cannot be lower than the energy difference between the initial and final structures; however, for cooperative migration, the barrier may be lower.

\subsection{Bulk t-LLZO phase}\label{sec:vac_bulk_t_llzo}

\begin{figure}[H]
\begin{center}
    \includegraphics[width=0.42\columnwidth]{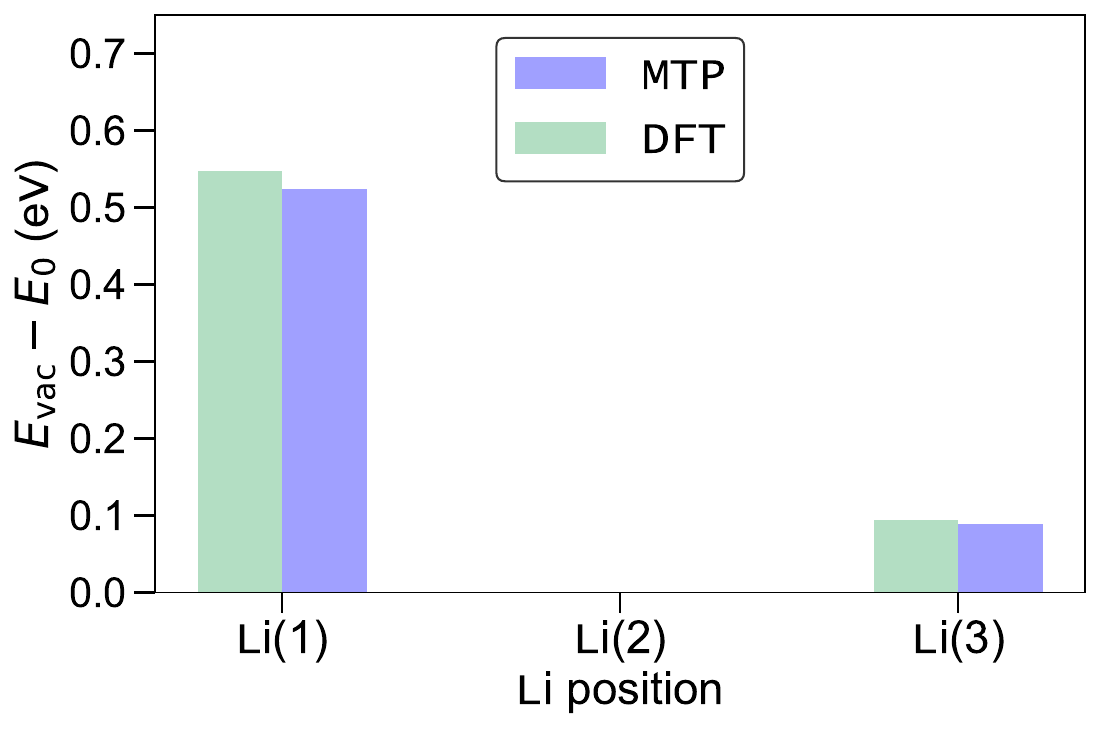}
\end{center}
\caption{Vacancy formation energies $E_{\mathrm{f}}$ for Li atoms at $\mu_{\mathrm{Li}} = 0$~eV, computed with DFT and MTP methods, for the three unique sites in t-LLZO: tetrahedral Li(1) and octahedral Li(2) and Li(3). Relative energies are reported with respect to the lowest-energy vacancy configuration obtained with each method. }
\label{fig:vac_bulk_llzo}
\end{figure}

We calculated the three unique vacancy formation energies $E_{\mathrm{f}}$ for Li atoms in t-LLZO at $\mu_{\mathrm{Li}} = 0$~eV. The results are presented in~\cref{fig:vac_bulk_llzo} and \cref{tab:vac_energy_tet}. The data show that MTP accurately reproduces the ranking of formation energies: $E_{\mathrm{f}}(\mathrm{Li1}) > E_{\mathrm{f}}(\mathrm{Li3}) > E_{\mathrm{f}}(\mathrm{Li2})$. In addition, the differences in the relative energies, which were normalized to the lowest-energy configuration, are less than 23~meV between the two methods.

The MTP reproduces the relative energy difference with an error of 23~meV, which is substantially larger than the MaxAE of 2~meV obtained for MTP-predicted energies of static configurations (see \cref{sec:errors_bulk_tet}). This discrepancy can be attributed to the fact that the MaxAE is determined from pairwise comparisons of DFT-calculated and MTP-predicted energies for static configurations, whereas vacancy formation energy calculations require full ionic relaxation and structural optimization after introducing a vacancy. This relaxation process samples a broader range of atomic environments, potentially involving extrapolation of some configurations, which may lead to lower calculated vacancy formation energies. Such behavior is also commonly observed for universal machine-learning interatomic potentials (uMLIPs)~\cite{deng2025systematic}. Nevertheless, the observed difference of 23~meV is significantly smaller than the discrepancies between different exchange-correlation functionals. For example, the typical difference in vacancy formation energies between the PBE and PBESol functionals is approximately 0.15~eV~\cite{bajdich2015surface}, which is nearly an order of magnitude larger than the MTP--DFT difference reported here.

\begin{table}[H]
\centering
\caption{Vacancy formation energies ($E_{\mathrm{f}}$) for Li atoms, calculated with DFT and MTP methods, for three unique sites in t-LLZO: tetrahedral Li(1), octahedral Li(2) and Li(3). We also provide relative energies, which are normalized to the lowest energy (DFT -- DFT(min) and MTP -- MTP(min)).  The difference between the minimum relative energies obtained from DFT and the MTP ($\Delta_{\mathrm{min}}^{\mathrm{DFT-MTP}}$) is also provided.  }
\label{tab:vac_energy_tet}
\renewcommand{\arraystretch}{1.2}
\begin{tabular}{l|ccccc}
\hline\hline
\multicolumn{1}{c|}{Position} & DFT & MTP & DFT -- DFT(min) & MTP -- MTP(min) &  $\Delta_{\mathrm{min}}$(DFT -- MTP) \\
\multicolumn{1}{c|}{} & (eV) & (eV) & (eV) & (eV) & (eV) \\ \hline
Li(1)                                  & 6.86     & 5.70  & 0.55  & 0.52 & 0.02  \\ 
Li(2)                                  & 6.31     & 5.17 & 0  & 0 & 0  \\ 
Li(3)                                  & 6.40     & 5.26 & 0.09 & 0.09 & 0.01 \\  \hline\hline
\end{tabular}%
\end{table}

\subsection{Effect of the strain on Li vacancy formation energies}\label{sec:vac_en_effect_strain}

\begin{figure}[H]
\centering
\includegraphics[width=0.42\columnwidth]{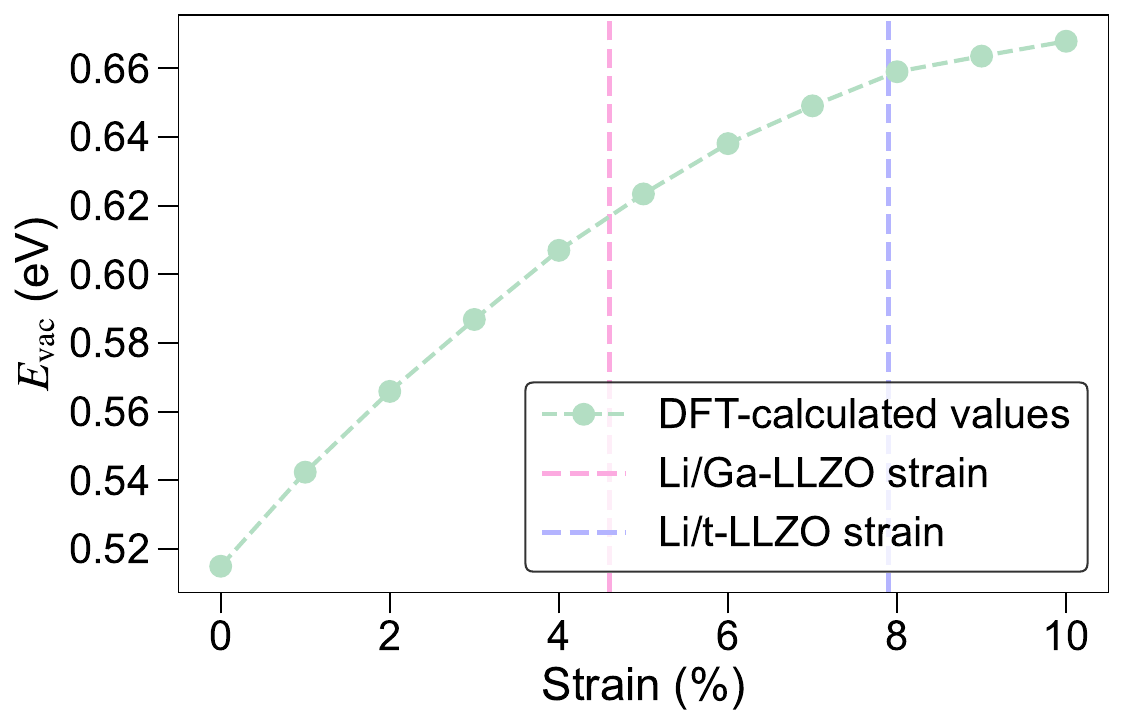}
\caption{Effect of isotropic strain on the Li vacancy formation energy ($E_{\mathrm{vac}}$) in a $3\times3\times3$ bcc Li supercell. The vertical dashed lines in blue and pink indicate the maximum in-plane strains in Li/t-LLZO and Li/Ga-LLZO supercells, respectively (see \cref{tb:cell_lattice_mismatch}).}
\label{fig:vac_en_strain_effect}
\end{figure}

To quantify the effect of in-plane strain on the Li vacancy formation energy in interface-containing supercells, we performed calculations on a $3\times3\times3$ bcc Li supercell for strain values spanning 0 to 10\%. The results are presented in~\cref{fig:vac_en_strain_effect}. At the maximum in-plane strain corresponding to the interface-containing structures, the vacancy formation energies change by approximately $0.1$~eV for Li/Ga-LLZO (4.9\% maximum strain, see \cref{tb:cell_lattice_mismatch}) and by about $0.15$~eV for Li/t-LLZO (7.9\% maximum strain, see \cref{tb:cell_lattice_mismatch}). The difference in vacancy formation energies between the two supercells at their respective strain levels is roughly $0.04$~eV. This magnitude is expected to be comparable in interface-containing supercells.

\subsection{Supercells with interfaces}\label{sec:vac_sup_cell_int}

\begin{figure}[H]
\begin{center}
    \includegraphics[width=0.9\columnwidth]{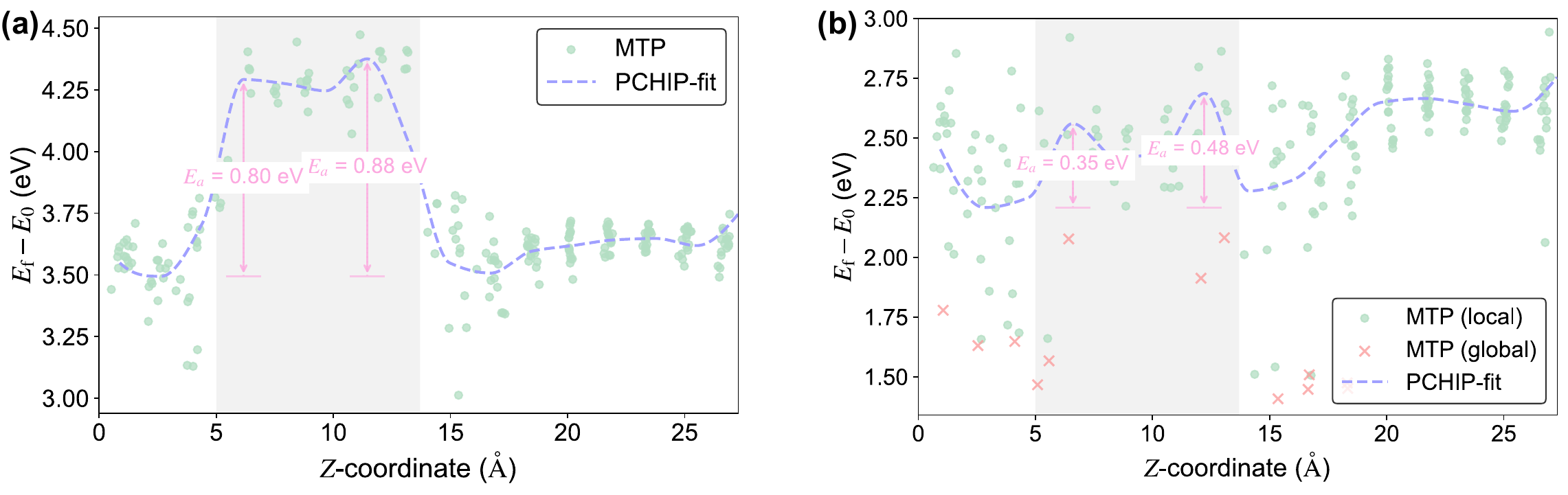}
\end{center}
\caption{Vacancy formation energies at $\mu_{\mathrm{Li}}=0$ ($E_{\mathrm{vac}}-E_0$) are shown as a function of the $z$-coordinate for Li atoms in a supercell containing Li/t-LLZO interfaces. Red crosses indicate vacancy sites for which atoms moved, on average, by more than $2~\text{\AA}$ during structural optimization. The peak positions, shapes, and heights are similar to those obtained in our previous DFT-obtained calculations~\cite{burov2024mechanism}. The data were fitted using the piecewise cubic Hermite interpolating polynomial (PCHIP) method~\cite{fritsch1984method}. The gray hatched region denotes the LLZO slab. (a) Single-point (unrelaxed) calculations; (b) fully relaxed structures.}
\label{fig:vac_int_tet}
\end{figure}

\begin{figure}[H]
\begin{center}
    \includegraphics[width=0.9\columnwidth]{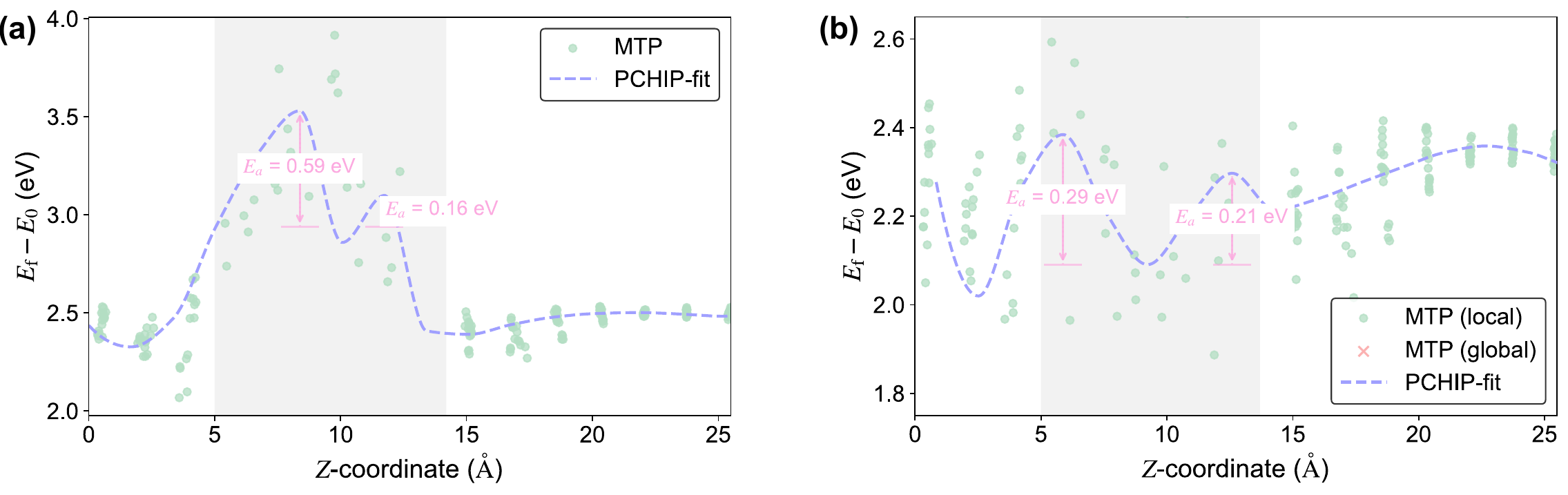}
\end{center}
\caption{Vacancy formation energies at $\mu_{\mathrm{Li}}=0$ ($E_{\mathrm{vac}}-E_0$) are shown as a function of the $z$-coordinate for Li atoms in a supercell containing Li/Ga-LLZO interfaces. The data were fitted using the piecewise cubic Hermite interpolating polynomial (PCHIP) method~\cite{fritsch1984method}. The gray hatched region denotes the LLZO slab. (a) Single-point (unrelaxed) calculations; (b) fully relaxed structures.}
\label{fig:vac_int_cub}
\end{figure}

Li vacancy formation energies were calculated in supercells containing two distinct interfaces: Li/t-LLZO and Li/Ga-LLZO. For each system, we computed both single-point (unrelaxed) energies and fully relaxed ionic structures. The results are presented in~\cref{fig:vac_int_tet} for Li/t-LLZO and in~\cref{fig:vac_int_cub} for Li/Ga-LLZO.

In both interface systems, the vacancy formation energy exhibits two maxima within the LLZO bulk region, while minima are consistently localized near the interface. Structural relaxation systematically lowers the vacancy formation energies across both systems. Specifically, the descriptor of activation energy for Li/t-LLZO decreases from 0.88 to 0.48~eV (a reduction of 0.40~eV), whereas for Li/Ga-LLZO it drops from 0.59 to 0.29~eV (a reduction of 0.30~eV).

\begin{figure}[H]
\begin{center}
    \includegraphics[width=0.35\columnwidth]{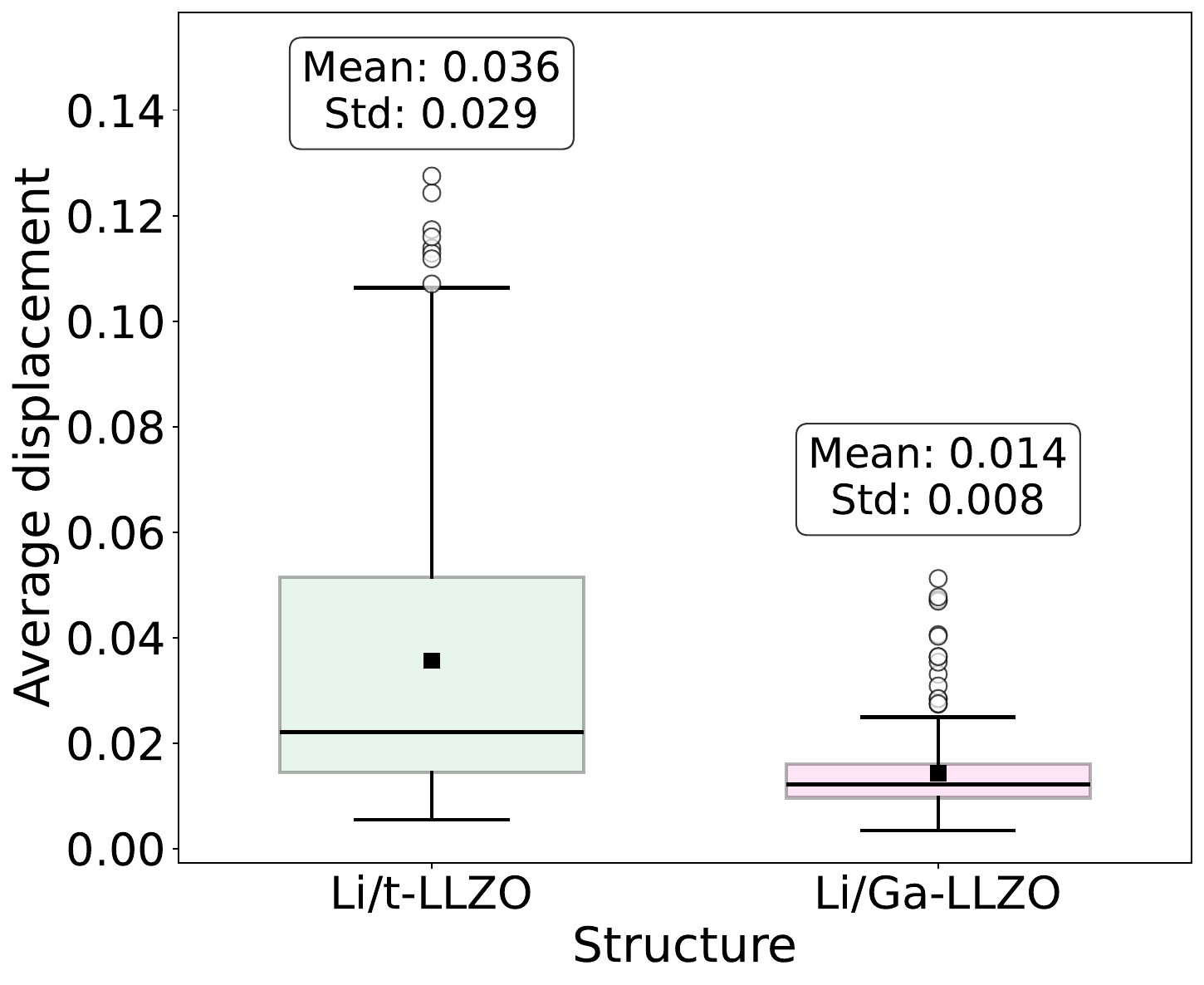}
\end{center}
\caption{Mean relative coordinate shifts with standard deviations for all Li atoms in Li/t-LLZO (green) and Li/Ga-LLZO (pink) interface supercells, as obtained from MTP calculations.}
\label{fig:displacements_int_mtp}
\end{figure}

A notable difference in the relaxation behavior is observed between the two systems. The atomic displacements during ionic relaxation are substantially smaller for Li/Ga-LLZO, with no vacancies triggering significant structural rearrangements. In contrast, for Li/t-LLZO (marked by red crosses in the figures), atoms move on average by approximately 2~\AA\ (see \cref{fig:displacements_int_mtp}), indicating pronounced local structural changes.

The unrelaxed vacancy formation energies differ by about 1~eV between the two interfaces (see \cref{fig:vac_int_tet}.a and \cref{fig:vac_int_cub}.a). This discrepancy cannot be explained by local strain effects, which are estimated to contribute only $\sim$0.04~eV, as discussed in~\cref{sec:vac_en_effect_strain}. Furthermore, after full relaxation, the global structural reordering in the interface-containing supercells brings the vacancy formation energies to a similar level of $\sim$2.5~eV (see \cref{fig:vac_int_tet}.b and \cref{fig:vac_int_cub}.b). In particular, the energies of equivalent Li regions become aligned upon relaxation, which is consistent with the local atomic environment concept underlying the calculation of energies in MLIPs. Therefore, neither strain nor size effects are responsible for the observed energy differences.

Instead, we attribute the variation in the local environment to changes in the electronic density extending far from the vacancy, as demonstrated in our previous work~\cite{burov2024mechanism}. A comprehensive investigation of this effect, requiring charge difference and Bader charge analyses, lies beyond the scope of the present work and warrants a dedicated systematic study.

\section{MTP-NEB calculations}\label{seq:s_mtp_neb} 

The most widespread method to estimate activation energy and find minimum-energy pathway (MEP) is nudged elastic band (NEB) method~\cite{jonsson1998nudged}. In some works, MTPs wrongly estimate energies of the initial and final states, whereas intermediate images are reproduced with high accuracy~\cite{kruglov2024surface}. Therefore, in order to evaluate the ability of MTPs to reproduce single-ion activation energies, we compared our previous DFT-NEB results with those values, obtained with MTP-NEB approach, as implemented in ASE software package~\cite{larsen2017atomic_ase}.

\subsection{Bulk phases: t-LLZO and bcc Li}

\begin{figure}[H]
\begin{center}
    \includegraphics[width=0.99\columnwidth]{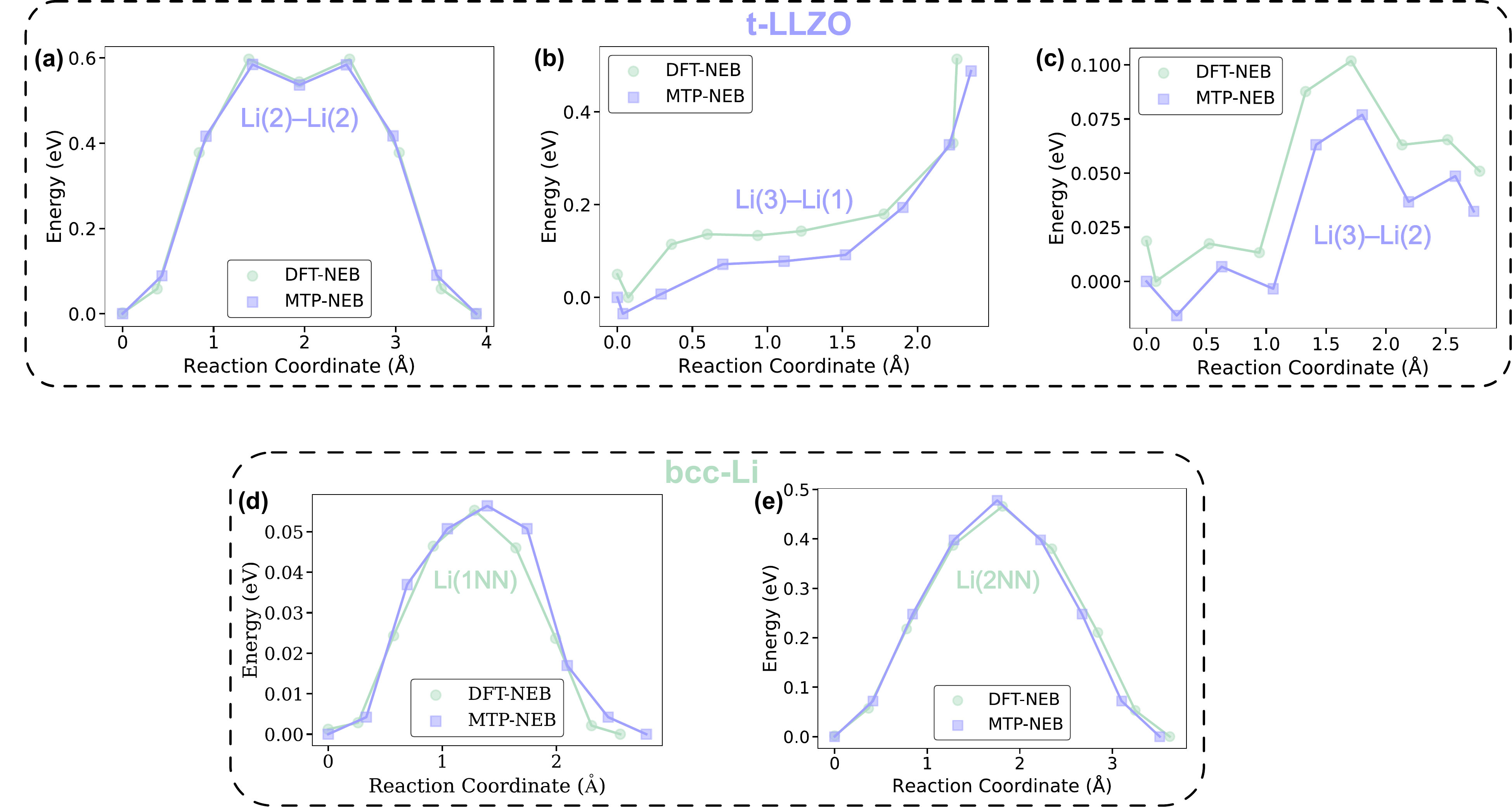}
\end{center}
\caption{The energy profiles and corresponding reaction coordinates for the unique Li single-ion diffusion events in the bulk phases: (a)--(c) t-LLZO; (d),(e) bcc Li. }
\label{fig:neb_llzo}
\end{figure}

\subsection{Supercells with interfaces: Li/t-LLZO and Li/Ga-LLZO}

\begin{figure}[H]
\begin{center}
    \includegraphics[width=0.99\columnwidth]{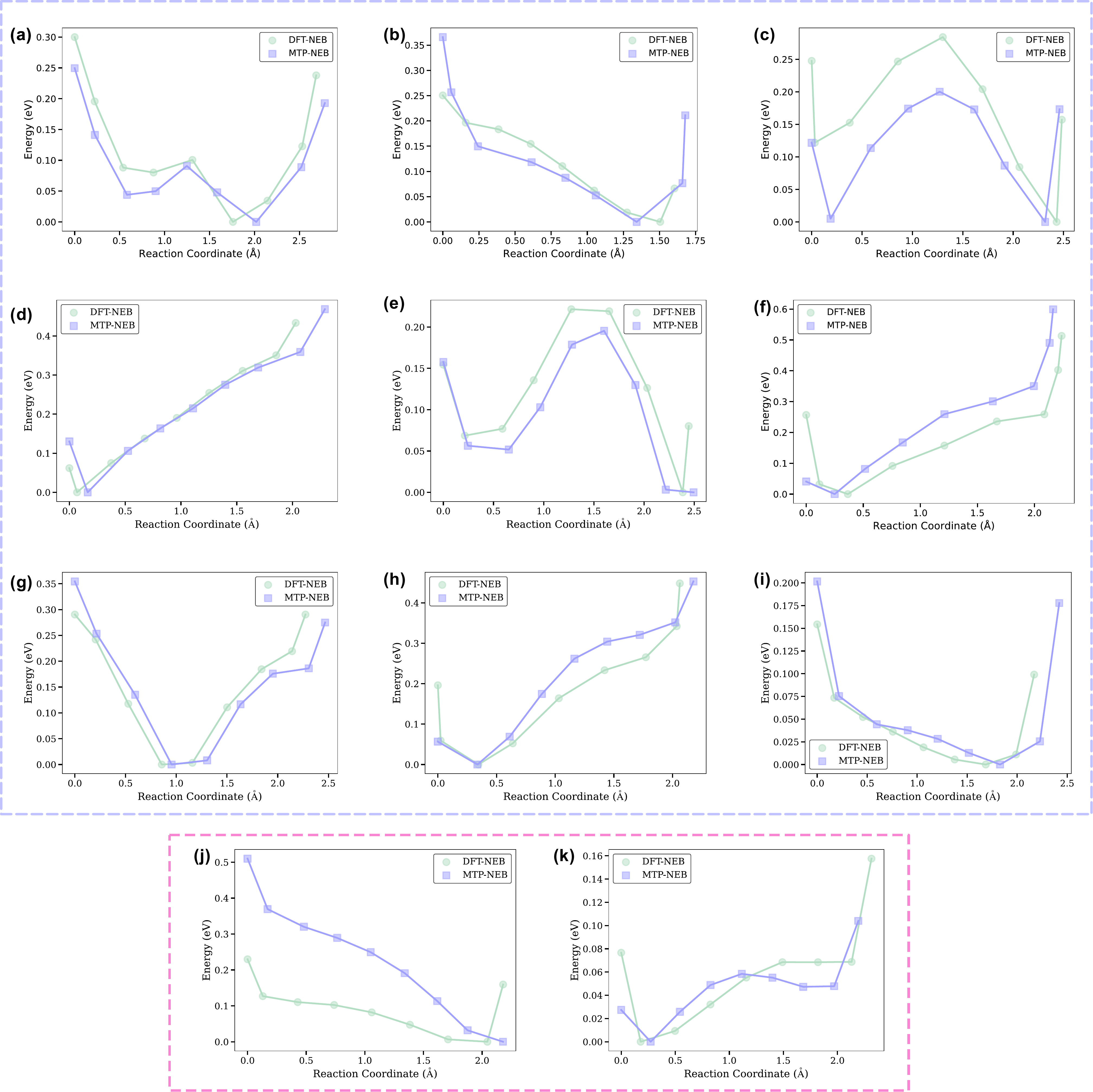}
\end{center}
\caption{ 
The energy profiles and corresponding reaction coordinates for the unique Li single-ion diffusion events in Li/t-LLZO: (a)-(i) paths with MTPs low extrapolation grades ($<30$); (j),(k) paths with MTPs high extrapolation grades ($>30$). }
\label{fig:neb_int_tet}
\end{figure}

The calculated migration Li-ion diffusion paths for Li/t-LLZO are shown in~\cref{fig:neb_int_tet}. When the extrapolation grades of MTPs are below 30, we get reliable MTP-NEB results, which accurately describe the geometry and pathways of DFT-NEB calculations (see~\cref{fig:neb_int_tet}.a-i). However, if the grades are above 30, the difference in activation energies can be up to 0.25~eV. 

The problem arises from the inability to sample all possible initial vacancy positions during MTP training.

\section{MSD of non-mobile ions}\label{seq:s_msd_nonli} 

\begin{figure}[H]
\begin{center}
    \includegraphics[width=0.85\columnwidth]{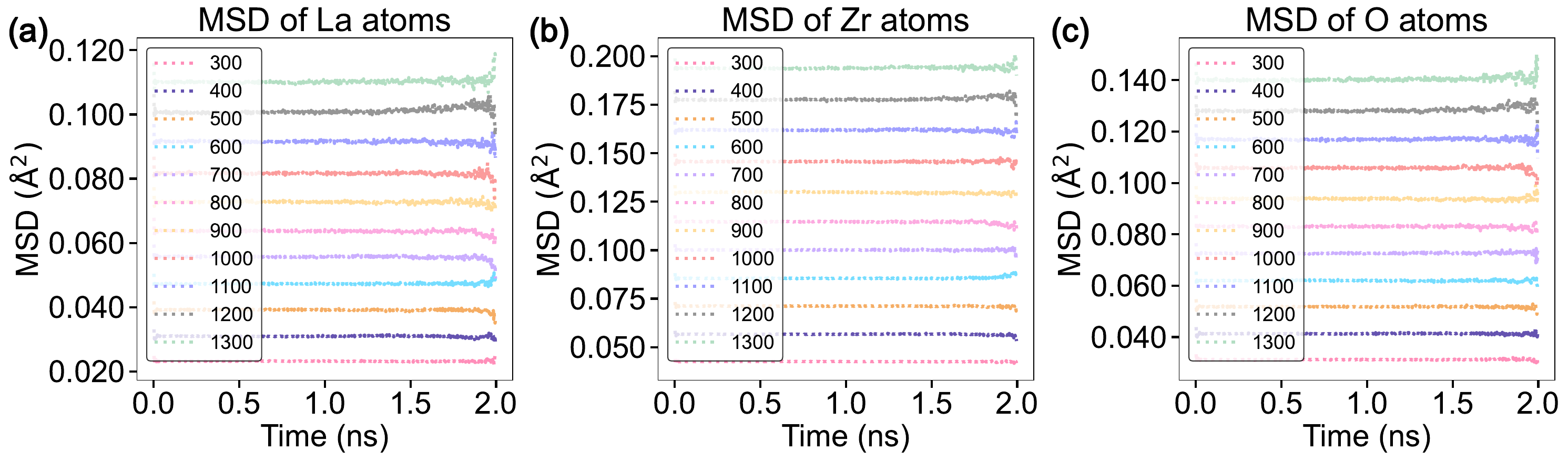}
\end{center}
\caption{MSD of La, Zr, and O atoms in the t-LLZO at various temperatures. Dotted lines indicate trajectories where the total MSD remains below 9~\AA$^2$.}
\label{fig:diff_non_tet_llzo}
\end{figure}

\begin{figure}[H]
\begin{center}
    \includegraphics[width=0.85\columnwidth]{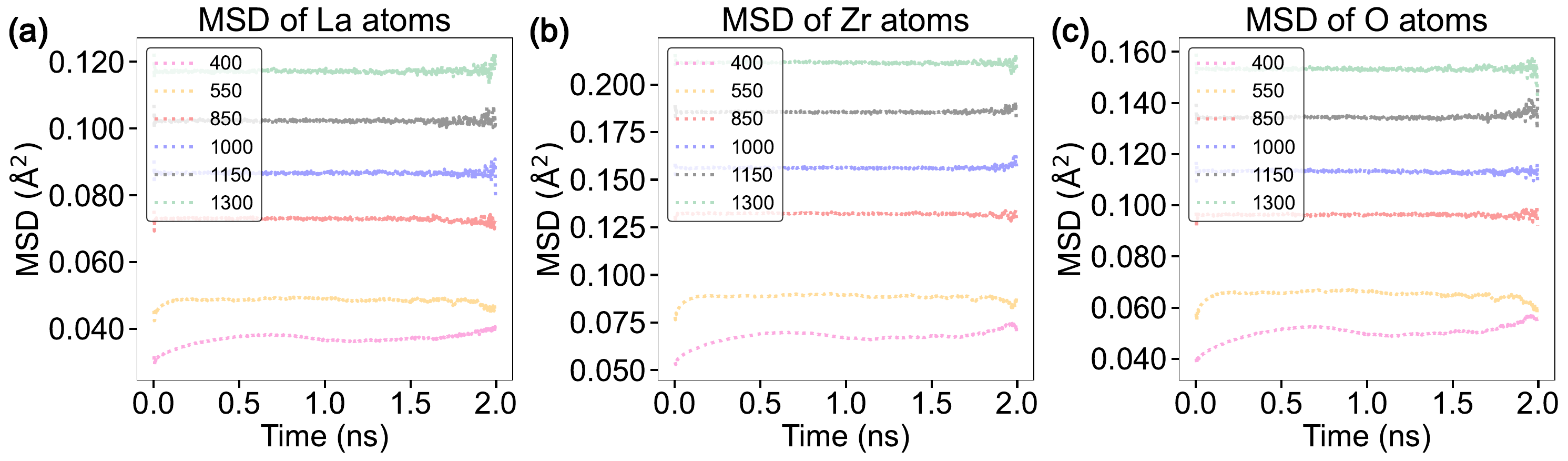}
\end{center}
\caption{MSD of La, Zr, and O atoms in the c-LLZO at various temperatures. Dotted lines indicate trajectories where the total MSD remains below 9~\AA$^2$.}
\label{fig:diff_non_cub_llzo}
\end{figure}

\begin{figure}[H]
\begin{center}
    \includegraphics[width=0.99\columnwidth]{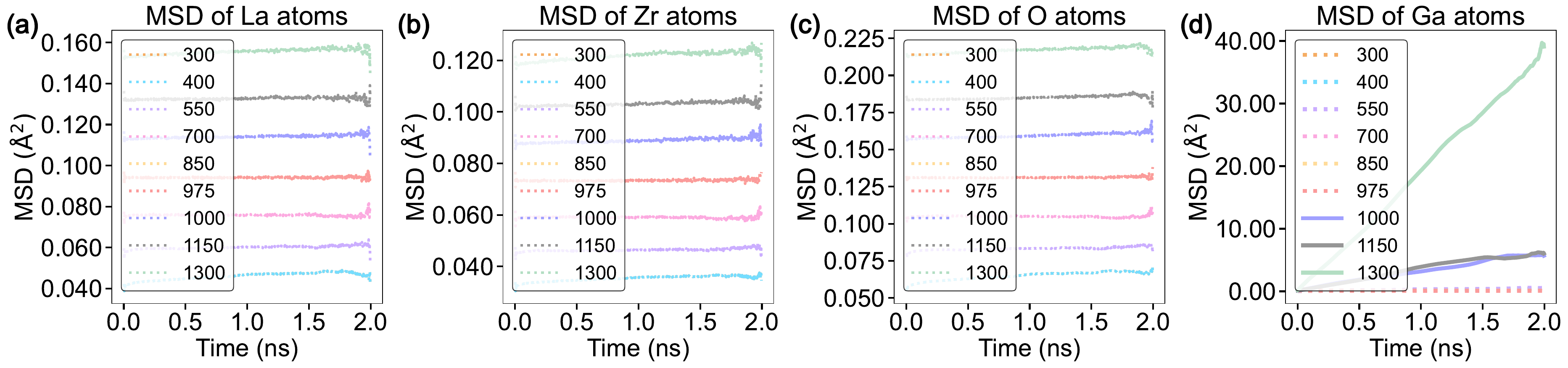}
\end{center}
\caption{MSD of La, Zr, Ga and O atoms in the Ga-LLZO at various temperatures. Dotted lines indicate trajectories where the total MSD remains below 9~\AA$^2$.}
\label{fig:diff_non_ga_llzo}
\end{figure}

\begin{figure}[H]
\begin{center}
    \includegraphics[width=0.9\columnwidth]{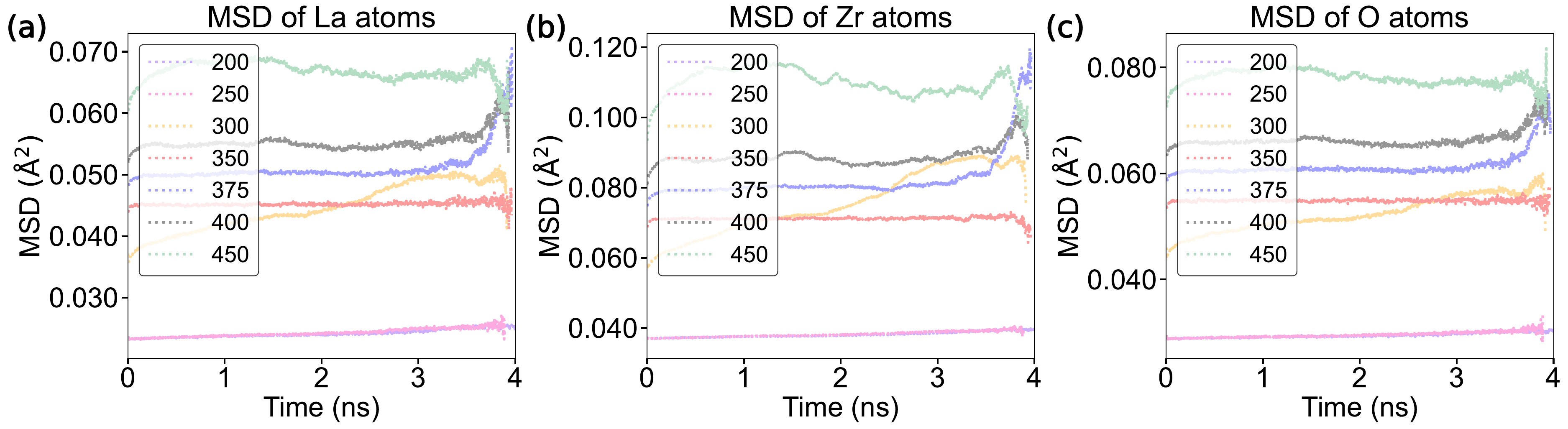}
\end{center}
\caption{MSD of La, Zr, and O atoms in the supercell with Li/t-LLZO interface at various temperatures. Dotted lines indicate trajectories where the total MSD remains below 20~\AA$^2$.}
\label{fig:diff_non_int_tet}
\end{figure}

\begin{figure}[H]
\begin{center}
    \includegraphics[width=0.99\columnwidth]{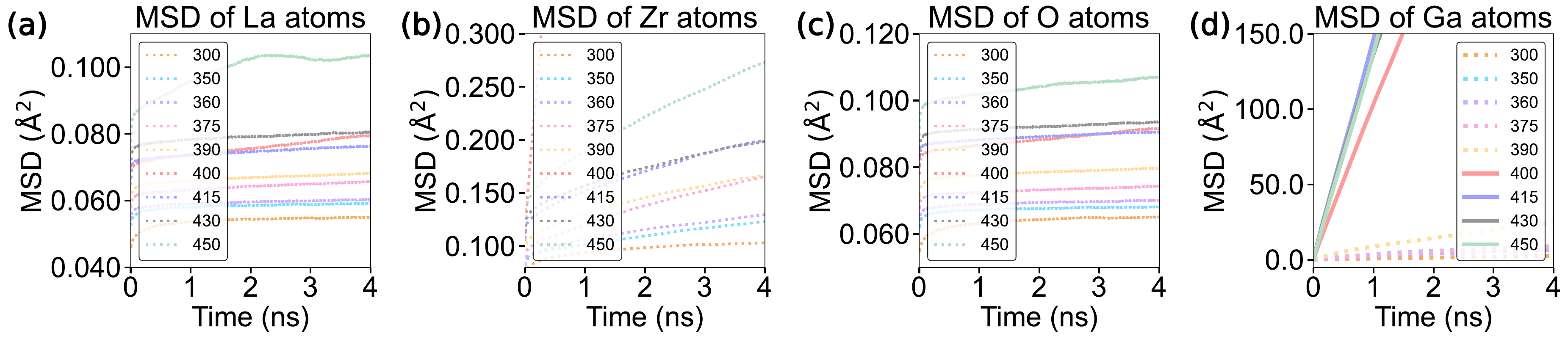}
\end{center}
\caption{MSD of La, Zr, Ga, and O atoms in the supercell with Li/Ga-LLZO interface at various temperatures. Dotted lines indicate trajectories where the total MSD remains below 20~\AA$^2$.}
\label{fig:diff_non_int_cub}
\end{figure}

To validate the integrity of the framework in structures, we calculated the MSD of non-Li atoms: La, Zr, Ga, and O. The results are shown in~\cref{fig:diff_non_tet_llzo} for t-LLZO, \cref{fig:diff_non_cub_llzo} for c-LLZO, \cref{fig:diff_non_ga_llzo} for Ga-LLZO, \cref{fig:diff_non_int_tet} for Li/t-LLZO, and \cref{fig:diff_non_int_cub} for Li/Ga-LLZO.

In each case, La, Zr, and O have small MSDs over all range of temperatures. The highest temperature for the bulk phases was 1300~K and for the supercells with interfaces, it was 450~K. However, Ga atoms in Ga-LLZO become mobile at $T>1000$~K with MSDs higher than 5~\AA$^{2}$. Such conditions are achieved during the Ga-LLZO synthesis at the annealing stage. In Li/Ga-LLZO, Ga atoms become mobile at 400~K, which can lead to the contact loss and transformation of the cubic phase to the tetragonal one.  

\subsection{Bulk phases}\label{seq:s_msd_nonli_bulk} 

\begin{figure}[H]
\begin{center}
    \includegraphics[width=0.85\columnwidth]{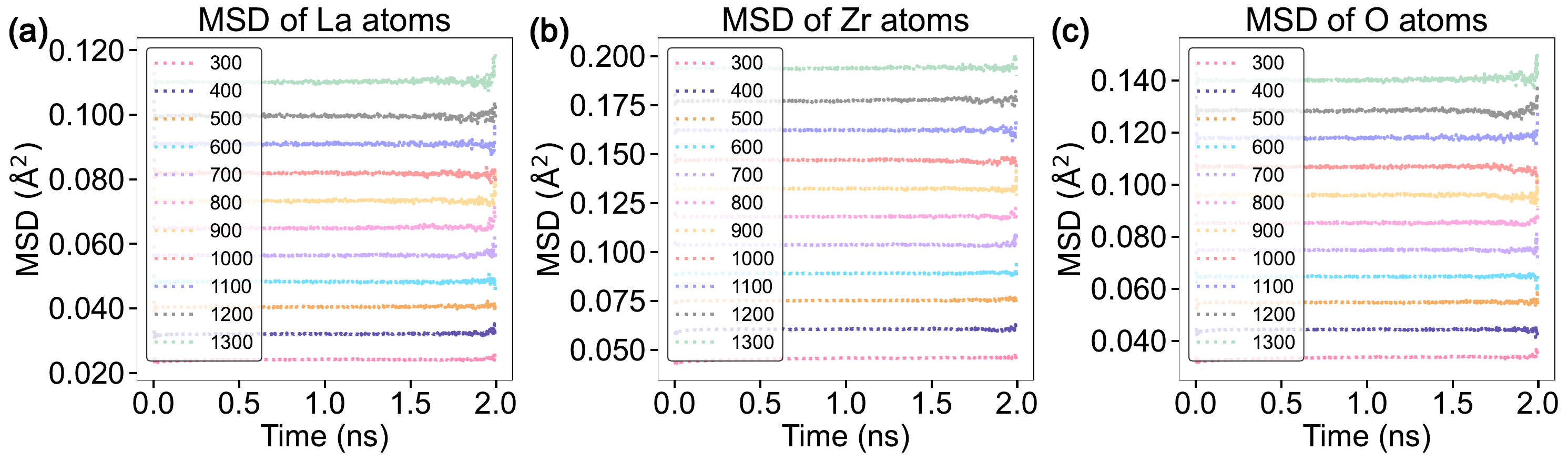}
\end{center}
\caption{MSD of La, Zr, and O atoms in the t-LLZO with additional 1.8\% vacancies at various temperatures. Dotted lines indicate trajectories where the total MSD remains below 9~\AA$^2$.}
\label{fig:diff_non_tet_llzo_with_vac}
\end{figure}

\subsection{Supercells with interfaces}\label{seq:s_msd_nonli_int}

\section{L\lowercase{i}-ion self-diffusion in additional bulk phases: \lowercase{t}-LLZO (1.8\% vacancies), \lowercase{c}-LLZO, and supercells with interfaces: L\lowercase{i}/\lowercase{t}-LLZO and L\lowercase{i}/G\lowercase{a}-LLZO}\label{seq:msd_li_other_phases}

\subsection{Tracer diffusion of Li vacancies in Li metal}\label{sec:diff_only_li_metal}

\begin{figure}[H]
\begin{center}
    \includegraphics[width=0.6\columnwidth]{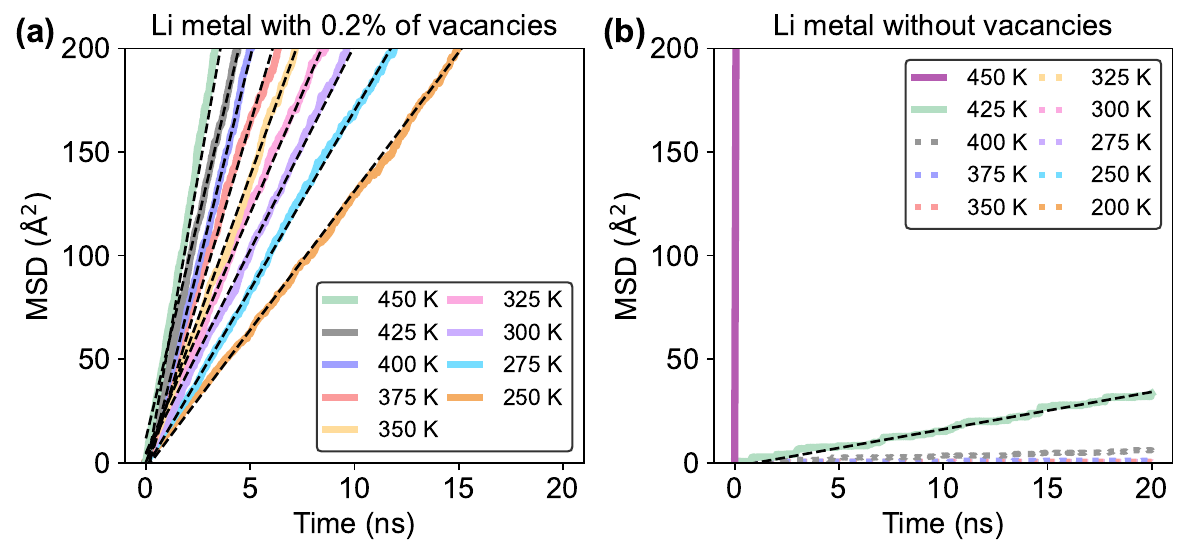}
\end{center}
\caption{MSD of bcc Li as a function of temperature: with 0.2\% vacancies (a) and without vacancies (b). To avoid trajectories with poor statistics, we included only those for which the MSD exceeded the squared typical bulk jump length, that is, $MSD > 9~\mathrm{\AA}^2$ (dotted lines in the plots). }
\label{fig:li_msd_both}
\end{figure}

\begin{figure}[H]
\begin{center}
    \includegraphics[width=0.4\columnwidth]{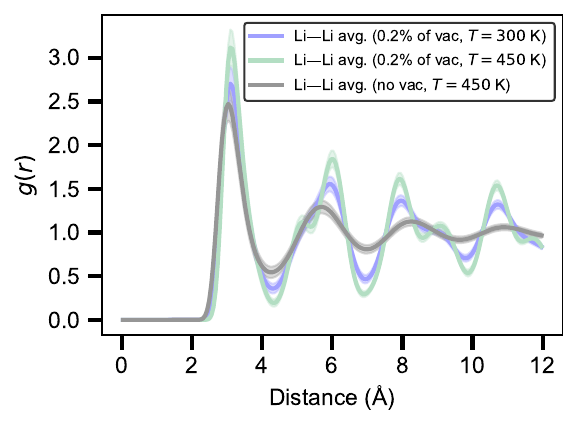}
\end{center}
\caption{RDFs for Li--Li distances in bcc Li: (a) structure with 0.2\% vacancies at low temperature ($T=300$~K), shown in blue; (b) structure with 0.2\% vacancies at high temperature ($T=450$~K), shown in green; and (c) vacancy-free structure at high temperature ($T=450$~K).  }
\label{fig:rdf_li_temp}
\end{figure}

\begin{figure}[H]
\begin{center}
    \includegraphics[width=0.8\columnwidth]{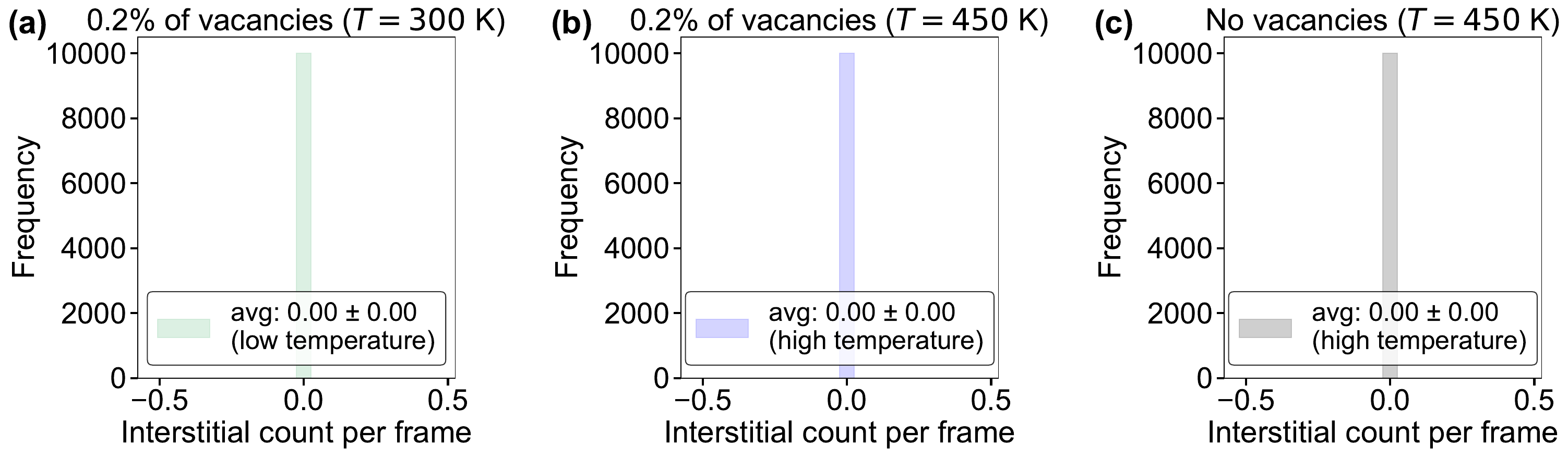}
\end{center}
\caption{Number of interstitials in bcc Li, calculated using the Wigner--Seitz method as implemented in the OVITO software package~\cite{stukowski2009visualization}. (a) Structure with 0.2\% vacancies at low temperature ($T=300$~K); (b) structure with 0.2\% vacancies at high temperature ($T=450$~K); and (c) vacancy-free structure at high temperature ($T=450$~K).  }
\label{fig:li_ws_analysis}
\end{figure}

We also performed diffusion MLMD simulations on the vacancy-free structure, as shown in~\cref{fig:li_msd_both}. No diffusion was observed up to $T>425$~K, where disordering of the crystal structure begins. This is reflected in the RDFs by peak broadening, as shown in~\cref{fig:rdf_li_temp}. However, no interstitials were observed in either the vacancy-containing or vacancy-free structures, as confirmed by the Wigner--Seitz analysis shown in~\cref{fig:li_ws_analysis}.

The concentration of Li vacancies is not accessible experimentally via positron annihilation spectroscopy~\cite{hautojarvi1979positrons}. Following the methodology of Frank \textit{et~al.}~\cite{frank1996first}, we therefore estimated the equilibrium vacancy concentration from DFT-calculated formation energies as follows:

\begin{equation}
    \label{eq:vac_form_eq}
    C_{\mathrm{vac}}^{\mathrm{eq}} 
= \exp\left(-\frac{G_{\mathrm{f}}}{k_{\mathrm{B}}T}\right) 
= \exp\left(\frac{S_{\mathrm{f}}}{k_{\mathrm{B}}} - \frac{E_{\mathrm{f}}}{k_{\mathrm{B}}T}\right) 
= \exp\left(\frac{S_{\mathrm{f}}}{k_{\mathrm{B}}}\right) \exp\left(-\frac{E_{\mathrm{f}}}{k_{\mathrm{B}}T}\right),
\end{equation}
where $G_{\mathrm{f}}$ is the Gibbs energy of the vacancy formation, $S_{\mathrm{f}}$ is the formation entropy, and $E_{\mathrm{f}}$ the formation energy of the defect at atmospheric pressure (calculated in DFT),  with the approximation that the formation enthalpy $H_{\mathrm{f}} \approx E_{\mathrm{f}}$ at ambient pressure.

Extracting data from the article figures, the calculated value of $S_{\mathrm{f}}(300~\mathrm{K})$ is approximately $2.4~k_{\mathrm{B}}$. The vacancy formation energy, corrected for elastic interaction, is 0.52~eV. 

Substituting this energy in~\cref{eq:vac_form_eq}, we get $C_{\mathrm{vac}}^{\mathrm{eq}} = 2.0 \times 10^{-7}$, which is in excellent agreement with the value of $2.2 \times 10^{-7}$, extracted from the plot in the article.

Using this concentration of vacancies, we can compare with experimentally measured vacancy diffusivities in bcc Li. The compiled data are provided in~\cref{tab:vac_mobility_works}. The values derived from the experiment lie in the range $(2.6-3.0)\times 10^{-4}$~cm$^2$\,s$^{-1}$~\cite{lodding1970isotope_li_diffusion, messer1975nuclear, mali19886li}, which is in good agreement with the value calculated from MD simulations of $2.8\times 10^{-4}$~cm$^2$\,s$^{-1}$~\cite{sergeev2023anomalous}. In contrast, AIMD simulations underestimate vacancy diffusivity with the values of $6.1\times 10^{-5}$~cm$^2$\,s$^{-1}$.

The experimental activation energy can be rationalized as the sum of the nearest-neighbor (NN) migration barrier (0.05~eV) and the vacancy formation energy (0.52~eV)~\cite{frank1996first}, yielding a total of 0.57~eV. This value falls within the experimentally reported range of $0.52-0.56$~eV, in excellent agreement.  

\begin{table}[H]

\renewcommand{\arraystretch}{1.2}
\centering
\setlength{\tabcolsep}{9pt}
\centering
\caption{Li self-diffusion parameters in bcc Li metal. $C_{\mathrm{vac}}$ denotes the equilibrium vacancy concentration (values in brackets correspond to the ratio of unoccupied to total sites), $E_{\mathrm{a}}$ is the self-diffusion activation energy, $D_{\mathrm{Li}}$~($300~\mathrm{K}$) is the Li diffusion coefficient, and $D_{\mathrm{vac}}$~($300~\mathrm{K}$) is the vacancy diffusion coefficient at room temperature. Experimental data were obtained by NMR; Sergeev \textit{et~al.} used modified embedded atom method (MEAM) molecular dynamics~\cite{sergeev2023anomalous}.}
\label{tab:vac_mobility_works}
\begin{tabular}{llllll}
\hline \hline
Work & Method & $C_{\mathrm{vac}}$ & $E_{\mathrm{a}}$ (meV) & $D_{\mathrm{Li}}(300~\mathrm{K})$ (cm$^2$/s) & $D_{\mathrm{vac}}(300~\mathrm{K})$ (cm$^2$/s) \\ \hline

Current work  &     MLMD  &  $2.3 \times 10^{-3}~\left(\frac{1}{432} \right)$     & $50\pm10$  &   $3.3 \times 10^{-7}$  &  $1.5 \times 10^{-4}$      \\

Sergeev, 2023~\cite{sergeev2023anomalous}   &   MD  &  $2.4 \times 10^{-4}~\left(\frac{1}{4096} \right)$      &   $44$     &     $^\dagger6.9 \times 10^{-8}$    & $2.8 \times 10^{-4}$   \\

Yang, 2021~\cite{yang2021interfacial}  &     AIMD       &   $7.8 \times 10^{-3}~\left(\frac{1}{128}\right)$     &  $45\pm23$    &   $^\dagger4.7\times10^{-7}$    &   $6.1\times10^{-5}$     \\

Lodding, 1970~\cite{lodding1970isotope_li_diffusion} &     NMR  &  $^{\S} 2.0 \times 10^{-7}$     &   $558\pm1$  &          $(5.9 \pm 2.9)\times10^{-11}$      &  $^\dagger(3.0 \pm 1.5) \times 10^{-4}$      \\
Messer, 1975~\cite{messer1975nuclear}  &   NMR      &   $^{\S} 2.0 \times 10^{-7}$     &  $521\pm13$    &   $(5.7 \pm 0.2)\times10^{-11}$   &   $^\dagger(2.9 \pm 0.1) \times 10^{-4}$     \\
Mali, 1988~\cite{mali19886li}   &   NMR  &  $^{\S} 2.0 \times 10^{-7}$     &   $561\pm2$     &     $(5.3 \pm 0.5)\times10^{-11}$        &  $^\dagger(2.6 \pm 0.3) \times 10^{-4}$ \\

\hline\hline
\end{tabular}
\begin{tablenotes}
      \small
      \item $^\S$ Experimental intrinsic Li vacancy equilibrium concentration derived from Frank \textit{et~al.}~\cite{frank1996first}. \\
        $^\dagger$ Data extracted from figures, as it was not explicitly reported in the article.
\end{tablenotes}

\end{table}

\subsection{Bulk phases}

\begin{figure}[H]
\begin{center}
    \includegraphics[width=0.99\columnwidth]{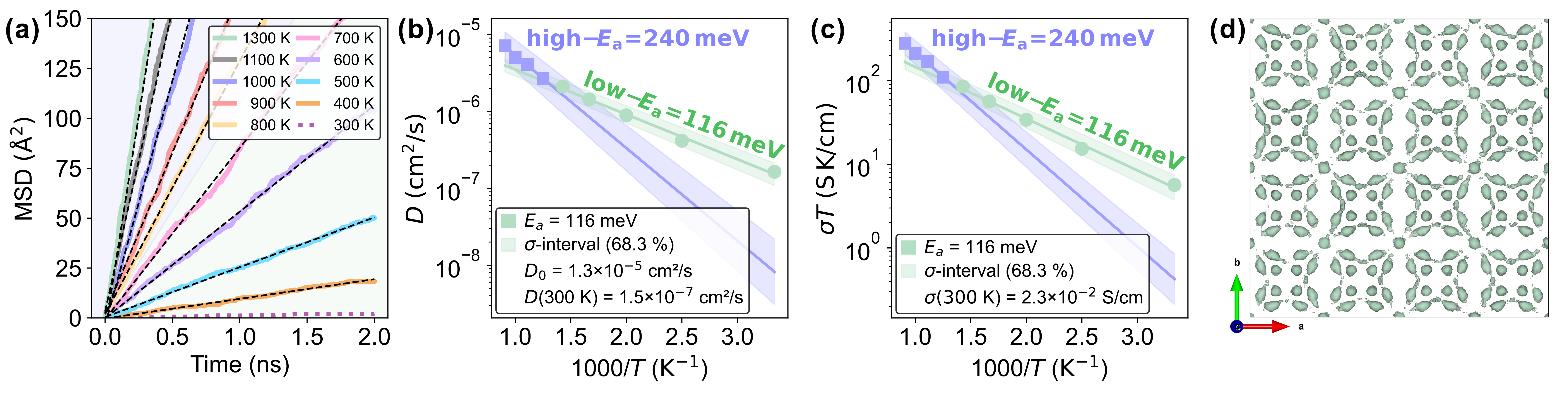}
\end{center}
\caption{Li-ion diffusion study at both low-temperature (vacancy-mediated) and high-temperature (Frenkel defect-mediated) regimes for t-LLZO with vacancies (55/56). (a) mean-squared displacement (MSD) over temperature; (b) diffusion coefficients and their Arrhenius fit; (c) ionic conductivities and their Arrhenius fit; (d) Li-ion probability density map (low-temperature regime at the isosurface level of $5 \cdot 10^{-4}$~$r_{\mathrm{Bohr}}^{-3}$). To avoid trajectories with poor statistics, we included only those for which the MSD exceeded the squared typical jump length in the bulk phases, that is, $MSD > 9~\mathrm{\AA}^2$ (dotted lines on plot).}
\label{fig:tet_vac_panel}
\end{figure}

\begin{figure}[H]
\begin{center}
    \includegraphics[width=0.99\columnwidth]{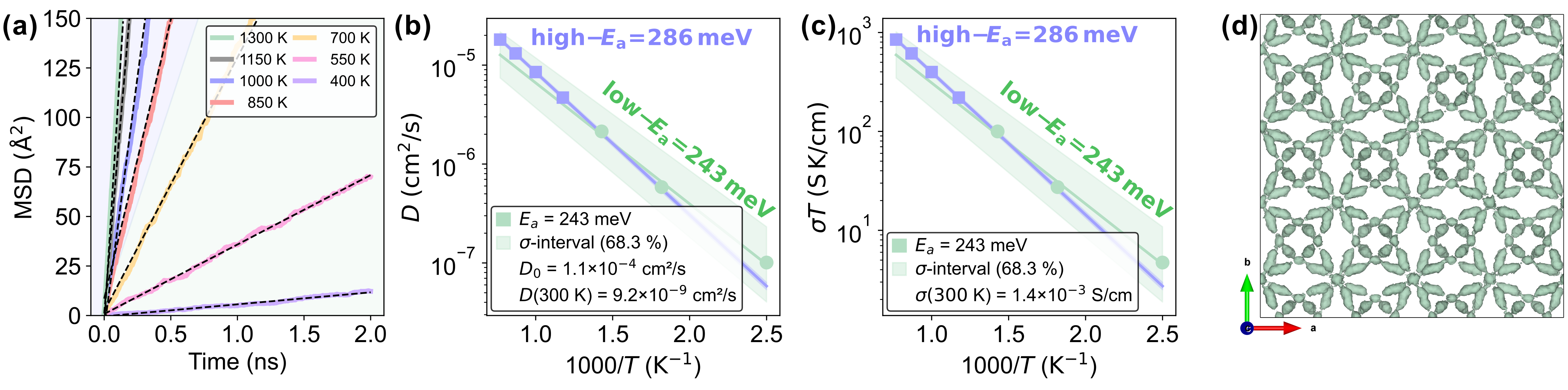}
\end{center}
\caption{Li-ion diffusion study at both low-temperature (vacancy-mediated) and high-temperature (Frenkel defect-mediated) regimes for c-LLZO. (a) mean-squared displacement (MSD) over temperature; (b) diffusion coefficients and their Arrhenius fit; (c) ionic conductivities and their Arrhenius fit; (d) Li-ion probability density map (low-temperature regime at the isosurface level of $5 \cdot 10^{-4}$~$r_{\mathrm{Bohr}}^{-3}$).}
\label{fig:cub_panel}
\end{figure}

\subsection{Supercells with interfaces}

\begin{figure}[H]
\begin{center}
    \includegraphics[width=0.7\columnwidth]{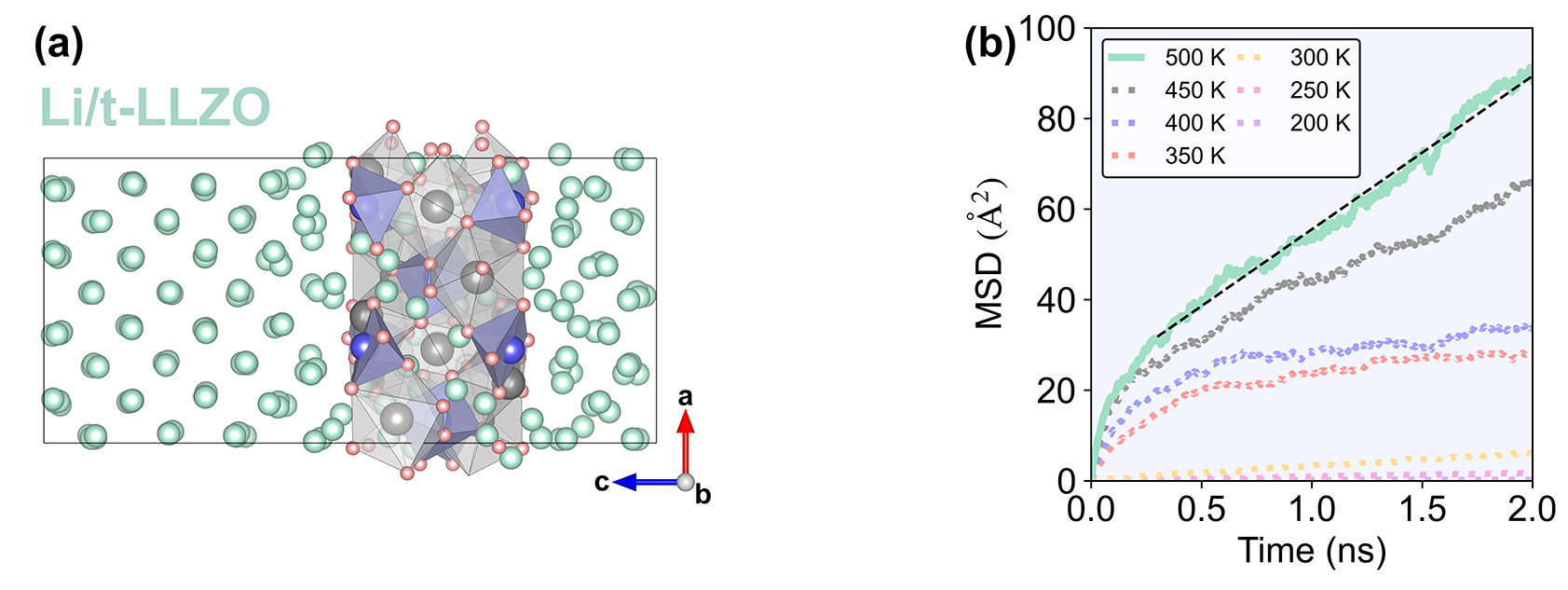}
\end{center}
\caption{Li-ion diffusion study at both low-temperature (vacancy-mediated) and high-temperature (Frenkel defect-mediated) regimes for Li/t-LLZO. (a) crystal structure; (b) mean-squared displacement (MSD) over temperature. To avoid trajectories with poor statistics, we included only those for which the MSD exceeded the squared typical jump length in the bulk phases, that is, $MSD_Z > 100~\mathrm{\AA}^2$ (dotted lines on plot).}
\label{fig:int_tet_diff_block}
\end{figure}

\begin{figure}[H]
\begin{center}
    \includegraphics[width=0.9\textwidth]{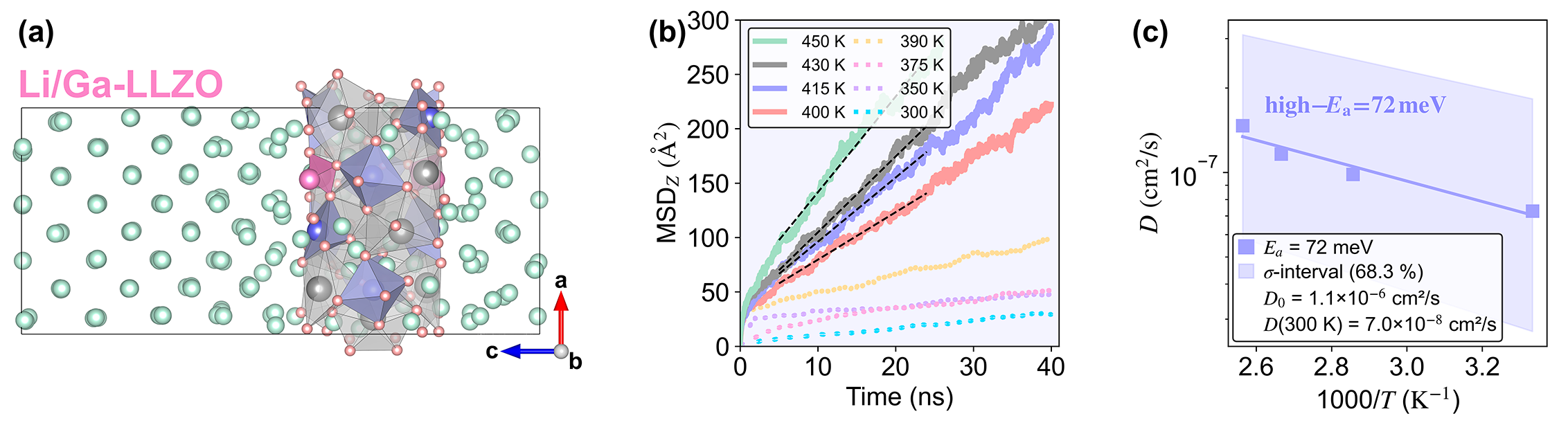}
\end{center}
\caption{Li-ion diffusion study at both low-temperature (vacancy-mediated) and high-temperature (Frenkel defect-mediated) regimes for Li/Ga-LLZO. (a) crystal structure; (b) mean-squared displacement (MSD) over temperature. To avoid trajectories with poor statistics, we included only those for which the MSD exceeded the squared typical jump length in the bulk phases, that is, $MSD_Z > 100~\mathrm{\AA}^2$ (dotted lines on plot). }
\label{fig:int_cub_diff_block}
\end{figure}

\section{L\lowercase{i} diffusion parameters}\label{seq:s_diff_parameters}

\begin{table}[H]
\centering
\setlength{\tabcolsep}{5pt}
\renewcommand{\arraystretch}{1.5}
\caption{For each structure, the table lists activation energies ($E_{\mathrm{a}}$), pre-exponential factors ($D_0$), room-temperature diffusivities ($D(300~\mathrm{K})$), and ionic conductivities ($\sigma(300~\mathrm{K})$), covering both low-temperature (vacancy-mediated) and high-temperature (Frenkel defect-mediated) regimes.}
\label{tab:si_diff_values_all}

\begin{tabularx}{\textwidth}{l ccccc}
\hline\hline
\multirow{2}{*}{Structure} 
  & \multicolumn{5}{c}{\text{low-$T$ limit (vacancy-mediated)}} \\
\cline{2-6}
  & $n_{\mathrm{vac}}$
  & $E_{\mathrm{a}}$ (meV)
  & $D^0_{\mathrm{Li}}$ (cm$^2$/s)
  & $D_{\mathrm{Li}}(300~\mathrm{K})$ (cm$^2$/s)
  & $\sigma(300~\mathrm{K})$ (mS/cm) \\
\hline
bcc-Li        & $\frac{1}{432}~(2.3\times10^{-3})$             & $50\pm10$       & $(2.33\pm0.28)\times10^{-6}$ & $(3.33\pm0.40)\times10^{-7}$ & --- \\
\mbox{t-LLZO (no vac)}   & 0   & ---             & 0         & 0   & 0 \\
\mbox{t-LLZO (with vac)}  & $\frac{1}{56}~(0.018)$   & $116\pm12$     & $(1.34\pm0.18)\times10^{-5}$ & $(1.53\pm0.21)\times10^{-7}$ & $23.41\pm0.73$ \\
c-LLZO          & $\frac{120-56}{120}~(0.53)$           & $243\pm33$     & $(1.11\pm0.42)\times10^{-4}$ & $(9.23\pm3.56)\times10^{-9}$ & $1.44\pm0.13$ \\
Ga-LLZO        & $\frac{118-50}{118}~(0.58)$            & $200\pm55$     & $(1.91\pm1.23)\times10^{-5}$ & $(8.21\pm5.28)\times10^{-9}$ & $1.14\pm0.17$ \\ \hline

\multirow{2}{*}{Structure} 
  & \multicolumn{5}{c}{\text{high-$T$ limit (Frenkel defect-mediated)}} \\
\cline{2-6}
  & $n_{\mathrm{vac}}$
  & $E_{\mathrm{a}}$ (meV)
  & $D^0_{\mathrm{Li}}$ (cm$^2$/s)
  & $D_{\mathrm{Li}}(300~\mathrm{K})$ (cm$^2$/s)
  & $\sigma(300~\mathrm{K})$ (mS/cm) \\
\hline

bcc-Li        & $\frac{1}{432}~(2.3\times10^{-3})$             & $99\pm9$       & $(2.62\pm0.28)\times10^{-6}$ & $(5.74\pm0.61)\times10^{-8}$ & --- \\
\mbox{t-LLZO (no vac)}  & 0   & $237\pm11$     & $(2.33\pm0.30)\times10^{-4}$ & $(2.39\pm0.31)\times10^{-8}$ & $(3.72\pm0.12)$ \\
\mbox{t-LLZO (with vac)}    & $\frac{1}{56}~(0.018)$    & $240\pm21$     & $(5.17\pm1.23)\times10^{-4}$ & $(4.90\pm1.17)\times10^{-8}$ & $7.51\pm0.41$ \\
c-LLZO         & $\frac{120-56}{120}~(0.53)$           & $286\pm2$      & $(7.75\pm0.18)\times10^{-3}$ & $(1.22\pm0.03)\times10^{-7}$ & $19.0\pm0.10$ \\
Ga-LLZO        & $\frac{118-50}{118}~(0.58)$            & $381\pm15$     & $(5.53\pm0.95)\times10^{-3}$ & $(2.19\pm0.38)\times10^{-9}$ & $0.33\pm0.01$ \\
\hline\hline
\end{tabularx}
\end{table}

\section{MSD anisotropy}\label{seq:msd_aniso}

We calculated diffusion anisotropy factor as:

\begin{equation}
    A_{\mathrm{aniso}} = \frac{D_{\perp}}{D_{\parallel}} = \frac{D_{z}}{D_{xy}} = \frac{2 MSD_{z}}{MSD_{x} + MSD_{y}},
\end{equation}
where $D_{\perp}$ is the diffusion coefficient in the direction perpendicular to an interface, $D_{\parallel}$ is the diffusion coefficient in the interface plane, and $MSD_{i}$ is MSD in the direction $i$. 

\begin{figure}[H]
\begin{center}
    \includegraphics[width=0.99\columnwidth]{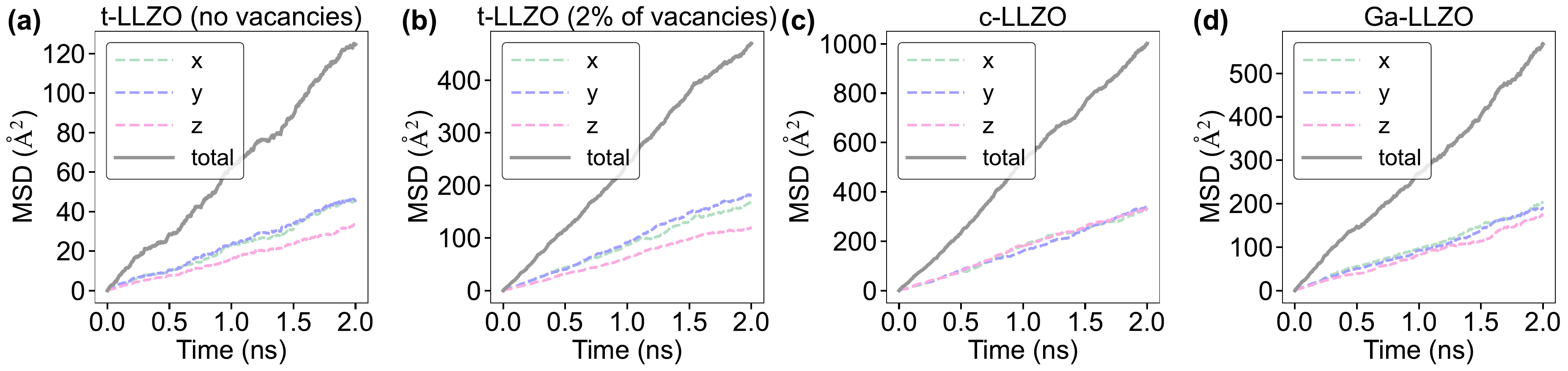}
\end{center}
\caption{Mean-squared displacements (MSD) at $T=1000$~K along the $x$, $y$, and $z$ directions, as well as the total MSD, for (a) t-LLZO without vacancies, (b) t-LLZO with 1.8\% vacancies, (c) c-LLZO, and (d) Ga-LLZO. }
\label{fig:msd_directions_llzo}
\end{figure}

\begin{figure}[H]
\begin{center}
    \includegraphics[width=0.85\columnwidth]{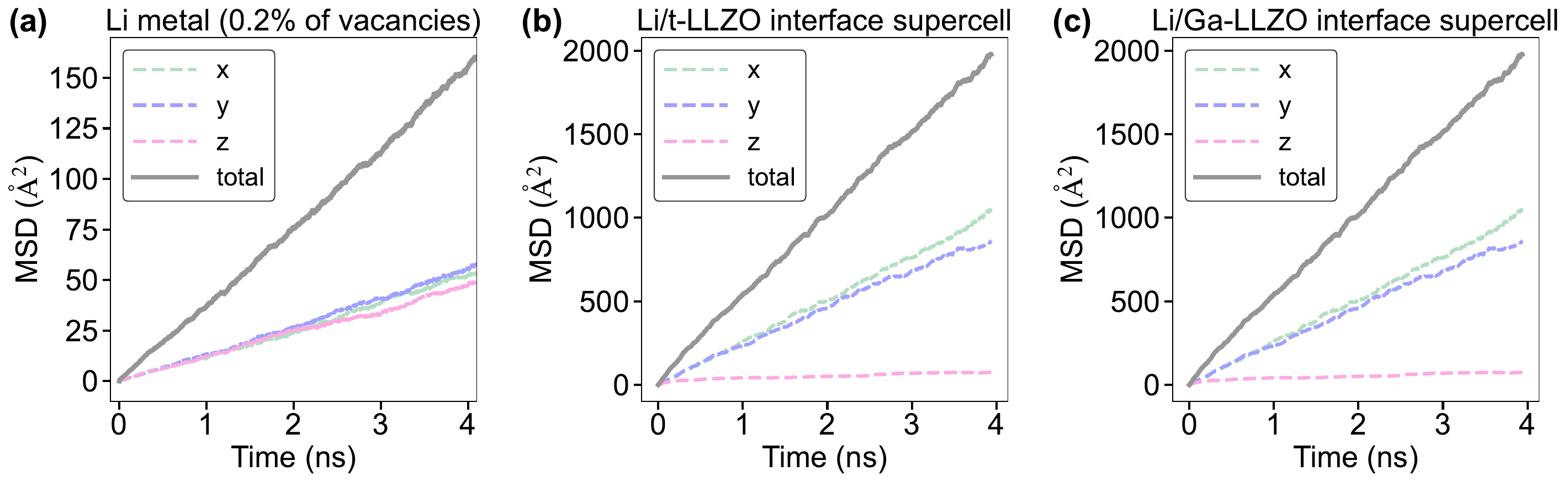}
\end{center}
\caption{Mean-squared displacements (MSD) at $T=400$~K along the $x$, $y$, and $z$ directions, as well as the total MSD, for (a) Li metal, (b) Li/t-LLZO interface supercell, and (c) Li/Ga-LLZO interface supercell. }
\label{fig:msd_directions_int}
\end{figure}

We calculated MSD in three directions for bulk phase and system with interface. The data for t-LLZO without vacancies and with 1.8\% vacancies, c-LLZO, and Ga-LLZO is shown in~\cref{fig:msd_directions_llzo} and for Li/t-LLZO and Li/Ga-LLZO interface supercells in~\cref{fig:msd_directions_int}. This factor is close to 1 for c-LLZO and Ga-LLZO, indicating nearly isotropic Li-ion diffusion. In contrast,  both t-LLZO structures with and without vacancies show $A_{\mathrm{aniso}} \approx 0.75$, reflecting slower diffusion along the $z$-direction than along $x$ and $y$, in agreement with previous MD simulations by Chen~\textit{et al.}~\cite{Chen2018}.

For the interfacial systems, $A_{\mathrm{aniso}}$ falls below 0.03, revealing strongly anisotropic transport due to rapid Li motion within the $xy$-plane. Diffusion normal to the interface is therefore much slower than diffusion within the interface plane.

\begin{figure}[H]
\begin{center}
    \includegraphics[width=0.5\columnwidth]{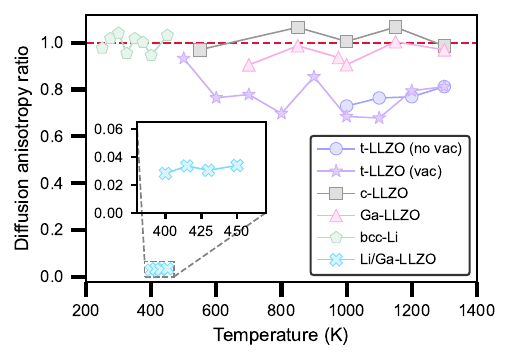}
\end{center}
\caption{Diffusion anisotropy factor ($A_\mathrm{aniso}$) as a function of temperature for the bulk phases (t-LLZO with and without vacancies, c-LLZO, Ga-LLZO, and bcc Li) and interface supercells (Li/t-LLZO and Li/Ga-LLZO). For the bulk phases, trajectories were included only if the total MSD exceeded $9$~\AA$^2$; for the interface supercells, the corresponding threshold was $100$~\AA$^2$. }
\label{fig:aniso_factor}
\end{figure}

The compiled data for anisotropy factors for all phases as a function of temperatures are provided in~\cref{fig:aniso_factor}.

\section{Charge transfer calculations}

\subsection{Number of jumps at the interface}\label{seq:num_jumps_methods}

We use a residence time, $t_{\mathrm{residence}}$, to avoid double-counting ion hops and calculate the total number of interface crossings. An ion is counted as having crossed the interface only if it remains in the LLZO or Li region for at least $t_{\mathrm{residence}}$; otherwise, such rapid back-and-forth rattling around the same site is not counted as contributing to the ionic flux. Ions, which cross interfaces and remain in either LLZO or Li region during $t_{\mathrm{residence}}$ are counted only once ($N_{\mathrm{int}}$). The transit ions, which crossed both interfaces are counted twice ($N_{\mathrm{transit}}$). 

Hence, the total number of ion jumps over simulation time $\Delta t$ is calculated as follows:

$$
    N_{\Sigma} = N_{\mathrm{int}} + 2 N_{\mathrm{transit}},
$$
where $N_{\Sigma}$ is the total crossings without double-counting, $N_{\mathrm{int}}$ is number of crossings that cross one interface only, $N_{\mathrm{transit}}$
is the number of through-slab transits, each contributes two crossings.

\begin{figure}[H]
\begin{center}
    \includegraphics[width=0.5\columnwidth]{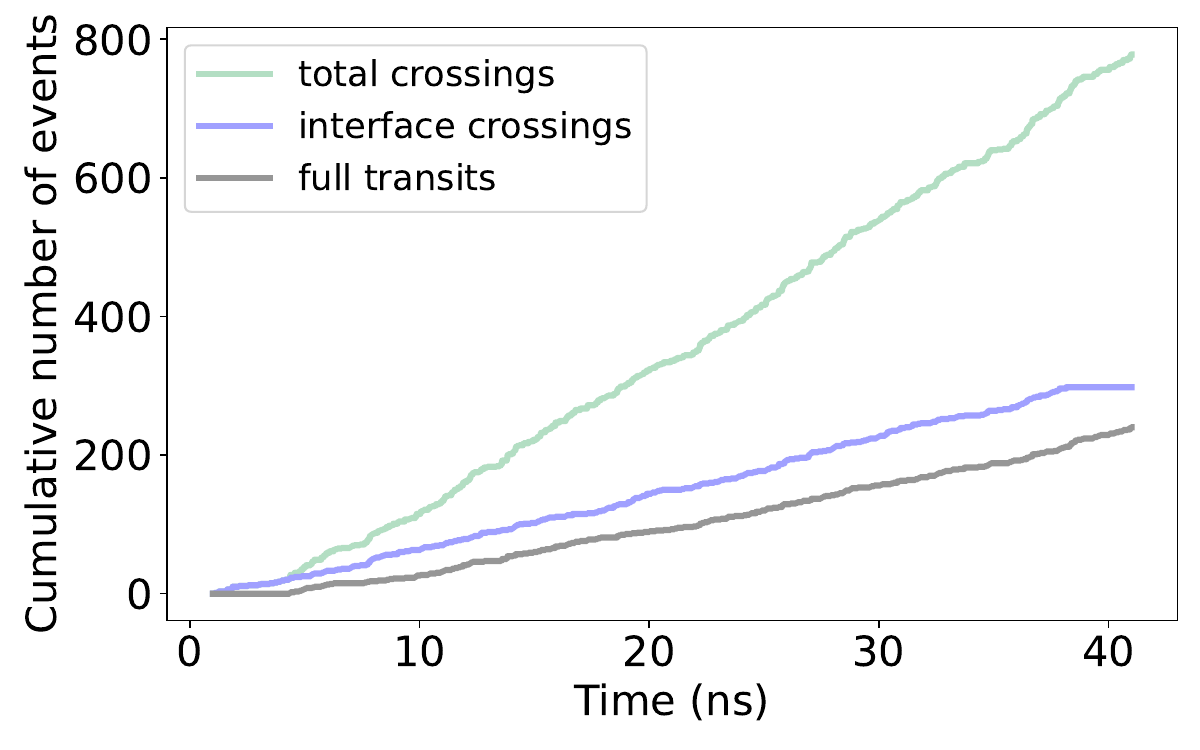}
\end{center}
\caption{Number of jumps across the Li/Ga-LLZO interface at 400~K. Green denotes the total number of crossings ($N_{\Sigma}$), blue denotes the number of interfacial crossings ($N_{\mathrm{int}}$,  counted once), and gray denotes the number of full transits ($N_{\mathrm{transit}}$, counted twice).}
\label{fig:crossing_example}
\end{figure}

The example of the number of $\ce{Li^{+}}$ jumps over the time is illustrated in~\cref{fig:crossing_example}.

\subsection{Current density}\label{seq:current_density_si}

Exchange-current density can be calculated from \cref{eq:number_of_jumps} as follows: 

$$
    j_z = \frac{eN_{\Sigma}}{2n_{\mathrm{int}}A\Delta t},
$$
where $A$ is the interface cross-sectional area, $n_{\mathrm{int}}$ is the number of interfaces included in the symmetry factor (2 in this work), and $\Delta t$ is the simulation time over which $N_{\Sigma}$ was accumulated.

\begin{figure}[H]
\begin{center}
    \includegraphics[width=0.55\columnwidth]{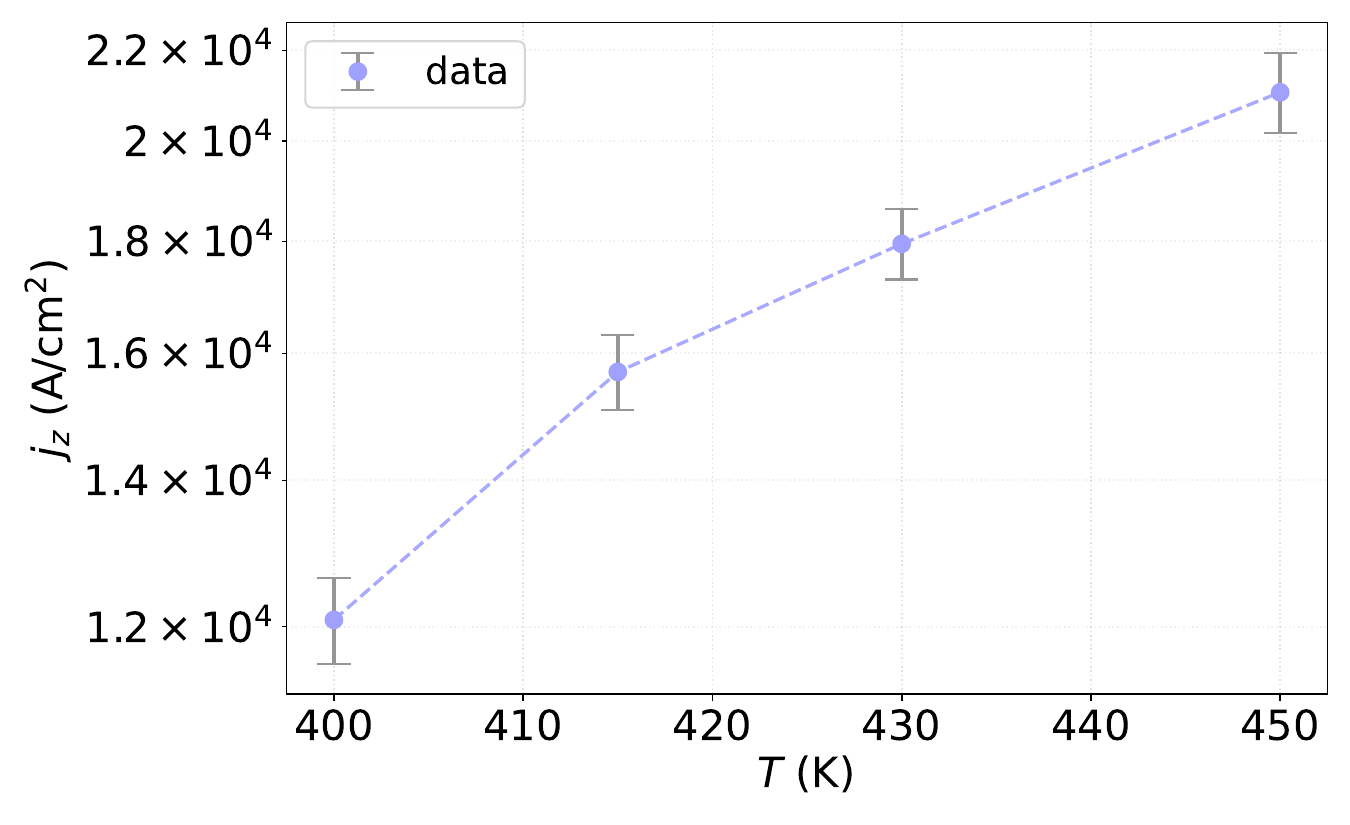}
\end{center}
\caption{Average current density, with error margins, as a function of 
temperature for the Li/Ga-LLZO interface supercell.}
\label{fig:current_density_over_t}
\end{figure}

The calculated average current density as a function of temperature 
is shown in~\cref{fig:current_density_over_t}.

We then used the same temperatures that provided sufficient jump statistics in the MSD analysis, from 400 to 450~K (see \cref{fig:int_cub_diff_block}), and fitted the current density as a function of $1/T$. For \cref{eq:current_density}, convergence with respect to the residence time is required to obtain a reliable fit. We therefore increased the residence time until the activation energy converged. The results are shown in~\cref{fig:min_residence_conv}, and the minimum required residence time was found to be 2.8~ns.

\section{Interface resistance derived from Butler--Volmer equation}\label{seq:interface_resistance}

To estimate the charge-transfer resistance from the activation energy, we use the Butler--Volmer equation:

\begin{equation}
     j = j_0 (T) \left[ \exp\left(\frac{\alpha_{\mathrm{a}} z F \eta}{RT}\right) - \exp\left(-\frac{\alpha_{\mathrm{c}} z F \eta}{RT}\right) \right] 
\end{equation}
where $j$ is electrode current density, $j_0$ is one-electron exchange current density, $T$ is absolute temperature, $z$ is number of electrons involved in electrode reaction, $F$ is Faraday constant, $R$ is universal gas constant, $\eta$ is activation overpotential, $\alpha_{\mathrm{a}}$ and $\alpha_{\mathrm{c}}$ are anodic and cathodic charge transfer coefficients, respectively.

At low overpotential limit $(|\eta| \ll RT/F)$, we can use Taylor expansion:
\begin{align*}
\exp\left(\frac{\alpha_\mathrm{a} z F \eta}{RT}\right) &= 1 + \frac{\alpha_\mathrm{a} z F \eta}{RT} + \frac{1}{2}\left(\frac{\alpha_\mathrm{a} z F \eta}{RT}\right)^2 + \cdots \\
\exp\left(-\frac{\alpha_\mathrm{c} z F \eta}{RT}\right) &= 1 - \frac{\alpha_\mathrm{c} z F \eta}{RT} + \frac{1}{2}\left(\frac{\alpha_\mathrm{c} z F \eta}{RT}\right)^2 - \cdots
\end{align*}

Substitute back:
\begin{align*}
j &= j_0 (T) \Bigg[ \left(1 + \frac{\alpha_\mathrm{a} z F \eta}{RT}\right) - \left(1 - \frac{\alpha_\mathrm{c} z F \eta}{RT}\right) + \mathcal{O}(\eta^2) \Bigg] = j_0 \frac{(\alpha_\mathrm{a} + \alpha_\mathrm{c}) z F \eta}{RT} + \mathcal{O}(\eta^2)
\end{align*}

Since $\alpha_\mathrm{a} + \alpha_\mathrm{c} = 1$ at microscopic reversibility:
\begin{equation}
j \approx j_0 (T) \frac{z F \eta}{RT} \quad \text{for} \quad |\eta| \ll \frac{RT}{z F}
\label{eq:bv_linear}
\end{equation}

Exchange current density for one-electron exchange reaction:
\begin{equation}
    \label{eq:current_density_si}
    j_0(T) = F k_{00} C_\mathrm{oxy}^{1-\beta} C_\mathrm{red}^\beta \exp\left(-\frac{E_\mathrm{a}}{k_\mathrm{B} T}\right)
\end{equation}
where $C_{\mathrm{oxy}}$ and $C_{\mathrm{red}}$ are the concentration of the oxidized and reduced species, respectively, $\beta$ is symmetry factor, $k_{00}$ is reaction rate constant, and $E_{\mathrm{a}}$ is activation energy of charge transfer for one ion at the interface in units of eV.

From Ohm's law, $j = \eta / R_\mathrm{ct}$ (where $\eta$ is the voltage drop across interface):

Substitute \cref{eq:bv_linear}:
$$ \frac{\eta}{R_\mathrm{ct}} = j_0 \frac{zF}{RT} \eta $$

Thus, we get the final equation:
\begin{equation*}
R_\mathrm{ct} = \frac{RT}{ z F j_0(T)} = \frac{RT}{ z F^2 k_{00} C_\mathrm{oxy}^{1-\beta} C_\mathrm{red}^\beta} \exp\left(\frac{E_\mathrm{a}}{k_\mathrm{B} T}\right)
\end{equation*}

\subsection{Calculated value of the rate constant}\label{sec:k00_from_calc}

Using \cref{eq:current_density_si} and data from \cref{fig:current_density_over_t}, we derived rate constant as:

\begin{equation*}
    k_{00} = \frac{j_0(T)}{F C_\mathrm{oxy}^{1-\beta} C_\mathrm{red}^\beta} \exp\left(\frac{E_\mathrm{a}}{k_\mathrm{B} T}\right) 
\end{equation*}

The calculated value equals $k_{00} = 339 \pm 166$~cm~s$^{-1}$.

\subsection{Experimental value of the rate constant}\label{sec:k00_from_experiment}

To obtain the experimental $k_{00}$, we use data from two works by Krauskopf~\textit{et al.}~\cite{krauskopf2020fast,krauskopf2019toward}, as few studies report activation energies, exchange currents, and charge transfer resistance for Li/LLZO interfaces. These works characterize $\ce{Li_{6.25}Al_{0.25}La_3Zr2O_{12}}$ interfaces, matching the dopant concentration of our Ga-LLZO model.

Using Butler--Volmer equations, the $k_{00}$ constant can be calculated from activation energy ($E_{\mathrm{a}}$) and charge transfer resistance ($R_{\mathrm{ct}}$) as follows:

\begin{equation}
    \label{eq:k00_from_ea}
    k_{00}(R_{\mathrm{ct}}) = \frac{RT}{z F^2 R_{\mathrm{ct}} C_\mathrm{oxy}^{1-\beta} C_\mathrm{red}^\beta } \exp{\left(\frac{E_{\mathrm{a}}}{k_{\mathrm{B}}T}\right)}
\end{equation}

\begin{table}[H]
\centering
\caption{Concentration of oxidizing and reducing species were derived from~\cref{tb:lat_const}.}
\label{tab:k00_exp_derived}
\begin{threeparttable}

\begin{tabular*}{0.7\columnwidth}{l @{\extracolsep{\fill}} ll}
\hline \hline
\textbf{Parameter} & \textbf{Value and Units} & Reference work \\
\hline
$z$ & $1$ & --- \\
$\beta$ & $0.5$ & --- \\

$E_{\mathrm{a}}$ & $0.37$~eV & Krauskopf, 2019~\cite{krauskopf2019toward} \\
$R_{\mathrm{ct}}$ & $0.08~\Omega~\mathrm{cm}^2$ & Krauskopf, 2020~\cite{krauskopf2020fast} \\
$C_{\mathrm{oxy}}^{\mathrm{Li^{+}}}$ & $38.0$~mmol cm$^{-3}$ &  Timusheva, 2025~\cite{timusheva2025chemical}  \\
$C_{\mathrm{red}}^{\mathrm{Li^{0}}}$ & $76.6$~mmol cm$^{-3}$ &  Feder, 1970~\cite{feder1970equilibrium} \\

\hline \hline
\end{tabular*}
\end{threeparttable}

\end{table}

To get $k_{00}$ from \cref{eq:k00_from_ea}, we had to use data both from two works: Krauskopf, 2019~\cite{krauskopf2019toward} and Krauskopf, 2020~\cite{krauskopf2020fast} as there is no available work on Li/LLZO interfaces, where both $E_{\mathrm{a}}$ and  $R_{\mathrm{ct}}$ are provided. However, both papers by Krauskopf use the same chemical composition, so this approach is viable. The calculated value is $k_{00}=102.1$~cm~s$^{-1}$.

\subsection{Charge transfer resistance for Li/LLZO interface from MLMD}

The oxidation and reduction reactions at Li/LLZO interface:
$$
\underbrace{\ce{Li^{+}_{(LLZO)} + e^{-}_{(interface)}}}_{\text{Oxidized species ($C_{\mathrm{oxy}}^{\mathrm{Li}^{+}}$)}} 
= 
\underbrace{\ce{Li^{0}_{(Li\ metal)}}}_{\text{Reducing species ($C_{\mathrm{red}}^{\mathrm{Li}^0}$)}}
$$

Using numerical values from \cref{tab:r_int_params_kJ}, we get:
\begin{multline*}
R_\mathrm{ct} = \frac{8.314~[\mathrm{J~mol^{-1}~K^{-1}}] \times 300~[\mathrm{K}]}{1\times 96485^2~[\mathrm{C}^2~\mathrm{mol}^{-2}]\times 3.39~[\mathrm{m~s}^{-1}]\times (3.73 \times 7.13 \times 10^{8} )^{1/2}~[\mathrm{mol~m}^{-3}] }  \\ \times
\exp{\left(\frac{0.167~[\mathrm{eV}]}{8.617 \times 10^{-5} [\mathrm{eV~K^{-1}}]\ \times 300~[\mathrm{K}]} \right)} = 1.01 \times 10^{-9}~\frac{\mathrm{J~s~m}^2}{\mathrm{C}^2}  = 1.01 \times 10^{-5}~\Omega~\mathrm{cm}^2
\end{multline*}

\begin{table}[H]
\centering
\caption{Parameters used to calculate the charge-transfer resistance at 300~K. Molar concentrations were derived from \cref{tb:lat_const}. Here, $z$ is the number of electrons transferred, $\beta$ is the symmetry factor, $R$ is the gas constant, $T$ is the absolute temperature, $F$ is the Faraday constant, and $k_{00}$ is the reaction rate constant. $C_{\mathrm{oxy}}^{\mathrm{Li}^{+}}$ and $C_{\mathrm{red}}^{\mathrm{Li}^{0}}$ are the concentrations of the oxidized and reduced species, respectively. $E_{\mathrm{a}}$ is the activation energy for charge transfer, and $j_0(300~\mathrm{K})$ is the exchange current density at 300~K. The values in parentheses are experimental values reported by Krauskopf~\textit{et al.}~\cite{krauskopf2019toward,krauskopf2020fast}. }
\label{tab:r_int_params_kJ}
\begin{threeparttable}

\renewcommand{\arraystretch}{1.15}

\begin{tabular*}{0.7\columnwidth}{l @{\extracolsep{\fill}} ll}
\hline \hline
\textbf{Parameter} & \textbf{Value} & \textbf{Units}   \\
\hline
\multicolumn{3}{c}{Kinetic metrics} \\ \hline
$j_{0}(300~\mathrm{K})$ & $2572~(0.3)$ &  A~cm$^{-2}$ \\ 
$k_{00}$ & $339~(102)$ &  cm~s$^{-1}$ \\ \hline
\multicolumn{3}{c}{Other values} \\ \hline
$z$ & $1$ & dimensionless \\
$\beta$ & $0.5$ & dimensionless \\
$R$ & $8.314$  & J~mol$^{-1}$~K$^{-1}$ \\
$T$ & $300$ & K \\
$F$ & $96485$ &  C mol$^{-1}$ \\
$k_{\mathrm{B}}$ & $8.617 \times 10^{-5}$  & eV~K$^{-1}$   \\
$C_{\mathrm{oxy}}^{\mathrm{Li^{+}}}$ & $37.3~(38.0)$ & mmol cm$^{-3}$ \\
$C_{\mathrm{red}}^{\mathrm{Li^{0}}}$ & $68.2~(76.6)$ &  mmol cm$^{-3}$ \\
$E_\mathrm{a}$ & $167~(370)$ &  meV \\ \hline

\multicolumn{3}{c}{Intermediate calculations} \\ \hline

$RT$ & $2494$ & J~mol$^{-1}$ \\
$k_\mathrm{B} T$ & $25.851 \times 10^{-3}$  & eV \\
$E_\mathrm{a}/k_{\mathrm{B}}T$ & $6.46$ &  dimensionless  \\
$\exp(E_\mathrm{a}/k_{\mathrm{B}}T)$ & $639.12$ &  dimensionless  \\
\hline

\multicolumn{3}{c}{Final value} \\ \hline
$R_\mathrm{ct}$ & $1 \times10^{-5}~(8 \times 10^{-2})$ &  $\Omega~\mathrm{cm}^2$  \\

\hline \hline
\end{tabular*}
\begin{tablenotes}
      \small
      \item Here, the reference value for $E_{\mathrm{a}}$ is taken from Krauskopf~\textit{et al.}~\cite{krauskopf2019toward}; $j_{0}(300~\mathrm{K})$ and $R_{\mathrm{ct}}$ are taken from the work~\cite{krauskopf2020fast}; and $k_{00}$ is derived from both works (see \cref{tab:k00_exp_derived}).
\end{tablenotes}

\end{threeparttable}

\end{table}

\section{Activation energy of charge transfer from effective activation energy for L\lowercase{i}/LLZO supercells }

\subsection{Derivation of Arrhenius diffusion across multiple regions}

For supercells containing Li/LLZO interfaces, $\ce{Li+}$ ions exhibit stationary motion along the $z$-direction (the diffusion direction). Consequently, the mean squared displacements in this direction are given by:

$$
MSD_z (t) \approx 2 D_{\text {eff }}(T) t
$$

We introduce response time as:

\begin{equation}
    \label{eq:response_time}
    \tau_j(T)=\tau_{0}^{j} \exp \left(\frac{E_{\mathrm{a}}^{j}}{k_{\mathrm{B}} T}\right),
\end{equation}
where ${E_{\mathrm{a}}^{j}}$ is activation energy of region $j$.

And effective diffusion energy is derived from the effective response time:
$$
D_{\mathrm{eff}}(T) = \frac{L^2}{2 \tau_{\mathrm{tot}}(T)},
$$

Or particularly for diffusion regions of Li, LLZO and effective:

\begin{equation}
    \label{eq:tau_introduced_regions}
    \tau_{\mathrm{Li}}(T)=\frac{l_{\mathrm{Li}}^2}{2 D_{\mathrm{Li}}(T)}, \quad \tau_{\mathrm{LLZO}}(T)=\frac{l_{\mathrm{LLZO}}^2}{2 D_{\mathrm{LLZO}}(T)}, \quad \tau_{\mathrm{eff}}(T)=\frac{L^2}{2 D_{\mathrm{eff}}(T)},
\end{equation}
where $l_{\mathrm{Li}}$ is the typical distance Li travels in the $z$-direction within the Li metal region (the thickness of the Li metal slab), $l_{\mathrm{LLZO}}$ is the typical distance within the LLZO region (the thickness of the LLZO slab), and $L$ is the total distance traveled in the $z$-direction during one charge transfer cycle (the thickness of the supercell).

Hence, we can introduce the total response times:

\begin{equation}
\tau_{\mathrm{tot}}(T) \approx \tau_{\mathrm{Li}}(T)+\tau_{\mathrm{ct}}(T)+\tau_{\mathrm{LLZO}}(T)
\end{equation}

In this approach, the effective energy is not a constant sum, rather, it is a temperature-dependent weighted average, with the weights given by the stage times themselves.

\begin{equation}
    \label{eq:ea_ct_introduced}
    E_{\mathrm{a}}^{\mathrm{eff}}(T)=k_{\mathrm{B}} \frac{d \ln \tau_{\mathrm{tot}}}{d(1 / T)}=\frac{\sum_j \tau_j(T) E_{\mathrm{a}}^{j}}{\sum_j \tau_j(T)}, \quad j \in\{\mathrm{Li}, \mathrm{ct}, \mathrm{LLZO}\}
\end{equation}

The charge transfer time per cycle is calculated as follows:

\begin{equation}
    \label{eq:tau_ct_introduced}
    \tau_{\mathrm{ct}}(T)=\tau_{\mathrm{eff}}(T) - \tau_{\mathrm{Li}}(T) - \tau_{\mathrm{LLZO}}(T) 
\end{equation}

Combining \cref{eq:ea_ct_introduced} with \cref{eq:tau_ct_introduced} yields:

$$
E_{\mathrm{a}}^{\mathrm{ct}}(T)=\frac{E_{\mathrm{a}}^{\mathrm{eff}} \tau_{\mathrm{eff}}-E_{\mathrm{a}}^{\mathrm{Li}} \tau_{\mathrm{Li}}-E_{\mathrm{a}}^{\mathrm{LLZO}} \tau_{\mathrm{LLZO}}}{\tau_{\mathrm{eff}}-\tau_{\mathrm{Li}}-\tau_{\mathrm{LLZO}}}
$$

Alternatively, by substituting $\tau_{j}(T)$ from \cref{eq:tau_introduced_regions} into the above equation, we obtain the final expression:

\begin{equation}
    \label{eq:e_a_ct_final}
    E_{\mathrm{a}}^{\mathrm{ct}}(T)=\frac{E_{\mathrm{a}}^{\mathrm{eff}}\,\frac{L^{2}}{2D_{\mathrm{eff}}}-E_{\mathrm{a}}^{\mathrm{Li}}\,\frac{l_{\mathrm{Li}}^{2}}{2D_{\mathrm{Li}}}-E_{\mathrm{a}}^{\mathrm{LLZO}}\,\frac{l_{\mathrm{LLZO}}^{2}}{2D_{\mathrm{LLZO}}}}{\frac{L^{2}}{2D_{\mathrm{eff}}}-\frac{l_{\mathrm{Li}}^{2}}{2D_{\mathrm{Li}}}-\frac{l_{\mathrm{LLZO}}^{2}}{2D_{\mathrm{LLZO}}}}
\end{equation}

If one of the stages is much faster than the other, then equation transforms to a simpler form. For instance, if the LLZO region hinders the $\ce{Li+}$-ion diffusion and the Li region is much faster, the equation transforms to the following:

$$
E_{\mathrm{a}}^{\mathrm{eff}} \approx E_{\mathrm{a}}^{\mathrm { LLZO }} 
$$

\subsection{Numerical estimations from our MLMD simulations}

We used the MLMD data summarized in~\cref{tab:ea_ct_estimations} to evaluate the sequential response times from \cref{eq:tau_introduced_regions,eq:tau_ct_introduced}. At $300$~K the LLZO transit already exceeds the whole-cell response time, $\tau_{\mathrm{LLZO}}>\tau_{\mathrm{eff}}$, so that \cref{eq:tau_ct_introduced} yields $\tau_{\mathrm{ct}}<0$. Within the sequential model, a rate-limiting charge-transfer stage would appear as a positive extra time, $\tau_{\mathrm{ct}}>0$, and would raise $\tau_{\mathrm{eff}}$ above $\tau_{\mathrm{Li}}+\tau_{\mathrm{LLZO}}$. The opposite inequality shows that interfacial charge transfer is not rate-limiting in this supercell. At the same time, $\tau_{\mathrm{ct}}<0$ places \cref{eq:e_a_ct_final} outside its domain of validity, so the construction cannot be used to extract a numerical charge-transfer activation energy. The comparison $\tau_{\mathrm{LLZO}}\gg\tau_{\mathrm{Li}}$ further indicates that, among the two bulk regions, $\ce{Li+}$ transport is limited by LLZO rather than by Li metal. The model assumes uncorrelated hops and strictly sequential stages, which is an additional approximation.

\begin{table}[H]
\centering
\caption{Parameters used to evaluate the sequential response times. The table includes: temperature ($T$); bulk-region (Li and LLZO) and effective activation energies ($E_{\mathrm{a}}^i$); diffusion coefficients at 300~K ($D_i(300~\mathrm{K})$); thicknesses of the Li region ($l_{\mathrm{Li}}$), LLZO region ($l_{\mathrm{LLZO}}$), and interface supercell ($L$). Intermediate calculations list the response times ($\tau_j$) for the Li, LLZO, and effective regions, and the derived charge-transfer time ($\tau_{\mathrm{ct}}$). Because $\tau_{\mathrm{ct}}<0$, \cref{eq:e_a_ct_final} is not used to report a charge-transfer activation energy. The reference data are given in Table~S22 and Figure~S38.}
\label{tab:ea_ct_estimations}
\begin{threeparttable}

\renewcommand{\arraystretch}{1.15}
\begin{tabular*}{0.7\columnwidth}{l @{\extracolsep{\fill}} ll}
\hline \hline
\textbf{Parameter} & \textbf{Value} & \textbf{Units} \\
\hline
$T$ & $300$ & K \\

$l_{\mathrm{Li}}$ & $15$ & \AA \\
$E_{\mathrm{a}}^{\mathrm{Li}}$ & $50$ & meV \\
$D_{\mathrm{Li}}(300~\mathrm{K})$ & $3.3 \times 10^{-7}$ & cm$^2$\,s$^{-1}$ \\

$l_{\mathrm{LLZO}}$ & $13.3$ & \AA \\
$E_{\mathrm{a}}^{\mathrm{LLZO}}$ & $200$ & meV \\
$D_{\mathrm{LLZO}}(300~\mathrm{K})$ & $8.2 \times 10^{-9}$ & cm$^2$\,s$^{-1}$ \\

$L$ & $28.3$ & \AA \\
$E_{\mathrm{a}}^{\mathrm{eff}}$ & $72$ & meV \\
$D_{\mathrm{eff}}(300~\mathrm{K})$ & $7.0 \times 10^{-8}$ & cm$^2$\,s$^{-1}$ \\ \hline

\multicolumn{3}{c}{Intermediate calculations} \\ \hline

$\tau_{\mathrm{Li}}$ & $3.41 \times 10^{-8}$ &  s  \\
$\tau_{\mathrm{LLZO}}$ & $1.08 \times 10^{-6}$ &  s  \\
$\tau_{\mathrm{eff}}$ & $5.72 \times 10^{-7}$ &  s  \\
$\tau_{\mathrm{ct}}$ & $-5.41 \times 10^{-7}$ &  s  \\

\hline \hline
\end{tabular*}
\end{threeparttable}

\end{table}

\clearpage
\section{Ensemble-averaged vacancy formation energies}\label{seq:vac_en_ensemble}

\subsection{Static calculations without relaxation of atomic position}

To show the consistency of vacancy formation energies at different MLMD snapshots, we calculated relative vacancy formation energies.

\begin{figure}[H]
\begin{center}
    \includegraphics[width=0.6\columnwidth]{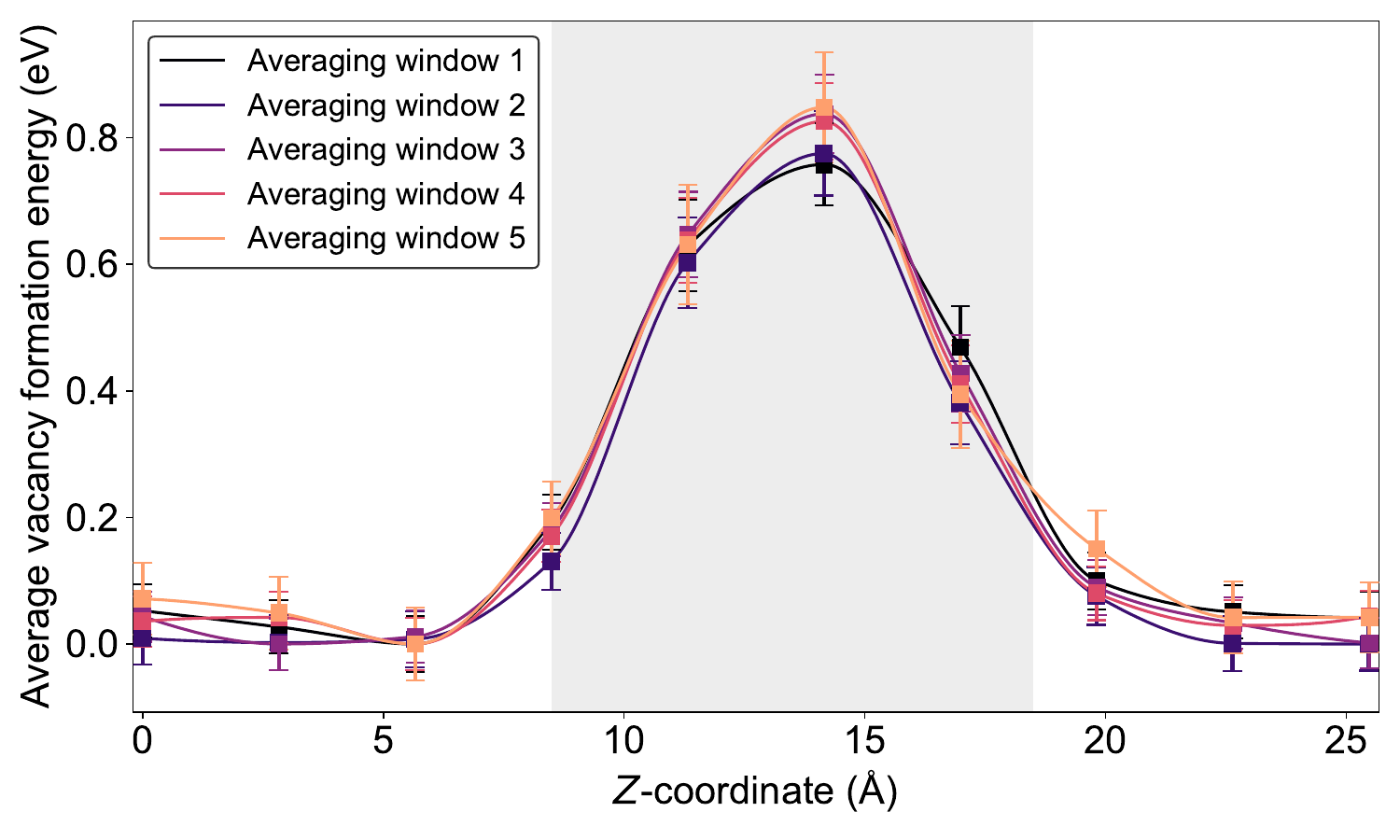}
\end{center}
\caption{Averaged lithium vacancy formation energies without relaxation at each timestep during MLMD simulations, shown at 0\%, 25\%, 50\%, and 100\% progress along the trajectory. The gray shaded area corresponds to the LLZO region. The bars represent the standard deviation within each data bin, and the filled squares denote the averaged values within each bin.}
\label{fig:vac_en_av_no_relax}
\end{figure}

\begin{figure}[H]
\begin{center}
    \includegraphics[width=0.6\columnwidth]{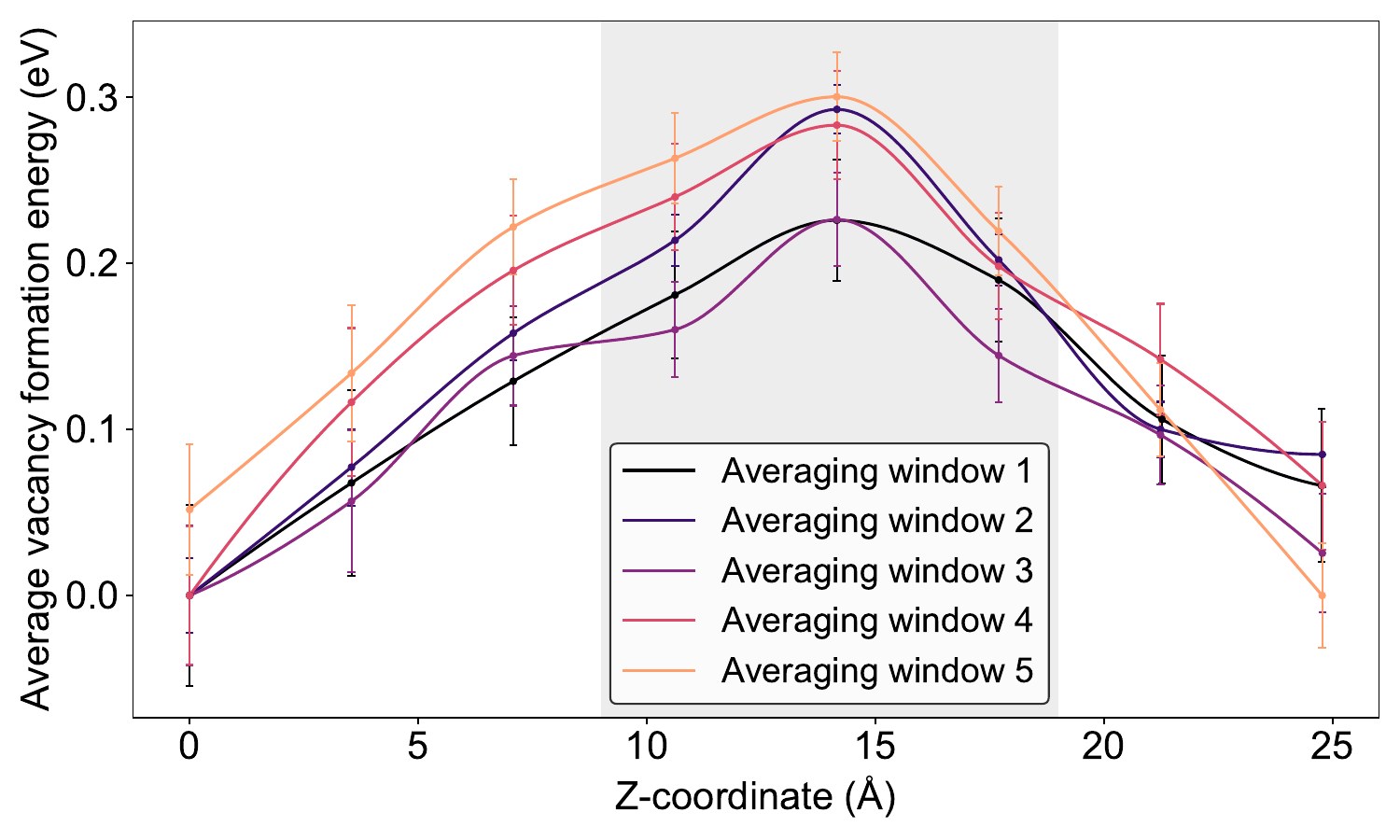}
\end{center}
\caption{Averaged lithium vacancy formation energies with relaxation at each timestep during MLMD simulations, shown at 0\%, 25\%, 50\%, and 100\% progress along the trajectory. The gray shaded area corresponds to the LLZO region. The bars represent the standard deviation within each data bin, and the filled squares denote the averaged values within each bin.}
\label{fig:vac_en_av_relax}
\end{figure}

\subsection{Ensemble-averaged for the whole trajectory}

\begin{figure}[H]
\begin{center}\includegraphics[width=0.92\textwidth]{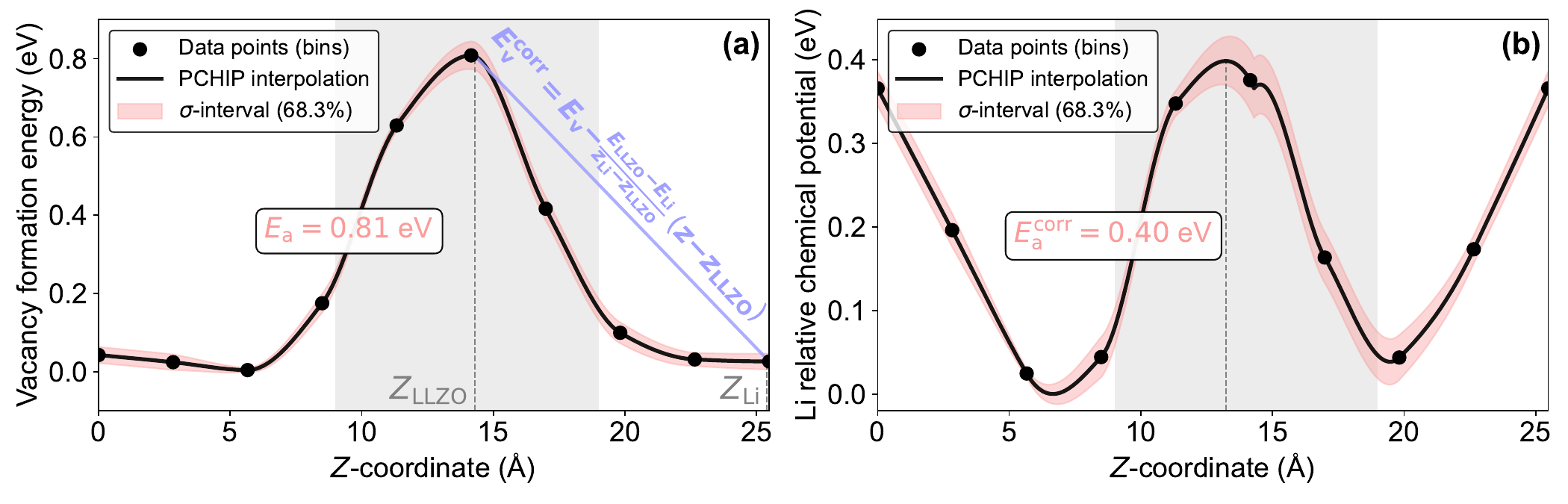}
\end{center}
\caption{(a) Vacancy formation energies, calculated with MTP and averaged over 200 snapshots for all Li within structure. Black circles are values averaged among data bin. Red shaded area is standard deviation error. (b) Relative chemical potentials of lithium in supercell with Li/Ga-LLZO interface.} 
\label{fig:vac_energies_all}
\end{figure}

We estimated the charge-transfer activation energy using static vacancy formation energies obtained from MTP calculations. To this end, we first computed the relative chemical potential of Li, $\mu_{\mathrm{Li}}^{\mathrm{rel}}$, by averaging the Li vacancy formation energies. Specifically, we sampled 200 snapshots uniformly from the linear regime of the MSD trajectory at 400~K in~\cref{fig:int_cub_diff_block}.b and performed single-point MTP calculations for every Li atom in the supercell. The resulting values were then averaged into 14 bins along the $z$-coordinate, as shown in~\cref{fig:vac_energies_all}.a.

We then applied a correction to account for the non-constant potential condition. Under operating conditions, the electrochemical potential of Li in the LLZO region and at the anode should be equal, and the interfacial energy barrier arises from the electric double layer. In this work, we computed the ensemble-averaged vacancy formation energy, $\left\langle E_{\mathrm{v}} \right\rangle$, which is proportional to the average electrochemical potential of Li, $\left\langle \mu_{\mathrm{Li}} \right\rangle$. Therefore, we applied a linear correction to $E_{\mathrm{v}}$, aligning $E_{\mathrm{v}}$ at the LLZO and Li reference sites and correcting intermediate values in proportion to their distance from the center of the LLZO region:

\begin{equation}
\label{eq:energy_correction}
\Delta E_{\mathrm{v}}^{\mathrm{corr}} =
\begin{cases}
0, & z = z_{\mathrm{LLZO}}, \\[4pt]
\dfrac{E_{\mathrm{LLZO}} - E_{\mathrm{Li}}}{z_{\mathrm{Li}} - z_{\mathrm{LLZO}}}
\left(z - z_{\mathrm{LLZO}}\right), &
z_{\mathrm{LLZO}} < z < z_{\mathrm{Li}}, \\[6pt]
0, & z = z_{\mathrm{Li}}.
\end{cases}
\end{equation}
where $z$ is the coordinate along the $z$-direction in the supercell, $z_{\mathrm{LLZO}}$ and $z_{\mathrm{Li}}$ denote the $z$-coordinates of the LLZO and Li region centers, respectively, and $E_{\mathrm{LLZO}}$ and $E_{\mathrm{Li}}$ are the corresponding average vacancy formation energies.

Next, we inverted the relation to obtain the averaged chemical potential of Li; the results are shown in~\cref{fig:vac_energies_all}.b. As a result, the activation energy in the interfacial region decreases from $0.81~\mathrm{eV}$ ($E_{\mathrm{a}}$) to $0.40~\mathrm{eV}$ ($E_{\mathrm{a}}^{\mathrm{corr}}$) after applying the correction. However, this correction is based on a simplified model, whereas the actual electrochemical potential of Li should be determined under constant-potential conditions~\cite{bonnet2012first}.

If we compare $E_{\mathrm{a}}^{\mathrm{corr}} = 0.4~\mathrm{eV}$ with the charge-transfer activation energy obtained from the number of jumps, $E_{\mathrm{ct}} = 0.17~\mathrm{eV}$ (see \cref{seq:current_density_si}), the former is substantially larger. This shows that the approach based on the ensemble-averaged chemical potential of Li is not applicable, because electron interactions are not captured in this model: Li is extracted together with its electron.

A similar overestimation is also observed in DFT-NEB calculations~\cite{burov2024mechanism}, where the migration of $\ce{Li^{0}}$ is enforced together with its electron. As a result, electron-charge redistribution at the LLZO interface is not properly accounted for, despite the reduction of LLZO. In our work, this effect is evident from the electron charge-density difference and the effective Bader charges.

\section{Estimation of the potential of mean force for interface supercells}\label{sec:pmf_interface}

The charge transfer resistance ($R_{\mathrm{ct}}$) calculated under constant-charge conditions represents a lower bound for the intrinsic resistance. To assess the possible correction associated with constant-potential boundary conditions at zero overpotential, we calculated the potential of mean force (PMF) from the Li density distribution within the interface Li/Ga-LLZO supercell. The equilibrium Li density along the stacking direction of the Li/LLZO/Li heterojunction encodes the thermodynamic driving force for interfacial transfer: Li is depleted where the local free energy is high and accumulates where it is low. Inverting the time-averaged number density yields a one-dimensional PMF, referenced separately to the bulk-metal and bulk-electrolyte reservoirs. Thus, $\Delta F(z)$ measures the excess free energy of Li relative to the coexisting phase at the same position $z$. Importantly, this profile is an equilibrium quantity and, unlike an Arrhenius barrier extracted from hop counts, does not by itself determine the kinetic rate of interface crossing. 

To obtain the one-dimensional PMF along the stacking axis, we inverted the time-averaged number density $\rho(z)$ relative to a piecewise-constant, phase-local reference $\rho_{\mathrm{ref}}(z)$:
\begin{equation}
  \Delta F(z)=-k_{\mathrm{B}}T\ln\bigl[\rho(z)/\rho_{\mathrm{ref}}(z)\bigr],
  \label{eq:deltaF_pmf}
\end{equation}
where $k_{\mathrm{B}}$ is Boltzmann's constant, $T$ is the temperature, $\rho(z)$ is the time-averaged Li number density, and $\rho_{\mathrm{ref}}(z)$ is the mean Li density of the coexisting bulk phase at the same $z$ (Li metal or LLZO).

\begin{figure}[H]
\centering
\includegraphics[width=0.50\textwidth]{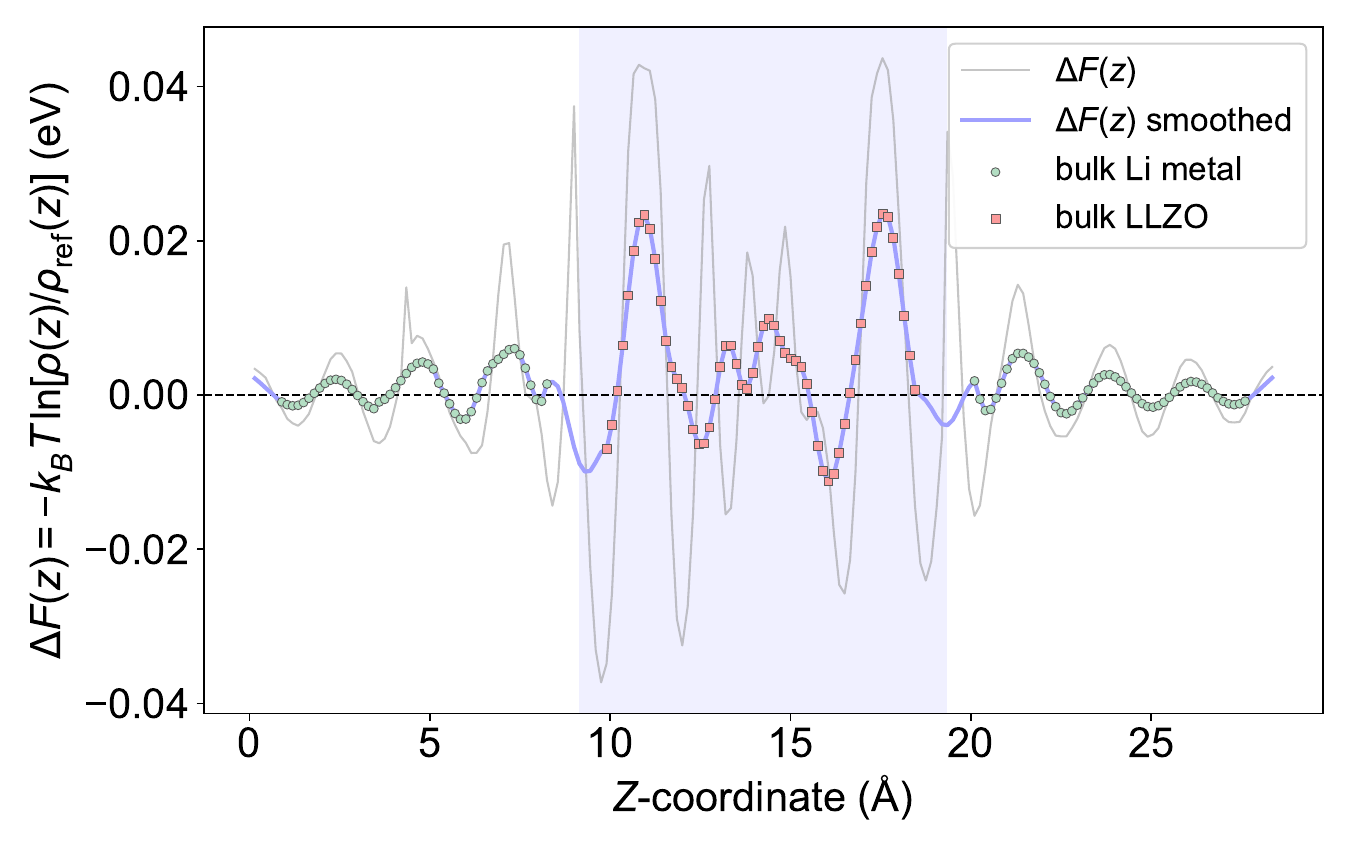}
\caption{One-dimensional potential of mean force (PMF, $\Delta F(z)$) across the supercell containing Li(100)/Ga-LLZO(100) interface, averaged over the full MLMD trajectory at $T = 450$~K and computed from \cref{eq:deltaF_pmf}. The raw PMF profile is shown in gray,  with the smoothed profile overlaid as a blue solid line. The green and pink markers denote the bulk Li metal and bulk LLZO regions, respectively, while the shaded area highlights the LLZO region.}
\label{fig:vac_gibbs_en}
\end{figure}

\cref{fig:vac_gibbs_en} shows $\Delta F(z)$ at $T=450~\mathrm{K}$ obtained in this way from the MLMD trajectory of Li/Ga-LLZO. The density $\rho(z)$ was histogrammed in bins of width $\Delta z=0.15~\mathrm{\AA}$. The LLZO slab was located from the smoothed La--Zr--O framework density (threshold $25\%$ of the maximum), wrapped so that the LLZO region lies at the center of the supercell. The reference $\rho_{\mathrm{ref}}(z)$ was set to the mean Li density in those windows ($\rho_{\mathrm{Li}}^{\mathrm{metal}}\approx 5.08\times 10^{-2}~\mathrm{\AA}^{-3}$, $\rho_{\mathrm{Li}}^{\mathrm{LLZO}}\approx 2.18\times 10^{-2}~\mathrm{\AA}^{-3}$). The bold curve is an averaging inside $11$ bins. Maxima at the interfaces correspond to Li depletion relative to $\rho_{\mathrm{ref}}$, and minima to accumulation.

The PMF exhibits similar maxima at the interface region to those obtained from static and ionically optimized MTP-calculated vacancy formation energies (see \cref{fig:vac_int_cub}). However, the barriers are substantially lower, with values of approximately 0.03~eV, compared to 0.17~eV from the residence-time window calculations. This suggests that corrections due to constant-potential conditions are likely to be of a similar order of magnitude as the PMF. Furthermore, this thermodynamic driving force suggests that kinetic barriers are also an intrinsic characteristic arising from the differing Li densities at Li and LLZO sites, rather than solely from migration between sites. Accordingly, the calculated $R_{\mathrm{ct}}$ should be interpreted as a lower-bound estimate of the intrinsic interfacial resistance under equilibrium conditions, rather than as a prediction applicable to high-rate or strongly non-equilibrium operating conditions.

\cref{eq:deltaF_pmf} should not be used as a standalone measure of interfacial Li transport and should be rather considered as a feasible correction to the constant-charge conditions, as $\Delta F(z)$ is an equilibrium PMF obtained from a one-dimensional occupancy projection: it indicates where Li is thermodynamically favored along the stacking axis, but does not determine the rate or mechanism of crossing. Regions with low density ($\rho \to 0$) are poorly sampled, making apparent barriers noisy, truncated, or dominated by the numerical floor on $\rho$. Additionally, the piecewise-constant $\rho_{\mathrm{ref}}(z)$ and the framework-density threshold used to assign bulk windows mix interfacial bins into the two reservoirs. Lateral heterogeneity, Ga-related defects, and distinct vacancy vs. interstitial pathways are averaged out, and $\Delta F(z)$ is not a vacancy-formation Gibbs energy. Kinetic quantities, such as hop counts with residence-time criteria, Arrhenius activation energies, and explicit transition-state searches are therefore essential to establish whether the thermodynamic profile controls Li transfer across the Li/Ga-LLZO interface. Finally, the raw data were bin-averaged and heavily smoothed to obtain a profile with small deviations, which may obscure some local effects.

\clearpage

\section{L\lowercase{i}-ion probability density}\label{seq:s_prob_density} 

\begin{figure}[H]
\begin{center}
    \includegraphics[width=0.99\columnwidth]{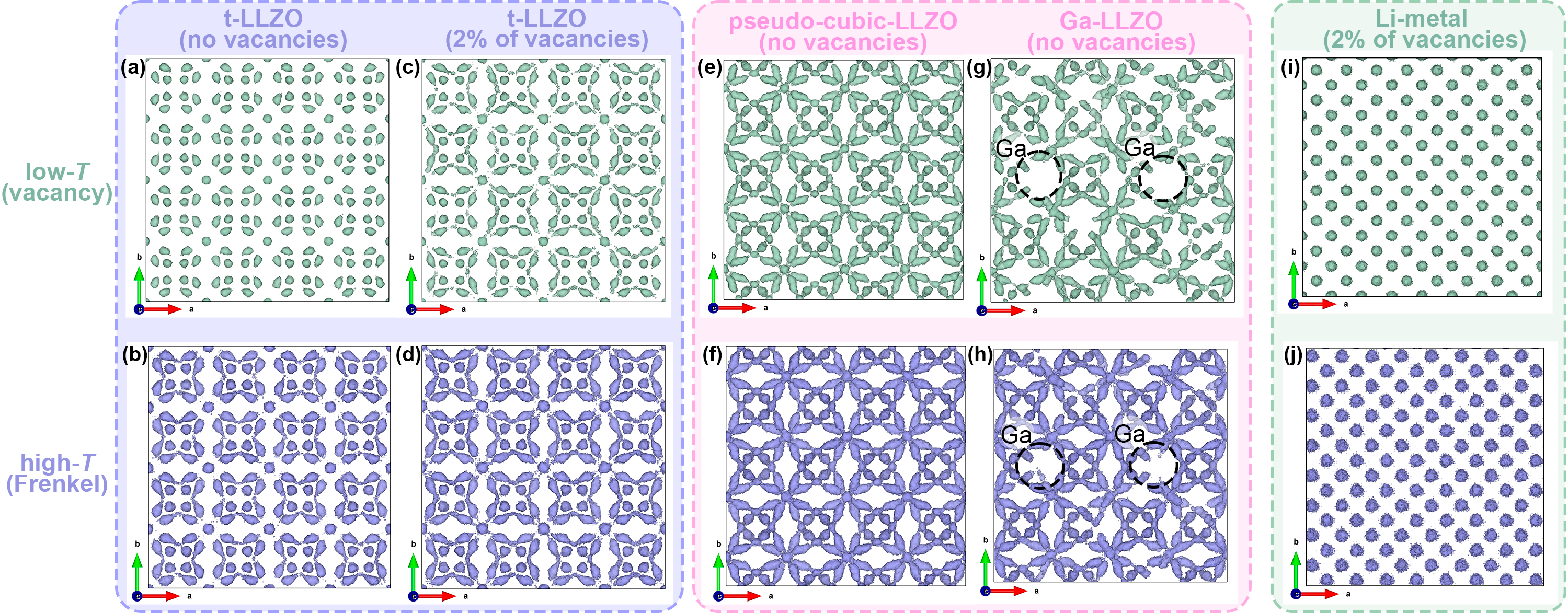}
\end{center}
\caption{Li-ion probability density for the low-temperature and high-temperature regimes. Isosurface level was of $5 \cdot 10^{-4}$ $r_{\mathrm{Bohr}}^{-3}$. (a), (b) t-LLZO without vacancies; (c), (d) t-LLZO with one vacancy per conventional cell (1/56); (e), (f) c-LLZO without non-intrinsic vacancies; (g), (h) Ga-LLZO without non-intrinsic vacancies;  (i), (j) bcc Li with one vacancy in the simulation cell (0.2\%, 1/432). }
\label{fig:li_prob_bulk}
\end{figure}

We calculated Li-ion probability density to visualize the dominant positions, which are mostly occupied during MLMD simulations. Here, we considered both low-temperature regime and high-temperature regimes. The results are shown in~\cref{fig:li_prob_bulk}.

For each structure, the probability density is localized at the lattice sites in the low-temperature regime. As the temperature increases, an increasing fraction of the probability density redistributes toward interstitial sites. In addition, some of the density in the migration channels spreads perpendicular to the migration pathways, indicating that Li ions can deviate from the primary migration pathways and move closer to the edges of the channels. This effect is particularly pronounced in t-LLZO, as shown in panels (a), (b), (c), and (d).

The analysis of the Li-ion probability density thus reveals that Li ions preferentially occupy interstitial sites in the high-temperature regime, a behavior not observed at low temperatures. This finding is consistent with the interstitial analysis presented in~\cref{sec:s_diffusion_mechanism}.

\clearpage

\section{L\lowercase{i}-ion diffusion mechanism in LLZO phases}\label{sec:s_diffusion_mechanism}

\subsection{Geometrical and crystal structure descriptors}

At room temperature, the dominant Li-ion self-diffusion mechanism in solid-state batteries should be vacancy-mediated, where an ion hops into a neighboring vacancy. However, at high temperatures, the dominant self-diffusion mechanism can switch to the Frenkel defect-mediated, where a cation leaves a lattice site and occupies an interstitial position~\cite{tufail2023design}.

\begin{figure}[H]
\begin{center}
    \includegraphics[width=0.99\columnwidth]{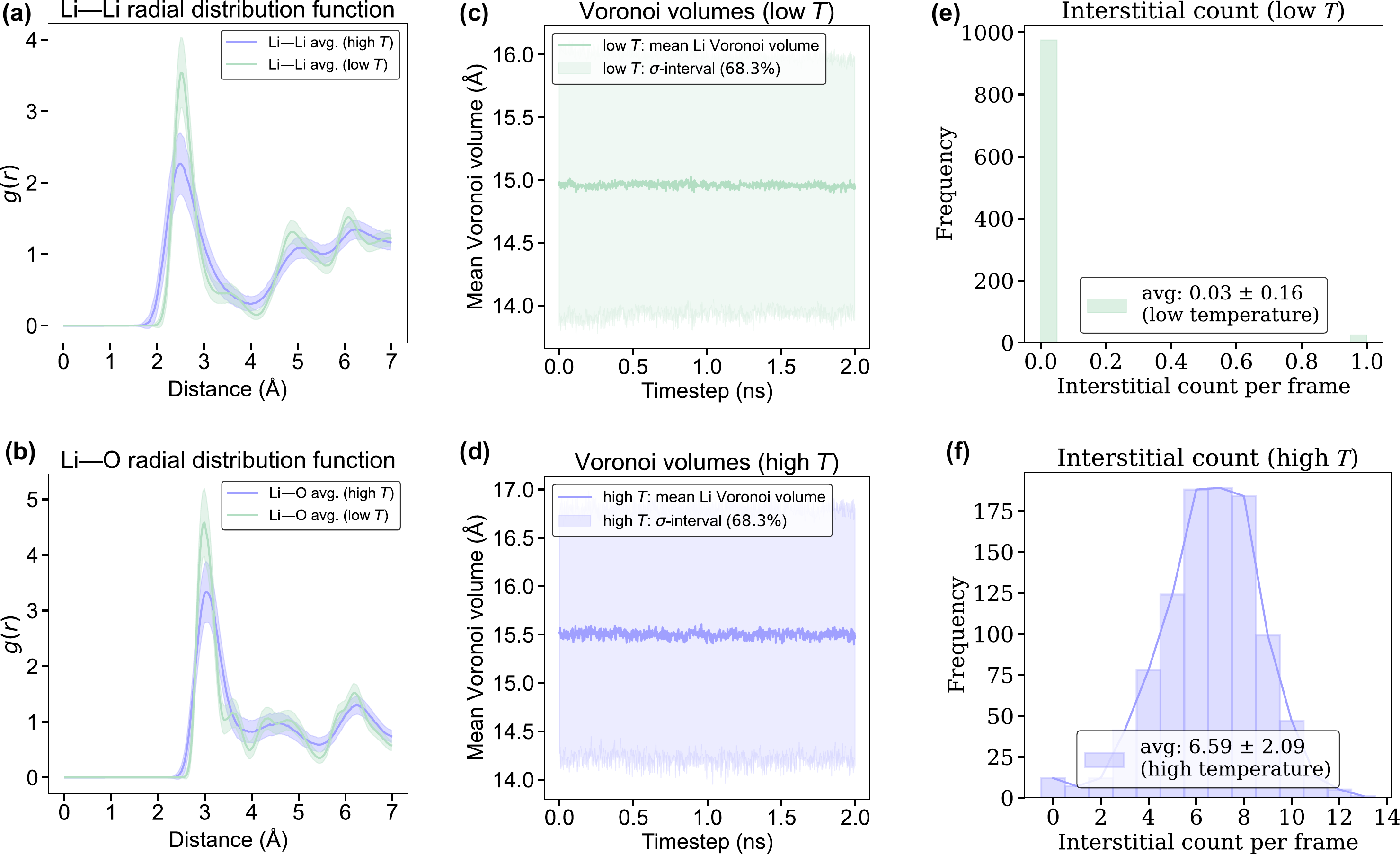}
\end{center}
\caption{Analysis of the diffusion mechanism for t-LLZO at 400~K (low-$T$ regime, vacancy mechanism) and 1000~K (high-$T$ regime, interstitial mechanism). Here, we present the radial distribution functions (RDFs) for Li--Li and Li--O distances, evaluated up to a cut-off radius of 7~\AA. Mean Voronoi volumes, calculated with the poly-disperse tessellation method. Number of interstitials, calculated with the Wigner--Seitz method. All values were averaged over simulation time and standard deviation energies were visualized. (a) RDFs for Li--Li distances; (b) RDFs for Li--O distances; (c) mean Voronoi volumes for the low-temperature regime; (d) mean Voronoi volumes for the high-temperature regime; (e) number of interstitials for the low-temperature regime; (f) number of interstitials for the high-temperature regime.   }
\label{fig:diff_type}
\end{figure}

To distinguish two Li-ion self-diffusion mechanisms, we first calculated radial distribution function (RDF) for Li--Li and Li--O distances, averaged over simulation time. The cut-off radius was set to 7~\AA. The results for t-LLZO at 400~K (low-temperature regime, vacancy mechanism) and 1000~K (high-temperature, interstitial mechanism) are shown in~\cref{fig:diff_type}. The broadening of peaks for Li--Li and Li--O distances are observed for high-temperature regime. The lattice sites are distinct with strongly pronounced peaks, whereas the peaks for interstitials have a broadening. The peaks with high intensities are provided in~\cref{tb:rdf_top_peaks}. The positions of peaks for both regimes are located at the same distances, though the intensities for high-temperature regimes are lower.

Second, we calculated mean Voronoi volumes, using poly-disperse tessellation method and averaging over simulation time. The data on mean values are provided in~\cref{tab:diff_type_param}. The mean Voronoi volume for the low-temperature regime is $(14.96 \pm 1.04)$~\AA$^3$, whereas for the high-temperature regime it is $(15.50 \pm 1.30)$~\AA$^3$. At lattice sites, the Voronoi volumes should be lower than that for interstitial sites.

Third, the Wigner--Seitz method was employed to calculate the number of interstitials, as implemented in OVITO software program~\cite{stukowski2009visualization}. The values were averaged over simulation time and provided in~\cref{tab:diff_type_param}. The mean number of interstitials for the low-temperature regime is $(0.03 \pm 0.16)$, whereas for the high-temperature regime it is $(6.59 \pm 2.09)$.

Combining three methods, one can conclude that the high-temperature regime has interstitials, which were absent in the low-temperature regime. As a result, the presence of Frenkel defects should affect the self-diffusion mechanism, which was vacancy-mediated for the low-temperature regime.

\begin{table}[H]
\centering
\caption{Highest peaks for Li--O and Li--Li radial distribution functions (RDFs) for the low-temperature and high-temperature regimes of diffusion. Peak positions are in units of \AA, intensities and their standard deviations are provided.}
\label{tb:rdf_top_peaks}
\setlength{\tabcolsep}{12pt} 
\renewcommand{\arraystretch}{1.3} 
\begin{tabular}{@{} c c c c c @{}}
\hline\hline
\multicolumn{2}{l}{Diffusion mechanism} & Peak & Position (\AA) & Intensity $\pm \: \sigma_{\mathrm{std}}$ \\
\hline
\multirow{8}{*}{\shortstack{Low-temperature regime \\ (vacancy mechanism)}} 
  & \multirow{5}{*}{Li--O} 
    & 1 & 2.992 & 4.584 $\pm$ 0.613 \\
  &  & 2 & 6.177 & 1.526 $\pm$ 0.166 \\
  &  & 3 & 3.587 & 1.165 $\pm$ 0.263 \\
  &  & 4 & 4.357 & 1.064 $\pm$ 0.198 \\
  &  & 5 & 4.742 & 1.031 $\pm$ 0.174 \\
\cmidrule(lr){2-5}
  & \multirow{3}{*}{Li--Li} 
    & 1 & 2.502 & 3.539 $\pm$ 0.492 \\
  &  & 2 & 6.072 & 1.521 $\pm$ 0.137 \\
  &  & 3 & 4.847 & 1.311 $\pm$ 0.167 \\
\midrule

\multirow{6}{*}{\shortstack{High-temperature regime \\ (Frenkel defect mechanism)}} 
  & \multirow{3}{*}{Li--O} 
    & 1 & 3.027 & 3.335 $\pm$ 0.548 \\
  &  & 2 & 6.247 & 1.297 $\pm$ 0.158 \\
  &  & 3 & 4.567 & 0.974 $\pm$ 0.193 \\
\cmidrule(lr){2-5}
  & \multirow{3}{*}{Li--Li} 
    & 1 & 2.502 & 2.268 $\pm$ 0.425 \\
  &  & 2 & 6.212 & 1.346 $\pm$ 0.129 \\
  &  & 3 & 5.092 & 1.091 $\pm$ 0.144 \\
\hline\hline
\end{tabular}

\end{table}

\begin{table}[H]
\centering
\caption{Mean Voronoi volumes and the mean number of interstitials during the simulation for the low-temperature and high-temperature regimes. The values were averaged over simulation time. }
\label{tab:diff_type_param}
\renewcommand{\arraystretch}{1.15} 
\begin{tabular}{lll}
\hline\hline
Calculated value & low-$T$ & high-$T$ \\ \hline
Mean Voronoi volume (\AA$^{3}$) &  $14.96 \pm 1.04$     &    $15.50 \pm 1.30$    \\
Mean number of interstitials &  $0.03 \pm 0.16$  &     $6.59 \pm 2.09$   \\ \hline\hline
\end{tabular}
\end{table}

\subsection{Cooperative migration from the connected graph}\label{sec:connected_graph}

To alternatively quantify cooperative migration, we adopted the method used by Dembitskiy~\textit{et al.}~\cite{dembitskiy2025new}. This approach analyzes simultaneous ion hops by constructing a connected graph between jumping ions over time intervals $\Delta t$.

First, each ion is assigned to its nearest lattice site in a static reference configuration. An ion is classified as a ``hopper'' if its site changes between frames. A neighbor list is then computed for both reference and current configurations using a cut-off radius $r_\mathrm{c} = 3$~\AA. Edges connect hoppers that remain in contact across frames, forming an undirected graph where connected components represent cooperative hop events.

The size of each component (number of vertices) gives the number of ions in a single cooperative hop. Smoothing trajectories with a moving average reduces noise before applying this graph analysis every $t_\mathrm{step}$ frames. Counting components by size across the trajectory yields statistics on cooperative jumps: events involving exactly $n$ simultaneous ions ($n \ge 1$) and total hopping events.

\begin{figure}[H]
\begin{center}\includegraphics[width=0.99\columnwidth]{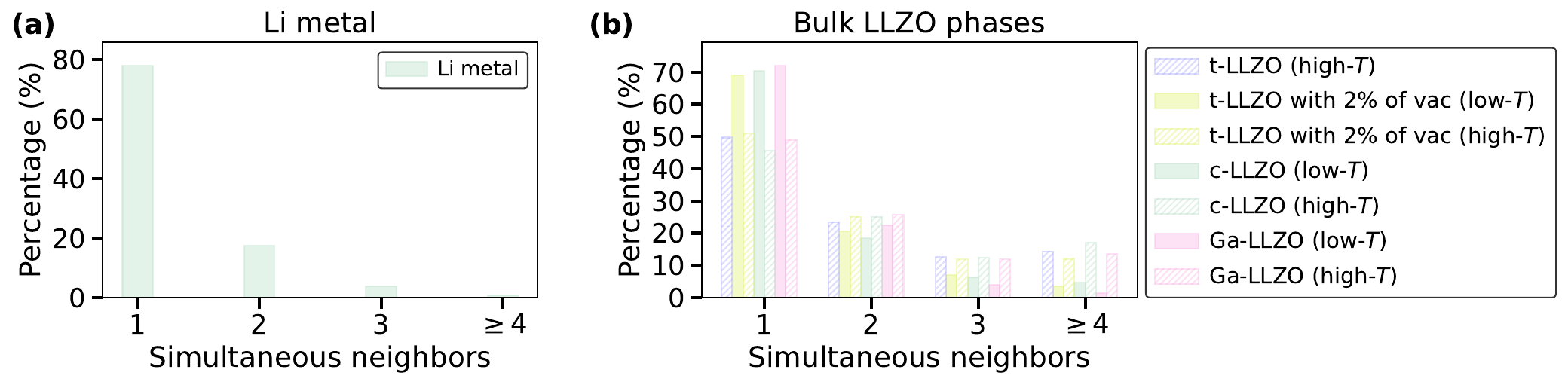}
\end{center}
\caption{Percent of cooperative jumps at low temperature (vacancy-mediated) and high temperature (Frenkel defect-mediated): (a) bcc Li with only low temperature regime at temperature in 250--450~K; (b) t-LLZO without and with 1.8\% vacancies, c-LLZO, Ga-LLZO at low temperature  ($T=600$~K) and high-temperature ($T=1000$~K) regimes. } 
\label{fig:diff_artem}
\end{figure}

\begin{figure}[H]
\begin{center}\includegraphics[width=0.5\columnwidth]{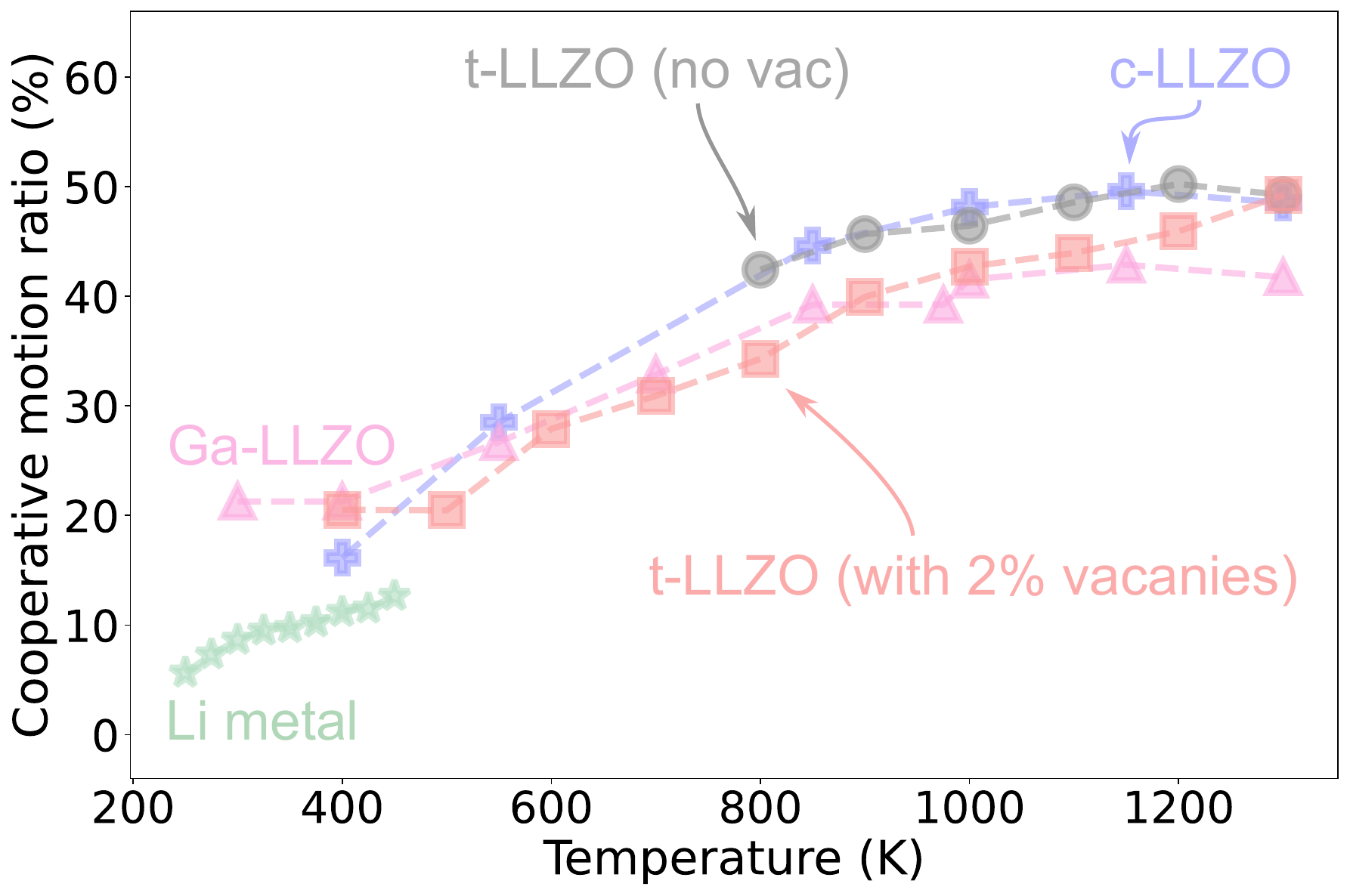}
\end{center}
\caption{Percent of cooperative jumps as a function of temperature for bulk phases: bcc Li, t-LLZO without and with 1.8\% vacancies, c-LLZO, Ga-LLZO. } 
\label{fig:diff_artem_temp}
\end{figure}

The example of analysis for bcc Li and LLZO systems are shown in~\cref{fig:diff_artem}. Here, we consider a migration as a cooperative if the number of simultaneously moving ions is greater than one. The effect of temperature on the ratio of cooperative jumps is shown in~\cref{fig:diff_artem_temp}. For all phases, the ratio increases with increasing temperature. For LLZO phases the ratio of cooperative jumps increases from $\sim20\%$ at 300~K and increases up to $\sim 50\%$ at 1200~K. For the bcc Li the corresponding ratio goes from 5 to 10\% at the temperature range of 250--450~K.

\section{van Hove correlation function}\label{sec:van_hove}

Self and distinct parts of the radial one-dimensional van Hove correlation functions represents the fraction of particles that have performed a given displacement $\mathbf{r}_i(0) - \mathbf{r}_i(t) = \mathbf{r}$ in a time $t$ and were calculated as follows:

$$
G_{\mathrm{s}}(\mathbf{r}, t) = \frac{1}{N} \left\langle \sum_{i=1}^N \delta[ \mathbf{r} + \mathbf{r}_i(0) - \mathbf{r}_i(t)] \right\rangle 
$$

$$
G_{\mathrm{d}}(\mathbf{r}, t) = \frac{1}{N} \left\langle \sum_{i=1}^N \sum_{j \neq i}^N \delta[\mathbf{r} + \mathbf{r}_i(0) - \mathbf{r}_j(t)] \right\rangle ,
$$
where $\delta$ is the Dirac delta function.

\subsection{Quantification of cooperative Li$^{+}$ jumps}\label{seq:van_hove_cooperative_method}

\begin{figure}[H]
\begin{center}
    \includegraphics[width=0.55\columnwidth]{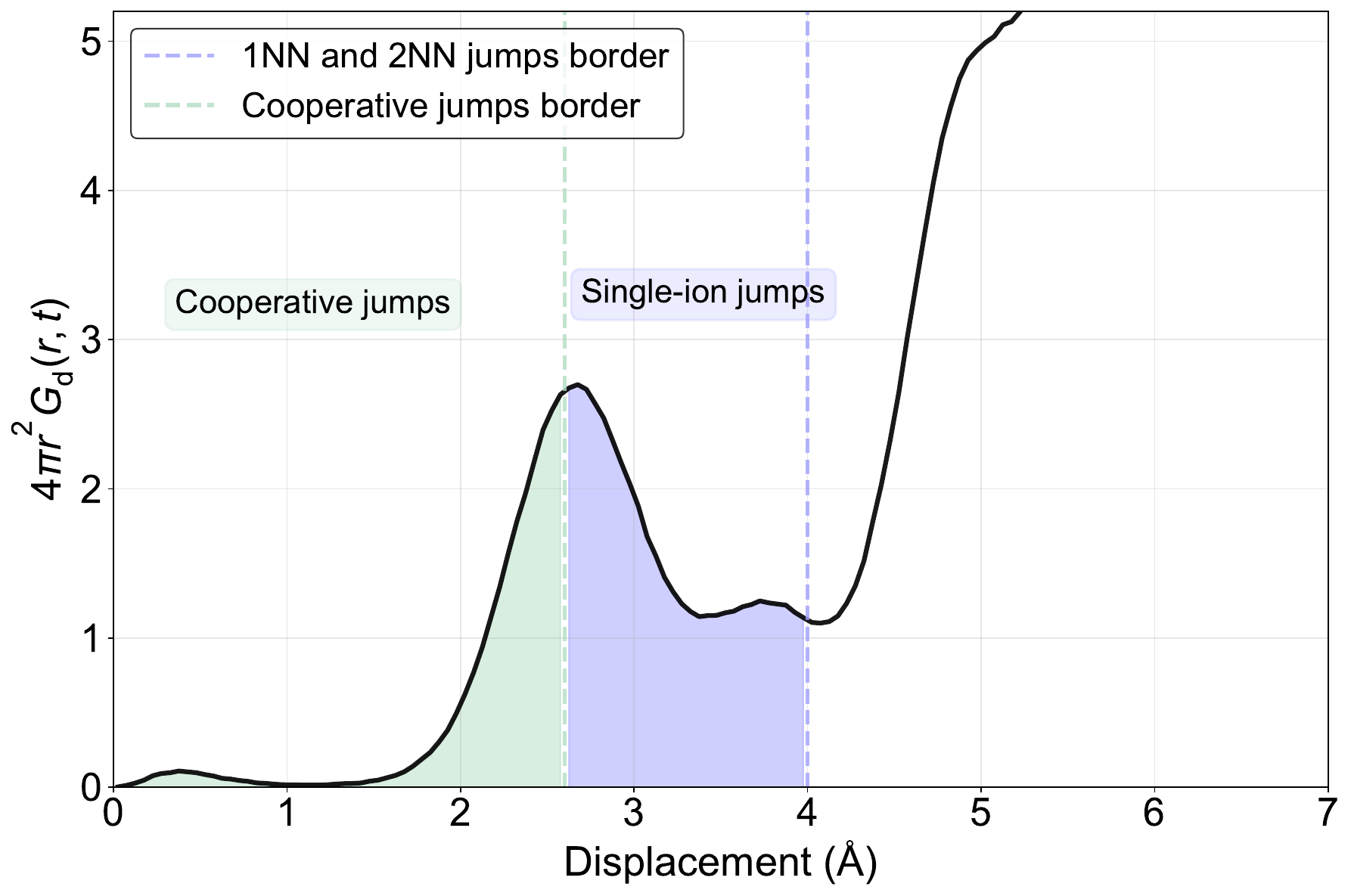}
\end{center}
\caption{Schematic illustration of different migration mechanisms in the distinct part of the radial one-dimensional van Hove correlation function ($G_{\mathrm{d}}$). Green represents cooperative jumps and blue represents single-ion jumps to nearest-neighbor (NN) positions.}
\label{fig:eta_method_distinct}
\end{figure}

The percentage of cooperative hops was calculated using the method shown in~\cref{fig:eta_method_distinct}. Migration distance was used to define the boundaries for single-ion diffusion (2.6--4.0~\AA) and cooperative diffusion ($<2.4$~\AA). The ratios were obtained from the integrated distinct van Hove functions for single-ion and cooperative jumps, as follows:

$$
\eta_\text{norm}(t) = \frac{ \int_{r_{\mathrm{coop}}^{\mathrm{low}}}^{r_{\mathrm{coop}}^{\mathrm{up}}}  r^2 G_\text{d}(r,t) \, dr }{ 
                       \int_{r_{\mathrm{single}}^{\mathrm{low}}}^{r_{\mathrm{single}}^{\mathrm{up}}} r^2 G_\text{d}(r,t) \, dr + 
                       \displaystyle \int_{r_{\mathrm{coop}}^{\mathrm{low}}}^{r_{\mathrm{coop}}^{\mathrm{up}}} r^2 G_\text{d}(r,t) \, dr }
$$
where $t$ denotes the lag time, $G_\mathrm{d}(r,t)$ the distinct van Hove function, $r$ is the radial displacement, and $r_i^j$ is the integration distance threshold ($i=$ single, coop; $j=$ low, up). Thresholds $r_i^j$ were determined from the migration pathway distances obtained via NEB calculations.

We did not analyze self part of van Hove functions as we did for the distinct part (\cref{fig:eta_method_distinct}), as distinguishing cooperative jumps from next-nearest-neighbor (NNN) jumps proves more challenging.

\begin{figure}[H]
\begin{center}
    \includegraphics[width=0.65\columnwidth]{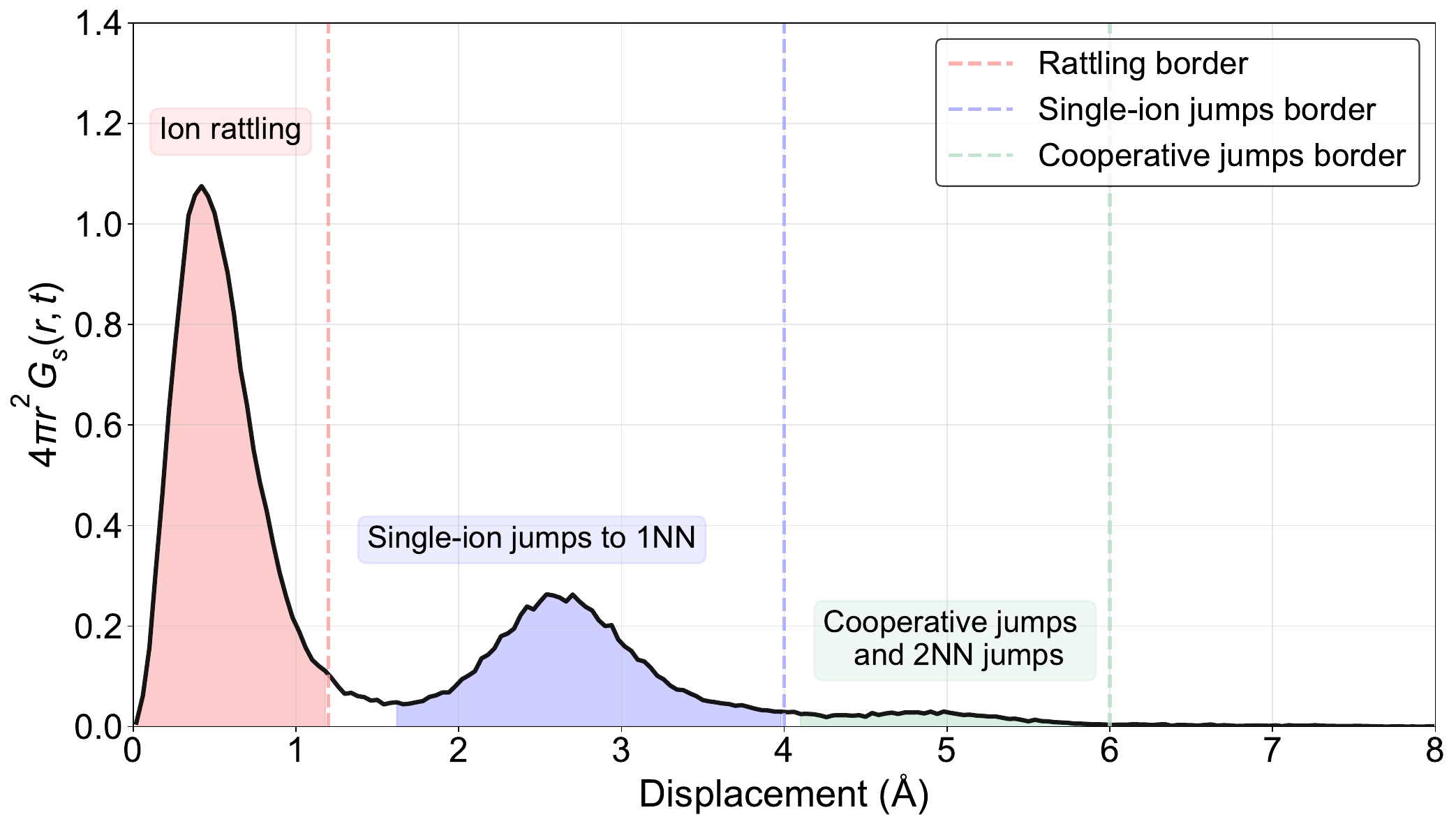}
\end{center}
\caption{Schematic illustration of different migration mechanisms in the self part of the radial one-dimensional van Hove correlation function ($G_{\mathrm{s}}$). Red highlights ion rattling, blue represents single-ion jumps to nearest-neighbor (NN) positions, and green denotes cooperative jumps along with a portion of single-ion jumps beyond the NN coordination shell of Li.}
\label{fig:eta_method}
\end{figure}

\subsection{Self part of the radial one-dimensional van Hove correlation function}

\begin{figure}[H]
\begin{center}
    \includegraphics[width=0.99\columnwidth]{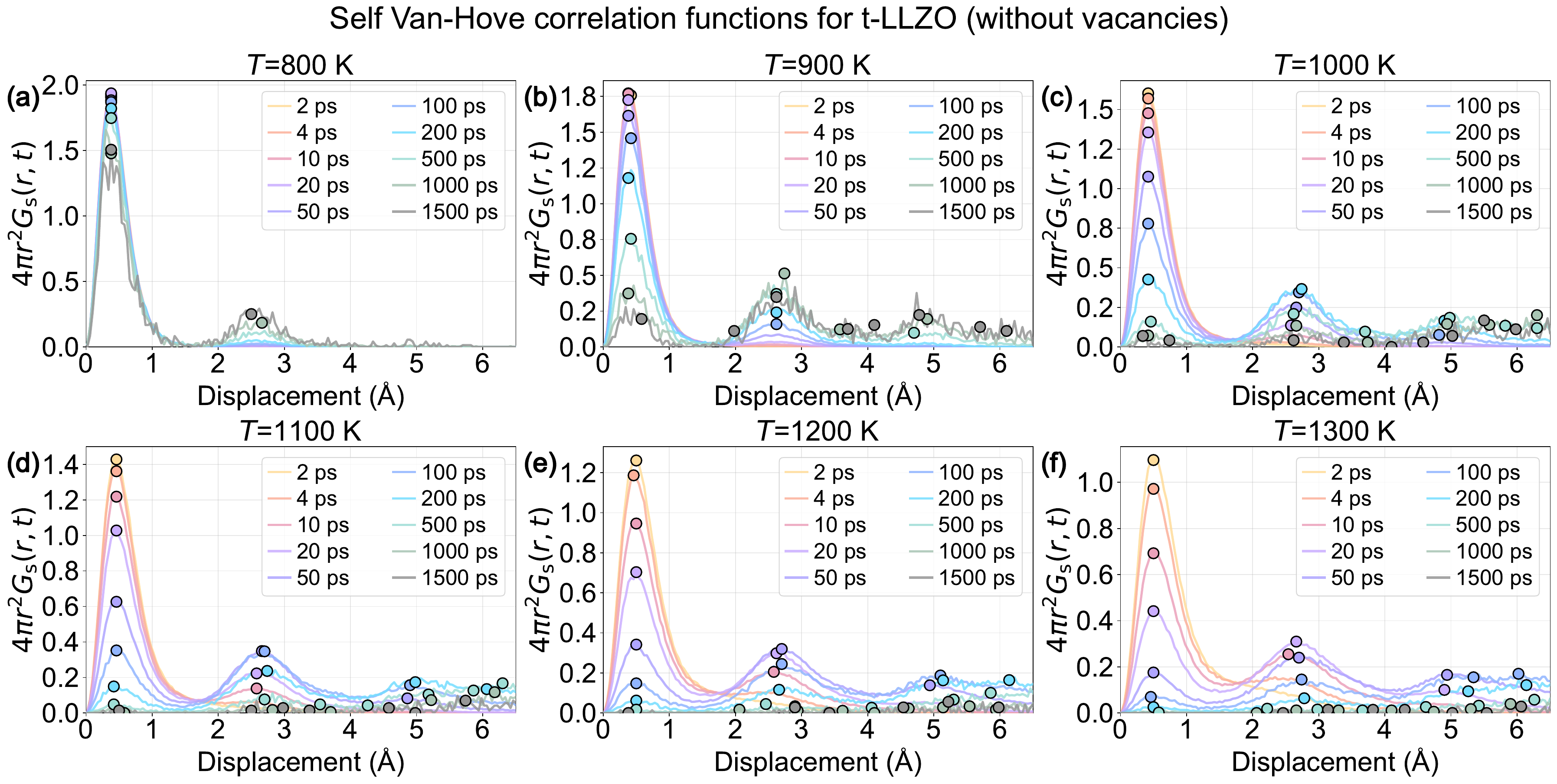}
\end{center}
\caption{Self part of the radial one-dimensional van Hove correlation function ($G_{\mathrm{s}}(r,t)$) at different lag times for various temperatures for t-LLZO without vacancies: (a) $T=300$~K, (b) $T=400$~K, (c) $T=500$~K, (d) $T=600$~K, (e) $T=700$~K, (f) $T=800$~K, (g) $T=900$~K, (h) $T=1000$~K, (i) $T=1100$~K, (j) $T=1200$~K, (k) $T=1300$~K. }
\label{fig:van_hove_tet_no_vac}
\end{figure}

\begin{figure}[H]
\begin{center}
    \includegraphics[width=0.99\columnwidth]{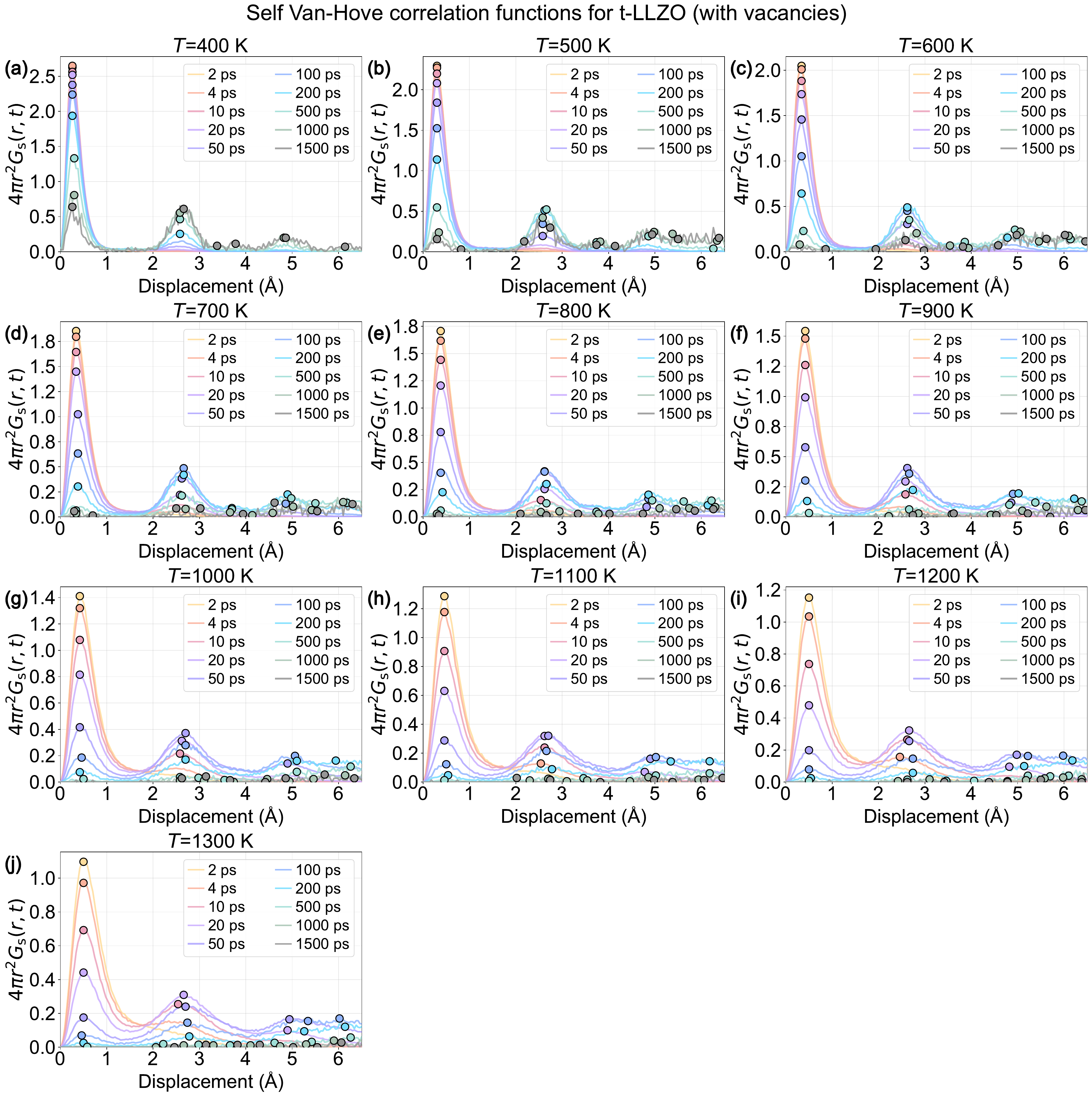}
\end{center}
\caption{Self part of the radial one-dimensional van Hove correlation function ($G_{\mathrm{s}}(r,t)$) at different lag times for various temperatures for t-LLZO with 1.8\% vacancies: (a) $T=300$~K, (b) $T=400$~K, (c) $T=500$~K, (d) $T=600$~K, (e) $T=700$~K, (f) $T=800$~K, (g) $T=1000$~K, (h) $T=1100$~K, (i) $T=1200$~K, (j) $T=1300$~K. }
\label{fig:van_hove_tet_vac}
\end{figure}

\begin{figure}[H]
\begin{center}
    \includegraphics[width=0.99\columnwidth]{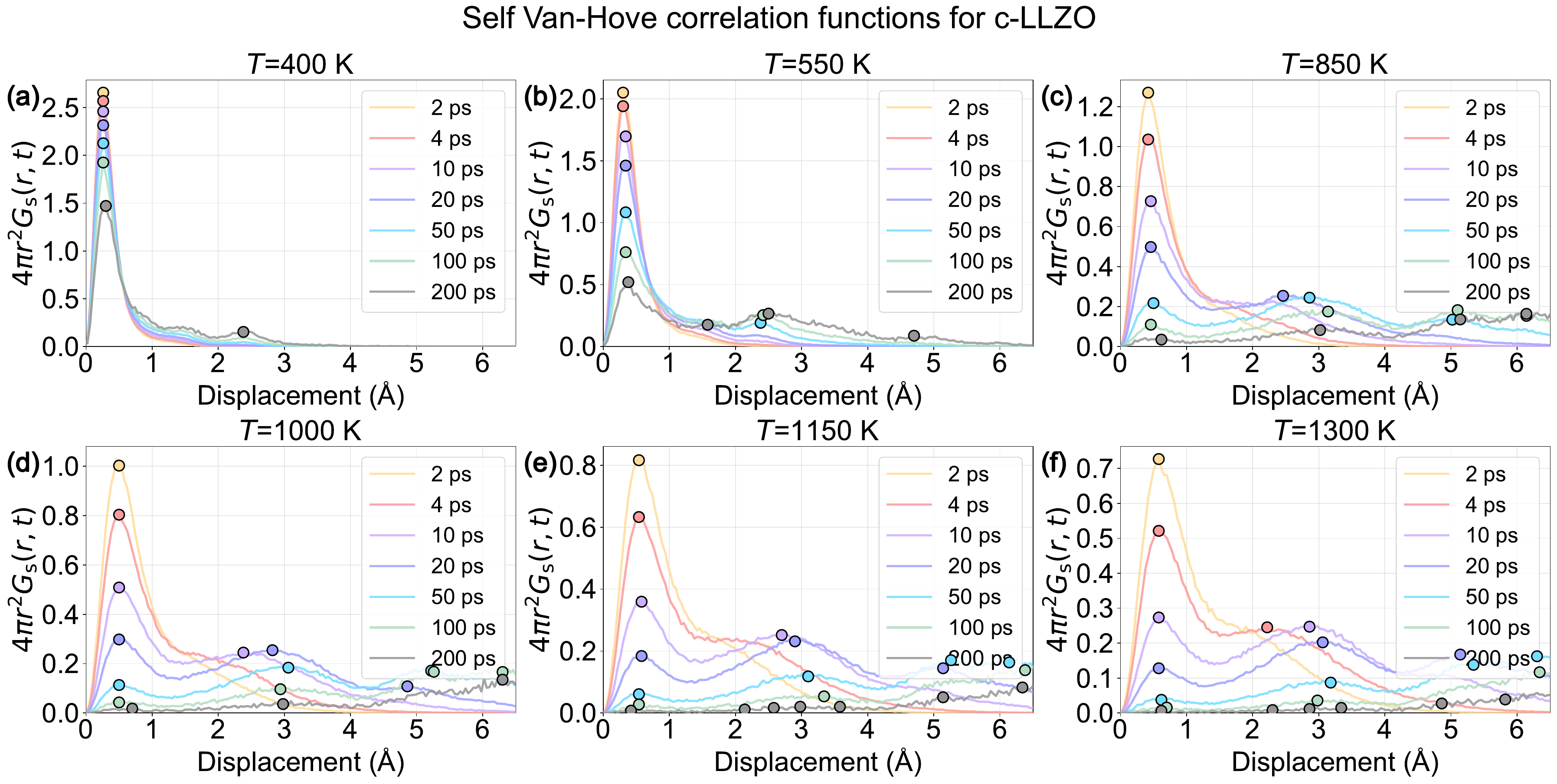}
\end{center}
\caption{Self part of the radial one-dimensional van Hove correlation function ($G_{\mathrm{s}}(r,t)$) at different lag times for various temperatures for c-LLZO: (a) $T=400$~K, (b) $T=550$~K, (c) $T=850$~K, (d) $T=1000$~K, (e) $T=1150$~K, (f) $T=1300$~K. }
\label{fig:van_hove_cub}
\end{figure}

\begin{figure}[H]
\begin{center}
    \includegraphics[width=0.99\columnwidth]{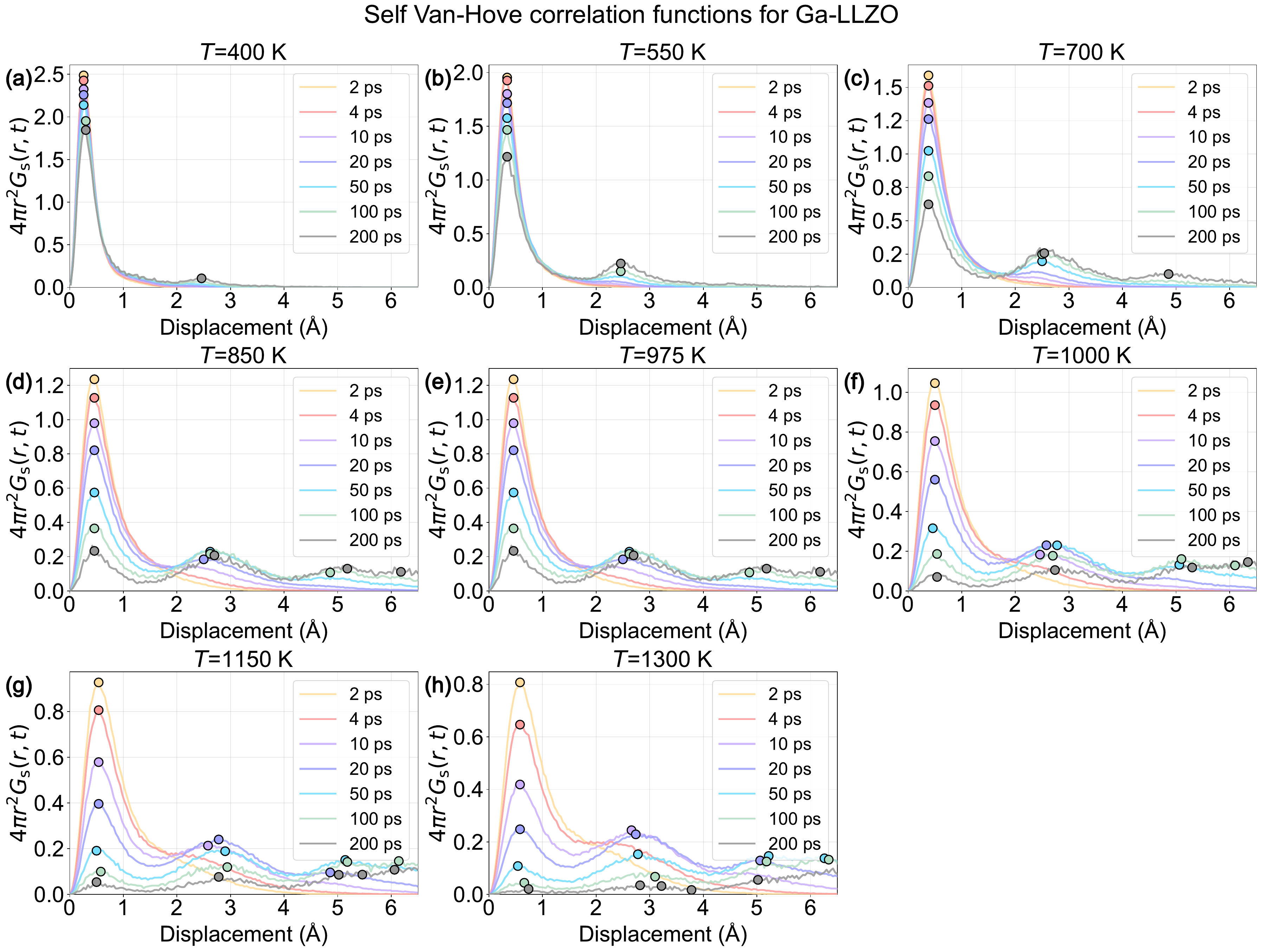}
\end{center}
\caption{Self part of radial one-dimensional van Hove correlation function ($G_{\mathrm{s}}(r,t)$) at different lag times for various temperatures for Ga-LLZO: (a) $T=300$~K, (b) $T=400$~K, (c) $T=550$~K, (d) $T=700$~K, (e) $T=850$~K, (f) $T=975$~K, (g) $T=1000$~K, (h) $T=1150$~K, (i) $T=1300$~K.  }
\label{fig:van_hove_llzo_ga}
\end{figure}

\begin{figure}[H]
\begin{center}
    \includegraphics[width=0.99\columnwidth]{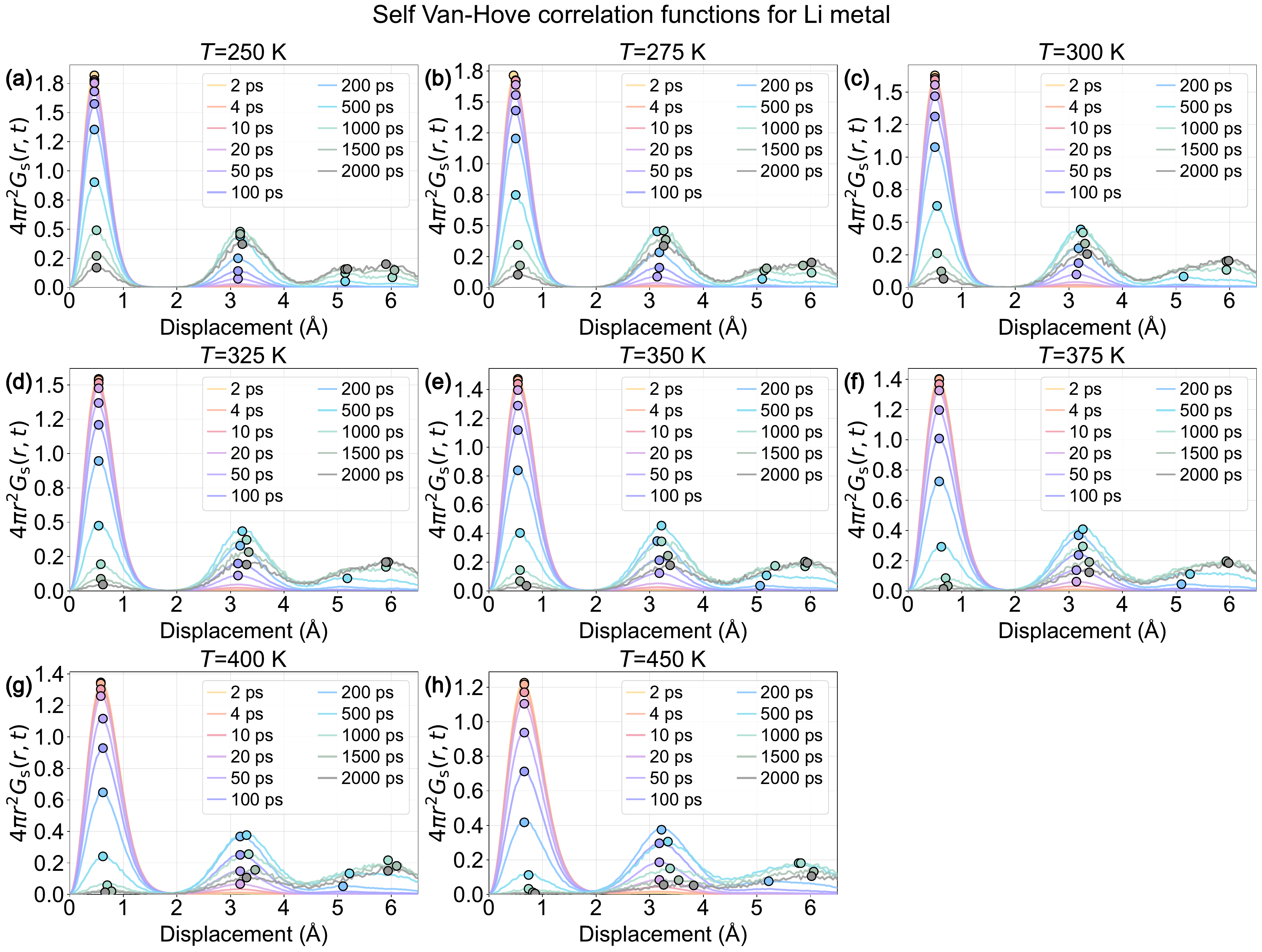}
\end{center}
\caption{Self part of the radial one-dimensional van Hove correlation function ($G_{\mathrm{s}}(r,t)$) at different lag times for various temperatures for Li metal:
(a) $T = 200$~K, (b) $T = 300$~K, (c) $T = 350$~K, (d) $T = 375$~K, (e) $T = 400$~K, (f) $T = 450$~K. }
\label{fig:van_hove_li_metal}
\end{figure}

\begin{figure}[H]
\begin{center}
    \includegraphics[width=0.8\columnwidth]{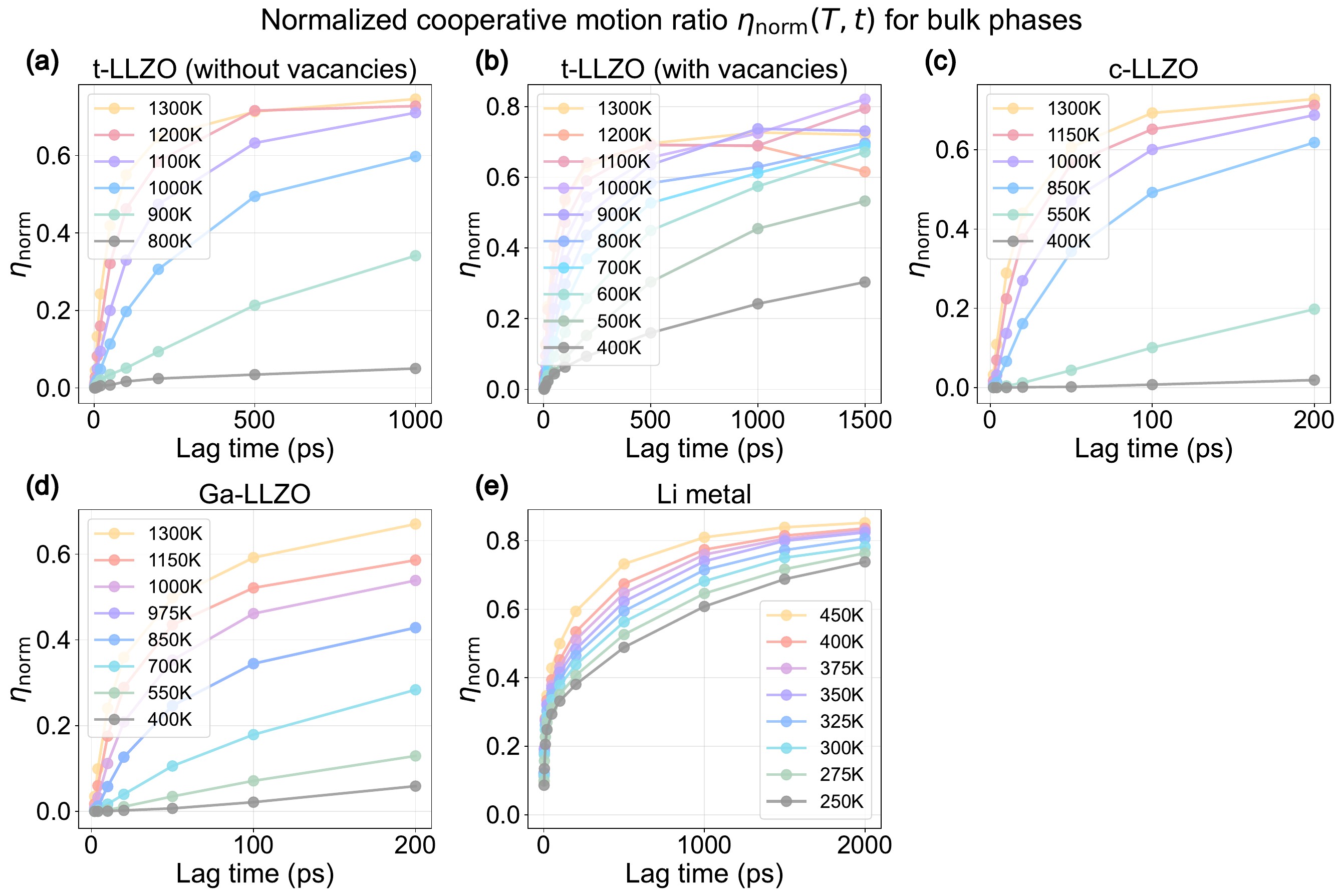}
\end{center}
\caption{Temperature dependence of the normalized ratio $\eta_{\mathrm{norm}}$ 
between cooperative and single-ion hops, extracted from the self part of the radial one-dimensional van Hove correlation function ($G_{\mathrm{s}}(r,t)$), for the bulk phases investigated:  (a) vacancy-free tetragonal LLZO (t-LLZO), (b) t-LLZO containing 1.8\% vacancies, (c) cubic LLZO (c-LLZO), (d) Ga-doped LLZO, and (e) pristine Li metal. The ratio provides a quantitative measure of the prevalence of cooperative ionic motion across different compositions and defect concentrations.}
\label{fig:van_hove_conv_bulk}
\end{figure}

\subsection{Distinct part of van Hove correlation function}

\begin{figure}[H]
\begin{center}
    \includegraphics[width=0.95\columnwidth]{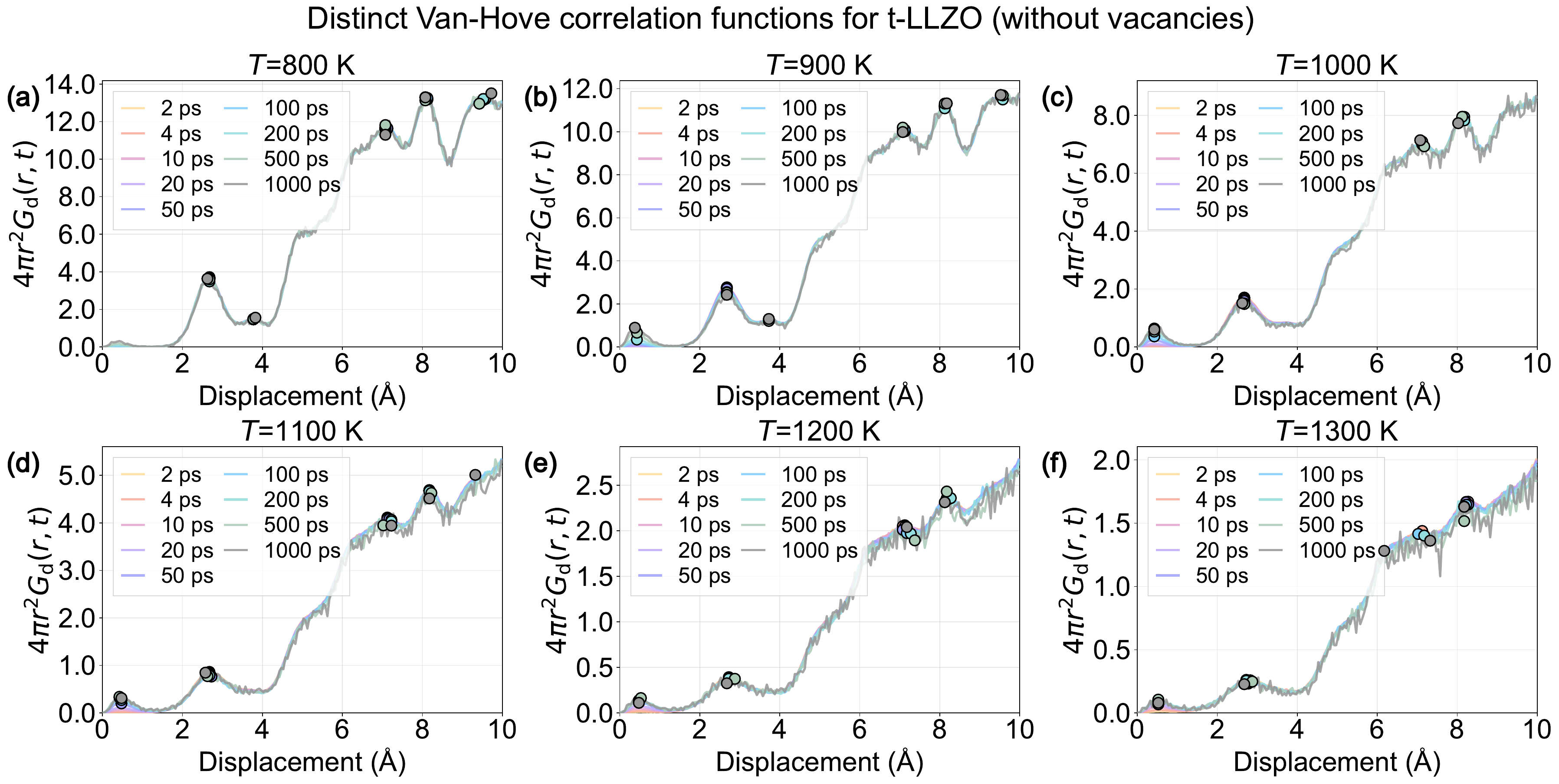}
\end{center}
\caption{Distinct part of the radial one-dimensional van Hove correlation function ($G_{\mathrm{d}}(r,t)$) at different lag times for various temperatures for t-LLZO without vacancies: (a) $T=300$~K, (b) $T=400$~K, (c) $T=500$~K, (d) $T=600$~K, (e) $T=700$~K, (f) $T=800$~K, (g) $T=900$~K, (h) $T=1000$~K, (i) $T=1100$~K, (j) $T=1200$~K, (k) $T=1300$~K. }
\label{fig:van_hove_tet_no_vac_distinct}
\end{figure}

\begin{figure}[H]
\begin{center}
    \includegraphics[width=0.99\columnwidth]{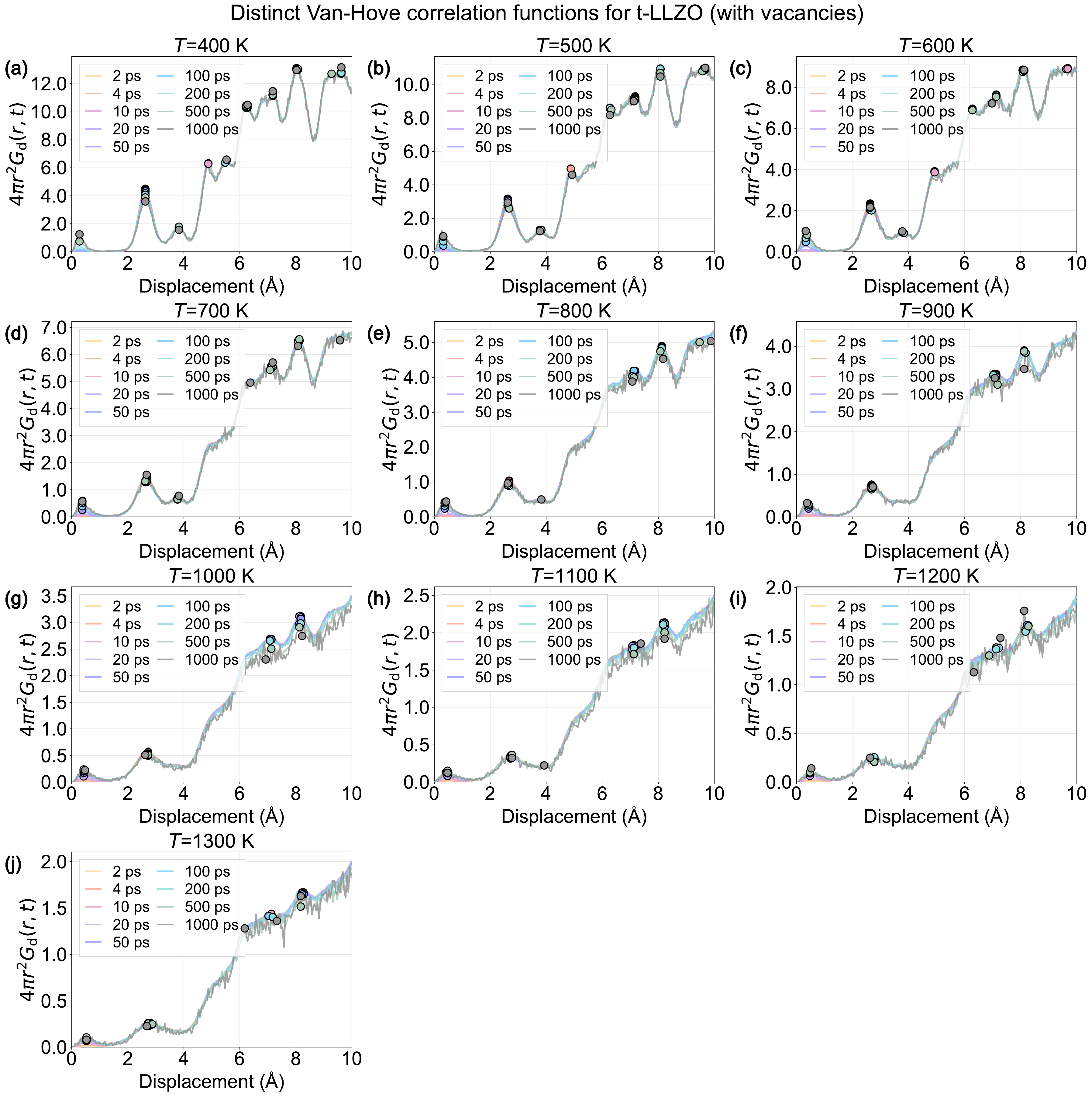}
\end{center}
\caption{Distinct part of the radial one-dimensional van Hove correlation function ($G_{\mathrm{d}}(r,t)$) at different lag times for various temperatures for t-LLZO with 1.8\% vacancies: (a) $T=300$~K, (b) $T=400$~K, (c) $T=500$~K, (d) $T=600$~K, (e) $T=700$~K, (f) $T=800$~K, (g) $T=1000$~K, (h) $T=1100$~K, (i) $T=1200$~K, (j) $T=1300$~K. }
\label{fig:van_hove_tet_vac_distinct}
\end{figure}

\begin{figure}[H]
\begin{center}
    \includegraphics[width=0.99\columnwidth]{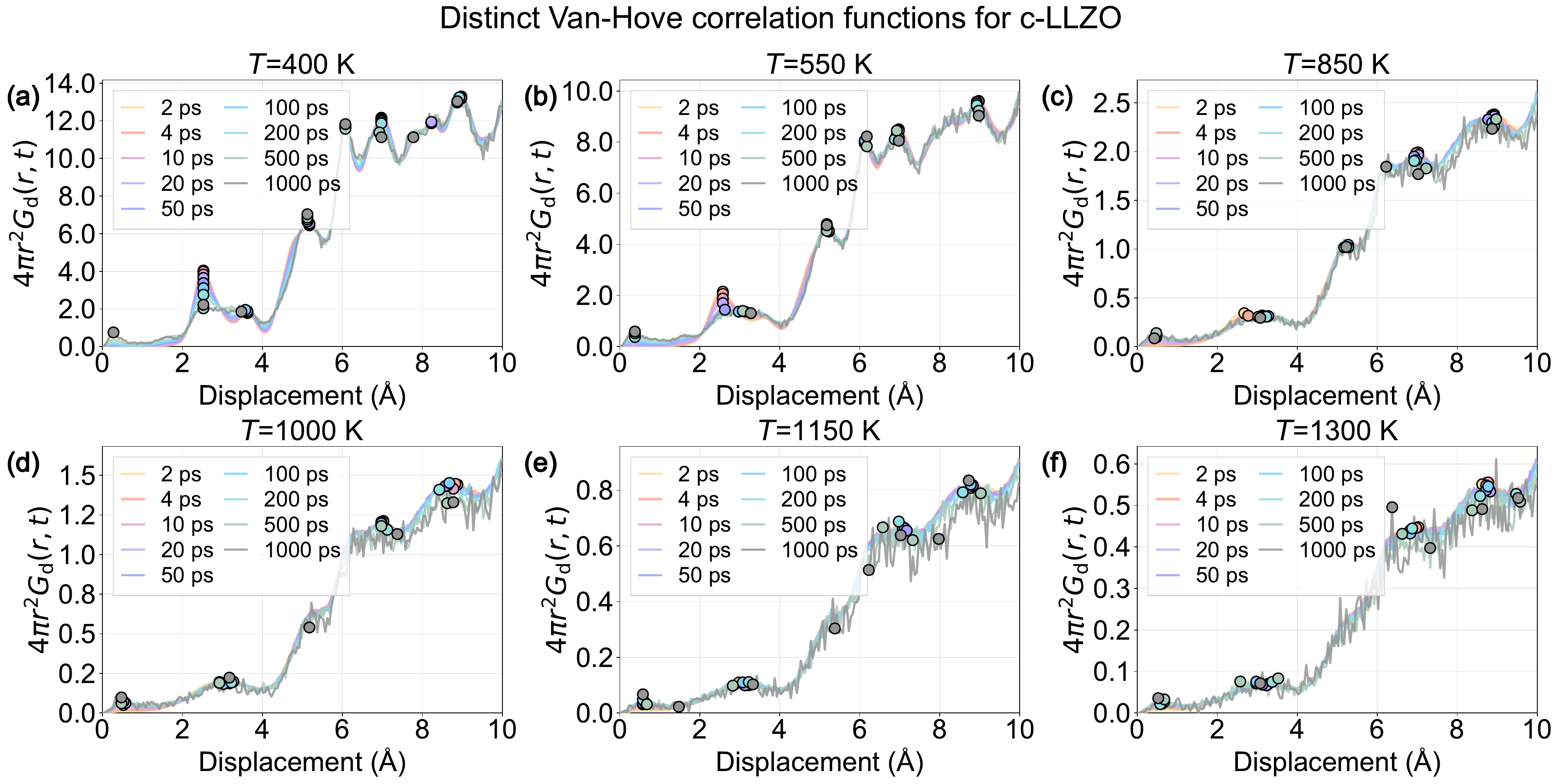}
\end{center}
\caption{Distinct part of the radial one-dimensional van Hove correlation function ($G_{\mathrm{d}}(r,t)$) at different lag times for various temperatures for c-LLZO: (a) $T=400$~K, (b) $T=550$~K, (c) $T=850$~K, (d) $T=1000$~K, (e) $T=1150$~K, (f) $T=1300$~K. }
\label{fig:van_hove_cub_distinct}
\end{figure}

\begin{figure}[H]
\begin{center}
    \includegraphics[width=0.99\columnwidth]{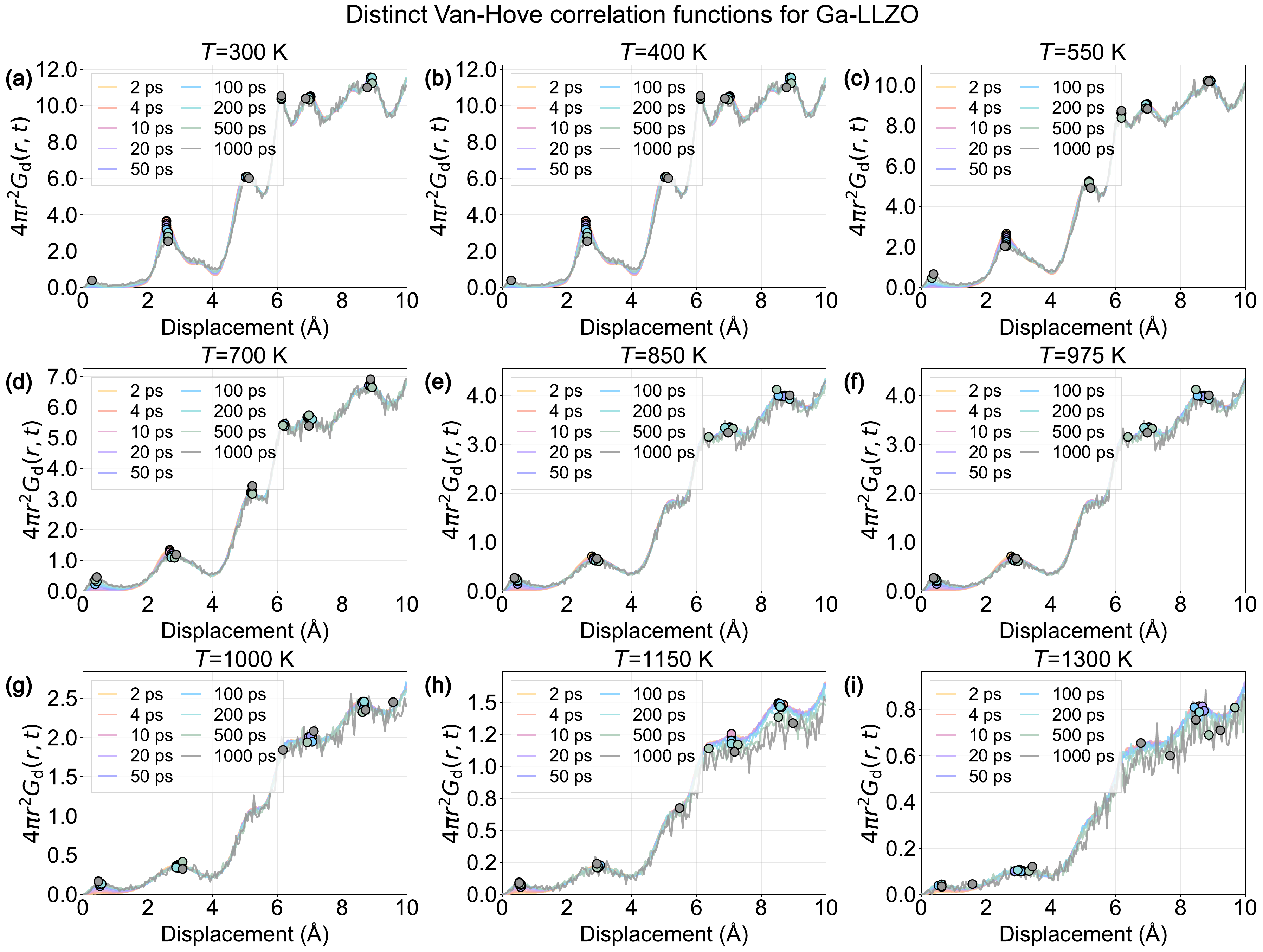}
\end{center}
\caption{Distinct part of the radial one-dimensional van Hove correlation function ($G_{\mathrm{d}}(r,t)$) at different lag times for various temperatures for Ga-LLZO: (a) $T=300$~K, (b) $T=400$~K, (c) $T=550$~K, (d) $T=700$~K, (e) $T=850$~K, (f) $T=975$~K, (g) $T=1000$~K, (h) $T=1150$~K, (i) $T=1300$~K. }
\label{fig:van_hove_llzo_ga_distinct}
\end{figure}

\begin{figure}[H]
\begin{center}
    \includegraphics[width=0.99\columnwidth]{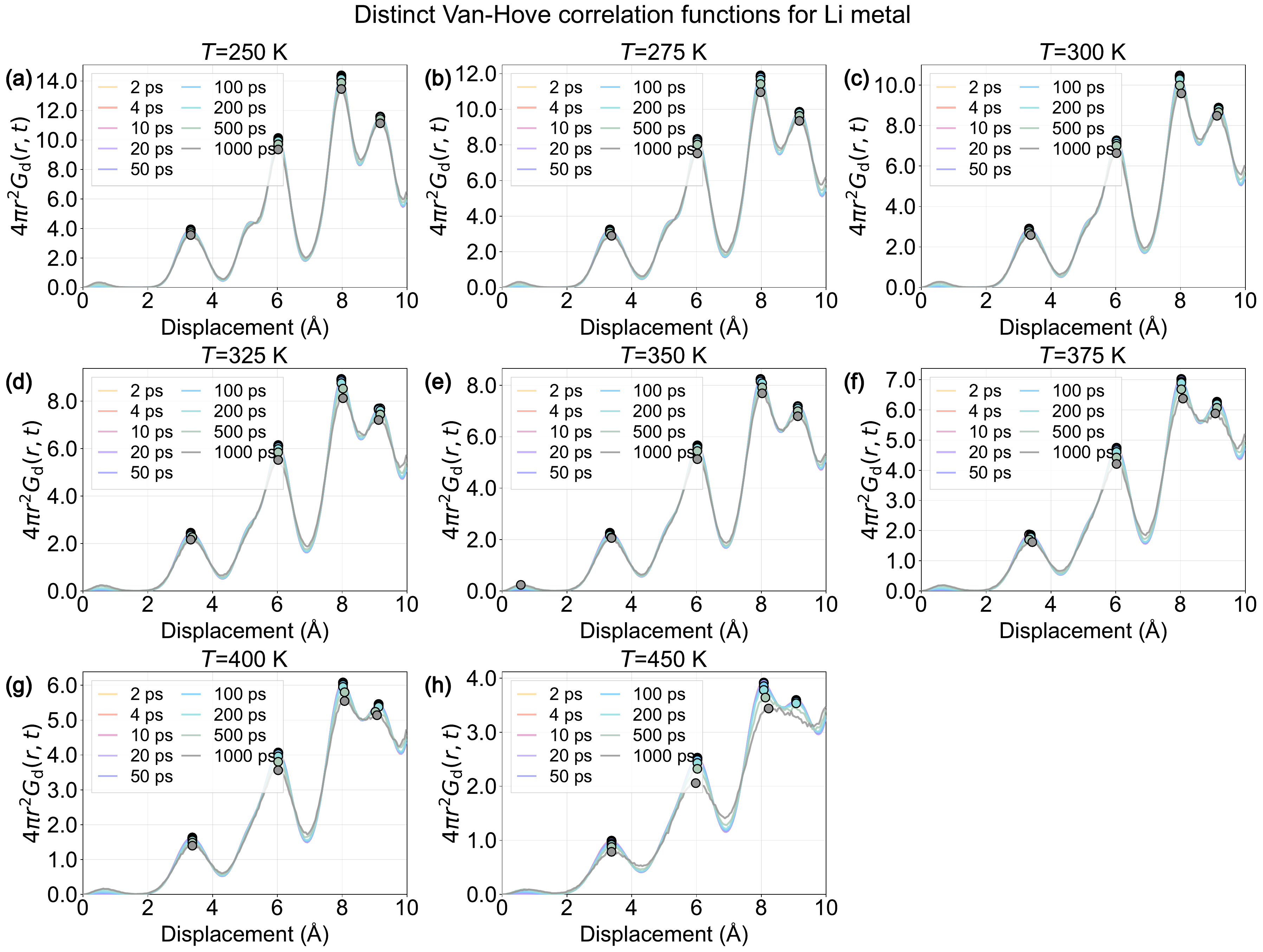}
\end{center}
\caption{Distinct part of the radial one-dimensional van Hove correlation function ($G_{\mathrm{d}}(r,t)$) at different lag times for various temperatures for Li metal:
(a) $T = 200$~K, (b) $T = 300$~K, (c) $T = 350$~K, (d) $T = 375$~K, (e) $T = 400$~K, (f) $T = 450$~K.
 }
\label{fig:van_hove_li_metal_distinct}
\end{figure}

\begin{figure}[H]
\begin{center}
    \includegraphics[width=0.8\columnwidth]{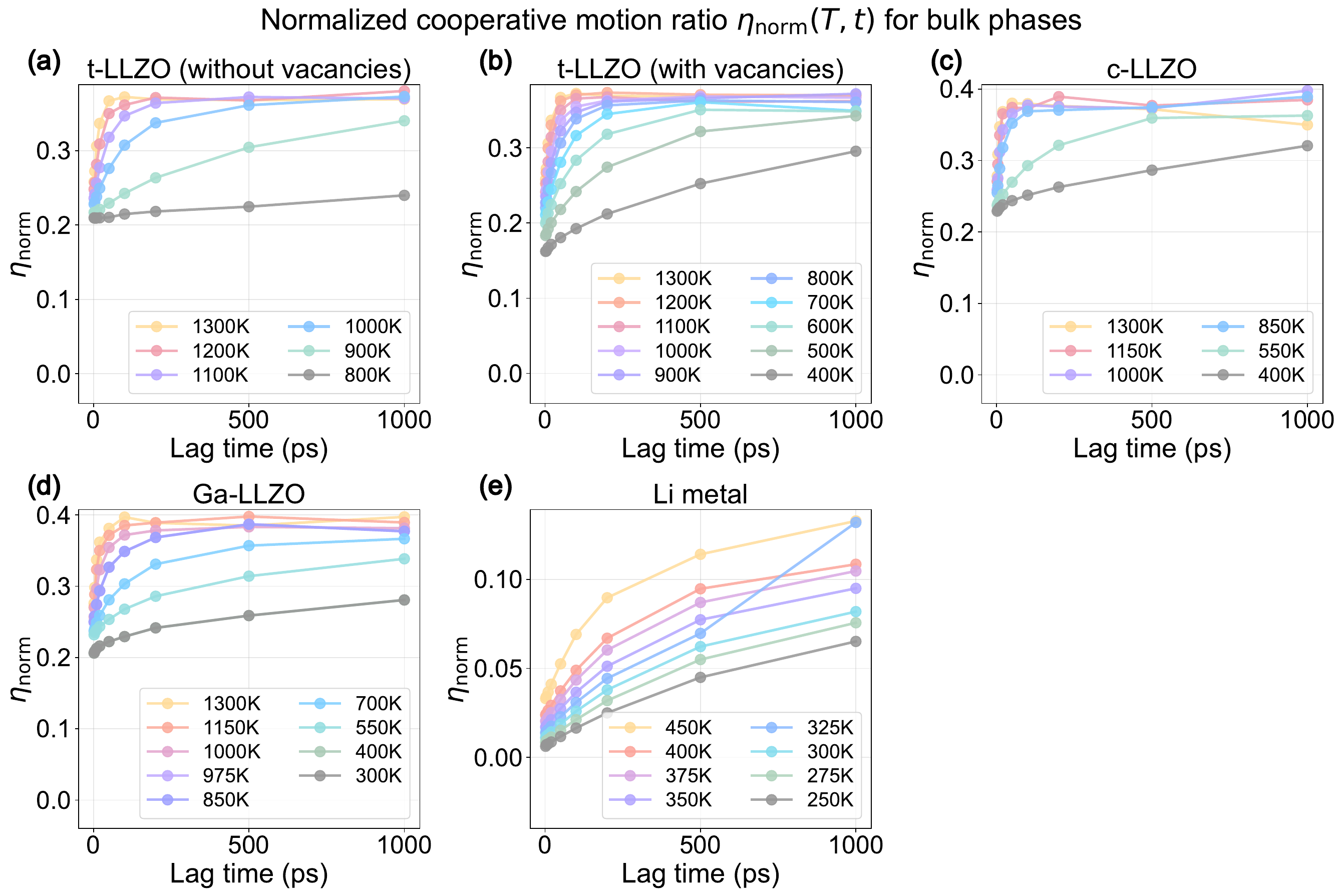}
\end{center}
\caption{Temperature dependence of the normalized ratio $\eta_{\mathrm{norm}}$ 
between cooperative and single-ion hops, extracted from the distinct part of the radial one-dimensional van Hove correlation function ($G_{\mathrm{d}}(r,t)$), for the bulk phases investigated:  (a) vacancy-free tetragonal LLZO (t-LLZO), (b) t-LLZO containing 1.8\% vacancies, (c) cubic LLZO (c-LLZO), (d) Ga-doped LLZO, and (e) pristine Li metal. The ratio provides a quantitative measure of the prevalence of cooperative ionic motion across different compositions and defect concentrations.}
\label{fig:van_hove_conv_bulk_distinct}
\end{figure}

\def\bibsection{\section*{Supplementary References}} 
\putbib[lib]
\end{bibunit}

\end{document}